\documentclass[11pt,a4paper,twoside]{article}
	\usepackage{a4wide}
  \usepackage{soul} 
  \usepackage{framed} 
  \usepackage{url}
  \usepackage[small,bf,hang]{caption} 
  \usepackage{footmisc}
  \usepackage{footnote}
  \usepackage{boxedminipage}
	\usepackage{graphicx} 
	\usepackage{wrapfig} 
	\usepackage{picins} 
	\usepackage{marvosym} 
	\usepackage{eurofont} 
	\usepackage{color} 
	\usepackage{amssymb} 
	\usepackage{extarrows} 
	\usepackage{amsthm} 
	\usepackage{amsmath}  
	\usepackage{bbm} 
	\usepackage{dsfont} 
	\usepackage{textcomp} 
	\usepackage[english]{babel}    
	\usepackage[all,cmtip]{xy} 
	\newcommand{\npar}{\par \vspace{2.3ex plus 0.3ex minus 0.3ex}}

	\DeclareMathOperator{\h}{\mathcal{H}} 
	\DeclareMathOperator{\A}{\mathcal{A}}

	\DeclareMathOperator{\een}{\mathds{1}}
	\DeclareMathOperator{\nul}{\mathds{O}}
	 
	\DeclareMathOperator{\dee}{d} 
	 
	\DeclareMathOperator{\Trace}{Tr} 
	\DeclareMathOperator{\Sp}{Sp}

	\DeclareMathOperator{\pee}{\mathds{P}}

	\DeclareMathOperator{\desda}{<\hspace{-0.8em}\textbf{=} 																														\hspace{-0.2em}\textbf{=}\hspace{-0.8em}>} 
	\DeclareMathAlphabet{\mathpzc}{OT1}{pzc}{m}{it} 	
	\DeclareMathOperator{\Qvee}{\vee_{\quad_{\quad_{\hspace{-2.8em}QL}}}}
	\DeclareMathOperator{\Qwedge}{\wedge_{\quad_{\quad_{\hspace{-2.75em}QL}}}}
	\DeclareMathOperator{\Qneg}{\neg_{\quad_{\quad_{\hspace{-2.75em}QL}}}}

	\newtheorem{stelling}{Theorem}[section]
	\newtheorem{lemma}[stelling]{Lemma}
	
	\newtheorem{propositie}[stelling]{Proposition}
	\newtheorem{gevolg}[stelling]{Corollary}
  \theoremstyle{definition}
	\newtheorem{voorbeeld}{Example}[section]
	\newtheorem{opmerking}{Remark}[section]
	\newtheorem{aanname}{Assumption}[section]
	\newtheorem{definitie}{Definition}[section] 
	
	\usepackage[T1]{fontenc} 
	    \makeatletter 
    \renewcommand\paragraph{\@startsection{paragraph}{4}{\z@}%
      {-3.25ex\@plus -1ex \@minus -.2ex}%
      {1.5ex \@plus .2ex}%
      {\normalfont\normalsize\bfseries}}
    \makeatother
\begin{document}

\setlength{\unitlength}{\textwidth}

\pagenumbering{roman}
\begin{titlepage}
\begin{center}
\hfill\\[15.0ex]
\textsc{\huge{\textbf{Quantum Mechanics:\\ From Realism to Intuitionism}}\\
\Large{A mathematical and philosophical investigation}\\[5.5ex]
Master's Thesis -- Mathematical Physics}\\[30.0ex]

\begin{minipage}{0.45\textwidth}
\begin{flushleft} \large
Author:\\
Ronnie Hermens\\
\npar
\npar
\npar
\end{flushleft}
\end{minipage}
%
\begin{minipage}{0.45\textwidth}
\begin{flushright} \large
Supervisor: \\
Prof. Dr. N. P. Landsman\\
\npar
Second Reader: \\
Dr. H. Maassen\\

\end{flushright}
\end{minipage}
\end{center}

\begin{figure}[b]
\begin{flushright}
\includegraphics[width=0.75\textwidth]{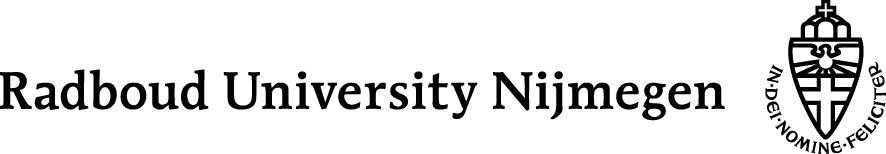}
\end{flushright}
\end{figure}

\end{titlepage}

\begin{center}
\hfill\\[15.0ex]
\textit{Aan de hoge blauwe hemel\\ zweeft een dapper wolkje,\\ in tegen de wind.\\ Hij komt niet ver,\\ maar probeert het tenminste.}\\
Jaap Robben
\end{center}

\clearpage

\section*{Preface}
\addcontentsline{toc}{section}{Preface}

As a student of physics I always felt that the difficulties I had with comprehending the explained theories were mostly due to my incapability. Pondering on questions like ``What \textit{is} an electric field?'' somehow prevented me from actually solving Maxwell's equations, which is in fact the thing that you have to do to pass your exam. But then working out the details of the actual solving leads to various difficulties again. Indeed, quite often in physics one encounters mathematical problems which one must march over to obtain the desired answer. Sometimes this results in peculiarities that seemed paradoxical to me like a discontinuous solution to a differential equation. As it turns out, I'm one of those persons who in many cases can't see the bigger picture until he's worked out a lot of the details. Fortunately, working out details is an activity praised in mathematics (the unfortunate thing for me was that it took me over four years to discover this).  

When I first came with the idea for this thesis I didn't know very much about the fundaments of quantum mechanics. Discussions on topics like ``hidden variables'', ``contextuality'' or ``locality'' always seemed gladly avoid by the teachers during the lectures on quantum theory. 
It was kind of a revelation (and a relief) for me to find out that the incomprehensibility of quantum mechanics is not easily stepped over. This became clearer to me when I followed a course on quantum probability taught by Dr. Maassen. One of the first topics discussed, was the violation of Bell inequalities by quantum mechanics, which demonstrates directly that the probability measures obtained in quantum mechanics are fundamentally different from those described by Kolmogorov's theory. This is interesting in particular for probability theorists but it wasn't the aim of Bell to advocate for a revision of probability theory. Rather, it was his aim to show that any hidden-variable theory that reproduces the predictions obtained from quantum mechanics must be non-local, i.e. it requires action at a distance. 

Roughly, a hidden-variable theory is a theory that allows a realist interpretation i.e., a theory in which observables can be interpreted to correspond to properties of systems that actually exist. Personally, I never considered that this should be taken as a demand for physical theories. Not because I have a strong opinion considering the realist/idealist question in philosophy, but more because I never considered it the task of physics to be judgmental about such philosophical problems. However, there were enough questions raised in my head to form a starting point for this thesis and luckily, none of them have been answered properly. These questions include the following. How can mathematics play a role in finding answers to metaphysical questions? How reliable is mathematics in this role? Why does quantum mechanics not allow (certain) realist interpretations? 

As it goes with such questions, trying out answers immediately leads to new questions. In particular it becomes of interest what the role of mathematics is in physics and even broader, what the nature of mathematics is in itself, i.e. what is mathematics actually about? Concerning this first problem I became particularly interested in probability theory, which, in my opinion, is one of the purest forms of physics.\footnote{This view doesn't seem very popular, but it is in fact in correspondence with Hilbert's vision who explained his sixth problem as follows: ``To treat in the same manner, by means of axioms, those physical sciences in which mathematics plays an important part; in the first rank are the theory of probabilities and mechanics'' \cite{Hilbert00}.}I remember a lecture during a course on statistical physics taught by Prof. Vertogen during which he gave a derivation of the notion of entropy from a Bayesian point of view making only use of philosophical and logical considerations (i.e., without resorting to the measure-theoretic approach). At that time I didn't recognize it as such, but it did make me realize how important logic is for the construction of physical theories. Unfortunately for me, at the same time I also followed a course on logic thought by Dr. Veldman. The unfortunate coincidence was that while Prof. Vertogen made extensive use of the logical law $\neg(\neg X)=X$ (it was actually the first formula in the accompanying reader), Dr. Veldman was advocating against the use of this law, which is abandoned in intuitionistic logic. Needless to say, I had my logical conceptions all mixed up and I ended up failing the exams for both the courses. 

Ever since I've had a sort of love-hate relationship with intuitionism. At first I didn't like it all and I tried to find a motivation for myself to see why the law of excluded middle should be true. From a physics point of view, all the motivations I could find were based on realism, which seemed to me to be a too strong assumption. On the other hand, it seemed to me that if one can doubt one specific logical law, one might as well doubt all of logic. This is also what Brouwer advocated and roughly, he proposed that not logic should be our guide to truth, but intuition. It never became clear to me why this approach would be more reliable when it comes to truth. But at least logic enables us to compare notes in an (almost) unambiguous way. I came to except logic as a tool for reasoning not because I believe it is true, but because I don't see a better alternative. Then what about the law of excluded middle? I think from a realist (or Platonist when it comes to mathematics) point of view it may be mandatory. Others may want to learn to use it with care. For me, sometimes its truth seems almost evident\footnote{I think this also must have been the case for Brouwer for I see no better way to motivate his continuity principle.} while on other occasions it seems very suspicious (and the same holds for the axiom of choice for that matter). And as for truth, perhaps truth is overrated.

I wouldn't have found a personally satisfactory view on physics and mathematics if it weren't for the aforementioned persons. In fact, I probably would have quit my study within the first three years without the down toned visions on physics Prof. Vertogen presented in his lectures and I would like to thank him for that. I also would like to thank Prof. Landsman and Dr. Maassen for making me enthusiastic about mathematical physics and Dr. Veldman for teaching me about the philosophy of mathematics and intuitionism in particular. Without these people I would never have guessed it to be possible to write a philosophical thesis on physics with the use of mathematical rigor that still makes sense. For this I must also thank Dr. Seevinck who taught me a lot about the foundations of physics and who was often willing to listen to my own ideas. Finally I'd like to thank my girlfriend Femke for supporting me in every step the past ten years and for being my philosophical sparring partner from time to time and above all for being my best friend.

\npar Nijmegen, October 2009
\clearpage
		
\tableofcontents
\cleardoublepage
\pagenumbering{arabic}

\markboth{Introduction}{Introduction}
\section{Introduction}
\begin{flushright}
\begin{minipage}[300pt]{0.6\linewidth}
\textit{Theoretical physicists live in a classical world, looking into a quantummechanical world.}
\end{minipage}
\end{flushright}
\begin{flushright}
- J. S. Bell
\end{flushright}

Quantum mechanics started as a counter-intuitive theory and has succeeded in preserving this status ever since.  
Most introductions to quantum mechanics start with Planck's radiation formula. This was the first formula that relied on the quantization of energy, which is a departure from the ``Natura non facit saltus''-principle. 
Proposing the theoretical interpretation of this radiation formula was described by Planck himself as an act of despair:
\begin{quote}
``Kurz zusammengefasst kann ich die ganze Tat als einen Akt der Verzweiflung bezeichnen, denn von Natur bin
ich friedlich und bedenklichen Abenteuern abgeneigt ..., aber eine Deutung musste um jeden Preis gefunden werden, und w\"are er noch so hoch. ... Im \"ubrigen war ich zu jedem Opfer an meinen bisherigen physikalischen \"Uberzeugungen bereit.'' \cite{Planck31}
\end{quote}
The act of despair in question was actually not the concept of allowing discontinuity in Nature (although this was related) but the reliance on Boltzmann's theory of statistical physics, which is based on atomism; i.e. the idea that all matter is made up of some smallest particles called atoms. Nowadays, the concept of atomism is part of the doctrine of natural science and nobody would question the existence of atoms. However, around 1900 there was no real consensus about the issue, and in fact Planck originally opposed it. 

In his later years, Planck came to accept the concept of atomism and thus conquered one of the (to him) counter-intuitive aspects of his radiation formula, and thus quantum mechanics. However, this acceptance took a large revision on what is to be expected of Nature and on what is to be expected of a physical theory. It seems that this has been characteristic for the discussion on the foundations of quantum mechanics ever since. 
Some of the revisions that have been proposed throughout the years will be discussed in this thesis. These include revisions of our view on: reality, causality, locality, free will, determinism and logic. Not the lightest of subjects, and it seems mind-boggling enough that quantum mechanics has led people to such considerations. 

An important motivation for the entire discussion is the search for an answer to the question: ``What is actually being measured when a measurement is performed?'' In classical physics there seemed to be an easy answer to this question; a measurement reveals some property possessed by the system under consideration. This is, roughly, the realist interpretation of physics. However, as it turns out, such an interpretation is not possible in quantum mechanics without making compromises. The proof of this statement is usually attributed to the Kochen-Specker Theorem and the violation of the Bell inequalities by quantum mechanics, which will be discussed in Chapter \ref{PostulatenSectie}. Both imply a compromise that has to be made if one wishes to maintain realism. The Kochen-Specker Theorem implies that one has to resort to contextuality\footnote{Of all the philosophical concepts that play a role in this thesis, this is probably the most peculiar one. Roughly, it states that \textit{what} is actually measured depends in a very strong sense on \textit{how} it is measured i.e., it depends on the measuring context. Of course, each of these concepts will be explained more carefully in the course of this thesis.}, and the Bell inequality argument implies that one has to resort to non-locality. Most people do not wish to make such compromises, and therefore these proofs are also known as `impossibility proofs' or `no-go theorems'.

There is a natural problem that arises in this situation. The statements derived are all of a philosophical nature, but on the other hand, rigorous proofs can only be made within mathematics. This is because when it comes to mathematical objects, most people agree on how these objects may be manipulated to obtain new objects.\footnote{It may be noted that there is no general consensus on what these objects \textit{are}. However, in many cases this doesn't influence what is considered a proof and what is not.} However, this also means that philosophical and mathematical concepts have to be linked to one another, and there is of course no rigorous way to do this. In fact, there is not even much consensus about what terms like reality, locality and free will \textit{mean} and what role they should play in a physical theory. This leaves a lot of room for discussion on what actually \textit{can} be proven about Nature and about physical theories in particular.

In Chapter \ref{NullificatieStuk} it will be shown that the Kochen-Specker Theorem is quite unstable considering speculations on what realism should entail. More specifically, it turns out that if one relaxes the view on what constitutes a physical observable, one may retain non-contextuality. 
The discussion becomes more philosophical in Chapter \ref{FWThoofdstuk}, where the Free Will Theorem of Conway and Kochen is discussed. 
This is also the first point where it becomes more clear that the strangeness of quantum mechanics does not just affect the realist interpretation of physics; the indeterminacy introduced by quantum mechanics seems unavoidable in any other proceeding theory (consistent with current experimental knowledge), irrespective of whether one has a realist or an anti-realist view. This leads to the question whether the earlier arguments used to point out problems in the realist interpretation can be extended to also point out problems that arise in other interpretations. In Chapter \ref{StrangenessSection} it is argued that this does indeed seem to be the case. 

Hoping to acquire a better understanding of these problems, I take on a short re-investigation of the Copenhagen interpretation.
It seems that the Copenhagen interpretation does provide certain conceptual tools to overcome some philosophical problems concerning quantum mechanics. 
However, most people, like myself, cannot help to feel some unease about this interpretation. This feeling is similar to the one sometimes encountered when studying a mathematical theorem; although the proof may convince one that the theorem should be true, it doesn't always provide the feeling that one understands what the theorem actually states. 
Often, a clarification is needed to explain why a theorem is stated the way it is, and what the idea behind the theorem is. 

Such a clarification appears to be missing for the Copenhagen interpretation. Bohr only recites some facts about the strangeness of quantum mechanics (at least, the facts as he sees them) and suggests how one should cope with them. 
The impossibility proofs show that these facts cannot easily be sidestepped and so indeed it seems that one must cope with them. 
However, structural philosophical arguments about what accounts for this strangeness are missing. 
There is no clear motivation for coping with the problems \textit{in the way suggested by Bohr}. 
In Chapter \ref{StrangenessSection} I will attempt to provide this motivation, by linking the philosophy of Bohr to some of the philosophical ideas behind intuitionistic logic. 
More precisely, my hope is that an abuse of language in the sense meant by Bohr, may be avoided by adopting a different form of logic. 
In particular, it seems that the law of excluded middle provides a devious way to introduce sentences that speak of phenomena that cannot be compared with one another. 
It is clear that Bohr wanted to avoid such sentences\footnote{In the words of Wittgenstein: ``Wovon man nicht reden kann, dar\"uber muss man schweigen.''}, and should have argued against this law, hence embracing intuitionistic logic. However, he insisted on the use of classical logic.

\clearpage
 
\markboth{Introduction to the Foundations of Quantum Mechanics}{Postulates of Quantum Mechanics}
\section{Introduction to the Foundations of Quantum Mechanics}\label{PostulatenSectie}

\begin{flushright}
\begin{minipage}[300pt]{0.6\linewidth}
\textit{There is a theory which states that if ever anyone discovers exactly what the Universe is for and why it is here, it will instantly disappear and be replaced by something even more bizarre and inexplicable. There is another which states that this has already happened.}
\end{minipage}
\end{flushright}
\begin{flushright}
-- D. Adams
\end{flushright}

\subsection{Postulates of Quantum Mechanics}\label{PostulatenZelf}
The goal of this section is to give a brief introduction to (some of) the counter-intuitive aspects of quantum theory and to show why they can't be resolved as easily as one might hope (namely, due to the impossibility proofs for hidden-variable theories).
First, (a version of) the postulates of quantum mechanics, as originally introduced by von Neumann \cite{Neumann55}, is formulated. The version I use here is the one that was presented to me by Michael Seevinck in a course on the foundations of quantum mechanics \cite{SeevinckLN}. Although probably most readers already know these postulates in some form, I think it is good to restate them to give a more complete overview. Moreover, it will make the discussion more precise, since there will now be less ambiguity on what I mean when I refer to a particular postulate.\footnote{This also holds more generally; a lot of confusion may be avoided if more authors took the time to restate important terms in their discussion.} More than once I found myself in a situation when I had an objection to some argument used in a text on foundations of quantum mechanics, only to find out that I was actually objecting to one of the postulates in a different form. Also, it seems a good occasion to introduce the notation used throughout this thesis.

\begin{itemize}
\item[\textbf{1.}] \textbf{State Postulate:} Every physical system can be associated with a (complex) Hilbert space\footnote{In our definition of a Hilbert space, the inner product will be linear in the second term and anti-linear in the first.} $\h$. 
Every nonzero vector $\psi\in\h$ gives a \textit{complete} description of the state of the system. For each $\lambda\in\mathbb{C}$, $\lambda\neq0$, the two vectors $\psi$ and $\lambda\psi$ describe the same state. 
If two systems are associated with spaces $\h_1$ and $\h_2$, then the composite system is described by the space $\h_1\otimes\h_2$.   
\end{itemize}
A more generalized notion of the state of a system is given by the language of density operators. 
The states in the form of vectors in a Hilbert space are then called \underline{pure states}. Notice that each pure state in fact corresponds with an entire `line' $(\lambda\psi)_{\lambda\in\mathbb{C}\backslash\{0\}}$ in $\h$, called a \underline{ray}. To each ray one associates the projection $P_\psi:\h\to\h$ on this line, given by
\begin{equation}\label{1dproj}
  P_\psi(\phi):=\frac{\langle\psi,\phi\rangle}{\langle\psi,\psi\rangle}\psi,\quad\forall\phi\in\h.
\end{equation}  
With a mixture of pure states one can then associate a convex combination of one-dimensional projection operators.\footnote{The interpretation of mixed states as an actual mixture of pure states is not entirely without problems. For example, two mixtures of different pure states may constitute the same mixed stated. Therefore, a mixed state doesn't give complete information about the pure states of which it is a mixture. Moreover, interpreting the mixed states as `not actually knowing what the pure state is' leads to problems when considering composite states, since a mixed state is in general a convex combination of pure states of the form $P_{\psi_1\otimes\psi_2}$. Such states are known as \underline{proper} mixtures, other mixtures are called \underline{improper}. This terminology is due to d'Espagnat. See also \cite{espagnat76}.}   
More formally, a mixed state is a positive trace-class operator with trace 1. The set of mixed states will be denoted by $\mathcal{S}(\h)$, and the set of pure states by $\mathcal{P}_1(\h)$ (which stands for the set of one-dimensional projections). The set $\mathcal{S}(\h)$ is a convex set. One may show that the set $\mathcal{P}_1(\h)$ corresponds to the set of all extreme points of $\mathcal{S}(\h)$ (i.e., the one-dimensional projections are precisely all the elements of $\mathcal{S}(\h)$ that cannot be written as a proper convex combination of other elements).  
 
\begin{itemize}
\item[\textbf{2.}] \textbf{Observable Postulate:} With each physical observable $\A$, there is associated a self-adjoint operator $A:D(A)\to\h$ with domain $D(A)$ dense in $\h$.\label{operatorpost} 
\end{itemize}

The theory of self-adjoint operators is noticeably more complex than most physics literature would lead one to believe, as is made clear in the following example.

\begin{framed}
\begin{voorbeeld}\label{plaats-impuls}
Consider the Hilbert space $\h=L^2(\mathbb{R})$, the space of all square integrable functions.
The position operator defined by $(X\psi)(x):=x\psi(x)$ does not map every $\psi\in\h$ to an element of $\h$. Therefore, the set $\h$ cannot be taken as its domain but instead, one must take some dense subset $D(X)$. Its adjoint operator $X^*$ is defined as the unique operator that satisfies
\begin{equation}
  \langle\psi,X\phi\rangle=\langle X^*\psi,\phi\rangle,\quad\forall\psi\in D(X^*),\phi\in D(X),
\end{equation}    
where the domain of $X^*$ is defined as
\begin{equation}
 D(X^*):=\{\psi\in\h\:;\:\phi\mapsto\langle\psi,X\phi\rangle\text{ is a bounded linear functional }\forall\phi\in D(X)\}.
\end{equation}
Intuitively, the larger one chooses $D(X)$, the smaller $D(X^*)$ becomes. It is therefore a delicate matter to choose $D(X)$ in such a way that $X$ is self-adjoint, i.e., $(X,D(X))=(X^*,D(X^*))$. It turns out that $X$ is self-adjoint on the domain $D(X)=\{\psi\in\h\:;\:X\psi\in\h\}$.

Note that for the specification of this domain it is necessary that $X\psi$ is actually \textit{defined} for all $\psi$. 
For $X$ this causes no problems, but for the momentum operator $P$, the expression $(P\psi)(x)=-i\hbar\frac{\dee}{\dee x}\psi(x)$ is not well-defined for most elements of $\h$ without introducing the notion of a generalized function (also called a distribution). First one introduces an injection $\psi\mapsto L_\psi$ of $\h$ into the set of all linear functionals on the vector space $C_c^\infty(\mathbb{R})$ (i.e. the set of all infinitely differentiable function with compact support):
\begin{equation}
	L_\psi(\phi):=\langle\psi,\phi\rangle,\quad\forall\phi\in C_c^\infty(\mathbb{R}).
\end{equation}
 On this sspace of linear functionals, one defines the derivative as
\begin{equation}
	\frac{\dee}{\dee x}L_\psi(\phi):=-\left\langle\psi,\frac{\dee}{\dee x}\phi\right\rangle,\quad\forall\phi\in C_c^\infty(\mathbb{R}).
\end{equation}
Now for any $\psi\in\h$ one takes the condition $\tfrac{\dee}{\dee x}\psi\in\h$ to mean that there exists a $\chi\in\h$ such that $\tfrac{\dee}{\dee x}L_\psi=L_\chi$. In this case, one defines $\tfrac{\dee}{\dee x}\psi=\chi$. In particular, if $\psi$ is differentiable (in the usual sense) with derivative in $\h$ one has
\begin{equation}
	\frac{\dee}{\dee x}L_\psi(\phi):=-\left\langle\psi,\frac{\dee}{\dee x}\phi\right\rangle
	=\left\langle\frac{\dee}{\dee x}\psi,\phi\right\rangle=L_{\frac{\dee}{\dee x}\psi}(\phi),\quad\forall\phi\in C_c^\infty(\mathbb{R}),
\end{equation}
so that this new notion of a derivative is a proper extension of the usual one. It then turns out that the momentum operator $P$ is self-adjoint on the domain
\begin{equation}
	D(P):=\{\psi\in AC(\mathbb{R})\:;\: P\psi\in\h\},
\end{equation}
where $AC(\mathbb{R})$ is the set of all functions that are absolutely continuous\footnotemark\ on each finite interval of $\mathbb{R}$.
A proof can be found in \cite[p. 198]{Yosida80}. In this book one can also find a proof of the peculiar fact that there is no self-adjoint momentum operator on the Hilbert space $L^2[0,\infty)$ (p. 353). A friendly text on these problems that also emphasizes the relevance for physics and chemistry is \cite{BFV01}.
\end{voorbeeld}
\end{framed}
\footnotetext[9]{A function $\psi$ is called \underline{absolutely continuous} on the interval $I$ if for each $\epsilon>0$ there exists a $\delta>0$ such that for each finite sequence $(a_n,b_n)$ of pairwise disjoint open sub-intervals of $I$, one has $\sum_{n}|\psi(b_n)-\psi(a_n)|<\epsilon$ whenever $\sum_{n}|b_n-a_n|<\delta$.}

These are examples of self-adjoint operators that play a large role in the theory of quantum mechanics. However, most self-adjoint operators don't play any role in quantum theory. One may, for example, consider the operator $X+P$ on the Hilbert space $L^2[0,1]$. In this case, the operator $X$ is self-adjoint on the domain $D(X)=\h$. The momentum operator is self-adjoint on the domain
\begin{equation}
	D(P)=\left\{\psi\in\h\:;\:\psi\text{ is absolutely continuous},-i\hbar\frac{\dee}{\dee x}\psi\in\h,\psi(0)=\psi(1)\right\}.
\end{equation}
It follows from the Kato-Rellich theorem (see for example \cite[Ch. 6]{Oliveira09}) that $X+P$ is self-\-adjoint on the domain $D(X+P)=D(P)$.

Though self-adjoint, the operator $X+P$ has no direct physical meaning. But even an indirect meaning (e.g. by adding the measuring results of position and momentum) is ambiguous, since one cannot measure both observables at the same time (because $X$ and $P$ do not commute). This leads one to question the converse of the observable postulate, i.e. the claim that every self-adjoint operator corresponds with an observable. It seems reasonable to deny this claim. On the other hand, it seems premature to exclude some operators (like $X+P$) \textit{a priori}, since it cannot be excluded that a meaning for such an observable will be found in the future. 

For a bounded operator $A$ on a Hilbert space $\h$ its spectrum is defined as the set 
\begin{equation}
	\sigma(A):=\{a\in\mathbb{C}\:;\:A-a\een\text{ is invertible}\}.
\end{equation} 
For an unbounded operator $A$ with domain $D(A)$ dense in $\h$ the spectrum $\sigma(A)$ can still be defined. In this case $a\in\sigma(A)$ if and only if there exists a bounded operator $B$ such that $(A-a\een)B=\een$ and $B(A-a\een)\psi=\psi$ for all $\psi\in D(A)$. 
The spectrum is a generalization of the notion of the set of eigenvalues of a matrix. As with matrices, the spectrum of a self-adjoint operator is always a subset of the real numbers. This physically justifies the following postulate.\footnote{This postulate is often seen as a part of the Born postulate. Indeed, the Born postulate implies that a measurement result almost surely (i.e., with probability one) is a value in the spectrum of $A$. The value postulate sharpens this by stating that measurement results outside $\sigma(A)$ can actually never be obtained.}

\begin{itemize}
\item[\textbf{3.}] \textbf{Value Postulate:} A measurement of a physical observable $\A$ yields a real number in the spectrum of the associated self-adjoint operator $A$. 
\end{itemize}

The value postulate does not make any statement about the actual \textit{result} of a specific measurement. To elaborate on this postulate, one of the wonderful results of functional analysis is needed: the spectral theorem.\footnote{See for example \cite{Conway90} or \cite{Rudin73} for proofs.}
\begin{stelling}\label{spectraalstelling}
For each densely-defined self-adjoint operator $A$, there is a spectral measure\footnote{A spectral measure is a map $\mu$ from the Borel subsets $\mathcal{B}$ of $\mathbb{R}$ to the projection operators $\mathcal{P}(\h)$ such that $\mu(\varnothing)=0$, $\mu(\mathbb{R})=\een$, $\mu(\Delta_1\cap\Delta_2)=\mu(\Delta_1)\mu(\Delta_2)$ $\forall\Delta_1,\Delta_2\in\mathcal{B}$ and for each countable set of disjoint subsets $(\Delta_i)_{i=1}^\infty$ in $\mathcal{B}$ one has $\mu(\cup_{i=1}^\infty\Delta_i)=\sum_{i=1}^\infty\mu(\Delta_i)$.} $\mu_A$ such that
\begin{enumerate}
\item $A=\int_{\mathbb{R}}z\dee\mu_A(z)$ (as a Stieltjes integral).
\item If $\Delta\cap\sigma(A)=\varnothing$, then $\mu_A(\Delta)=0$ for each Borel set $\Delta$.
\item For each open subset $U\subset\mathbb{R}$ with $U\cap\sigma(A)\neq\varnothing$, one has $\mu_A(U)\neq0$.
\item If $B$ is a bounded operator such that $BA\subset AB$\footnote{This means that $D(BA)\subset D(AB)$ and $AB\psi=BA\psi$ for all $\psi\in D(BA)$.}, then $B$ also commutes with $\mu_A(\Delta)$ for every Borel set $\Delta$.
\end{enumerate}  
\end{stelling}
This mathematically justifies the following postulate.   

\begin{itemize}
\item[\textbf{4.}] \textbf{Born Postulate:} The probability of finding a result $a\in\Delta$ upon measurement of the observable $\A$ on a system in the state $\psi$ for a Borel set $\Delta$ is given by
\begin{equation}\label{Born}
	\pee_{\psi}[\A\in\Delta]=\frac{1}{\|\psi\|^2}\langle\psi,\mu_A(\Delta)\psi\rangle.
\end{equation} 
\end{itemize}
The notation $\pee_{\psi}[\A\in\Delta]$ is probably more familiar to probability theorists than to physicists, but I think it is a convenient one. It is to be read as the probability for the event $\A\in\Delta$, given the state $\psi$. Similarly, I write
\begin{equation}\label{qmverwacht}
	\mathbb{E}_{\psi}(\A)=\int_{\mathbb{R}}z\pee_{\psi}[\A\in\{\dee z\}]=\int_{\mathbb{R}}\frac{1}{\|\psi\|^2}\langle\psi,z\mu_A(\dee z)\psi\rangle
	=\frac{\langle\psi,A\psi\rangle}{\langle\psi,\psi\rangle}
\end{equation}
for the expectation value, instead of the often seen notation $\langle A\rangle_\psi$. Note that it is more common to take the right-hand side of (\ref{qmverwacht}) as the definition of the quantum-mechanical expectation value. Its relation with the probabilities for measurement results (the left hand side of (\ref{qmverwacht})) may then be seen to be a consequence of the spectral theorem (i.e., according to this theorem (\ref{qmverwacht}) is equivalent to (\ref{Born})).

In the generalized case where mixed states are considered, the Born rule generalizes to
\begin{equation}\label{BornRuleTrace}
	\pee_{\rho}[\A\in\Delta]=\Trace(\rho\mu_A(\Delta))
\end{equation}
for the mixed state $\rho$, where $\Trace$ denotes the trace operation. One easily checks that this results in (\ref{Born}) in case that $\rho=P_{\psi}$ (i.e., whenever $\rho$ is a pure state).

\begin{opmerking}\label{janee}
 Observables associated with projection operators (in particular, those of the form $\mu_A(\Delta)$) are usually regarded as the ``yes-no''-questions. Since their spectrum is $\{0,1\}$, a measurement of such an observable always yields one of these numbers. For an observable $\mathcal{P}$ associated with the operator $P$, the number 1 corresponds to the answer ``The state of the system after the measurement lies in the space $P\h$.'' and the number 0 corresponds to the answer ``The state of the system after the measurement lies in the space $(\een-P)\h$.'' In particular, a measurement of the observable associated with an operator of the form $\mu_A(\Delta)$ can be associated with the question ``Does the value of $\A$ lie in the set $\Delta$?'' The precise meaning of these questions (and their answers) will play an underlying role in this thesis.
\end{opmerking}

Many physicists may find this use of mathematics overwhelming and maybe even unnecessary. In most physics literature one simply refers to the spectral decomposition of an operator without explicitly defining what this means. Operators are often treated as if they are matrices and their spectra are then referred to as the set of eigenvalues with corresponding eigenstates and eigenspaces. As a student of physics I became confused when first realizing that, for example, the position operator $X$ does not have any eigenstates. Moreover, it didn't become clear to me why the probability of finding a particle in some (Borel) subset $\Delta$ was given by
\begin{equation}\label{phys-born}
	\pee_{\psi}[\mathcal{X}\in\Delta]=\frac{1}{\|\psi\|^2}\int_{\Delta}|\psi(x)|^2\dee x,
\end{equation} 
until I learned (in a mathematics course)  that the spectral measure for the position operator is simply given by\footnote{Here, $1_{\Delta}$ denotes the indicator function for the set $\Delta$.} $\mu_X(\Delta)\psi=1_{\Delta}\psi$ (because $\sigma(X)=\mathbb{R}$), so that (\ref{phys-born}) is a special case of (\ref{Born}).

Finally, it has to be specified how the state of the system changes in time. Actually, two postulates are needed for this.
\begin{itemize}
\item[\textbf{5.}] \textbf{Schr\"odinger Postulate:} When no measurement is performed on the system, the change of the state in time is described by a unitary transformation. That is,
\begin{equation*}
	\psi(t)=U(t)\psi(0)
\end{equation*}
for some strongly continuous unitary one-parameter group\footnote{This means that the set $\{U(t)\:;\:t\in\mathbb{R}\}$ forms a group of unitary operators satisfying $U(t+s)=U(t)U(s)$ $\forall s,t$, where the map $t\mapsto U(t)$ is continuous in the sense that $\lim_{s\to t}U(s)\psi=U(t)\psi$ for all $t\in\mathbb{R},\psi\in\h$.} $t\mapsto U(t)$.
\end{itemize}
Note that no distinction in notation is made between the map $\psi:\mathbb{R}\to\h$, $t\mapsto\psi(t)$ and the vector $\psi$ in $\h$ as is standard in most literature. This postulate is in fact equivalent to the one found in more standard physics literature. Namely, because of Stone's Theorem there exists a self-adjoint operator $H$ such that $U(t)=e^{-iHt}$ $\forall t$, which brings one back to the original Schr\"odinger equation
\begin{equation*}
	i\frac{\dee}{\dee t}\psi(t)=H\psi(t),\quad\forall\psi\in D(H).
\end{equation*}
For a mixed state $\rho$ the time evolution is given by $\rho(t)=U(t)\rho U^*(t)$, or $i\tfrac{\dee\rho(t)}{\dee t}=[H,\rho(t)]$.

\begin{itemize}
\item[\textbf{6.}] \textbf{Von Neumann Postulate (Projection Postulate):} When an observable $\A$ corresponding to an operator $A$ with discrete spectrum is measured and the measurement yields some $a\in\sigma(A)$, the state of the system changes discontinuously from $\psi$ to $\mu_A(\{a\})\psi$. 
\end{itemize}
Note that the state of the system after measurement is indeed always a state (i.e. $\mu_A(\{a\})\psi\neq0$) because of the Born postulate.

The motivation for introducing this postulate is that it ensures that if a measurement of an observable yields some value $a$, an immediate second measurement of the same observable will yield exactly the same result. It is founded on experimental experience (von Neumann based it on the Compton-Simons experiment \cite{Neumann55}) and therefore seems a necessary claim. However, the postulate as stated here only applies for discrete observables (i.e., those whose corresponding operators have a discrete spectrum). It has, in fact, been shown that a similar postulate for continuous spectra cannot be formulated: repeatable measurements are only possible for discrete observables \cite{Ozawa84}. A more extensive discussion can be found in \cite{Busch91}.

The von Neumann postulate is probably the most controversial postulate of quantum mechanics. Because the time evolution of the state of a system depends so greatly on whether or not there is a measurement being performed on the system, one is tempted to ask what exactly constitutes a measurement. No satisfactory answer to this question exists in my opinion, and it is one of the underlying questions of what is known as ``the measurement problem'' (see also \cite{Bell90}). Compared to the difficulty of this problem, the problem of repeatability for observables with a continuous spectrum seems a rather small one. It seems likely to me that a philosophically satisfying solution of the measurement problem may also solve the latter.\footnote{There have been proposals for introducing generalizations of the von Neumann postulate that are rich enough to incorporate observables with a continuous spectrum. However, there are great consequences involved. In the scenario described in \cite{Srinivas80} it requires a generalization of the state postulate to a point where the original states (the density operators) are only associated with the probability functions they generate through the Born postulate. The generalized states also allow probability functions that are no longer of the form (\ref{BornRuleTrace}) and are no longer $\sigma$-additive in general.}

Although it certainly is an interesting topic for research, the measurement problem will not be the focus of this paper. My problem is rather related to one of the earliest objections against quantum mechanics made clear for the first time by Einstein, namely its possible incompleteness. 

       \markboth{Introduction to the Foundations of Quantum Mechanics}{The (In)completeness of Quantum Mechanics}
       \subsection{The (In)completeness of Quantum Mechanics (Part I)}\label{The (In)completeness of Quantum Mechanics (Part I)}

It follows from the Born postulate that, in general, the state of the system does not determine what the result of the outcome of a measurement is. This in itself was not the only problem Einstein had with quantum theory. An even more serious problem for him was that certain observables like energy and momentum, which are even supposed to be conserved, are not attributed a particular value at all in quantum mechanics. That is, if one knows the state of the system, one cannot in general say what the momentum of a particle in the system is. In \cite{EPR} Einstein, Podolsky and Rosen also gave a seemingly convincing reason why a complete theory \textit{should} attribute a value to such observables at all times. Below, an experiment is discussed that is quite similar to the one discussed in \cite{EPR}, but has the advantage that there are no unbounded operators involved. It was first introduced by Bohm \cite{Bohm57} and has played a central role in many discussions ever since.    

\begin{framed}
\begin{voorbeeld}[The EPRB-experiment]\label{EPRB}
Consider a system of two spin-$\tfrac{1}{2}$ particles (say, two electrons). Each particle taken by itself can be described in a Hilbert space $\mathbb{C}^2$, where a basis is choosen such that $(1,0)$ stands for spin up and $(0,1)$ for spin down. Physicists would use $|\uparrow\rangle$ and $|\downarrow\rangle$ to denote those basis vectors. Traditionally the spin is considered along the $z$-axis and the corresponding observable for the spin along this axis is
\begin{subequations}
\begin{equation}
	\sigma_z:=\begin{pmatrix} 1 & 0 \\ 0 &-1\end{pmatrix}.
\end{equation}
For the $x$ and $y$ axes one has
\begin{equation}
	\sigma_x:=\begin{pmatrix} 0 & 1 \\ 1 &0\end{pmatrix},\quad\sigma_y:=\begin{pmatrix} 0 & -i \\ i &0\end{pmatrix}.
\end{equation}
\end{subequations}
Consequently, for a measurement of the spin along the axis 
\begin{equation}
	r=(\cos(\vartheta)\sin(\varphi),\sin(\vartheta)\sin(\varphi),\cos(\varphi))
\end{equation} 
one has the observable
\begin{equation}
\begin{split}
	\sigma_r&:=\cos(\vartheta)\sin(\varphi)\sigma_x+\sin(\vartheta)\sin(\varphi)\sigma_y+\cos(\varphi)\sigma_z\\
	&=
	\begin{pmatrix}	\cos(\varphi) & \cos(\vartheta)\sin(\varphi)-i\sin(\vartheta)\sin(\varphi)\\
									\cos(\vartheta)\sin(\varphi)+i\sin(\vartheta)\sin(\varphi) & -\cos(\varphi)
	\end{pmatrix}\\
	&=P_{r+}-P_{r-},
\end{split}
\end{equation}
where $P_{r+}=\tfrac{1}{2}(\een+\sigma_r)$ and $P_{r-}=\tfrac{1}{2}(\een-\sigma_r)$ are one-dimensional projections (this is easily checked using $\sigma_r^2=\een$). Thus, the projection $P_{r+}$ corresponds to the question if one will find spin up along the $r$-axis. Let's denote the corresponding observable by $\mathcal{P}_{r}$ (see Remark \ref{janee}). For a spin-$\tfrac{1}{2}$ particle in a state $\psi$ it then follows that
\begin{equation}
	\pee_\psi[\mathcal{P}_{r}=1]=\langle\psi,P_{r+}\psi\rangle,\quad  
	\pee_\psi[\mathcal{P}_{r}=0]=\langle\psi,(\een-P_{r+})\psi\rangle=\langle\psi,P_{r-}\psi\rangle.
\end{equation}
The combined system is then described by the Hilbert space $\mathbb{C}^2\otimes\mathbb{C}^2\simeq\mathbb{C}^4$ where the following connection is made between the two descriptions of this space:
\begin{equation}
\begin{aligned}
	\begin{pmatrix} 1\\0\end{pmatrix}\otimes\begin{pmatrix} 1\\0\end{pmatrix}&=(1,0,0,0)\:\left(=|\uparrow\uparrow\rangle\right)& 
	\begin{pmatrix} 1\\0\end{pmatrix}\otimes\begin{pmatrix} 0\\1\end{pmatrix}&=(0,1,0,0)\:\left(=|\uparrow\downarrow\rangle\right)\\
	\begin{pmatrix} 0\\1\end{pmatrix}\otimes\begin{pmatrix} 1\\0\end{pmatrix}&=(0,0,1,0)\:\left(=|\downarrow\uparrow\rangle\right)& 
	\begin{pmatrix} 0\\1\end{pmatrix}\otimes\begin{pmatrix} 0\\1\end{pmatrix}&=(0,0,0,1)\:\left(=|\downarrow\downarrow\rangle\right)
\end{aligned}
\end{equation}
An observable $\mathcal{A}$ corresponding with the operator $A$ for the one-particle system extends to an observable for the first particle (the one `on the left') in the combined system with the operator $A\otimes\een$. For the second particle $\A$ extends to the operator $\een\otimes A$. Such observables always commute, since
\begin{equation}
	(A\otimes\een)(\een\otimes B)=A\otimes B=(\een\otimes B)(A\otimes\een).
\end{equation}
Therefore, one can simultaneously perform measurements on particle one and on particle two. Also, note that for two projections $P_1$ and $P_2$, the operator $P_1\otimes P_2$ is again a projection. 
Now suppose the system is prepared in the state
\begin{equation}
	\psi=\frac{1}{\sqrt{2}}(0,1,-1,0)\:\left(=\frac{1}{\sqrt{2}}\left(|\uparrow\downarrow\rangle-|\downarrow\uparrow\rangle\right)\right).
\end{equation}
If the spin along the $z$-axis for the first particle is measured one finds that the probabilities for finding spin up or spin down are respectively
\begin{equation}
\begin{split}
	\pee_\psi[\mathcal{P}_z=1]
	&=
	\langle\psi,P_{z+}\otimes\een\psi\rangle=\frac{1}{2}\langle(1,0),P_{z+}(1,0)\rangle=\frac{1}{2};\\
	\pee_\psi[\mathcal{P}_z=0]
	&=
	\langle\psi,P_{z-}\otimes\een\psi\rangle=\frac{1}{2}\langle(0,1),P_{z-}(0,1)\rangle=\frac{1}{2}.
\end{split}
\end{equation}
The reasoning of Einstein, Podolsky and Rosen now is as follows.
According to the von Neumann postulate, after the measurement the state of the system will be $(0,1,0,0)$ upon finding spin up and $(0,0,1,0)$ upon finding spin down. In either case, a measurement of the spin along the $z$-axis on the \textit{second} particle will yield a particular result (the opposite of the spin of the first particle) with absolute certainty. Since one can make a prediction of the spin of the second particle without in any way disturbing this particle (the distance between the two particles may be arbitrarily large) one may state that the spin along the $z$-axis has a \textit{real} meaning. That is, the spin along the $z$-axis appears to be an observable that is actually meaningful for the observed system. Einstein, Podolsky and Rosen would say that there exists \textit{an element of physical reality} that corresponds to this observable.   

Now, if one would measure the spin along the $y$-axis on particle one instead, the state of the system would be projected to the state $(-1,1,-1,1)$ if one would find spin up, and to the state $(1,1,-1,-1)$ if one would find spin down. Each happens with probability $\tfrac{1}{2}$. In either case, a measurement of the spin along the $y$-axis on the second particle yields a particular result (the opposite of the spin of the first particle) with absolute certainty. Following the same reasoning, one concludes that also the spin along the $y$-axis of the second particle should correspond to an element of physical reality. 

Now the problem is the following: in the states $(0,1,0,0)$ and $(0,0,1,0)$ one can assign a value to the spin along the $z$-axis for the second particle, but the spin along the $y$-axis does not have a definite value. \textit{Vice versa} for the states $(-1,1,-1,1)$ and $(1,1,-1,-1)$. It turns out that there is no state that can assign a definite value to both the observables at the same time and hence there is no state in quantum mechanics that can give a complete description of the system, because there is always at least one observable that has no definite value in that state. 
\end{voorbeeld}
\end{framed}

The standard literature uses the terminology of Einstein, Podolsky and Rosen, which is more formal. The crucial terms they use are (quotations are taken from \cite{EPR}):
\begin{itemize}
	\item \textbf{EOPR:} ``If, without in any way disturbing a system, we can predict with certainty (i.e., with probability equal to unity) the value of a physical quantity, then there exists an element of physical reality [EOPR] corresponding to this physical quantity.''
	\item \textbf{Comp:} A necessary condition for the completeness of a physical theory is that ``every element of the physical reality must have a counterpart in the physical theory.''
	\item \textbf{Loc:} The performance of a measurement on a physical system does not have an instantaneous influence on elements of the physical reality that are located at some distance of this system.
\end{itemize}
In these terms the example above now reads as follows. Since without in any way disturbing the second particle (because of Loc) either its spin along the $y$-axis or along the $z$-axis can be predicted, both these observables must correspond to elements of the physical reality (EOPR). Since no state of the system can describe simultaneously the values of both these observables, the theory of quantum mechanics is not complete (because of Comp).

These terms may sound a bit metaphysical, if only because they hinge upon a particular definition of physical reality. However, the argument still holds if one takes the criteria of completeness not to be about physical reality, but about possible observations, eliminating the objections that instrumentalists (or idealists) may have against this argument. One could argue that a physical theory should be local in the sense that, if one can make predictions about observables of system 1 by performing measurements on some system 2 separated from system 1, the theory should be able to make those predictions also without the use of system 2. In addition, the theory is complete if, in this situation, it actually \textit{does} make these predictions. This seems at least \textit{sensible}.

Bohr, as the great defender of the completeness of quantum mechanics, published his own response to this experiment \cite{Bohr35}. His main objection is undoubtedly to be sought in the following passage.
\begin{quote}
	``From our point of view we now see that the wording of the above-mentioned criterion of physical reality proposed by Einstein, Podolsky and Rosen contains an ambiguity as regards the meaning of the expression ``without in any way disturbing the system.'' Of course there is in a case like that just considered no question of a mechanical disturbance of the system under investigation during the last critical stage of the measuring procedure. But even at this stage there is essentially the question of \textit{an influence on the very conditions which define the possible types of predictions regarding the future behavior of the system.} Since these conditions constitute an inherent element of the description of any phenomenon to which the term ``physical reality'' can properly be attached, we see that the argumentation of the above-mentioned authors [Einstein, Podolsky and Rosen] does not justify their conclusion that the quantum-mechanical description is incomplete.'' \cite{Bohr35}
\end{quote}
In terms of example \ref{EPRB}, it seems that Bohr finds an ambiguity in the reasoning used to obtain the conclusion that both the spin along the $z$-axis, as the spin along the $y$-axis correspond to elements of physical reality. Apparently, some form of disturbance must be at hand. In my understanding, there is an ambiguity in the fact that in \cite{EPR} it is demanded that the theory gives a \textit{simultaneous} description of a pair of observables that \textit{cannot} be measured simultaneously. Bohr declares such observables to be \textit{complementary}. No doubt, I do think that Bohr's reply might tell us that we may consider quantum mechanics a complete theory in a certain sense (although I think more clarification is needed), but it doesn't really tell us why we \textit{cannot} consider it to be incomplete in a different sense. Thus a search for a theory that is complete in the sense of Einstein, Podolsky and Rosen (i.e., a so-called hidden-variable theory) seems to me justified, certainly back in 1935, but even today. However, it appears that attempts to find such a theory that is also local are doomed to fail.

       \markboth{Introduction to the Foundations of Quantum Mechanics}{Impossibility Proofs for Hidden Variables}
       \subsection{Impossibility Proofs for Hidden Variables}\label{NOGOS}

What constitutes a hidden-variable theory? Thus far, it has only been argued that quantum mechanics does not satisfy the criteria because of its alleged incompleteness (that is, according to Einstein, Podolsky and Rosen). Let's make some seemingly reasonable assumptions on the structure of a theory that \textit{is} supposedly complete (or at least, more complete than quantum mechanics).
 
As in any contemporary approach to physics, suppose there is a set $\Lambda$ called the state-space. The completeness claim now implies that there exist states $\lambda\in\Lambda$ that, for each observable $\A$,  determine the value $\lambda(\A)$ of that observable. 
Such a state will be called a \underline{pure state} and it is supposed that $\Lambda$ only consists of pure states. 
As a result, for each observable $\A$, a function $f_{\A}:\Lambda\to\mathbb{V}_{\A}$ can be constructed, assigning to each state the \underline{value of the observable in that state}:
\begin{equation}\label{classicobs}
	f_{\A}(\lambda):=\lambda(\A)   
\end{equation}
Here $\mathbb{V}_{\A}$ denotes the set of all possible values that $\A$ may have. It is common belief that one can take this to be the set of real numbers (or an $n$-tuple of real numbers, e.g. in the case of position or momentum).\footnote{It is a remarkable accomplishment of modern science that everything is described by numbers; even phenomena like colors. However, it seems good to point out that we are holding on to a dogma here, and that one day it may appear that using numbers isn't an appropriate way to describe all phenomena.} 

Now, the interpretation of (\ref{classicobs}) is that if a system in a state $\lambda$ is considered, and the observable $\A$ is measured, then one will find the value $f_{\A}(\lambda)$ with probability one. 
This implies that measurements reveal properties possessed by the system prior to the measurement. In particular, the outcomes of experiments are pre-determined (unlike in quantum mechanics). Furthermore, if one assumes that a measurement does not disturb the state of the system, one automatically gains repeatability of measurements (i.e., successive measurements of the same observable will yield the same result). There is no need for a discontinuous state change like the one introduced by the von Neumann postulate.

The statistics of quantum mechanics should be recovered by the hidden-variable model by introducing appropriate probability measures $\pee$ on the set $\Lambda$ (which should be turned into a measurable space by an appropriate choice of some $\sigma$-algebra). The expectation value of the observable $\A$ for the ensemble $\pee$ would then be given by
\begin{equation}\label{vwwaarde}
	\mathbb{E}(\A)=\int_{\Lambda}f_{\A}(\lambda)\dee\pee(\lambda).
\end{equation}

\subsubsection{Von Neumann's Theorem}\label{neumann}
The impossibility proof of von Neumann as presented in \cite{Neumann55} is quite extensive and complex (it spans about ten pages, preceded by about fifteen pages of introductory discussion). In fact, a good understanding of the proof is hard to acquire and even in recent years explanations of it have been put online \cite{Rosinger04}, \cite{Singh05}, \cite{Dmitriev05}.\footnote{This last article actually originates from 1974, but has only been published recently.}

It is not surprising that the original proof appears to be somewhat vague at first sight. It is concerned with the question whether or not the stochastic behavior of quantum mechanics can be reproduced by a classical theory. However, von Neumann's book (from 1932) dates from before the time the mathematical axioms of classical probability were properly introduced by Kolmogorov \cite{Kolmogorov56} in 1933. The clear structure as presented above therefore wasn't available to von Neumann at that time.\footnote{Most likely, von Neumann was acquainted with the recent developments made in probability theory, since he himself was also working on measure theory. Still, even Kolmogorov's work seems sometimes less formal from a modern perspective.} In fact, von Neumann doesn't explicitly speak about assigning definite values to observables at all and doesn't make use of the notion of a probability space (like $\Lambda$). Instead, the focus is on ensembles of systems and properties of the expectation values for observables. Von Neumann investigates what kind of properties ensembles \textit{do} have from the point of view of quantum mechanics, and \textit{should} have from the point of view of hidden-variable theories. A discrepancy between these two leads von Neumann to conclude that no completion of quantum mechanics in terms of hidden variables is possible. 

In terms of the above described structure, one may think of an ensemble as a function $\mathbb{E}:\mathcal{O}\to\mathbb{R}$ on the set $\mathcal{O}$ of all observables, defined by equation (\ref{vwwaarde}).
 Although this is a good concept to keep in mind when von Neumann talks about an expectation-value function, it should be emphasized that von Neumann actually refers to a broader notion. In fact, von Neumann almost proves that the expectation-value functions that appear in quantum mechanics cannot be of the form (\ref{vwwaarde}) (this is proven more explicitly by the violation of the Bell inequalities, see Section \ref{The Bell Inequality}).

To accomplish this, von Neumann relies on four axioms for a hidden-variable theory:
\begin{itemize}
\item[\textbf{vN1}] For each observable $\A$ corresponding to the operator $A$, and for each polynomial $f:\mathbb{R}\to\mathbb{R}$, the observable $f(\A)$ (corresponding to applying the function $f$ to each measurement result of $\A$) corresponds to the operator $f(A)$.
\item[\textbf{vN2}] If $\A$ is an observable that only takes positive values, then for each ensemble of systems one has $\mathbb{E}(\A)\geq0$.
\item[\textbf{vN3}] For each sequence of observables $\A_1,\A_2,\ldots$ corresponding to operators $A_1,A_2,\ldots$, there is an observable $\A_1+\A_2+\ldots$ corresponding to the operator $A_1+A_2+\ldots$.
\item[\textbf{vN4}] For each sequence of observables $\A_1,\A_2,\ldots$, each sequence of real numbers $c_1,c_2,\ldots$ and each ensemble of systems it should hold that
\begin{equation}
	\mathbb{E}(c_1\A_1+c_2\A_2+\ldots)=c_1\mathbb{E}(\A_1)+c_2\mathbb{E}(\A_2)+\ldots.
\end{equation}
\end{itemize}
Axiom vN3 is a bit ambiguous. At first sight, it is not clear if von Neumann allows the sums to be infinite. It turns out in the proof that this assumption is indeed necessary. In that case, the following difficulty arises. In general, a sequence of operators $\sum_{i=1}^nA_i$ will not converge to any operator (neither uniformly, nor strongly, nor weakly). In fact, it is not even clear that if $A_1$ and $A_2$ are self-adjoint, that their sum is too (since it is not clear how to choose $D(A_1+A_2)$). For sake of simplicity, one may consider only observables whose corresponding operators are bounded. Also, von Neumann nowhere uses vN3 in this form in his proof. Instead, one may introduce the following modified axiom.
\begin{itemize}
\item[\textbf{vN3'}] If $\A$ is an observable corresponding to the bounded operator $A$ and $A_1,A_2,\ldots$ is a sequence of bounded self-adjoint operators such that $\sum_{i=1}^nA_i$ converges strongly to $A$ (as $n\to\infty$), then each of the operators $A_i$ corresponds to a certain observable $\A_i$.
\end{itemize}
For the same reasons, vN4 will also be modified:
\begin{itemize}
\item[\textbf{vN4'}] If $\A$ is an observable corresponding to the bounded operator $A$, and $A_1,A_2,\ldots$ is a sequence of bounded self-adjoint operators and $c_1,c_2,\ldots$  a sequence of real numbers such that $\sum_{i=1}^nc_iA_i$ converges strongly to $A$ (as $n\to\infty$), then 
\begin{equation}
	\mathbb{E}(\A)=\mathbb{E}(c_1\A_1+c_2\A_2+\ldots)=c_1\mathbb{E}(\A_1)+c_2\mathbb{E}(\A_2)+\ldots.
\end{equation}
\end{itemize}
Note that vN3' and vN4' are in a sense the axioms vN3 and vN4 reversed. Indeed, vN3 postulates the existence of a single observable given the existence of an entire sequence of observables, whereas vN3' postulates the existence of an entire sequence of observables, given the existence of a single observable.
From the axioms presented in this way, it follows that each bounded self-adjoint operator should correspond to an observable. 
Also, it follows from vN4' that $\mathbb{E}(c\A)=c\mathbb{E}(\A)$. For this reason, one can always normalize any expectation-value function $\mathbb{E}$ such that $\mathbb{E}(\mathpzc{1})=1$ (except for the pathological case where $\mathbb{E}(\A)=0$ for all $\A$, or $\mathbb{E}(\mathpzc{1})=\infty$). Therefore, mainly normalized ensembles will be considered.

Besides these axioms, von Neumann introduces two definitions.
\begin{definitie}
An expectation-value function $\mathbb{E}:\mathcal{O}\to\mathbb{R}$ is called \underline{dispersion free} if 
\begin{equation}
  \mathbb{E}(\A^2)=\mathbb{E}(\A)^2,\quad\forall\A.
\end{equation}
\end{definitie}
This definition expresses the idea that for every observable $\A$, its variance in a dispersion-free state is zero. That is, in such an ensemble a measurement of any observable $\A$ will yield a particular result almost surely. A hidden-variable state, then, would have to be dispersion free. 
\begin{definitie}\label{puur}
An expectation-value function $\mathbb{E}:\mathcal{O}\to\mathbb{R}$ is called \underline{pure} or \underline{homogeneous} if for all expectation-value functions $\mathbb{E}',\mathbb{E}''$  the condition
\begin{equation}\label{homo1}
	\mathbb{E}(\A)=\mathbb{E}'(\A)+\mathbb{E}''(\A), \quad\forall\A,
\end{equation}
implies that there exist positive constants $c',c''$ (independent of $\A$, with $c'+c''=1$), such that 
\begin{equation}
	\mathbb{E}'(\A)=c'\mathbb{E}(\A)\text{ and }\mathbb{E}''(\A)=c''\mathbb{E}(\A),\quad\forall\A.
\end{equation}
\end{definitie}
This definition expresses that a homogeneous ensemble is not the mixture of two other ensembles. That is, every split made in the ensemble only gives two versions of the original ensemble.

The main mathematical result by von Neumann may now be formulated as follows:
\begin{stelling}\label{spoorneumann}
	If vN3' holds, then for every expectation-value function $\mathbb{E}$ that satisfies vN2, vN4' and $\mathbb{E}(\mathpzc{1})<\infty$, there exists a positive trace-class operator $U$ such that
	\begin{equation}
		\mathbb{E}(\A)=\Trace(UA),\quad\forall \A\in\mathcal{O}.
	\end{equation}
	Conversely, if $U$ is a positive trace-class operator $U$, then the expectation-value function $\mathbb{E}(\A)=\Trace(UA)$ satisfies vN2 and vN4'.
\end{stelling}

\noindent
\textit{Proof:}\hspace*{\fill}\\
For a unit vector $e$, let $P_e$ denote the projection on the ray spanned by $e$, and let $\mathcal{P}_e$ denote the corresponding observable (which exists according to vN3'). For any pair of unit vectors $e_1,e_2$, the operators $F_{e_1,e_2}$ and $G_{e_1,e_2}$ are defined to be
\begin{equation}
	F_{e_1,e_2}\psi:=\langle e_2,\psi\rangle e_1+\langle e_1,\psi\rangle e_2,\quad 
	G_{e_1,e_2}\psi:=i\langle e_2,\psi\rangle e_1-i\langle e_1,\psi\rangle e_2=F_{ie_1,e_2},
\end{equation}
or, equivalently,
\begin{equation}
	F_{e_1,e_2}=P_{(e_1+e_2)/\sqrt{2}}-P_{(e_1-e_2)/\sqrt{2}},\quad
	G_{e_1,e_2}=P_{(ie_1+e_2)/\sqrt{2}}-P_{(ie_1-e_2)/\sqrt{2}}.
\end{equation}
\indent

One easily checks that, if $e_1$ and $e_2$ are either orthogonal or identical (i.e., if $P_{e_1}$ and $P_{e_2}$ commute), then these operators are self-adjoint. The corresponding observables are denoted by $\mathcal{F}_{e_1,e_2}$ and $\mathcal{G}_{e_1,e_2}$. Note that one has $F_{e,e}=2P_e$ and $G_{e,e}=0$ for all unit vectors $e$.

Let $\mathbb{E}:\mathcal{O}\to\mathbb{R}$ be given. The operator $U$ can now be defined in the following way. Let $e$ be an arbitrary unit vector and let $(e_n)_{n=1}^\infty$ be an orthonormal basis of the Hilbert space such that there is an $i$ with $e=e_i$ (note that Hilbert spaces are by definition separable in \cite{Neumann55}). Now consider the functional
\begin{equation}
	f_e:\psi\mapsto
	\sum_{n=1}^\infty\langle\psi,e_n\rangle
	\left(\frac{1}{2}\mathbb{E}(\mathcal{F}_{e_n,e})+\frac{i}{2}\mathbb{E}(\mathcal{G}_{e_n,e})\right).
\end{equation}
It must be checked that this limit indeed exists for each $\psi$. Note that projection operators correspond to positive observables. From vN2 and vN4' it then follows that
\begin{equation}
\begin{split}
	\mathbb{E}(\mathcal{F}_{e_1,e_2})
	&=\mathbb{E}(\mathcal{P}_{(e_1+e_2)/\sqrt{2}})-\mathbb{E}(\mathcal{P}_{(e_1-e_2)/\sqrt{2}})\\
	&\leq\mathbb{E}(\mathcal{P}_{(e_1+e_2)/\sqrt{2}})=\mathbb{E}(1)-\mathbb{E}(1-\mathcal{P}_{(e_1+e_2)/\sqrt{2}})\\
	&\leq\mathbb{E}(1)<\infty,
\end{split}
\end{equation}
and similarly, $\mathbb{E}(\mathcal{G}_{e_1,e_2})\leq\mathbb{E}(1)<\infty$ for all unit vectors $e_1,e_2$. Therefore,
\begin{equation}\label{functionaalafschatting}
	\lim_{N\to\infty}\left|\sum_{n=1}^N\langle\psi,e_n\rangle
	\left(\frac{1}{2}\mathbb{E}(\mathcal{F}_{e_n,e})+\frac{i}{2}\mathbb{E}(\mathcal{G}_{e_n,e})\right)\right|
	\leq
	\mathbb{E}(1)\lim_{N\to\infty}\left|\sum_{n=1}^N\langle\psi,e_n\rangle\right|=\|\psi\|\mathbb{E}(1)<\infty.
\end{equation}
It is straightforward, though tedious, to show that the value of $f_e(\psi)$ does not depend on the choice of the basis in which $e$ appears. I will omit this part of the proof here. From (\ref{functionaalafschatting}) it follows that the functional $f_e$ is in fact bounded and hence, according to Riesz' representation theorem, there is a unique vector in $\h$, which will be denoted by $Ue$, such that
\begin{equation}
	f_e(\psi)=\langle\psi,Ue\rangle,\quad\forall\psi\in\h.
\end{equation}
This defines the operator $U$. From this definition it follows that	
\begin{equation}
	\langle e_1,Ue_2\rangle:=\frac{1}{2}\mathbb{E}(\mathcal{F}_{e_1,e_2})+\frac{i}{2}\mathbb{E}(\mathcal{G}_{e_1,e_2}),\quad\text{whenever }P_{e_1}\text{ and }P_{e_2}\text{ commute.}
\end{equation}
In particular, one has
\begin{equation}\label{potieveU}
	\langle e,Ue\rangle=\frac{1}{2}\mathbb{E}(\mathcal{F}_{e,e})+\frac{i}{2}\mathbb{E}(\mathcal{G}_{e,e})=\mathbb{E}(\mathcal{P}_e).
\end{equation}
It is now easy to show that $U$ is self-adjoint. For each pair of unit vectors $e_1,e_2$ with $[P_{e_1},P_{e_2}]=0$ one has
\begin{equation}
\begin{split}
	\langle Ue_1,e_2\rangle=&
	\langle e_2,Ue_1\rangle^*=\frac{1}{2}\mathbb{E}(\mathcal{F}_{e_2,e_1})-\frac{i}{2}\mathbb{E}(\mathcal{G}_{e_2,e_1})\\
	=&
	\frac{1}{2}\mathbb{E}(\mathcal{F}_{e_1,e_2})+\frac{i}{2}\mathbb{E}(\mathcal{G}_{e_1,e_2})=\langle e_1,Ue_2\rangle.\\
\end{split}
\end{equation}
The general result
\begin{equation}
	\langle\psi,U\phi\rangle=\langle U\psi,\phi\rangle,\quad\forall \psi,\phi\in\h,
\end{equation}
then follows by expanding both vectors with respect to the same basis.

Now let $\A$ be an arbitrary observable and let $A$ be its corresponding bounded self-adjoint operator. For any orthonormal basis $(e_n)_{n=1}^\infty$ of $\h$, write $a_{nm}:=\langle e_n,Ae_m\rangle$. Then 
\begin{equation}
	A=\sum_{n=1}^\infty\sum_{m=1}^{n-1} \left(a_{nn}P_{e_n}+\mathrm{Re}(a_{nm})F_{e_n,e_m}+\mathrm{Im}(a_{nm})G_{e_n,e_m}\right),
\end{equation}
where the right hand side converges strongly to $A$. Indeed, for $\psi\in \h$ one has
\begin{multline}
\lim_{N\to\infty}\sum_{n=1}^N\sum_{m=1}^{n-1} \left(a_{nn}P_{e_n}+\mathrm{Re}(a_{nm})F_{e_n,e_m}+\mathrm{Im}(a_{nm})G_{e_n,e_m}\right)\psi\\
\begin{split}
	=&
	\lim_{N\to\infty}\sum_{n=1}^N\sum_{m=1}^{n-1}\sum_{j=1}^\infty
	\left(a_{nn}P_{e_n}+\mathrm{Re}(a_{nm})F_{e_n,e_m}+\mathrm{Im}(a_{nm})G_{e_n,e_m}\right)\langle e_j,\psi\rangle e_j\\
	=&
	\lim_{N\to\infty}\sum_{n=1}^N\sum_{m=1}^{n-1}\Bigr(
	\langle e_n,\psi\rangle a_{nn} e_n+\mathrm{Re}(a_{nm})\left(\langle e_m,\psi\rangle e_n+\langle e_n,\psi\rangle e_m\right)\\
	&+i\mathrm{Im}(a_{nm})\left(\langle e_m,\psi\rangle e_n-\langle e_n,\psi\rangle e_m\right)\Bigl)\\
	=&
	\lim_{N\to\infty}\sum_{n=1}^N\sum_{m=1}^{n-1}\langle e_n,A e_n\rangle\langle e_n,\psi\rangle e_n
	+\langle e_n,A e_m\rangle\langle e_m,\psi\rangle e_n+\langle e_m,A e_n\rangle\langle e_n,\psi\rangle e_m\\
	=&
	\lim_{N\to\infty}\sum_{n,m=1}^N\langle e_n,A e_m\rangle\langle e_m,\psi\rangle e_n=A\psi.
\end{split}
\end{multline}
Finally, using vN4', it follows that
\begin{equation}
\begin{split}
	\mathbb{E}(\A)
	=&
	\sum_{n=1}^\infty\sum_{m=1}^{n-1} \left(a_{nn}\mathbb{E}(\mathcal{P}_{e_n})
	+\mathrm{Re}(a_{nm})\mathbb{E}(\mathcal{F}_{e_n,e_m})+\mathrm{Im}(a_{nm})\mathbb{E}(\mathcal{G}_{e_n,e_m})\right)\\
	=&
	\sum_{n=1}^\infty\sum_{m=1}^{n-1} \left(a_{nn}\Trace(UP_{e_n})
	+\mathrm{Re}(a_{nm})\Trace(UF_{e_n,e_m})+\mathrm{Im}(a_{nm})\Trace(UG_{e_n,e_m})\right)\\
	=&
	\sum_{n=1}^\infty\sum_{m=1}^{n-1}
	\Trace\left(U\left(a_{nn}P_{e_n}+\mathrm{Re}(a_{nm})F_{e_n,e_m}+\mathrm{Im}(a_{nm})G_{e_n,e_m}\right)\right)\\
	=&
	\Trace(UA),
\end{split}
\end{equation}
where the second step almost immediately follows from the definition of $U$. The positivity of $U$ follows from equation (\ref{potieveU}) together with vN2. From the same equation together with vN4' it also follows that $U$ is trace-class. Indeed, for any orthonormal basis $(e_i)_{i=1}^\infty$ one finds
\begin{equation}
	\sum_{i=1}^\infty\langle e_i,U e_i\rangle=\sum_{i=1}^\infty\mathbb{E}(\mathcal{P}_{e_i})=\mathbb{E}(\mathpzc{1})<\infty.
\end{equation}
The proof of the converse statement is straightforward and is omited here.
\hfill $\square$\\[0.5ex]

From this theorem, the following corollaries are obtained.
\begin{gevolg}\label{homogeneousgevolg}
	If in addition to the conditions of Theorem \ref{spoorneumann} the ensemble $\mathbb{E}$ is pure and $U\neq0$, there is a unit vector $e$ and real number $\lambda$ such that $U=\lambda P_e$. Conversely, for each unit vector $e$, the expectation-value function $\mathbb{E}(\A)=\Trace(P_eA)=\langle e,Ae\rangle$ is pure.
\end{gevolg}

\noindent
\textit{Proof:}\hspace*{\fill}\\
To prove the first statement, let $\psi_0\in\h$ such that $U\psi_0\neq0$. Then define the operators
\begin{equation}
	U':\psi\mapsto\frac{\langle U\psi_0,\psi\rangle}{\langle\psi_0,U\psi_0\rangle}U\psi_0,\quad U'':\psi\mapsto U\psi-U'\psi.
\end{equation}
It follows from the self-adjointness of $U$ that these are both self-adjoint too. Moreover, they are positive\footnote{These inequalities follow from using the Cauchy-Schwartz inequality for the mapping $(\psi,\phi)\mapsto\langle\psi,U\phi\rangle$, which is an inner product because $U$ is positive. This also shows that $\langle\psi_0,U\psi_0\rangle>0$.}:
\begin{equation}
	\langle\psi,U'\psi\rangle=\frac{|\langle\psi,U\psi_0\rangle|^2}{\langle\psi_0,U\psi_0\rangle}\geq0,\quad
	\langle\psi,U''\psi\rangle
	=\frac{\langle\psi,U\psi\rangle\langle\psi_0,U\psi_0\rangle-|\langle\psi,U\psi_0\rangle|^2}{\langle\psi_0,U\psi_0\rangle}\geq0.
\end{equation}
Therefore, the expectation-value functions $\mathbb{E}'$ and $\mathbb{E}''$ associated with $U'$ and $U''$ satisfy $\mathbb{E}(\A)=\mathbb{E}'(\A)+\mathbb{E}''(\A)$. Then, because $\mathbb{E}$ is pure, it follows that there are $c',c''$ such that $U'=c'U$, $U''=c''U$ and $c',c''>0$. Because $U''\psi_0=0$, it follows that $c'=1$ and $c''=0$.
\indent

Now set $e:=\tfrac{1}{\|U\psi_0\|}U\psi_0$. For every $\psi\in\h$, it then holds that
\begin{equation}
	U\psi=U'\psi=\frac{\langle U\psi_0,\psi\rangle}{\langle\psi_0,U\psi_0\rangle}U\psi_0=
	\frac{\langle e,\psi\rangle \|U\psi_0\|^2}{\langle\psi_0,U\psi_0\rangle}e=
	\frac{\|U\psi_0\|^2}{\langle\psi_0,U\psi_0\rangle}P_{e}\psi.
\end{equation}

For the converse, let $U=P_e$ for some unit vector $e$ and let $\mathbb{E}$ denote the expectation-value function associated with $U$. Suppose $\mathbb{E}'$ and $\mathbb{E}''$ satisfy (\ref{homo1}), and let $U'$ and $U''$ be the positive semi-definite self-adjoint operators associated with these ensembles. It follows that $U=U'+U''$. Now let $\psi\in\h$ be any vector and set $\psi^{\parallel}=P_e\psi$ and $\psi^{\perp}=(\een-P_e)\psi$. Then 
\begin{equation}
	0\leq\langle\psi^{\perp},U'\psi^{\perp}\rangle\leq\langle\psi^{\perp},U'\psi^{\perp}\rangle+\langle\psi^{\perp},U''\psi^{\perp}\rangle
	=\langle\psi^{\perp},U\psi^{\perp}\rangle=0.
\end{equation}
Thus, $U'\psi^{\perp}=0$ and also $U''\psi^{\perp}=0$. Furthermore,
\begin{equation}
	\langle(\een-P_e)U'\psi,(\een-P_e)U'\psi\rangle=\langle(\een-P_e)U'\psi,U'\psi\rangle=\langle U'(\een-P_e)U'\psi,\psi\rangle
	=\langle0,\psi\rangle=0.
\end{equation}
That is, $U'\psi\in P_e\h$ $\forall\psi\in\h$ (and similarly for $U''$). Now set $c'=\langle e,U'e\rangle$. Then $U'e=c'e$ and
\begin{equation}
	U'\psi=U'\psi^{\parallel}=\langle e,\psi\rangle c'e=c'U\psi,\quad\forall\psi\in\h.
\end{equation}
In the same way set $c''=\langle e,U''e\rangle$, and it follows that $c'+c''=1$. This shows that $U$ is pure.
\hfill $\square$\\[0.5ex]

In other words, this corollary states that the only possible pure states, as defined in Definition \ref{puur}, are in fact the ones already given by quantum mechanics. The other corollary is the following.

\begin{gevolg}
	If the axioms vN1 and vN3' are satisfied, there are no normalisable dispersion-free expectation-value functions that satisfy vN2 and vN4'. 
\end{gevolg}

\noindent
\textit{Proof:}\hspace*{\fill}\\
Let $\mathbb{E}$ be an expectation-value function that satisfies vN2 and vN4' and let $U$ be the self-adjoint operator defined by this function according to the previous theorem. Let $e$ be any unit vector. Because $\mathbb{E}$ is dispersion free, and because of vN1 one has  
\begin{equation}
	\Trace(UP_e)^2=\mathbb{E}(\mathcal{P}_e)^2=\mathbb{E}(\mathcal{P}_e^2)=\Trace(UP_e^2)=\Trace(UP_e).
\end{equation}
Since $\mathbb{E}(\mathcal{P}_e)\leq\mathbb{E}(1)<\infty$, it follows that $\Trace(UP_e)\in\{0,1\}$ for all $e$. Then, since $e\mapsto\Trace(UP_e)$ is a continuous function on the unit vectors, it must be constant. Consequently, either $U=\nul$ or $U=\een$ (where $\nul$ denotes the zero operator and $\een$ the unit operator). But if $U=\nul$ one has $\mathbb{E}(\A)=\Trace(\nul\cdot A)=0$ for every observable $\A$ and the function is not normalisable\footnote{This is also physically unacceptable. Using the words of von Neumann: ``$U=\nul$ furnishes no information.''}. On the other hand, if $U=\een$, for each pair of orthonormal vectors $e_1,e_2$ the operator $P_{e_1}+P_{e_2}$ is again a projection, and
\begin{equation}
\begin{split}
	2&=\Trace(U(P_{e_1}+P_{e_2}))=\Trace(U(P_{e_1}+P_{e_2})^2)\\
	&=\mathbb{E}((\mathcal{P}_{e_1}+\mathcal{P}_{e_2})^2)=\mathbb{E}(\mathcal{P}_{e_1}+\mathcal{P}_{e_2})^2
	=\Trace(U(P_{e_1}+P_{e_2}))^2=4,
\end{split}
\end{equation}
which is again a contradiction. Therefore, there are no normalisable dispersion-free ensembles.
\hfill $\square$\\[0.5ex]
\indent

The conclusion drawn by von Neumann is the following. Since there are no states that are dispersion free, there are no hidden-variable states that can counter the indeterminism of quantum mechanics. In fact, since all pure states are given by unit vectors in the Hilbert space, no extension of quantum mechanics is possible. In his own words:

\begin{quote}
``There would still be the question [\ldots] as to whether the dispersions of the homogeneous ensembles characterized by the wave functions [\ldots] are not due to the fact that these are not the real states, but only mixtures of several states [\ldots] which together would determine everything causally, i.e., lead to dispersion free ensembles. The statistics of the homogeneous ensemble [the ones given by the unit vectors] would then have resulted from the averaging over that region of values of the ``hidden parameters'' which is involved in those states. But this is impossible for two reasons: First, because then the homogeneous ensemble in question could be represented as a mixture of two different ensembles, contrary to its definition. Second, because the dispersion free ensembles, which would have to correspond to the ``actual'' states [\ldots], do not exist. It should be noted that we need not go any further into the mechanism of the ``hidden parameters,'' since we now know that the established results of quantum mechanics can never be re-derived with their help.'' \cite{Neumann55}
\end{quote}

So, if I understand von Neumann correctly, any search for a hidden-variable model will be doomed to fail. However, the general rule that the more complex an argument becomes, the more likely it will be that it is flawed, turns out to apply once again.

\subsubsection{A Counterexample}\label{catalogtheory}
Despite von Neumann's proof it turns out to be easy to show that hidden-variable theories that reproduce the quantum-mechanical statistics do exist. The simplest way of accomplishing this will be called the catalog theory (for reasons made clear later on). Let $\mathcal{O}$ denote the set of all self-adjoint operators on the Hilbert space $\h$ associated with the system in question.
Since quantum mechanics has proven to be very accurate in describing phenomena, 
and because a theory is desired that \textit{completes} quantum mechanics and doesn't \textit{replace} it, 
take $\mathbb{V}_{\A}=\sigma(A)$ for the set of possible values of $\A$, where $A$ is the self-adjoint operator associated with the 
observable $\A$ in quantum mechanics.
Now the set of hidden variables is defind to be
\begin{equation}\label{groteL}
	\Lambda_{cat}:=\{\lambda:\mathcal{O}\to\mathbb{R}\:;\:\lambda(A)\in\sigma(A)\:\forall A\in\mathcal{O}\}.
\end{equation}
A $\sigma$-algebra on this set can be formed in the following way. For each Borel subset $\Delta\subset\mathbb{R}$ and observable $\A$, let $[\A\in\Delta]$ denote the set of all states for which the value of $\A$ lies in $\Delta$:
\begin{equation}
	[\A\in\Delta]=\{\lambda\in\Lambda_{cat}\:;\:\lambda(A)\in\Delta\}.
\end{equation}
Then $\Sigma_{cat}$ will be the $\sigma$-algebra generated by all these sets. For each quantum-mechanical state $\psi$, define the probability measure\footnote{No distinction in notation is made between this measure and the probability function in (\ref{Born}).} $\pee_\psi$ to be
\begin{equation}
	\pee_\psi[\A\in\Delta]=\Trace(P_\psi\mu_A(\Delta)),
\end{equation}
which ensures that (\ref{vwwaarde}) holds.
It follows from Kolmogorov's extension theorem that this defines a probability measure on $(\Lambda_{cat},\Sigma_{cat})$ (I omit the details to keep the argument clear). It follows from von Neumann's proof that this probability measure is not dispersion free. However, it is a mixture of ensembles that \textit{are} dispersion free (each of the $\lambda$ may be associated with a dispersion free ensemble). And although it was shown that this ensemble is homogeneous, it still is a mixture of ensembles. So, accepting the validity of von Neumann's proof, there must be a conflict with his assumptions. 

In the proof of the second part of Corollary \ref{homogeneousgevolg} it was stated that if the ensemble given by $P_\psi$ can be written as the mixture of two ensembles, these ensembles can again be associated with a positive trace-class operator. However, this step uses Theorem \ref{spoorneumann}, which only holds if these ensembles satisfy vN2 and vN4'. And although vN2 is satisfied in this model (positive observables can indeed only have positive values), vN4' is not. In fact, if one considers an ensemble that picks out $\lambda$ almost surely, the axiom reads
\begin{equation}\label{vn4cat}
	\lambda(\A)=\mathbb{E}(\A)=c_1\mathbb{E}(\A_1)+c_2\mathbb{E}(\A_2)+\ldots=c_1\lambda(\A_1)+c_2\lambda(\A_2)+\ldots,
\end{equation}
whenever $A=c_1A_1+c_2A_2+\ldots$.

Obviously, this property doesn't hold for any of the $\lambda\in\Lambda_{cat}$. It is, in a certain sense, the \textit{absence} of such a property that makes the theory $(\Lambda_{cat},\Sigma_{cat})$ meaningless. It is indeed just a big catalog of all possible measurement results, and a state is only determined if one measures \textit{all} observables. If this is done, then one can look in the catalog to find what the state is. There are no laws in this theory. That is, having measured some observables, together with this theory, one cannot make a prediction about measurement results of any other observable that is any more precise then the prediction that can be made in case the earlier measurements hadn't been performed. For example, measuring the momentum $p$ of a particle does not help to predict its kinetic energy $p^2/2m$. Thus such a theory is completely meaningless, for there is no causal structure. We want a theory that at least says that a ball will move if we kick it. However, the importance of the catalog theory lies in the fact that it exposes the assumption (\ref{vn4cat}) made by von Neumann as a rather dubious one.

The realization that the proof of von Neumann doesn't exclude all possible hidden-variable theories has resulted in heavy critique on the theorem (see for example \cite{Mermin93}). I think this is unreasonable. As a proof for the non-existence of hidden-variable theories it may perhaps depend on unnecessarily strong assumptions, but as an investigation of the possible existence of hidden-variable theories it remains valuable.  

How strong is, in fact, assumption vN4'? It is a natural assumption that any hidden-variable theory should be empirically equivalent to quantum mechanics. Now the two postulates of quantum mechanics that relate the mathematical structure to empirical statements are the value postulate and the Born postulate. In a hidden-variable theory the first could easily be accounted for by demanding that $\mathbb{V}_{\A}=\sigma(A)$ for all observables. The second one is harder to pinpoint. The Born postulate does in fact imply that for any three observables $\A,\A_1,\A_2$ corresponding to self-adjoint operators $A,A_1,A_2$ such that $A=c_1 A_1+c_2 A_2$ one has 
\begin{equation}\label{Neumannannnameklein}
	\mathbb{E}(\A)=c_1\mathbb{E}(\A_1)+c_2\mathbb{E}(\A_2)
\end{equation}
for every empirically admissible ensemble (i.e., for every quantum-mechanical ensemble). Is it then necessary that this relation should also hold for all sub-ensembles? In other words, is a violation of $\A=c_1\A_1+c_2\A_2$ in a sub-ensemble susceptible to empirical investigation?

To test the relation $A=c_1 A_1+c_2 A_2$ empirically, one simply measures the three observables $\A$, $\A_1$ and $\A_2$ and checks if the relation holds. If the operators $A_1$ and $A_2$ commute\footnote{In that case they automatically also commute with $A$.}, then according to (property (iv) of) Theorem \ref{spectraalstelling}, all the projection operators they generate by their spectral measure also commute. 
Together with the von Neumann postulate, this implies that a measurement of one of the observables alters the probability distribution over the possible outcomes of the other observables in such a way that the relation $A=c_1 A_1+c_2 A_2$ will be satisfied. 
The procedure to test the relation $A=c_1 A_1+c_2 A_2$ described above is therefore a meaningful one, and it seems a reasonable assumption that equation (\ref{Neumannannnameklein}) should hold for all sub-ensembles. 
However, in the case that $A_1$ and $A_2$ do not commute, a measurement of one observable in general does not alter the probability distribution over the possible outcomes in such a way that $A=c_1 A_1+c_2 A_2$ is satisfied. In that case, a measurement of $\A_1$ and $\A_2$ may best be performed by again splitting a set of systems in the same state in two parts and measuring $\A_1$ on one part and $\A_2$ on the second. But this is not expressed by the relation $\A=c_1\A_1+c_2\A_2$. In fact, Bell \cite{Bell66} pointed out that if one considers the observables associated with the operators $\sigma_z$, $\sigma_y$ and $\sigma_r$ from example \ref{EPRB} with $r=\frac{1}{2}\sqrt{2}(1,0,1)$, one has
\begin{equation}
	\sigma_r=\frac{1}{2}\sqrt{2}\left(\sigma_z+\sigma_y\right).
\end{equation}
However, this relation can never be satisfied empirically if each of the observables is only allowed to take the values -1 or 1 upon measurement (i.e., eigenvalues are not linear for non-commuting operators\footnote{Already in 1935 a related objection was made by Grete Hermann \cite{Hermann35} which, however, remained ignored for many years. The interested reader may consult \cite{Seevinck04} and \cite{Herzenberg08} for more information.}). 

The conclusion is now drawn that (\ref{Neumannannnameklein}) may only seem a reasonable demand if the corresponding self-adjoint operators commute. The arguments used to come to this conclusion appear to me to be reasonably involved and I therefore disagree with Bell and Mermin who stated that assumption vN4' (and even von Neumann's proof in general) is ``silly''.\footnote{In \cite{Mermin93}, Mermin uses this term and defends it referring to a quote from Bell in an interview. Ever since, vN4' has become known as von Neumann's ``silly assumption''.} Furthermore, the question is left open of what \textit{would} be a reasonable relation between observables for which the corresponding self-adjoint operators do \textit{not} commute. Certainly, the classical relation between energy, momentum and position,
\begin{equation}
	H=\frac{p^2}{2m}+V(x),
\end{equation} 
should have some meaning in a hidden-variable theory, especially since it plays an important role in quantum physics (as an operator equation). Should the assumption that, in a hidden-variable theory, (\ref{Neumannannnameklein}) only holds for quantum-mechanical ensembles (i.e. those described by a positive trace-class operator) really be the remnant of this (once so fundamental) equation?

At this point, it is not at all clear that a weakening of vN4', e.g., by stating that it only has to hold for observables whose corresponding operators commute, enables one to find a possible hidden-variable theory that \textit{is} satisfactory. The catalog theory presented above is of course unsatisfactory, if only because it is empirically in violation with quantum mechanics (e.g. most states will violate a relation like $\A=c_1\A_1+c_2\A_2$ even if the corresponding operators commute). The catalog theory may only be saved by introducing an ad hoc state change upon the performance of a measurement to ensure the preservation of physical laws. Bell showed in \cite{Bell66} that a hidden-variable model that is completely consistent with quantum mechanics is in fact possible for the system of a single spin $\tfrac{1}{2}$-particle; in his model equation (\ref{Neumannannnameklein}) is satisfied for all commuting observables in all states. In that case, the dimension of the Hilbert space is 2. The question then arises if such a model can be extended to cover systems described by Hilbert spaces of arbitrary dimension. The next theorem will show that this cannot be done.

\subsubsection{The Kochen-Specker Theorem}\label{Kochen-Specker-sectie}

For a hidden-variable theory, it is demanded that there exists a set $\Lambda$ consisting of all pure states. Appealing to the completeness condition of Einstein, Podolsky and Rosen, it is demanded that for each observable $\A$, there must be a map $f_{\A}:\Lambda\to\mathbb{V}_{\A}$, assigning to each state the value of the observable $\A$ in that state. To obtain empirical equivalence with quantum mechanics, the set of possible values for $\A$ is taken to be $\mathbb{V}_{\A}=\sigma(A)$, where $\sigma(A)$ is the spectrum of the self-adjoint operator $A$ associated to the observable $\A$ in quantum mechanics. 

Thus far, the set $\Lambda$ resembles the state space of the catalog theory in the sense that it is completely lawless. From the proof by von Neumann it follows that demanding 
\begin{equation}\label{lineairfunc}
	\lambda(\A)=c_1\lambda(\A_1)+c_2\lambda(\A_2),\quad\forall\lambda\text{ whenever }A=c_1A_1+c_2A_2,
\end{equation}
is too strict, and in fact not even necessary to obtain empirical equivalence when $A_1$ and $A_2$ do not commute. 

It seems good to point out again that von Neumann never explicitly required (\ref{lineairfunc}) to hold for the hidden-variable states. In fact, von Neumann explicitly appealed only to the statistical form of this law (i.e., requiring it only to hold for the expectation values), as explained earlier. Because it follows from the Born postulate, this is in fact a necessary requirement for the quantum-mechanical ensembles. In my opinion, the only flaw in the reasoning of von Neumann is to be found in the proof of the homogeneousity of the quantum-mechanical ensembles described by the unit vectors in the Hilbert space (Corollary \ref{homogeneousgevolg}). It was here that the unreasonable assumption was made that \textit{if} the quantum-mechanical ensemble was a mixture of two ensembles, \textit{then} these sub-ensembles should also satisfy vN4'.

In contrast with von Neumann, Kochen and Specker do explicitly work with pure states $\lambda$. Of course, linear laws like (\ref{lineairfunc}) (for commuting observables) are not the only ones to be satisfied in the hidden-variable model. For any Borel function $f$ and every observable $\A$ one can introduce the observable $f(\A)$ which coincides with applying the function $f$ to the measurement result of $\A$. Kochen and Specker made the following assumption in \cite{KS67}:
\begin{quote}
	\textbf{FUNC':} If the observable $\A$ is associated with the self-adjoint operator $A$, then the observable $f(\A)$ is associated with the self-adjoint operator $f(A)$ (if it exists).
\end{quote}
Note that this is a generalization of vN1.
In the precise form given here, this assumption is somewhat implicit in \cite{KS67}. There, it automatically results since there is no distinction in notation between observables and self-adjoint operators. The definition of the operator $f(A)$ is then given by the Borel functional calculus.\footnote{See for example \cite{Conway90}.} It follows that $A$ and $f(A)$ commute, and hence their corresponding observables can be measured simultaneously (their measurement results are always related by $f$, no matter in what order they are measured). This motivates the following assumption.
\begin{quote}
	\textbf{FUNC:} For each observable $\A$ and each Borel function $f$, any hidden-variable state $\lambda$ satisfies
	\begin{equation}
		\lambda(f(\A))=f(\lambda(\A)),
	\end{equation}
	where $f(\A)$ is defined as stated above.
\end{quote}
Note that vN2 is a special case of this assumption. From FUNC and FUNC' together it also follows that (\ref{lineairfunc}) holds whenever $A_1$ and $A_2$ commute.
Now, if for a certain physical system $\mathpzc{Obs}$ denotes the set of observables\footnote{Each observable may be associated with a self-adjoint operator according to the observable postulate. The set $\mathpzc{Obs}$ can thus be viewed as a subset of all self-adjoint operators.}, the set of hidden variables that satisfy the criteria imposed by Kochen and Specker is given by
\begin{equation}
	\Lambda_{KS}:=\{\lambda:\mathpzc{Obs}\to\mathbb{R}\:;\:\lambda(\A)\in\sigma(A), \lambda(f(\A))=f(\lambda(\A)),\forall\text{ Borel functions }f\}.
\end{equation} 
An element of this set is often also called a \underline{valuation function}. Using this definition, the theorem Kochen \& Specker can now be formulated in the following way:

\begin{stelling}[Kochen-Specker Theorem]
If the Hilbert space associated with the system has dimension greater than 2, then the set $\Lambda_{KS}$ is empty. 
\end{stelling}
The proof is actually fairly long and I'll try to present it here in the form of a story, discussing the technical details along the way. 
The story focuses on special observables only, namely those corresponding to projection operators $P$ in quantum mechanics, which are regarded as the ``yes-no''-questions (see Remark \ref{janee}). 
For such operators, the following lemma can be proven in a very direct way.\footnote{In their article \cite{KS67}, Kochen \& Specker refer to theorem 6 in \cite{Neumark54} to state that whenever two operators $A_1,A_2$ commute, there exists an operator $A$ and functions $f_1,f_2$ such that $f_i(A)=A_i$ for $i=1,2$. This result appears also in a lot of textbooks. See for example proposition 1.21 in \cite{Takesaki79-1}, which states that every Abelian von Neumann algebra is generated by a single self-adjoint operator. In a lot of literature, one simply refers to this as ``a well known mathematical fact'' without specifying exactly what this fact is (it took me quite some time to figure out what it was). To clarify the discussion, I have chosen to only prove the part that is necessary for this discussion.}

\begin{lemma}\label{KSlemma1}
For any $n$-tuple $P_1,\ldots,P_n$ of mutually orthogonal (and hence commuting) projection operators, there exists a self-adjoint operator $A_n$ and functions $f_{n,1},\ldots,f_{n,n}$ such that $P_i=f_{n,i}(A_n)$, for $i=1,\ldots,n$.
\end{lemma}  

\noindent
\textit{Proof:}\hspace*{\fill}\\
With induction in $n$. For $n=1$ the assertion is trivial, just take $A_1=P_1$. For arbitrary $n$, take 
\begin{equation}
A_n=\alpha_1P_1+\alpha_2P_{2}+\ldots+\alpha_nP_n,\quad\text{with }
\alpha_1=1,\quad\alpha_k=\frac{1}{2}\left(1-\sqrt{1-\alpha_{k-1}}\right).
\end{equation}
Consider the function $h:\mathbb{C}\to\mathbb{C}$, $h:x\mapsto4(x-x^2)$. One then has
\begin{equation}
\begin{split}
	h(A_n)=4(A_n-A_n^2)
	&=
	4(\alpha_1-\alpha_1^2)P_1+4(\alpha_2-\alpha_2^2)P_2+\ldots+4(\alpha_n-\alpha_n^2)P_n\\
	&=
	\alpha_1P_2+\alpha_2P_3+\ldots+\alpha_{n-1}P_n.
\end{split}
\end{equation}
Therefore, applying $h$ $n-1$ times to $A_n$ gives $h^{n-1}(A_n)=P_n$. Now suppose the lemma is true for certain $n$, for the case $n+1$, take $f_{n+1,n+1}=h^n$ (applying $h$ $n$ times) and for $k=1,\ldots,n$ define
\begin{equation}
\begin{gathered}
	f_{n+1,k}:\mathbb{C}\to\mathbb{C};\\
	f_{n+1,k}(x):=f_{n,k}(x-\alpha_{n+1}f_{n+1,n+1}(x)).
\end{gathered}
\end{equation}
This proves the lemma since
\begin{equation}
\begin{split}
	f_{n+1,k}(A_{n+1})&=f_{n,k}(A_{n+1}-\alpha_{n+1}f_{n+1,n+1}(A_{n+1}))\\
	&=f_{n,k}(A_{n+1}-\alpha_{n+1}P_{n+1})=f_{n,k}(A_n)=P_k.
\end{split}
\end{equation}
\hfill $\square$\\[0.5ex]
\indent

\begin{gevolg}[Finite Sum Rule]\label{eindigesomregel}
For any $n$-tuple $P_1,\ldots,P_n$ of mutually orthogonal projection operators, every $\lambda\in\Lambda_{KS}$ must satisfy
\begin{equation}\label{sumrule}
	\lambda(\mathcal{P}_1+\ldots+\mathcal{P}_n)=\lambda(\mathcal{P}_1)+\ldots+\lambda(\mathcal{P}_n),
\end{equation}
where $\mathcal{P}_i$ is the observable associated with the operator $P_i$.
\end{gevolg}
\noindent
\textit{Proof:}\hspace*{\fill}\\
Let $P_1,\ldots,P_n$ be an $n$-tuple of orthogonal projection operators. Take $A_n$ and $f_{n,1},\ldots,f_{n,n}$ as constructed in Lemma \ref{KSlemma1} and take $g=f_{n,1}+\ldots+f_{n,n}$. Let $\A_n$ denote the observable associated with the operator $A_n$. One then has for every $\lambda\in\Lambda_{KS}$ 
\begin{equation}
\begin{split}
	\lambda(\mathcal{P}_1+\ldots+\mathcal{P}_n)
	&=
	\lambda(g(\mathcal{A}_n))=g(\lambda(\A_n)\\
	&=
	f_{n,1}(\lambda(\A_n))+\ldots+f_{n,n}(\lambda(\A_n))=\lambda(f_{n,1}(\A_n))+\ldots+\lambda(f_{n,n}(\A_n))\\
	&=
	\lambda(\mathcal{P}_1)+\ldots+\lambda(\mathcal{P}_n).
\end{split}
\end{equation}
\hfill $\square$\\[0.5ex]
\indent

Here it shows that the apparently innocent FUNC rule has serious consequences. In fact, the finite sum rule enables one to prove the following lemma, which, together with its corollary, already almost proves the Kochen-Specker Theorem. The following lemma is about four-dimensional Hilbert spaces. Later on, I will present a similar result for the three-dimensional case (Lemma \ref{KS3dim}). Lemma \ref{KS4dim} may be seen as an appetizer for the one presented later.

\begin{lemma}\label{KS4dim}
If $\Lambda_{KS}$ is a set of hidden variables to describe a system associated with a four-dimensional Hilbert space, then $\Lambda_{KS}$ is empty.
\end{lemma} 

\noindent
\textit{Proof:}\hspace*{\fill}\\
For any 4-tuple of orthogonal 1-dimensional projection operators $P_1,P_2,P_3,P_4$, the finite sum rule implies that
\begin{equation}\label{Cabellosumrule}
	\lambda(\mathcal{P}_1)+\lambda(\mathcal{P}_2)+\lambda(\mathcal{P}_3)+\lambda(\mathcal{P}_4)=
	\lambda(\mathcal{P}_1+\mathcal{P}_2+\mathcal{P}_3+\mathcal{P}_4)=\lambda(\mathpzc{1})=1,\quad\forall\lambda\in\Lambda_{KS},
\end{equation}
where $\mathpzc{1}$ denotes the observable associated with the unit operator $\een$. 
\indent

Now let $\lambda$ be any element of $\Lambda_{KS}$. 
By (\ref{Cabellosumrule}), for any 4-tuple of 1-dimensional projection operators $P_1,P_2,P_3,P_4$, the state $\lambda$ must assign the value 1 to exactly one of the corresponding observables, and the value 0 to all the others. 

\begin{table}[ht]\caption{The 18 vectors appearing in the proof of Lemma \ref{KS4dim}. Each vector appears exactly two times.}\label{cabellotabel}
\scriptsize
\begin{center}
\begin{tabular}{|c|c|c|c|c|c|c|c|c|}
\hline
	(0,0,0,1) &	(0,0,0,1) &	(1,-1,1,-1) &	(1,-1,1,-1) &	(0,0,1,0) &	(1,-1,-1,1) &	(1,1,-1,1) &	(1,1,-1,1) &	(1,1,1,-1) \\
 	(0,0,1,0) &	(0,1,0,0) &	(1,-1,-1,1) &	(1,1,1,1) &	(0,1,0,0) &	(1,1,1,1) &	(1,1,1,-1) &	(-1,1,1,1) &	(-1,1,1,1) \\
 	(1,1,0,0) &	(1,0,1,0) &	(1,1,0,0) &	(1,0,-1,0) &	(1,0,0,1) &	(1,0,0,-1) &	(1,-1,0,0) &	(1,0,1,0) &	(1,0,0,1) \\
 	(1,-1,0,0) &	(1,0,-1,0) &	(0,0,1,1) &	(0,1,0,-1) &	(1,0,0,-1) &	(0,1,-1,0) &	(0,0,1,1) &	(0,1,0,-1) &	(0,1,-1,0) \\
\hline			
\end{tabular}
\end{center}
\end{table}
\normalsize

Now consider the 18 vectors in table \ref{cabellotabel}. This table of vectors was first introduced by Cabello in \cite{Cabello96}. Each column of the table constitutes an orthogonal basis of the Hilbert space $\mathbb{C}^4$. With each of the vectors one can associate the observable that corresponds to the projection on the line spanned by that vector. Thus for every column in the table, the state $\lambda$ associates the value 1 to exactly one of the vectors in the column. In total, since there are nine columns, the value 1 would appear nine times. On the other hand, since there are 18 different vectors in total --and since once a vector is associated with the number 1 in one column, it must also be associated with the number 1 in the other column in which it appears-- the number 1 should appear an even number of times. Since 9 is odd, the state $\lambda$ cannot exist. 

In terms more common in this discussion, the state $\lambda$ would lead to an impossible coloring of the vectors in table \ref{cabellotabel} (e.g. associating 1 with the color black, and 0 with the color white).
\hfill $\square$\\[0.5ex]

\begin{gevolg}\label{KSeindig}
The set $\Lambda_{KS}$ is empty if the associated Hilbert $\h$ space has finite dimension $n\geq4$.
\end{gevolg}

\noindent
\textit{Proof:}\hspace*{\fill}\\
Suppose $P_1,\ldots,P_n$ is an $n$-tuple of 1-dimensional orthogonal projection operators on $\h$ and let $\lambda$ be any state in $\Lambda_{KS}$. By the finite sum rule, the state $\lambda$ must assign the value 1 to precisely one of the observables $\mathcal{P}_1,\ldots,\mathcal{P}_n$. Suppose $\lambda({P}_j)=1$ and let $\h'$ be any four-dimensional linear subspace of $\h$ containing $P_j\h$ as a subspace (i.e. $P_j\h\subset\h'\subset\h$). 
\indent

Let $P_{\h'}:\h\to\h$ denote the projection on the subspace $\h'$. Now for any observable $\A'$ associated with the operator $A'$ acting on the space $\h'$ (i.e., $\A'$ is an observable for the system associated with a four-dimensional Hilbert space), let $\A$ denote the observable (for the system associated with the Hilbert space $\h$) associated with the operator $A'P_{\h'}$. 

Now the state $\lambda$ generates a state $\lambda'$ in the set $\Lambda_{KS}'$ of hidden variables associated with $\h'$ according to
\begin{equation}
	\lambda'(\A'):=\lambda(\A),
\end{equation}
where $\A$ is the observable generated by $\A'$ as described above. It is easy to check that this indeed gives an element of $\Lambda_{KS}'$. In particular, for any 4-tuple of 1-dimensional orthogonal projection operators $P_1',P_2',P_3',P_4'$ (acting on $\h'$) one has 
\begin{equation}
	\lambda'(\mathcal{P}'_1)+\lambda'(\mathcal{P}'_2)+\lambda'(\mathcal{P}'_3)+\lambda'(\mathcal{P}'_4)
	=
	\lambda'(\mathcal{P}'_1+\mathcal{P}'_2+\mathcal{P}'_3+\mathcal{P}'_4)=\lambda(\mathcal{P}_{\h'})=1.
\end{equation}
This is sufficient to show that $\lambda'$ (and therefore $\lambda$) does not exist, using the proof of Lemma \ref{KS4dim}.
\hfill $\square$\\[0.5ex]

Thus far it is accomplished that a hidden-variable theory (that has the structure of the set $\Lambda_{KS}$), cannot be used to describe a system whose associated Hilbert space has finite dimension $n\geq4$. 
The generalization to separable Hilbert spaces is more technical and in fact, I haven't found any article (about the Kochen-Specker Theorem) that handles the complications involved in detail. 
This is a bit surprising, since the Kochen-Specker Theorem would loose some of its power if it would not apply to separable Hilbert spaces. One may, for example, argue that every physical system is in fact associated with a separable Hilbert space. 
Even the simple example of a spin-$\tfrac{1}{2}$ particle usually described by the Hilbert space $\mathbb{C}^2$, is in fact only correctly described by the Hilbert space $L^2(\mathbb{R}^3)\otimes\mathbb{C}^2$ (since one cannot exclude the possibility that the particle always has some freedom to move in space). From this point of view it may seem possible (though unlikely) that the extra structure provided by the infinite-dimensional Hilbert space does allow for the existence of hidden variables.     

The problems involved in the infinite-dimensional case become clear when looking at the proof of Corollary \ref{KSeindig}.  To construct an appropriate four-dimensional subspace, it is necessary that for each $\lambda$, one can choose a projection operator $P_\lambda$ for which $\lambda(\mathcal{P}_\lambda)=1$ and $P_\lambda\h$ has dimension at most 4. But although for every projection operator on the separable space $\h$ one can establish whether $\lambda(\mathcal{P})$ equals 1 or 0, it is not trivial to find a \textit{finite}-dimensional projection operator $P$ for which $\lambda(\mathcal{P})=1$. Fortunately, that such a projection can indeed be found is proved by the following lemma.

\begin{lemma}[Infinite Sum Rule]
For each sequence of 1-dimensional mutually orthogonal projection operators $P_1,P_2,\ldots$ with $\sum_{n=1}^\infty P_n=\een$ (strongly) and for each $\lambda\in\Lambda_{KS}$, there is exactly one $n$ such that $\lambda(\mathcal{P}_n)=1$.
\end{lemma}

\noindent
\textit{Proof:}\hspace*{\fill}\\
Consider the (bounded) self-adjoint operator $A:=\sum_{n=1}^\infty\frac{1}{n}P_n$ and define functions $(f_n)_{n=1}^\infty$ as follows
\begin{equation}
	f_n(x):=\begin{cases} 1,&\text{if }x=\frac{1}{n};\\0,&\text{else}.\end{cases}
\end{equation}
The Borel functional calculus then implies that $f_n(A)=P_n$ for each $n$. 
It then follows from the FUNC rule that
\begin{equation}
	\sum_{n=1}^\infty \lambda(\mathcal{P}_n)=\sum_{n=1}^\infty \lambda(f_n(A))=\sum_{n=1}^\infty f_n(\lambda(A))=1=\lambda(\mathpzc{1})=\lambda(\sum_{n=1}^\infty\mathcal{P}_n),
\end{equation}
where I used $\sum_{n=1}^\infty f_n(\lambda(A))=1$, since
\begin{equation}
	\lambda(\A)\in\sigma(A)=\{\tfrac{1}{n}\:;\:n\in\mathbb{N}\backslash\{0\}\}.
\end{equation}
Therefore, there is exactly one $n$ such that $\lambda(\mathcal{P}_n)=1$.
\hfill $\square$\\[0.5ex]
\indent

\begin{gevolg}\label{KSoneindig}
The set $\Lambda_{KS}$ is empty if the associated Hilbert $\h$ space is either separable, or has finite dimension $n\geq4$.
\end{gevolg}

So the only thing left to prove is that the set $\Lambda_{KS}$ is also empty in case $\mathrm{dim}\h=3$. As seen in the proof of Lemma \ref{KS4dim}, it is sufficient to show that there exists no map that assigns either the value 0 or 1 to each observable associated with a projection operator, such that for any triplet of mutually orthogonal 1-dimensional projections $P_1,P_2,P_3$ (often called a triad) precisely one is assigned the value 1. That is, only the following lemma has to be proven.

\begin{wraptable}[26]{I}{0.5\textwidth}\caption{The 33 vectors appearing in the proof of Lemma \ref{KS3dim}.}\label{Peresvectors}
\scriptsize
\begin{tabular}{|rrr|}
\hline
	$e_1=\begin{pmatrix}1\\0\\0\end{pmatrix}$,& 
	$e_2=\begin{pmatrix}0\\1\\0\end{pmatrix}$,& 
	$e_3=\begin{pmatrix}0\\0\\1\end{pmatrix}$ 
	\\
	\hline
	$f_1^1=\frac{1}{\sqrt{2}}\begin{pmatrix}0\\1\\1\end{pmatrix}$,&
	$f_2^1=\frac{1}{\sqrt{2}}\begin{pmatrix}1\\0\\1\end{pmatrix}$,&
	$f_3^1=\frac{1}{\sqrt{2}}\begin{pmatrix}1\\1\\0\end{pmatrix}$\\
	$f_1^2=\frac{1}{\sqrt{2}}\begin{pmatrix}0\\-1\\1\end{pmatrix}$,&
	$f_2^2=\frac{1}{\sqrt{2}}\begin{pmatrix}-1\\0\\1\end{pmatrix}$,& 
	$f_3^2=\frac{1}{\sqrt{2}}\begin{pmatrix}1\\-1\\0\end{pmatrix}$\\
	\hline
	$g_1^1=\frac{1}{3}\begin{pmatrix}0\\\sqrt{3}\\\sqrt{6}\end{pmatrix}$, &
	$g_2^1=\frac{1}{3}\begin{pmatrix}\sqrt{6}\\0\\\sqrt{3}\end{pmatrix}$, &
	$g_3^1=\frac{1}{3}\begin{pmatrix}\sqrt{3}\\\sqrt{6}\\0\end{pmatrix}$, \\
	$h_1^1=\frac{1}{2}\begin{pmatrix}\sqrt{2}\\-1\\1\end{pmatrix}$, &
	$h_2^1=\frac{1}{2}\begin{pmatrix}-1\\\sqrt{2}\\1\end{pmatrix}$, &
	$h_3^1=\frac{1}{2}\begin{pmatrix}-1\\1\\\sqrt{2}\end{pmatrix}$
	\\
	$g_1^2=\frac{1}{3}\begin{pmatrix}0\\\sqrt{6}\\-\sqrt{3}\end{pmatrix}$, &
	$g_2^2=\frac{1}{3}\begin{pmatrix}-\sqrt{3}\\0\\\sqrt{6}\end{pmatrix}$, &
	$g_3^2=\frac{1}{3}\begin{pmatrix}\sqrt{6}\\-\sqrt{3}\\0\end{pmatrix}$, \\
	$h_1^2=\frac{1}{2}\begin{pmatrix}\sqrt{2}\\1\\-1\end{pmatrix}$, &
	$h_2^2=\frac{1}{2}\begin{pmatrix}1\\\sqrt{2}\\-1\end{pmatrix}$, &
	$h_3^2=\frac{1}{2}\begin{pmatrix}1\\-1\\\sqrt{2}\end{pmatrix}$
	\\
	$g_1^3=\frac{1}{3}\begin{pmatrix}0\\\sqrt{6}\\\sqrt{3}\end{pmatrix}$, &
	$g_2^3=\frac{1}{3}\begin{pmatrix}\sqrt{3}\\0\\\sqrt{6}\end{pmatrix}$, &
	$g_3^3=\frac{1}{3}\begin{pmatrix}\sqrt{6}\\\sqrt{3}\\0\end{pmatrix}$, \\ 
	$h_1^3=\frac{1}{2}\begin{pmatrix}\sqrt{2}\\-1\\-1\end{pmatrix}$, &
	$h_2^3=\frac{1}{2}\begin{pmatrix}-1\\\sqrt{2}\\-1\end{pmatrix}$, &
	$h_3^3=\frac{1}{2}\begin{pmatrix}-1\\-1\\\sqrt{2}\end{pmatrix}$
	\\
	$g_1^4=\frac{1}{3}\begin{pmatrix}0\\-\sqrt{3}\\\sqrt{6}\end{pmatrix}$, &
	$g_2^4=\frac{1}{3}\begin{pmatrix}\sqrt{6}\\0\\-\sqrt{3}\end{pmatrix}$, &
	$g_3^4=\frac{1}{3}\begin{pmatrix}-\sqrt{3}\\\sqrt{6}\\0\end{pmatrix}$, \\
	$h_1^4=\frac{1}{2}\begin{pmatrix}\sqrt{2}\\1\\1\end{pmatrix}$, &
	$h_2^4=\frac{1}{2}\begin{pmatrix}1\\\sqrt{2}\\1\end{pmatrix}$, &
	$h_3^4=\frac{1}{2}\begin{pmatrix}1\\1\\\sqrt{2}\end{pmatrix}$
	\\
	\hline
\end{tabular} 
\end{wraptable}
\normalsize

\begin{lemma}\label{KS3dim}
Let $\mathcal{P}_1(\mathbb{C}^3)$ denote the set of all 1-dimensional projection operators on $\mathbb{C}^3$. There exists no map $v:\mathcal{P}_1(\mathbb{C}^3)\to\{0,1\}$ such that for each triad $P_1,P_2,P_3$, there is precisely one $j\in\{1,2,3\}$ with $v(P_j)=1$.
\end{lemma}

\noindent
\textit{Proof:}\hspace*{\fill}\\
For the proof a specific finite subset of $\mathcal{P}_1(\mathbb{C}^3)$ is considered, and it is shown that no map satisfying the desired properties on this domain can exist. In the original proof \cite{KS67}, Kochen and Specker used a subset consisting of 117 elements. The proof discussed here is based on a modification due to Peres \cite{Peres90}. It starts with 33 projections, taken to be the projections on the lines spanned by the unit vectors in table \ref{Peresvectors}.
\indent

With these vectors, 16 triads in total can be constituted. In each triad, precisely one of the vectors must be assigned the value one. Besides these triads, there are 24 pairs of orthogonal vectors. For a pair of orthogonal vectors $v_1,v_2$ consider the following modification of the sum rule:
\begin{multline}\label{suminequality}
	v(\mathcal{P}_{v_1})+v(\mathcal{P}_{v_2})\\
	\leq v(\mathcal{P}_{v_1})+v(\mathcal{P}_{v_2})+v(\mathcal{P}_{v_1\times v_2})\\
	=1,
\end{multline}
where $\times$ denotes the exterior product. This implies that whenever one of the observables $\mathcal{P}_{v_1}$, $\mathcal{P}_{v_2}$ is assigned the value 1 by $v$, the other must be assigned the value 0. The 16 triads, together with the 24 pairs of orthogonal vectors\footnote{In \cite{PMMM05} it is argued (among other things) that the 24 extra vectors needed to accomplish (\ref{suminequality}) are essential to make the proof constructive. This would imply that the finite subset of $\mathcal{P}_1(\mathbb{C}^3)$ for which the map $v$ doesn't exist should actually consists of 57 elements instead of 33. An interesting consequence is that, from this point of view, the leading score in the competition of finding the smallest set for which $v$ does not exist, is no longer held by Conway and Kochen \cite[p. 114]{Peres02} (requiring a set of 31/51 elements) but by Bub \cite{Bub96} (requiring a set of 33/49 elements).} are listed in table \ref{Perestriads}.

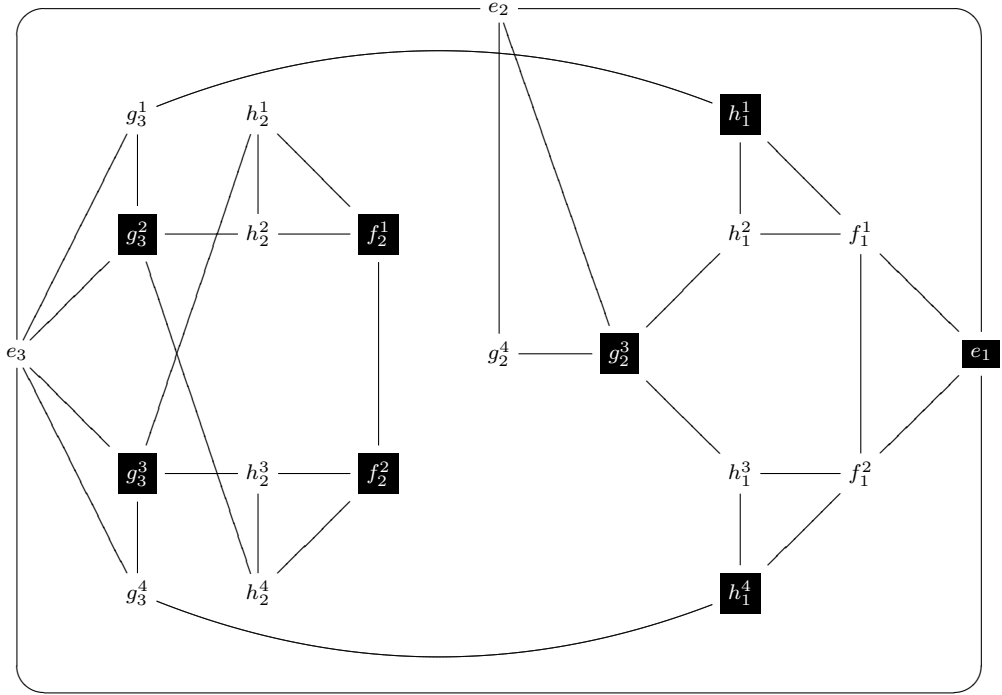
\begin{figure}[t]
	\scriptsize
	\centerline{
	\xymatrix@!C{
		&&&& e_2\ar@{-}[ddd]\ar@{-}[dddr] \ar @{-} `[rrrrddd] [rrrrddd]
	\ar @{-} `l[dddllll] [dddllll]
	&&&& \\
		&g^1_3 \ar@{-}[d]\ar@{-} @/^2pc/ [rrrrr] & h^1_2\ar@{-}[d]\ar@{-}[dr]
	&&&& \colorbox{black}{\textcolor{white}{$h^1_1$}}\ar@{-}[dr]\ar@{-}[d]	&&\\
		& \colorbox{black}{\textcolor{white}{$g^2_3$}}\ar@{-}[r]\ar@{-}[dddr] & h^2_2\ar@{-}[r]
	&\colorbox{black}{\textcolor{white}{$f^1_2$}}\ar@{-}[dd]
	&&& h^2_1\ar@{-}[r] & f^1_1\ar@{-}[dr]\ar@{-}[dd] &\\
		e_3  \ar @{-}`d[dddrrrrrrrr]`[rrrrrrrr][rrrrrrrr] \ar@{-}[uur]\ar@{-}[ur]\ar@{-}[dr]\ar@{-}[ddr]
	&&&& g^4_2\ar@{-}[r] & \colorbox{black}{\textcolor{white}{$g^3_2$}}\ar@{-}[ur]\ar@{-}[dr]
	&&& \colorbox{black}{\textcolor{white}{$e_1$}} \\
		& \colorbox{black}{\textcolor{white}{$g^3_3$}}\ar@{-}[r]\ar@{-}[uuur] & h^3_2\ar@{-}[r]
	&\colorbox{black}{\textcolor{white}{$f^2_2$}}
	&&& h^3_1\ar@{-}[r] & f^2_1\ar@{-}[ur] &\\
		&g^4_3 \ar@{-}[u]\ar @/_2pc/ @{-}[rrrrr] & h^4_2\ar@{-}[u]\ar@{-}[ur]
	&&&& \colorbox{black}{\textcolor{white}{$h^4_1$}}\ar@{-}[ur]\ar@{-}[u]	&&\\
	&&&&&&&&
	}}
	\caption{A schematic view of the consequences for assigning the value 1 to the vectors $e_1$ and $g^3_2$. A black marking depicts the assignment of the value 1 to the vector, whilst an absent marking depicts the assignment of the value 0. A line between two vectors denotes that these two vectors are orthogonal (only the lines used in the proof are drawn).}\label{kleuring1}
\end{figure}
\normalsize

\begin{wraptable}[17]{I}{0.5\textwidth}\caption{The 16 triads constituted with the vectors of table \ref{Peresvectors}, together with the remaining 24 orthogonal vector pairs.}\label{Perestriads}
\scriptsize
\renewcommand{\arraystretch}{1.4}
\begin{center}
\begin{tabular}{|c|c|c|c|}
\hline
$e_1,e_2,e_3$ & $e_1,f^1_1,f^2_1$ & $e_2,f^1_2,f^2_2$ & $e_3,f^1_3,f^2_3$ \\
\hline
 & $e_1,g^1_1,g^2_1$ & $e_2,g^1_2,g^2_2$ & $e_3,g^1_3,g^2_3$ \\
\hline
 & $e_1,g^3_1,g^4_1$ & $e_2,g^3_2,g^4_2$ & $e_3,g^3_3,g^4_3$ \\
\hline
 & $f^1_1,h^1_1,h^2_1$ & $f^1_2,h^1_2,h^2_2$ & $f^1_3,h^1_3,h^2_3$ \\
\hline
 & $f^1_1,h^3_1,h^4_1$ & $f^1_2,h^3_2,h^4_2$ & $f^1_3,h^3_3,h^4_3$ \\
\hline
\hline
$g_1^1, h_2^2$ & $g_1^2, h_3^1$ & $g_1^3, h_3^2$ & $g_1^4, h_2^1$\\
\hline
$g_1^1, h_2^3$ & $g_1^2, h_3^4$ & $g_1^3, h_3^3$ & $g_1^4, h_2^4$\\
\hline
$g_2^1, h_3^1$ & $g_2^2, h_1^1$ & $g_2^3, h_1^2$ & $g_2^4, h_3^2$\\
\hline
$g_2^1, h_3^3$ & $g_2^2, h_1^4$ & $g_2^3, h_1^3$ & $g_2^4, h_3^4$\\
\hline
$g_3^1, h_1^1$ & $g_3^2, h_2^2$ & $g_3^3, h_2^1$ & $g_3^4, h_1^2$\\
\hline
$g_3^1, h_1^3$ & $g_3^2, h_2^4$ & $g_3^3, h_2^3$ & $g_3^4, h_1^4$\\
\hline
\end{tabular}
\end{center}
\end{wraptable}
\normalsize

Because the set of vectors from table \ref{Peresvectors} is invariant under permutations of the $x$, $y$ and $z$ axes, any of the vectors $e_1,e_2,e_2$ may be chosen to be assigned the value 1 without loss of generality. Let $e_1$ be assigned the value 1 (for any $v$ one may label the axes such that $v(\mathcal{P}_{e_1})=1$). 

As a consequence, $e_2$ will be assigned the value 0. Therefore, from each of the pairs $g_2^1,g_2^2$ and $g_2^3,g_2^4$ exactly one must be assigned the value 1. I will first show that assigning the value 1 to $g_2^3$ leads to a contradiction. The line of reasoning is also depicted in figure \ref{kleuring1}.

When $g_2^3$ is assigned the value 1, it immediately follows that $h_1^2$ and $h_1^3$ should be assigned the value 0. Also, the vectors $f^1_1$ and $f^2_1$ must be assigned the value 0, since they form a triad together with $e_1$. Consequently, from the triads $h^1_1,h^2_1,f^1_1$ and $h^3_1,h^4_1,f^2_1$, the vectors $h^1_1$ and $h^4_1$ must be assigned the value 1.  This results in $g_3^1$ and $g_3^4$ being assigned the value 0.
As a next step, note that $g_3^2$ and $g_3^3$ must be assigned the value 1 (because they appear in the triads $e_3,g_3^1,g_3^2$ and $e_3,g_3^3,g_3^4$). This implies that $h_2^1,h_2^2,h_2^3,h_2^4$ are assigned the value 0. One then infers that both $f_2^1$ and $f_2^2$ must be assigned the value 1. But this is a contradiction, since $f_2^1$ and $f_2^2$ are orthogonal. Thus, $g_2^3$ cannot be assigned the value 1.

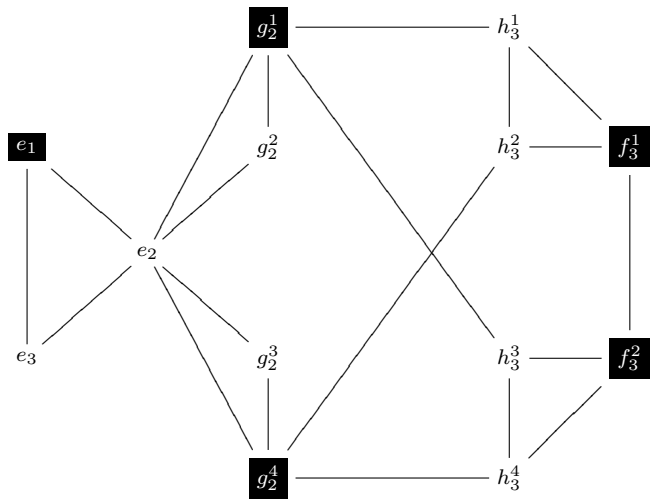
\begin{figure}[hb]
	\scriptsize
	\centerline{
	\xymatrix@!C{
		&& \colorbox{black}{\textcolor{white}{$g_2^1$}} \ar@{-}[rr]\ar@{-}[dddrr]\ar@{-}[d] 
	&& h_3^1 \ar@{-}[dr]\ar@{-}[d] 
	& \\
		\colorbox{black}{\textcolor{white}{$e_1$}} \ar@{-}[dr]\ar@{-}[dd] 
	&& g_2^2 & & h_3^2 \ar@{-}[r]        
	& \colorbox{black}{\textcolor{white}{$f_3^1$}} \ar@{-}[dd] \\
		& e_2 \ar@{-}[uur]\ar@{-}[ur]\ar@{-}[dr]\ar@{-}[ddr] &       
	&&& \\
		e_3 \ar@{-}[ur] 
	&& g_2^3\ar@{-}[d] 
	&& h_3^3 \ar@{-}[r] & \colorbox{black}{\textcolor{white}{$f_3^2$}} \\
		&& \colorbox{black}{\textcolor{white}{$g_2^4$}} \ar@{-}[rr]\ar@{-}[uuurr] 
	&& h_3^4 \ar@{-}[ur]\ar@{-}[u] &  
	}}
	\caption{A schematic view of the consequences for assigning the value 1 to the vectors $e_1$, $g^1_2$ and $g^4_2$. The meaning of the markings and lines is the same as in figure \ref{kleuring1}.}\label{kleuring2}
\end{figure}

Because of the symmetry around the $x$-axis, the same line of reasoning can be used to show that $g_2^2$ cannot be assigned the value 1. The only option left is to assign the value 1 to both $g_2^1$ and $g_2^4$. This will also lead to a contradiction. Indeed, in this case it follows that $h_3^1,h_3^2,h_3^3$ and $h_3^4$ must be assigned the value 0 (see figure \ref{kleuring2}). Consequently, both $f_3^1$ and $f_3^2$ must be assigned the value 1 which is in violation with (\ref{suminequality}). 

This completes the proof.\hfill $\square$\\[0.5ex]

\normalsize

So Lemma \ref{KS3dim} and Corollary \ref{KSoneindig} together constitute a proof of the Kochen-Specker Theorem. Lemma \ref{KS3dim} is often seen as the key element of the proof, named a \underline{coloring theorem}. A set for which a function $v$ does not exist is called not KS-colorable (or simply not colorable). 

As with von Neumann's theorem, the Kochen-Specker Theorem has also attracted some criticism (see for example\footnote{The criticism is in fact much more moderate than the criticism von Neumann attracted. Note that the article by Bell dates from before the Kochen-Specker Theorem. In fact, in this article Bell criticizes a similar result proven by him in the same article. The theorem is therefore also often called the Bell-Kochen-Specker Theorem.} \cite{Bell66},\cite{HeywoodRedhead83},\cite{Mermin93}). The arguments used however, are more involved. 

A technical objection may be raised against the FUNC assumption. In itself, this seems a very natural assumption to construct a theory that is not lawless, but it relies on the more complex assumption FUNC' together with some mathematical considerations. In the case of the finite sum rule, it was shown that the technicalities could be dealt with easily. Indeed, for a simple polynomial function $f$, the observable $f(\A)$ is quite naturally associated with the operator $f(A)$ since $f(A)$ has a clear explicit definition. But in the case of the infinite sum rule, the assumption FUNC' appears to me to be more artificial. 

Not every text that presents the Kochen-Specker Theorem uses the FUNC rule, so objections related to precisely this rule aren't very strong. For example, in \cite{Cabello96} one simply starts with the assumption of the infinite sum rule. I will now present a possible (though not very strong) direct motivation for this rule.\footnote{A more extensive investigation of possible starting assumptions and how they are related can be found in \cite{Fine78}.}

Consider an observable $\A$ corresponding to the operator $A$. A measurement of $\A$ can be interpreted as asking the system the question: ``What is the value of $\A$?'' This formulation is a bit sloppy, since the discussion is exactly about whether or not a value can be assigned to the system at all times. One may, instead, read this question as ``What is the value of $\A$ after the measurement?'', at least, in case one accepts that ``the value of $\A$'' has a meaning if the system is in an eigenstate of the operator $A$. Otherwise, one may read this question as ``What will be the value yielded when I perform a measurement of $\A$?'' or, if one rejects the possibility of infinite precise measurements: ``What subset of $\sigma(A)$ will emerge when I perform a measurement that I associate with the measurement of $\A$?'' Whatever question the reader may prefer, let's agree to abbreviate it with ``What is the value of $\A$?'' 

Conversely, it seems reasonable that similar, physical meaningful, questions could be associated with operators. Suppose $\Delta\subset\sigma(A)$. One may ask the question ``Does the value of $\A$ lie in $\Delta$?'' In fact, one can associate an operator with this question, namely the projection operator $\mu_A(\Delta)$ from Theorem \ref{spectraalstelling}. Now let $(\Delta_i)_{i\in I}$ be any partition of $\sigma(A)$. A measurement of $\A$ may then be interpreted as simultaneously asking all the yes-no questions\footnote{See also Remark \ref{janee}.} ``Does the value of $\A$ lie in $\Delta_i$?'' for $i\in I$. Since a measurement will yield only one result, precisely one of the questions must be answered yes, and all the others must be answered no. This motivates the (infinite) sum rule.

A possible objection against this line of reasoning is that it is in general not clear whether or not a sequence of mutually orthogonal projection operators $(P_i)_{i\in I}$ can be associated with a single observable such that $P_i=\mu_A(\Delta_i)$ for all $i$ for some partition $(\Delta_i)_{i\in I}$ of $\sigma(A)$. This exposes a second problem: Why should the self-adjoint operators mentioned in the proof indeed correspond to observables\footnote{In the von Neumann proof, it was simply postulated (though implicitly) that the converse of the observable postulate should hold. One may, of course, postulate the same for the Kochen-Specker Theorem, but that seems unreasonable to me (see earlier considerations when introducing the observable postulate on page \pageref{operatorpost}).}? 

A short investigation of this problem will reveal yet another objection when considering the proof of Lemma \ref{KS4dim}. In this setting, for each column of table \ref{cabellotabel} one would have to find an observable $\A_i$ and a partition $(\Delta_i^k)_{k=1}^4$ of $\sigma(A_i)$ , $i=1,\ldots,9$ (since there are nine columns) such that each vector in the $i$th column is associated with the observable corresponding to $\mu_{A_i}(\Delta_i^k)$ for some $k$. However, from this point of view, it is not at all clear why, whenever $\mu_{A_i}(\Delta_i^k)=\mu_{A_j}(\Delta_j^l)=P$ ($i\neq j$), they should both be associated with the same observable $\mathcal{P}$ and why they should be assigned the same value. One may argue that the observable associated with $\mu_{A_i}(\Delta_i^k)$ is in fact not the same as the one associated with $\mu_{A_j}(\Delta_j^l)$, especially since the observables $\A_i$ and $\A_j$ cannot be measured at the same time. This leads to the following definition.

\begin{definitie}
A theory is called \underline{contextual} if the value of an observable may depend on the measuring context. That is, it may depend on what other observables are being measured. 
\end{definitie}

In the terms of the observables $\A_i,\A_j$ and $\mathcal{P}$, contextuality states that the value of $\mathcal{P}$ may depend upon whether it is measured simultaneously with $\A_i$ or with $\A_j$. This seems a natural assumption if $\mathcal{P}$ is considered to be the observable corresponding to $\mu_{A_i}(\Delta_i^k)$ in one measuring context, whilst corresponding to $\mu_{A_j}(\Delta_j^l)$ in the other. But what if $\mathcal{P}$ has a meaning all of its own, independently of the observables $\A_i$ and $\A_j$? In that case, contextuality becomes a stronger assumption. It then states that the value of $\mathcal{P}$ may depend on the way we (as experimenters look) at it. 

It turns out that if one wishes to use the contextuality argument to criticize the Kochen-Specker Theorem, one necessarily has to adopt this stronger form. This is because the operators that play a role in the proof of Lemma \ref{KS3dim} \textit{can} each be associated with a single observable that has a clear physical meaning independently of its measuring context. Indeed, every vector can be associated with the squared spin of a spin-1 particle along that axis. The operators associated with these observables commute if and only if the associated axes are orthogonal. Thus ever triad can be associated with a set of three observables that can be measured at the same time. It is predicted by quantum mechanics that such triplet measurements always yield twice the value 1, and once the value 0. The proof of Lemma \ref{KS3dim} thus also shows that a (non-contextual) definite value assignment to these observables is impossible. 

Bell \cite{Bell66} argued that the value assigned to any observable may well depend on which other observables are to be measured. This is because different sets of observables must be measured using different measuring devices. This is, in fact, in line with (and partly inspired by) the philosophy of Bohr, who thought that observables have no real meaning at all without specifying the measuring context. But, for me, the idea that this may be so, i.e. that the value of an observable depends on the way we measure it, is a substantial compromise to what is to be expected from a hidden-variable theory. It parts from the idea that observables should have a definite value independent of the observer. Quite a step for someone who was `against `measurement'' \cite{Bell90}.

\subsubsection{The Bell Inequality}\label{The Bell Inequality}

Instead of studying the possibility of hidden-variable theories in a broad sense\footnote{The proof by von Neumann focuses on ensembles for arbitrary physical systems and also the Kochen-Specker Theorem is a statement about all physical statements that are described by a Hilbert space of dimension greater than 2.}, the argument of Bell focuses on one specific physical system: that of a pair of spin-$\tfrac{1}{2}$ particles. The original first paper \cite{Bell64} does in fact not account for contextual theories (i.e., it assumes definite values for observables independent of the measuring context). The argument was improved in \cite{Bell71}, where Bell incorporated the possibility that the actual value obtained when measuring an observable may also depend on the setting of the measurement device. Later, Bell's argument has also been extended to incorporate stochastic hidden-variable theories (i.e. theories in which measurement results are not necessarily pre-determined, but do obey certain other requirements). To make the argument for stochastic theories more comprehensible, I will first discuss a Bell-type argument for deterministic theories. 
A more extensive account of the variety of Bell arguments can be found in \cite{Clauser78} and \cite{Seevinck08}.

\paragraph{Deterministic Hidden Variables}

I will present the argument here in a way that is slightly different from what is common in most literature. The differences will become more apparent along the way. The main difference is that I start from the hidden state of the system $\lambda$ that is supposed to determine all outcomes of all possible experiments for all possible contexts, whereas one usually argues from the value of a specific observable as a function of several hidden variables, which may then be `local' or `non-local'.  

The system under consideration is again that of the two spin-$\tfrac{1}{2}$ particles of Example \ref{EPRB}. The question will be whether or not a hidden-variable theory can reproduce the statistics that quantum mechanics dictates if the system is prepared in the state
\begin{equation}\label{correlatietoestand}
	\psi=\frac{1}{\sqrt{2}}(0,1,-1,0).
\end{equation}
First, an examination of the structure of a contextual hidden-variable theory is necessary. Since it is now allowed that values of observables may depend on measuring contexts, one should reconsider the notion of a state in a hidden-variable theory. Now a state $\lambda$ should assign a value in $\sigma(A)$ to each observable $\A$, for each measuring context $\mathcal{C}$ in which it is possible to measure $\A$. To give a precise meaning to this requirement, the following definition is helpful.
\begin{definitie}
A \underline{measuring context} $\mathcal{C}$ is a set $\{\A_i\:;\:i\in I\}$ of observables for which all corresponding self-adjoint operators $\{A_i\:;\:i\in I\}$ commute.
\end{definitie} 

By an appeal to the FUNC' rule, each measuring context $\{\A_i\:;\:i\in I\}$ might be represented by the commutative W*-algebra\footnote{A W*-algebra (or von Neumann algebra) is a set of bounded operators that is closed under taking adjoints and that is equal to its double commutant, where the commutant of a set of bounded operators $V$ is the set of all bounded operators that commute with every element of $V$. There are also more abstract definitions that do not require the notion of an operator on a Hilbert space (see for example \cite{Davidson96}.} of operators generated by the set $\{A_i\:;\:i\in I\}$. Indeed, this construction exactly adds all the observables $\A$ to the measuring context for which there exists an $i\in I$ and a Borel function $f$ such that $\A=f(\A_i)$ for some $i$. However, since no special structure (like that of a W*-algebra) is required to make the argument of this paragraph work, it seems redundant to demand it. In fact, the FUNC' rule may better be avoided (in order not to make unnecessary assumptions) and an observable of the form $f(\A)$ may only be interpreted as applying the function $f$ to the measurement result of $\A$ (that is, it is not required that it is associated with a self-adjoint operator). 

\begin{definitie}
A \underline{contextual pure state} $\lambda$ is a rule that to each measuring context $\mathcal{C}$ assigns a function $\lambda_{\mathcal{C}}:\mathcal{C}\to\mathbb{R}$ such that $\lambda_{\mathcal{C}}(\A)\in\sigma(A)$ for all $\A\in\mathcal{C}$. 
\end{definitie}

It turns out that the set of all contextual pure states is large enough to avoid a conflict with quantum mechanics. In fact, contextual hidden-variable theories do exist, Bohmian mechanics probably being the most popular one (\cite{Bohm52}, see \cite{Tumulka04} for a friendly introduction). The main objection against Bohmian mechanics is that it is non-local i.e., it allows action at a distance. This is what inspired Bell to write his article \cite{Bell64}; he wanted to investigate whether this non-local behavior is a necessary property of any hidden-variable theory. It turns out that it is, at least if one adopts the following definition of locality.\footnote{One may easily check that also the catalog theory of Section \ref{catalogtheory} must indeed be non-local if it is modified to be empirically equivalent with quantum mechanics.}

\begin{definitie}\label{LOCBell1}
A contextual pure state is called \underline{local} if for any pair of observables $\A_1,\A_2$ corresponding to commuting operators $A_1,A_2$ that can be measured using measuring devices that may be separated at an arbitrary distance from each other, and for every measuring context $\mathcal{C}$ containing $\A_1$ and $\A_2$, one has
\begin{equation}\label{LOCBell}
	\lambda_{\mathcal{C}}(\A_1)=\lambda_{\mathcal{C}\backslash\{\A_2\}}(\A_1)\text{ and }
	\lambda_{\mathcal{C}}(\A_2)=\lambda_{\mathcal{C}\backslash\{\A_1\}}(\A_2).
\end{equation}
Such a pair of observables will be called \underline{separable}. The set of all local contextual pure states is called $\Lambda_{Bell}$.
\end{definitie}

One may think of separable observables as observables of separated systems that are now combined into one system. That is, the system is described by a Hilbert space of the form $\h_1\otimes\h_2$. Two observables may then be called separable if their corresponding operators are of the form $A_1\otimes\een$ and $\een\otimes A_2$. Thus $\A_1$ is an observable of the separate subsystem described by the space $\h_1$ and $\A_2$ is an observable of the subsystem described by the space $\h_2$. Consequently, a local pure state must assign a value to the observable $\A_1$ in the context $\mathcal{C}$ independently of any observable in $\mathcal{C}$ associated with a spatially separated subsystem. In other words, equation (\ref{LOCBell}) may be itterated an arbitrary number of times until the context $\mathcal{C}$ no longer contains any observables associated with a spatially separated subsystem.

The theorem due to Bell may now be formulated as follows.

\begin{stelling}\label{Bell1}
  There is no local, contextual hidden-variable theory (and consequently no non-contextual one) that can reproduce the statistics of the system of two spin-$\tfrac{1}{2}$ particles prepared in the state $\psi=\frac{1}{\sqrt{2}}(0,1,-1,0)$ (that are space-like separated). Therefore, any such theory is empirically in disagreement with quantum mechanics.
\end{stelling}

Due to the extra structure acquired (compared to $\Lambda_{cat}$ or $\Lambda_{KS}$) by introducing contextuality, an examination of how the statistics of quantum mechanics are to be reproduced is required before the theorem can be proven. This is done in the same way as in classical statistical physics (as was also roughly sketched in the beginning of paragraph \ref{NOGOS}). 

A \underline{macro state} in the hidden-variable theory is a probability distribution $\mu$ over the set $\Lambda_{Bell}$.\footnote{The term `macro state' is taken from statistical physics where it is also used to describe probability distributions over pure states (also called micro states). The term may seem a bit awkward, since it describes what may be called a micro state in quantum mechanics. However, it is precisely this viewpoint (i.e., that the quantum states are micro states) that is criticized by Einstein, Podolsky and Rosen.} One usually thinks of the system as an ensemble of systems, or one can think of a system whose pure state fluctuates rapidly in time.\footnote{The philosophy of the interpretation of statistical mechanics is a topic of its own \cite{Sklar09}.} To talk sensibly about a probability measure on the set $\Lambda_{Bell}$, a $\sigma$-algebra $\Sigma_{Bell}$ of subsets of $\Lambda_{Bell}$ is required. 

This $\sigma$-algebra is constructed as follows. For each observable $\A$ corresponding to some operator $A$, for each measuring context $\mathcal{C}$ such that $\A\in\mathcal{C}$ and for each Borel set $\Delta\subset\sigma(A)$, consider the event that the measurement of $\A$ yields a result in $\Delta$ in the context $\mathcal{C}$. This event is identified with the set 
\begin{equation}\label{hakenset}
	[\A_{\mathcal{C}}\in\Delta]:=\{\lambda\in\Lambda_{Bell}\:;\:\lambda_{\mathcal{C}}(\A)\in\Delta\}.
\end{equation}
For fixed $\mathcal{C}$, let $\Sigma_{\mathcal{C}}$ be the $\sigma$-algebra generated by all such sets. Then take $\Sigma_{Bell}$ to be the $\sigma$-algebra generated by all these $\sigma$-algebras. For a fixed measuring context $\mathcal{C}$, the following equivalence relation on $\Lambda_{Bell}$ is introduced: 
\begin{equation}
	\lambda\sim_{\mathcal{C}}\lambda' \quad\desda\quad \lambda_{\mathcal{C}}(\A)=\lambda'_{\mathcal{C}}(\A),\forall \A\in\mathcal{C}.
\end{equation}
This defines a set of equivalence classes
\begin{equation}\label{modulo}
	\Lambda_{\mathcal{C}}:=\Lambda_{Bell}/\sim_{\mathcal{C}},\quad
	[\lambda]_{\mathcal{C}}:=\{\lambda'\in\Lambda_{Bell}\:;\:\lambda'\sim_{\mathcal{C}}\lambda\}.
\end{equation}
For any $S\in\Sigma_{\mathcal{C}}$, it follows that if $\lambda\in S$, then $[\lambda]_{\mathcal{C}}\subset S$. Therefore, the $\sigma$-algebra $\Sigma_{\mathcal{C}}$ extends in a natural way to a $\sigma$-algebra on $\Lambda_{\mathcal{C}}$ by taking
\begin{equation}
	\{\{[\lambda]_{\mathcal{C}}\:;\:\lambda\in S\}\:;\:S\in\Sigma_{\mathcal{C}}\}.
\end{equation}
For notational convenience, both $\sigma$-algebras will be denoted by the same symbol. Also, no distinction in notation will be made for their elements. Any macro state $\mu$ now extends to a probability measure $\pee_{\mathcal{C}}$ on $(\Lambda_{\mathcal{C}},\Sigma_{\mathcal{C}})$ for any measuring context $\mathcal{C}$ by taking
\begin{equation}\label{contextmaat}
	\pee_{\mathcal{C}}(S):=\mu(S)=\int_{\Lambda_{Bell}}1_S(\lambda)\dee\mu(\lambda),\quad\forall S\in\Sigma_{\mathcal{C}}.
\end{equation}
These restricted probability spaces $(\Lambda_{\mathcal{C}},\Sigma_{\mathcal{C}},\pee_{\mathcal{C}})$ have the advantage that any observable $\A\in\mathcal{C}$ can be viewed as a stochastic variable on $\Lambda_{\mathcal{C}}$ by taking
\begin{equation}
	\A([\lambda]_{\mathcal{C}}):=\lambda_{\mathcal{C}}(\A),\quad\lambda\in[\lambda]_{\mathcal{C}},
\end{equation}
which is well defined (i.e. independent of the choice of $\lambda$).

Here one finds a relation between the usual way in which the argument is presented with $\A$ as a function of the hidden parameters on the left-hand side, and the way I present the argument on the right-hand side. From this equation, the difference between the two approaches can also be explained. In the more common discussions, $\A$ is seen as a function of the state of the system and the measurement context separately. One would write something like $\A(\lambda,m_{\mathcal{C}})$ where $\lambda$ is a hidden parameter associated with the preparation of the system and $m_{\mathcal{C}}$ is a hidden parameter associated with the preparation of the measurement device. Both elements are here incorporated in the use of the state $\lambda$, which may of course depend on much more than only the system preparation and measurement preparation.

Now consider two separable observables $\A_1,\A_2$. Because of the locality assumption (\ref{LOCBell}), it follows that
\begin{equation}
\begin{split}
	\A_1([\lambda]_{\{\A_1,\A_2\}})
	&=
	\lambda_{\{\A_1,\A_2\}}(\A_1)
	=
	\lambda_{\{\A_1\}}(\A_1)=\A_1([\lambda]_{\{\A_1\}});\\
	\A_2([\lambda]_{\{\A_1,\A_2\}})
	&=
	\lambda_{\{\A_1,\A_2\}}(\A_2)
	=
	\lambda_{\{\A_2\}}(\A_2)=\A_2([\lambda]_{\{\A_2\}}).
\end{split}
\end{equation}

Now return to the example of the two spin-$\tfrac{1}{2}$ particles. Let $\sigma_{r_1}$ be the spin of the first particle along the $r_1$-axis (one should actually write $\sigma_{r_1}\otimes\een$) and $\sigma_{r_2}$ the spin of the second particle along the $r_2$-axis. Clearly, these are separable observables. Viewing them as stochastic variables in the measuring context $\{\sigma_{r_1},\sigma_{r_2}\}$, the expectation value of their product\footnote{Sometimes this is called a correlation, but I will refrain from using that term so as to avoid confusion with the mathematical term.} for a fixed macro state $\mu$ is given by
\begin{equation}\label{correlatie}
\begin{split}
	\mathbb{E}_{\{\sigma_{r_1},\sigma_{r_2}\}}(\sigma_{r_1}\sigma_{r_2})
	&=
	\int_{\Lambda_{Bell}}
	\sigma_{r_1}([\lambda]_{\{\sigma_{r_1},\sigma_{r_2}\}})
	\sigma_{r_2}([\lambda]_{\{\sigma_{r_1},\sigma_{r_2}\}})\dee\mu(\lambda)\\
	&=
	\int_{\Lambda_{Bell}}\lambda_{\{\sigma_{r_1}\}}(\sigma_{r_1})\lambda_{\{\sigma_{r_2}\}}(\sigma_{r_2})\dee\mu(\lambda).
\end{split}
\end{equation}  
This equation is the key to the proof of the following lemma.

\begin{lemma}\label{Bellongelijkheid1}
Let $\sigma_{r_1}$ and $\sigma_{r'_1}$ denote the spin along the $r_1$ and $r'_1$ axes of the first particle and let $\sigma_{r_2}$ and $\sigma_{r'_2}$ denote the spin along the $r_2$ and $r'_2$ axes of the second particle. Then any macro state $\mu$ of the local, contextual hidden-variable theory satisfies the following Bell inequality:
\begin{equation}\label{inequality1}
	\left|\mathbb{E}_{\{\sigma_{r_1},\sigma_{r_2}\}}(\sigma_{r_1}\sigma_{r_2})
	-\mathbb{E}_{\{\sigma_{r_1},\sigma_{r'_2}\}}(\sigma_{r_1}\sigma_{r'_2})\right|+
	\left|\mathbb{E}_{\{\sigma_{r'_1},\sigma_{r_2}\}}(\sigma_{r'_1}\sigma_{r_2})
	+\mathbb{E}_{\{\sigma_{r'_1},\sigma_{r'_2}\}}(\sigma_{r'_1}\sigma_{r'_2})\right|
	\leq2.
\end{equation} 
\end{lemma}

\noindent
\textit{Proof:}\hspace*{\fill}\\
The first term in the inequality can be estimated in the following way:
\begin{multline}\label{BellAfschatting1}
 \left|\mathbb{E}_{\{\sigma_{r_1},\sigma_{r_2}\}}(\sigma_{r_1}\sigma_{r_2})
	-\mathbb{E}_{\{\sigma_{r_1},\sigma_{r'_2}\}}(\sigma_{r_1}\sigma_{r'_2})\right|=  \cdots \\
\begin{split}
	\cdots &=
	\left|\int_{\Lambda_{Bell}}\lambda_{\{\sigma_{r_1}\}}(\sigma_{r_1})
	\left(\lambda_{\{\sigma_{r_2}\}}(\sigma_{r_2})-\lambda_{\{\sigma_{r'_2}\}}(\sigma_{r'_2})\right)\dee\mu(\lambda)\right| \\
	&\leq
	\int_{\Lambda_{Bell}}\left|\lambda_{\{\sigma_{r_1}\}}(\sigma_{r_1})\right|
	\left|\lambda_{\{\sigma_{r_2}\}}(\sigma_{r_2})-\lambda_{\{\sigma_{r'_2}\}}(\sigma_{r'_2})\right|\dee\mu(\lambda) \\
	&=
	\int_{\Lambda_{Bell}}
	\left|\lambda_{\{\sigma_{r_2}\}}(\sigma_{r_2})-\lambda_{\{\sigma_{r'_2}\}}(\sigma_{r'_2})\right|\dee\mu(\lambda). 	
\end{split}
\end{multline}
Similarly,
\begin{multline}\label{BellAfschatting2}
 \left|\mathbb{E}_{\{\sigma_{r'_1},\sigma_{r_2}\}}(\sigma_{r'_1}\sigma_{r_2})
	+\mathbb{E}_{\{\sigma_{r'_1},\sigma_{r'_2}\}}(\sigma_{r'_1}\sigma_{r'_2})\right|=  \cdots \\
\begin{split}
	\cdots &=
	\left|\int_{\Lambda_{Bell}}\lambda_{\{\sigma_{r'_1}\}}(\sigma_{r'_1})
	\left(\lambda_{\{\sigma_{r_2}\}}(\sigma_{r_2})+\lambda_{\{\sigma_{r'_2}\}}(\sigma_{r'_2})\right)\dee\mu(\lambda)\right| \\
	&\leq
	\int_{\Lambda_{Bell}}\left|\lambda_{\{\sigma_{r'_1}\}}(\sigma_{r'_1})\right|
	\left|\lambda_{\{\sigma_{r_2}\}}(\sigma_{r_2})+\lambda_{\{\sigma_{r'_2}\}}(\sigma_{r'_2})\right|\dee\mu(\lambda) \\
	&=
	\int_{\Lambda_{Bell}}
	\left|\lambda_{\{\sigma_{r_2}\}}(\sigma_{r_2})+\lambda_{\{\sigma_{r'_2}\}}(\sigma_{r'_2})\right|\dee\mu(\lambda). 	
\end{split}
\end{multline}
Next, note that $\Lambda_{Bell}$ can be split to four disjoint pieces
\begin{equation}
\begin{gathered}
	\Lambda_{++}
	=\{\lambda\in\Lambda_{Bell}\:;\:\lambda_{\{\sigma_{r_2}\}}(\sigma_{r_2})=\lambda_{\{\sigma_{r'_2}\}}(\sigma_{r'_2})=1\};\\
	\Lambda_{+-}
	=\{\lambda\in\Lambda_{Bell}\:;\:\lambda_{\{\sigma_{r_2}\}}(\sigma_{r_2})=-\lambda_{\{\sigma_{r'_2}\}}(\sigma_{r'_2})=1\};\\
	\Lambda_{-+}
	=\{\lambda\in\Lambda_{Bell}\:;\:\lambda_{\{\sigma_{r_2}\}}(\sigma_{r_2})=-\lambda_{\{\sigma_{r'_2}\}}(\sigma_{r'_2})=-1\};\\
	\Lambda_{--}
	=\{\lambda\in\Lambda_{Bell}\:;\:\lambda_{\{\sigma_{r_2}\}}(\sigma_{r_2})=\lambda_{\{\sigma_{r'_2}\}}(\sigma_{r'_2})=-1\},
\end{gathered}
\end{equation}
so that $\Lambda_{Bell}=\Lambda_{++}\cup\Lambda_{+-}\cup\Lambda_{-+}\cup\Lambda_{--}$.
One easily checks that on each of these pieces one has
\begin{equation}
  \left|\lambda_{\{\sigma_{r_2}\}}(\sigma_{r_2})-\lambda_{\{\sigma_{r'_2}\}}(\sigma_{r'_2})\right|+
  \left|\lambda_{\{\sigma_{r_2}\}}(\sigma_{r_2})+\lambda_{\{\sigma_{r'_2}\}}(\sigma_{r'_2})\right|=2.
\end{equation}
This leads to the estimate
\begin{multline}
	\left|\mathbb{E}_{\{\sigma_{r_1},\sigma_{r_2}\}}(\sigma_{r_1}\sigma_{r_2})
	-\mathbb{E}_{\{\sigma_{r_1},\sigma_{r'_2}\}}(\sigma_{r_1}\sigma_{r'_2})\right|+
	\left|\mathbb{E}_{\{\sigma_{r'_1},\sigma_{r_2}\}}(\sigma_{r'_1}\sigma_{r_2})
	+\mathbb{E}_{\{\sigma_{r'_1},\sigma_{r'_2}\}}(\sigma_{r'_1}\sigma_{r'_2})\right|
	\\
 \leq
 \int_{\Lambda_{Bell}}2\dee\mu(\lambda)=2, 
\end{multline}
which proves the lemma.
\hfill $\square$\\[0.5ex]
\indent

With the use of this lemma, Theorem \ref{Bell1} can now be proven.

\noindent
\textit{Proof of Theorem \ref{Bell1}:}\hspace*{\fill}\\
All that has to be shown, is that the inequality (\ref{inequality1}), can be violated in quantum mechanics. Using the notation of Example \ref{EPRB}, setting $e_1=(1,0)$ and $e_2=(0,1)$ and taking $\psi$ as in (\ref{correlatietoestand}), the quantum-mechanical equivalent of (\ref{correlatie}) can be calculated:
\begin{equation}
\begin{split}
	\mathbb{E}_{\psi}(\sigma_{r_1}\sigma_{r_2})
	=&
	\langle\psi,\sigma_{r_1}\otimes\sigma_{r_2}\psi\rangle\\
	=&
	\frac{1}{2}\left\langle e_1\otimes e_2-e_2\otimes e_1,\sigma_{r_1}\otimes\sigma_{r_2}\left(e_1\otimes e_2-e_2\otimes e_1\right)\right\rangle\\
	=&
	\frac{1}{2}\left\langle e_1\otimes e_2,\sigma_{r_1}\otimes\sigma_{r_2}e_1\otimes e_2\right\rangle
	-\frac{1}{2}\left\langle e_1\otimes e_2,\sigma_{r_1}\otimes\sigma_{r_2}e_2\otimes e_1\right\rangle\\
	&+
	\frac{1}{2}\left\langle e_2\otimes e_1,\sigma_{r_1}\otimes\sigma_{r_2}e_2\otimes e_1\right\rangle
	-\frac{1}{2}\left\langle e_2\otimes e_1,\sigma_{r_1}\otimes\sigma_{r_2}e_1\otimes e_2\right\rangle.
\end{split}
\end{equation}
Writing out these inner products and using some trigonometry leads to the following relation:
\begin{equation}
\begin{split}
	\mathbb{E}_{\psi}(\sigma_{r_1}\sigma_{r_2})
	=&
	-\frac{1}{2}\cos(\varphi_1)\cos(\varphi_2)\\
	&-\frac{1}{2}(\cos(\vartheta_1)\sin(\varphi_1)-i\sin(\vartheta_1)\sin(\varphi_1))
	             (\cos(\vartheta_2)\sin(\varphi_2)+i\sin(\vartheta_2)\sin(\varphi_2))\\
	&-\frac{1}{2}\cos(\varphi_1)\cos(\varphi_2)\\
  &-\frac{1}{2}(\cos(\vartheta_1)\sin(\varphi_1)+i\sin(\vartheta_1)\sin(\varphi_1))
	             (\cos(\vartheta_2)\sin(\varphi_2)-i\sin(\vartheta_2)\sin(\varphi_2))\\
	=&
	-\cos(\varphi_1)\cos(\varphi_2)
	-\left(\cos(\vartheta_1)\cos(\vartheta_2)+\sin(\vartheta_1)\sin(\vartheta_2)\right)\sin(\varphi_1)\sin(\varphi_2)\\
	=&
	-\cos(\varphi_1)\cos(\varphi_2)
	-\cos(\vartheta_1-\vartheta_2)\sin(\varphi_1)\sin(\varphi_2).
\end{split}
\end{equation}
To simplify, take $\varphi_1=\varphi_2=\pi/2$, so that $\mathbb{E}_{\psi}(\sigma_{r_1}\sigma_{r_2})=-\cos(\vartheta_1-\vartheta_2)$. With this expression, (\ref{inequality1}) becomes
\begin{equation}
	\left|\cos(\vartheta_1-\vartheta_2)-\cos(\vartheta_1-\vartheta'_2)\right|+
	\left|\cos(\vartheta'_1-\vartheta_2)+\cos(\vartheta'_1-\vartheta'_2)\right|\leq2.
\end{equation} 
It is not hard to see that this inequality can be violated. For example, if one takes 
\begin{equation}
	\vartheta_1=0,\vartheta'_1=\tfrac{\pi}{2},\vartheta_2=\tfrac{7}{4}\pi,\vartheta'_2=\tfrac{5}{4}\pi,
\end{equation}
this inequality would result in
\begin{equation}\label{violatieongelijkheid}
	2\sqrt{2}=4\tfrac{1}{2}\sqrt{2}=\left|\cos(\vartheta_1-\vartheta_2)-\cos(\vartheta_1-\vartheta'_2)\right|+
	\left|\cos(\vartheta'_1-\vartheta_2)+\cos(\vartheta'_1-\vartheta'_2)\right|\leq2,
\end{equation} 
which is a contradiction. This completes the proof.
\hfill $\square$\\[0.5ex]
\indent

\paragraph{Stochastic Hidden Variables}

The starting point for a stochastic hidden-variable theory is again a set $\Lambda$ of pure states that may be contextual in nature. However, instead of assigning a value to each observable, a pure state $\lambda$ assigns to each observable $\A$ a probability measure on the space $(\sigma(A),\Sigma_A)$  for each measuring context $\mathcal{C}$ with $\A\in\mathcal{C}$, where  $\Sigma_A$ is the $\sigma$-algebra of Borel-sets in $\sigma(A)$. This measure will be denoted by
\begin{equation}
	\Delta\mapsto\pee_{\mathcal{C}}[\A\in\Delta|\lambda].
\end{equation}

It is further assumed that there is a way to conditionalize. That is, if the measurement of the observable $\A_1$ yields some subset $\Delta_1\in\Sigma_{A_1}$ as a measurement result, the pure state of the system may be altered going from $\lambda$ to $\lambda'$. Instead of introducing some notation for the altered state, I introduce the following notation to denote that the state may be altered by a previous measurement:
\begin{equation}\label{conditionering}
	\pee_{\mathcal{C}}[\A_2\in\Delta_2|\A_1\in\Delta_1,\lambda]:=\pee_{\mathcal{C}}[\A_2\in\Delta_2|\lambda'],
\end{equation}
for any $\A_2\in\mathcal{C}$, $\Delta_2\in\Sigma_{A_2}$. 

The reader may notice an ambiguity in this notation. If the state $\lambda$ changes into the state $\lambda'$, all probability distributions may be altered. 
That is, (\ref{conditionering}) is supposed to hold for arbitrary observables $\A_2$ in arbitrary contexts $\mathcal{C}$ that do not necessarily contain $\A_1$. It then seems natural to assume that the new state may depend on the context in which the result $\A_1\in\Delta_1$ was obtained. 
To incorporate this possibility, the notation should in fact look something like
\begin{equation} 
	\pee_{\mathcal{C}}[\A_2\in\Delta_2|\A_1\in_{\mathcal{C}_1}\Delta_1,\lambda],
\end{equation} 
where $\A_1\in_{\mathcal{C}_1}\Delta_1$ denotes that the result $\A_1\in\Delta_1$ was obtained in the measuring context $\mathcal{C}_1$. 
However, in what follows it will be the case that, once a context is chosen, it remains fixed for all subsequent measurements and one may think of $\pee_{\mathcal{C}}[\A_2\in\Delta_2|\A_1\in\Delta_1,\lambda]$ as shorthand for $\pee_{\mathcal{C}}[\A_2\in\Delta_2|\A_1\in_{\mathcal{C}}\Delta_1,\lambda]$. It may also be noted that this way of conditionalizing is different from the one in standard probability theory. For example, in general it is not to be expected that, whenever $\A_1,\A_2\in\mathcal{C}$, the equality
\begin{equation}\label{KlassiekConditioneren}
	\pee_{\mathcal{C}}[\A_2\in\Delta_2|\A_1\in\Delta_1,\lambda]\pee_{\mathcal{C}}[\A_1\in\Delta_1|\lambda]
	=
	\pee_{\mathcal{C}}[\A_1\in\Delta_1|\A_2\in\Delta_2,\lambda]\pee_{\mathcal{C}}[\A_2\in\Delta_2|\lambda]
\end{equation}
will hold for all $\lambda$.

A \underline{macro state} of the system will again correspond to a probability distribution over the pure states. At first glance this may look like a distribution over distributions. However, the pure states are not probability distributions themselves but rather collections of distributions. Also, one is encouraged to think of the pure states as \textit{complete} descriptions of the system (in the sense that they contain maximal information about the system), and to interpret the macro states as descriptions of the system based on incomplete information (hence as not exactly knowing what the pure state is). Strictly speaking, this definition only makes sense if the space $\Lambda$ is endowed with a $\sigma$-algebra $\Sigma$ of subsets. This is introduced in the following way. For a fixed context $\mathcal{C}$, observable $\A\in\mathcal{C}$ and Borel set $\Delta\in\Sigma_A$ there is a map $\Lambda\to[0,1]$ given by $\lambda\mapsto\pee_{\mathcal{C}}[\A\in\Delta|\lambda]$. $\Sigma$ is taken to be the smallest $\sigma$-algebra that makes all these maps measurable.

A macro state $\mu$ can be used to assign probabilities to actual events by means of the stochastic variables $\lambda\mapsto\pee_{\mathcal{C}}[\A\in\Delta|\lambda]$ by taking
\begin{equation}
	\pee_{\mathcal{C}}[\A\in\Delta|\mu]:=\int_\Lambda \pee_{\mathcal{C}}[\A\in\Delta|\lambda]\dee\mu(\lambda).
\end{equation}
This is to be read as the probability to find a value in $\Delta$ if one measures the observable $\A$ in the context $\mathcal{C}$, given that the state is $\mu$.
It is good to note that this is not the probability of an event, that is, it is not the measure of some subset of $\Lambda$, but rather the expectation value of the variable $\lambda\mapsto\pee_{\mathcal{C}}[\A\in\Delta|\lambda]$. This does indeed seem the only natural way to make sense of probabilities of experimental events in this abstract mathematical context in such a way that
\begin{equation}\label{SHVTDirac}
	\pee_{\mathcal{C}}[\A\in\Delta|\delta_\lambda]:=\int_\Lambda \pee_{\mathcal{C}}[\A\in\Delta|\lambda']\dee\delta_\lambda(\lambda')
	=\pee_{\mathcal{C}}[\A\in\Delta|\lambda],
\end{equation}
where $\delta_\lambda$ is the macro state given by 
\begin{equation}
	\delta_\lambda(S)=\begin{cases} 1,& \lambda\in S;\\ 0 ,& \lambda\notin S,\end{cases}\quad\forall S\in\Sigma.
\end{equation}
In the same way, conditionalized probabilities may be introduced by taking
\begin{equation}
	\pee_{\mathcal{C}}[\A_2\in\Delta_2|\A_1\in\Delta_1,\mu]:=
	\int_\Lambda \pee_{\mathcal{C}}[\A_2\in\Delta_2|\A_1\in\Delta_1,\lambda]\dee\mu(\lambda),
\end{equation}
and the expectation value of the observable $\A$ in the context $\mathcal{C}$ given the state $\mu$ is defined to be
\begin{equation}
\begin{split}
	\mathbb{E}_{\mathcal{C}}[\A|\mu]
	&:=\int_\Lambda \mathbb{E}_{\mathcal{C}}[\A|\lambda]\dee\mu(\lambda) \\
	&=\int_\Lambda \int_{\sigma(A)}x\pee_{\mathcal{C}}[\A\in\dee x|\lambda]\dee\mu(\lambda).
\end{split}
\end{equation}

As was done for deterministic theories, a condition of locality is now introduced. For the deterministic models it was implicitly assumed that measurements do not disturb the state of the system, whereas here disturbances are allowed. However, in order to proceed one needs to assume that such disturbances may only have a local effect. Consequently, to obtain a Bell-inequality, two locality assumptions are made for the stochastic model.
\begin{definitie}\label{Jarrettlocal} $\quad$ 
\begin{enumerate}
\item A state $\mu$ is called \underline{outcome independent} (OILOC) if for any pair of separable observables $\A_1,\A_2$, corresponding to commuting operators $A_1,A_2$, for every measuring context $\mathcal{C}$ containing $\A_1$ and $\A_2$, and for every pair of measurable sets $\Delta_1\subset\sigma(A_1)$ and $\Delta_2\subset\sigma(A_2)$, one has
\begin{equation}\label{uitkomstonafh}
	\pee_{\mathcal{C}}[\A_1\in\Delta_1|\A_2\in\Delta_2,\mu]=\pee_{\mathcal{C}}[\A_1\in\Delta|\mu].
\end{equation}
\item A state $\mu$ is called \underline{context independent} (CILOC) if for any pair of separable observables $\A_1,\A_2$, corresponding to commuting operators $A_1,A_2$, for every measuring context $\mathcal{C}$ containing $\A_1$ and $\A_2$, and for every measurable set $\Delta\subset\sigma(A_1)$, one has
\begin{equation}\label{parameteronafh}
	\pee_{\mathcal{C}}[\A_1\in\Delta|\mu]=\pee_{\mathcal{C}\backslash\{\A_2\}}[\A_1\in\Delta|\mu].
\end{equation}
\end{enumerate}
\end{definitie}

These two locality criteria are due to Jarrett \cite{Jarrett84} (similar criteria are also discussed in \cite{Fraassen82}). The first condition states that the measurement result is not allowed to depend on the result of a (simultaneous) measurement made far away. That is, measurement results may not influence future measurement results \textit{instantaneously}. The second condition expresses the idea that the measurement result of an observable may not depend on the experimental setup used far away. Note that for OILOC states and separable observables, relation (\ref{KlassiekConditioneren}) holds. For such observables one can thus sensibly talk about a joint probability distribution by taking
\begin{equation}
	\pee_{\mathcal{C}}[(\A_1\in\Delta_1)\wedge(\A_2\in\Delta_2)|\lambda]:=
	\pee_{\mathcal{C}}[\A_2\in\Delta_2|\A_1\in\Delta_1,\lambda]\pee_{\mathcal{C}}[\A_1\in\Delta_1|\lambda].
\end{equation}

To finally obtain a Bell-type inequality, a third implicit assumption is made, namely, that the state $\mu$ of the system does not depend on the measuring context $\mathcal{C}$. The idea is that the state may have been prepared long before the choice of the measuring context was made. This is not so much an assumption of locality, but an assumption of free choice: the experimenter is free to choose the experimental set-up, independent of the history of the system. A truely deterministic theory may deny this assumption (although advocates of such a theory would probably not hope to find a solution in stochastic hidden variables anyway).
This leads to the following formulation of the Bell inequality, again in terms of the experiment of Example \ref{EPRB}:
	
\begin{lemma}\label{SHVT-Bell}
	Let $\sigma_{r_1}$ and $\sigma_{r'_1}$ denote the spin of the first particle along the $r_1$ and $r'_1$ axes and let $\sigma_{r_2}$ and $\sigma_{r'_2}$ denote the spin of the second particle along the $r_2$ and $r'_2$ axes. Consider the measuring contexts
\begin{equation}
	\mathcal{C}_{12}=\{\sigma_{r_1},\sigma_{r_2}\},\quad \mathcal{C}_{12'}=\{\sigma_{r_1},\sigma_{r'_2}\},\quad
	\mathcal{C}_{1'2}=\{\sigma_{r'_1},\sigma_{r_2}\},\quad \mathcal{C}_{1'2'}=\{\sigma_{r'_1},\sigma_{r'_2}\}.
\end{equation}
	Then any macro state $\mu$ of the contextual stochastic hidden-variables theory that satisfies the locality claims (i) and (ii) of Definition \ref{Jarrettlocal} satisfies the following Bell-inequality:
\begin{equation}\label{inequality2}
	\left|\mathbb{E}_{\mathcal{C}_{12}}[\sigma_{r_1}\sigma_{r_2}|\mu]
	-\mathbb{E}_{\mathcal{C}_{12'}}[\sigma_{r_1}\sigma_{r'_2}|\mu]\right|+
	\left|\mathbb{E}_{\mathcal{C}_{1'2}}[\sigma_{r'_1}\sigma_{r_2}|\mu]
	+\mathbb{E}_{\mathcal{C}_{1'2'}}[\sigma_{r'_1}\sigma_{r'_2}|\mu]\right|
	\leq2.
\end{equation} 
\end{lemma} 

\noindent
\textit{Proof:}\hspace*{\fill}\\
First, an expression for the expectation values is obtained using both locality assumptions:
\begin{multline}
	\mathbb{E}_{\mathcal{C}_{12}}[\sigma_{r_1}\sigma_{r_2}|\mu]
	=
	\int_{\Lambda}\mathbb{E}_{\mathcal{C}_{12}}[\sigma_{r_1}\sigma_{r_2}|\lambda]\dee\mu(\lambda) \\
\begin{split}
	=&
	\int_{\Lambda}\pee_{\mathcal{C}_{12}}[\sigma_{r_1}=1\wedge\sigma_{r_2}=1|\lambda]
	+\pee_{\mathcal{C}_{12}}[\sigma_{r_1}=-1\wedge\sigma_{r_2}=-1|\lambda]\dee\mu(\lambda) \\
	&
	-\int_{\Lambda}\pee_{\mathcal{C}_{12}}[\sigma_{r_1}=1\wedge\sigma_{r_2}=-1|\lambda]
	+\pee_{\mathcal{C}_{12}}[\sigma_{r_1}=-1\wedge\sigma_{r_2}=1|\lambda]\dee\mu(\lambda) \\
	=&
	\pee_{\mathcal{C}_{12}}[\sigma_{r_1}=1\wedge\sigma_{r_2}=1|\mu]+\pee_{\mathcal{C}_{12}}[\sigma_{r_1}=-1\wedge\sigma_{r_2}=-1|\mu] \\
	&
	-\pee_{\mathcal{C}_{12}}[\sigma_{r_1}=1\wedge\sigma_{r_2}=-1|\mu]-\pee_{\mathcal{C}_{12}}[\sigma_{r_1}=-1\wedge\sigma_{r_2}=1|\mu] \\
	=&
	\pee_{\mathcal{C}_{12}}[\sigma_{r_1}=1|\sigma_{r_2}=1,\mu]\pee_{\mathcal{C}_{12}}[\sigma_{r_2}=1|\mu]+
	\pee_{\mathcal{C}_{12}}[\sigma_{r_1}=-1|\sigma_{r_2}=-1,\mu]\pee_{\mathcal{C}_{12}}[\sigma_{r_2}=-1|\mu] \\
	&
	-\pee_{\mathcal{C}_{12}}[\sigma_{r_1}=1|\sigma_{r_2}=-1,\mu]\pee_{\mathcal{C}_{12}}[\sigma_{r_2}=-1|\mu]
	-\pee_{\mathcal{C}_{12}}[\sigma_{r_1}=-1|\sigma_{r_2}=1,\mu]\pee_{\mathcal{C}_{12}}[\sigma_{r_2}=1|\mu] \\
	\stackrel{(\ref{uitkomstonafh})}{=}&
	\pee_{\mathcal{C}_{12}}[\sigma_{r_1}=1|\mu]\pee_{\mathcal{C}_{12}}[\sigma_{r_2}=1|\mu]+
	\pee_{\mathcal{C}_{12}}[\sigma_{r_1}=-1|\mu]\pee_{\mathcal{C}_{12}}[\sigma_{r_2}=-1|\mu] \\
	&
	-\pee_{\mathcal{C}_{12}}[\sigma_{r_1}=1|\mu]\pee_{\mathcal{C}_{12}}[\sigma_{r_2}=-1|\mu]
	-\pee_{\mathcal{C}_{12}}[\sigma_{r_1}=-1|\mu]\pee_{\mathcal{C}_{12}}[\sigma_{r_2}=1|\mu] \\
	\stackrel{(\ref{parameteronafh})}{=}&
	\pee_{\{\sigma_{r_1}\}}[\sigma_{r_1}=1|\mu]\pee_{\{\sigma_{r_2}\}}[\sigma_{r_2}=1|\mu]+
	\pee_{\{\sigma_{r_1}\}}[\sigma_{r_1}=-1|\mu]\pee_{\{\sigma_{r_2}\}}[\sigma_{r_2}=-1|\mu] \\
	&
	-\pee_{\{\sigma_{r_1}\}}[\sigma_{r_1}=1|\mu]\pee_{\{\sigma_{r_2}\}}[\sigma_{r_2}=-1|\mu]
	-\pee_{\{\sigma_{r_1}\}}[\sigma_{r_1}=-1|\mu]\pee_{\{\sigma_{r_2}\}}[\sigma_{r_2}=1|\mu] \\
	=&
	\left(\pee_{\{\sigma_{r_1}\}}[\sigma_{r_1}=1|\mu]-\pee_{\{\sigma_{r_1}\}}[\sigma_{r_1}=-1|\mu]\right)
	\left(\pee_{\{\sigma_{r_2}\}}[\sigma_{r_2}=1|\mu]-\pee_{\{\sigma_{r_2}\}}[\sigma_{r_2}=-1|\mu]\right).
\end{split}
\end{multline}
Similar relations hold for the other expectation values. Next, introduce the functions
\begin{equation}
\begin{split}
	f_i(\mu)&:=\pee_{\{\sigma_{r_i}\}}[\sigma_{r_i}=1|\mu]-\pee_{\{\sigma_{r_i}\}}[\sigma_{r_i}=-1|\mu];\\
	f'_i(\mu)&:=\pee_{\{\sigma_{r'_i}\}}[\sigma_{r'_i}=1|\mu]-\pee_{\{\sigma_{r'_i}\}}[\sigma_{r'_i}=-1|\mu],
\end{split}
\end{equation}
for $i=1,2$. In terms of these functions, the inequality (\ref{inequality2}) reads
\begin{equation}
	\left|f_1(\mu)f_2(\mu)-f_1(\mu)f'_2(\mu)\right|+
	\left|f'_1(\mu)f_2(\mu)+f'_1(\mu)f'_2(\mu)\right|
	\leq2.
\end{equation} 
That this inequality is satisfied follows almost immediately from $|f_i(\mu)|\leq1$ and $|f'_i(\mu)|\leq1$ for $i=1,2$. Indeed, one has
\begin{multline}
  \left|f_1(\mu)f_2(\mu)-f_1(\mu)f'_2(\mu)\right|+
	\left|f'_1(\mu)f_2(\mu)+f'_1(\mu)f'_2(\mu)\right|\\
	\leq
	\left|f_2(\mu)-f'_2(\mu)\right|+
	\left|f_2(\mu)+f'_2(\mu)\right|
	\leq2.
\end{multline}
This completes the proof.
\hfill $\square$ \\[0.5ex]
\indent

\begin{gevolg}\label{Bell2}
  There is no local, contextual stochastic hidden-variable theory that can reproduce the statistics of the system of two spin-$\tfrac{1}{2}$ particles prepared in the state $\psi=\frac{1}{\sqrt{2}}(0,1,-1,0)$ (that are space-like separated). Therefore, any such theory is empirically in violation with quantum mechanics.
\end{gevolg}

\begin{opmerking}
It has been shown (see for instance \cite{Tsirelson80}, \cite{Landau87}) that the maximal violation of the Bell-inequality is with a factor $\sqrt{2}$ (like in equation (\ref{violatieongelijkheid})). That is, for any 4-tuple of operators $A_1,A_2,A_3,A_4$ each of the form $A_i=2P_i-\een$ with $P_i$ a projection, such that $[P_1,P_3]=[P_2,P_4]=0$ and $[P_1,P_2]\neq0$ and $[P_3,P_4]\neq0$, it holds for every state $\psi$ that
\begin{equation}
	\left|\mathbb{E}_{\psi}(A_1A_3)-\mathbb{E}_{\psi}(A_1A_4)\right|+\left|\mathbb{E}_{\psi}(A_2A_3)+\mathbb{E}_{\psi}(A_2A_4)\right|
	\leq2\sqrt{2}.
\end{equation}
Thus quantum mechanics itself satisfies an inequality similar to the Bell inequality.  
\end{opmerking}

\paragraph{Discussion}
Theorem \ref{Bell1} and Corollary \ref{Bell2} are remarkable results. Whereas the theorems of von Neumann and Kochen \& Specker prove the incompatibility of hidden variables with quantum mechanics at an abstract level, the Bell inequality provides a distinction between local realist theories and quantum mechanics that is susceptible to experimental investigation. No wonder Shimony coined the term ``experimental metaphysics'' to describe related experimental research. 
The first experimental tests were performed by Freedman and Clauser\footnote{In fact, this experiment tested a modification of the Bell inequality that is due to Clauser, Horne, Shimony and Holt (known as the CHSH inequality) \cite{Clauser69}.} \cite{Freedman72}, but often the ones performed by Aspect \cite{Aspect82} are seen as the decisive ones (in favor of quantum mechanics).

Although it is the general consensus that the violation of the Bell inequality excludes the possibility of local realism, there are possible objections against this claim. For example, besides expressing a certain notion of locality, equation (\ref{LOCBell}) may also be viewed as a consequence of free will. Indeed, it expresses the idea that the measurement result of $\A_1$ does not depend on whether or not one chooses to use a measuring context in which $\A_2$ can be measured. This choice may therefore be considered to be free. Conversely, if one denies this form of free will, a violation of (\ref{LOCBell}) is possible without allowing action at a distance. This is because from this point of view, the actual measuring context was already determined before the experiment was even thought of.\footnote{See also the implicit assumption made just above the formulation of Lemma \ref{SHVT-Bell}.} There is therefore no need to demand certain relations between different measuring contexts, since only one measuring context is actual. In particular, for the equation $\lambda_{\mathcal{C}}(\A_1)=\lambda_{\mathcal{C}\backslash\{\A_2\}}(\A_1)$ there is no reason to assume that both sides of (\ref{LOCBell}) should be defined from the determinist point of view i.e., $\lambda_{\mathcal{C}}$ need only be defined for the actual measuring context that is determined to be chosen. Some recent development made following this philosophy can be found in \cite{Hooft02} and \cite{Blasone09}. However, most scientists find this viewpoint `conspirational' and don't wish to abandon free will

for the sake of local realism. I will come back to this discussion in Chapter \ref{FWThoofdstuk}.

Certain other explicit and implicit assumptions made in the derivation of the Bell inequality have also been subjected to criticism\footnote{Some of them are not directly translatable to the notation used in this paper. It might be interesting to investigate whether or not they still apply here or if the notation used here leads to other possible objections.}; see for example \cite{Hajek92}, \cite{Butterfield07}, \cite{Gisin07} and \cite{Shimony09}. The experimental tests that show the violation of Bell inequalities in Nature have also been criticized (c.f. \cite{Feoli03}, \cite{Gill03}, \cite{Santos05}). Furtermore, for an investigation and criticism of the Bell inequalities from a probability theoretic point of view, one may refer to \cite{Khrennikov08}. 
The entire criticism may perhaps best be summarized in the following way.

\begin{quote}
``Between the metaphysical statement ``a local realistic theory is impossible'' and the actual experimental set-up there is a huge gap, which can only be bridged with the aid of many auxiliary hypotheses. Any one of these could be wrong. Proceeding from the experimental side, we can, for example, point out that there are ``experimental loopholes'' [\ldots] in Aspect's experiments, which, if investigated further, might turn out to be responsible for the result. We can also suspect the existence of some ``selection effect'' which influences the detection probabilities, so that Aspect's experiments do not actually test Bell's inequality [\ldots]. We can accept the possibility that some additional implicit assumption has entered into the mathematical derivation of Bell's inequality, or doubt that the mathematical criteria used in this derivation are accurate and complete translations of the metaphysical concepts ``realism'' and ``locality''. And surely, there are many more possibilities, which we cannot see from within the network of present-day physical concepts.'' \cite{BenDov94}
\end{quote}

Indeed, one must always be careful about what conclusions are drawn from mathematical theorems. As will be shown in the next chapter, there may be creative ways to escape plausible assumptions, thereby rendering them implausible. And in Chapter \ref{FWThoofdstuk} it will also be argued that no mathematical theorem at all can ever decide on the true nature of reality. 

However, as long as we are stuck with these present-day physical concepts, it seems that the only way to maintain the possibility of a hidden-variable theory is to accept either non-locality or absolute determinism (thereby denying free will). These two options are also unacceptable for me (as for most people) and the only remaining option seems to be to accept the strangeness of quantum mechanics, and try to make sense of the Copenhagen interpretation, or to find a better interpretation.
In the end, it will turn out that a choice has to be made on what is to be \textit{expected} from a physical theory and what is to be \textit{demanded} from it. I will argue that if one wishes to make as few metaphysical assumptions as possible, a view must be adopted that explicitly incorporates the idea that not everything can be known.

\clearpage

\markboth{The Alleged Nullification of the Kochen-Specker Theorem}{Introduction}
\section[The Alleged Nullification of the Kochen-Specker Theorem]{The Alleged Nullification of the Kochen-Specker\\ Theorem}\label{NullificatieStuk}

\begin{flushright}
\begin{minipage}[300pt]{0.6\linewidth}
\textit{What is proved by impossibility proofs is lack of imagination.}
\end{minipage}
\end{flushright}
\begin{flushright}
-- J. S. Bell
\end{flushright}

\subsection{Introduction}
It is a common phenomenon in any discussion on the foundations of some subject that the stronger a statement, the more creative the theories that oppose this statement. This is no different in the hidden-variable discussion. In 1999 Meyer \cite{Meyer99} unleashed a discussion on the alleged `nullification' of the Kochen-Specker Theorem, which seems to have resulted in a stalemate between Appleby on one side \cite{Appleby05} and Barrett and Kent on the other \cite{Barrett-Kent04}. The aim of this chapter is to give an overview of this discussion and assess whether or not the Kochen-Specker Theorem has been `nullified'. Although \cite{Barrett-Kent04} already gives an overview of the discussion, this overview is obviously prejudiced, and hence it seems worthwhile to present the story from the perspective of an outsider.

\subsection{The Nullification}
\markboth{The Alleged Nullification of the Kochen-Specker Theorem}{The Nullification}
The loophole in the Kochen-Specker Theorem that Meyer found lies neither in the mathematical proof of the theorem, nor in the somewhat abstract FUNC rule, but in its use of the observable postulate. All proofs of the Kochen-Specker Theorem are based on finding a set of self-adjoint operators that cannot all be assigned a definite value in such a way that the FUNC rule is satisfied. A contradiction with quantum mechanics then arises if one assumes that these operators correspond to actual observables. This dubious assumption is the reversal of the observable postulate (Section \ref{PostulatenZelf}). In fact, it is not clear at all why every self-adjoint operator (or any specific self-adjoint operator) should correspond to an observable. On the other hand, the operators that appear in the proof of Lemma \ref{KS3dim} \textit{are} generally accepted to correspond to observables (see also the discussion at the end of Section \ref{Kochen-Specker-sectie}).

But is it really \textit{necessary} to consider these operators to be observables? More precisely, are all operators corresponding to squared spins along some axis empirically distinguishable? Meyer thinks they are not. According to him, ``no experimental arrangement could be aligned to measure spin projections along coordinate axes specified within more than finite precision'' \cite{Meyer99}. I tend to agree with this. For small enough $\epsilon$, the squared spin along some axis $r$ may well be indistinguishable from the squared spin along the $r+\epsilon$ axis. Then Meyer continues to argue that all that has to be done to `nullify' the Kochen-Specker Theorem is to find a set of observables that can be assigned values in accordance with the FUNC rule, such that the squared spin along any axis is indistinguishable from some observable in this set.

The choice of the squared spin along some axis $r$ is commonly identified with the point on the unit sphere $S^2$ in $\mathbb{R}^3$ where the axis intersects with this sphere.\footnote{In fact, for each axis there are two points on the sphere. Such points are considered to be equivalent.} The most natural subset of $S^2$ that results in a set of observables that are empirically indistinguishable from the set of observables when taking the entire $S^2$, is $S^2\cap\mathbb{Q}^3$. Note that this is not the same as the set of all points in $S^2$ whose corresponding axes go through a point in $\mathbb{Q}^3$. For example, the axis through $(1,1,0)$ enters the sphere in the points $\pm1/\sqrt{2}(1,1,0)\notin S^2\cap\mathbb{Q}^3$. It is, however, true that $S^2\cap\mathbb{Q}^3$ is indeed dense in $S^2$. It is in fact also true for higher dimensions (i.e. $S^n\cap\mathbb{Q}^{n+1}$ is dense in $S^n$, see for example \cite{Schmutz08}).

To complete the argument of Meyer, the following proposition must be proven. 
\begin{propositie}\label{MeyerLemma}
The set $\mathpzc{Obs}_{FP}$ of all squared spin observables along axes that intersect the set $S^2\cap\mathbb{Q}^3$ can be assigned definite values in accordance with the FUNC rule. More explicitly, there exists a map $f:S^2\cap\mathbb{Q}^3\to\{0,1\}$ such that for all $x,x',x''\in S^2\cap\mathbb{Q}^3$ one has
\begin{subequations}\label{func-null}
\begin{equation}\label{func-null-a}
	f(x)=f(-x);
\end{equation}
\begin{equation}\label{func-null-b}
	\text{if }x\bot x',\text{ then }f(x)+f(x')\geq1;
\end{equation}
\begin{equation}\label{func-null-c}
	\text{if }x\bot x'\bot x''\bot x,\text{ then } f(x)+f(x')+f(x'')=2.
\end{equation}
\end{subequations} 
\end{propositie}
The proof I will present here is based on the one given by Havlicek, Krenn, Summhammer and Svozil in \cite{Havlicek99}. It is based on the following lemma, which is also proven in \cite{Havlicek99}.

\begin{lemma}
An axis intersects $S^2\cap\mathbb{Q}^3$ if and only if it intersects the set of all triples of integers that satisfy the Pythagorean property;
\begin{equation}
 \mathbb{Z}^3_{Pyth}:=\{(x,y,z)\in\mathbb{Z}^3\backslash\{0\}\:;\:x^2+y^2+z^2=n^2,\:n\in\mathbb{N}\}. 
\end{equation}
\end{lemma}

\noindent
\textit{Proof:}\hspace*{\fill}\\
For the `only if', suppose $\left(\frac{n_1}{m_1},\frac{n_2}{m_2},\frac{n_3}{m_3}\right)\in S^2\cap\mathbb{Q}^3$. Let $n\in\mathbb{N}$ be the least common multiple (lcm) of $m_1,m_2,m_3$. Then $n\cdot\left(\frac{n_1}{m_1},\frac{n_2}{m_2},\frac{n_3}{m_3}\right)\in\mathbb{Z}^3$  lies on the same axis as $\left(\frac{n_1}{m_1},\frac{n_2}{m_2},\frac{n_3}{m_3}\right)$ and
\begin{equation}
 \left(n\frac{n_1}{m_1}\right)^2+\left(n\frac{n_2}{m_2}\right)^2+\left(n\frac{n_3}{m_3}\right)^2=
		n^2\left(\left(\frac{n_1}{m_1}\right)^2+\left(\frac{n_2}{m_2}\right)^2+\left(\frac{n_3}{m_3}\right)^2\right)=n^2 
\end{equation}
\indent
For the `if', suppose $(x,y,z)\in\mathbb{Z}^3_{Pyth}$. Then $\frac{1}{n}(x,y,z)\in S^2\cap\mathbb{Q}^3$, and it lies on the same axis as $(x,y,z)$.
\hfill $\square$\\[0.5ex]

\noindent
\textit{Proof of Proposition \ref{MeyerLemma}:}\hspace*{\fill}\\
For any point $(x,y,z)\in\mathbb{Z}^3_{Pyth}$, let $(x',y',z')=\left(\frac{x}{\mathrm{gcd}(x,y,z)},\frac{y}{\mathrm{gcd}(x,y,z)},\frac{z}{\mathrm{gcd}(x,y,z)}\right)$ where gcd stands for greatest common divisor. This is again a point in $\mathbb{Z}^3_{Pyth}$. It follows that precisely one of the numbers $x',y'$ or $z'$ must be odd. This can be seen as follows. Let $n^2=x'^2+y'^2+z'^2$. 
If $n$ is even, then $n^2=0[\mathrm{mod}4]$. Because not all $x',y'$ and $z'$ can be even (since then their greatest common divisor would equal 1), precisely two must be odd. Suppose $y'$ and $z'$ are odd. Since $y'^2-1=(y'-1)(y'+1)$ is the product of two even numbers, $y'^2=1[\mathrm{mod}4]$, and similarly $z'^2=1[\mathrm{mod}4]$. But this implies that 
\begin{equation}
	2=n^2-x'^2-(y'^2-1)-(z'^2-1)=0[\mathrm{mod}4],
\end{equation} 
which is a contradiction. Therefore, $n$ cannot be even.
If $n$ is odd, then either $x',y',z'$ are all odd, or precisely one of them is odd. If they are all odd, then $x'^2+y'^2+z'^2=3[\mathrm{mod}4]$, which leads to a contradiction, since $n^2=1[\mathrm{mod}4]$. Therefore, precisely one of the $x',y',z'$ must be odd.
\indent

This leads to the definition of the function
\begin{equation}
	f:S^2\cap\mathbb{Q}^3\to\{0,1\},\quad f\left(\frac{n_1}{m_1},\frac{n_2}{m_2},\frac{n_3}{m_3}\right):=
	\begin{cases}
		0,& \frac{\mathrm{lcm}(m_1,m_2,m_3)}{\mathrm{gcd}(n_1,n_2,n_3)}\frac{n_3}{m_3}\text{ is odd};\\
		1,& \frac{\mathrm{lcm}(m_1,m_2,m_3)}{\mathrm{gcd}(n_1,n_2,n_3)}\frac{n_3}{m_3}\text{ is even}.
	\end{cases}
\end{equation}
To show that this function satisfies (\ref{func-null}), note that the map
\begin{equation}
	S:\left(\frac{n_1}{m_1},\frac{n_2}{m_2},\frac{n_3}{m_3}\right)\mapsto
	\frac{\mathrm{lcm}(m_1,m_2,m_3)}{\mathrm{gcd}(n_1,n_2,n_3)}\left(\frac{n_1}{m_1},\frac{n_2}{m_2},\frac{n_3}{m_3}\right) 
\end{equation}
takes elements of $S^2\cap\mathbb{Q}^3$ to elements $(x,y,z)\in\mathbb{Z}^3_{Pyth}$, with $\mathrm{gcd}(x,y,z)=1$. Condition (\ref{func-null-a}) immediately follows from the definition of $f$. Condition (\ref{func-null-b}) will follow from (\ref{func-null-c}), since if $x,x'\in S^2\cap\mathbb{Q}^3$ with $x\bot x'$, then $x\times x'\in S^2\cap\mathbb{Q}^3$, where $\times$ denotes the exterior product. So the only thing left to show is (\ref{func-null-c}). 

Suppose $x,x',x''$ is a triple of mutually orthogonal vectors in $S^2\cap\mathbb{Q}^3$. If $f(x)=0$, it must be shown that $f(x')=f(x'')=1$. For this, set
\begin{equation}
	 	x=\left(\frac{n_1}{m_1},\frac{n_2}{m_2},\frac{n_3}{m_3}\right),\quad
		x'=\left(\frac{n_1'}{m_1'},\frac{n_2'}{m_2'},\frac{n_3'}{m_3'}\right),\quad
		x''=\left(\frac{n_1''}{m_1''},\frac{n_2''}{m_2''},\frac{n_3''}{m_3''}\right), 
\end{equation}
and 
\begin{equation} 
	S(x)=(x_1,x_2,x_3),\quad S(x')=(x_1',x_2',x_3'),\quad S(x'')=(x_1'',x_2'',x_3''). 
\end{equation}
If $f(x)=0$, then $x_3$ is odd and $x_1$ and $x_2$ are both even. Furthermore, since $x_1x_1'+x_2x_2'+x_3x_3'=0$, the number
\begin{equation}
	x_3'=-\frac{x_1x_1'+y_1y_1'}{x_3} 
\end{equation} 
must be even too, and so $f(x')=1$. Similarly, $f(x'')=1$.  

Secondly, it must be shown that if $f(x)=f(x')=1$, then $f(x'')=0$. Note that
\begin{equation}
	f(x'')=
		\begin{cases}
			0,& x_3''\text{ is odd};\\
			1,& x_3''\text{ is even}.\\
		\end{cases}
\end{equation}
However, because $f(x)=f(x')=1$, $x_3$ and $x_3'$ are even and $x_1,x_2,x_1'$ and$x_2'$ are odd. Suppose now that $x_3''$ is even. If $x_1''$ is even, it follows that $x_2''=\frac{-1}{x_2}\left(x_1x_1''+x_3x_3''\right)$ is also even. Similarly, if $x_2''$ is even, $x_1''$ must be even. Thus if $x_3''$ is even, $x_1'',x_2'',x_3''$ must all be even which is a contradiction. Therefore, $x_3''$ must be odd and hence $f(x'')=0$.

This completes the proof.
\hfill $\square$\\[0.5ex]

Meyer suggests that any measurement of a squared spin along some axis $r$ eventually results in the measurement of the squared spin along some axis $r'$, close to $r$, that intersects with $S^2\cap\mathbb{Q}^3$. Also, the selection of the direction $r'$ occurs in such a way that if one measures the squared spin along three orthogonal axes $x,y,z$, actually the spin along some orthogonal axes $x',y',z'$ are measured, where $x',y',z'$ intersect with $S^2\cap\mathbb{Q}^3$ and are close to the axes $x,y,z$. This makes the actual measurement empirically indistinguishable from the intended measurement, and (\ref{func-null-c}) ensures that the measurement result is in agreement with predictions made by quantum mechanics. This is Meyer's alleged `nullification' of the Kochen-Specker Theorem.

Of course, the fact that only rational vectors are considered in the discussion above is not that relevant. The same arguments can be used for any other dense, so-called colorable subset of $S^2$. Not much later, Kent \cite{Kent99} showed that dense colorable sets not only exist in the three-dimensional case, but in fact for any finite-dimensional Hilbert space. He also generalized the scheme to so-called positive operator valued measurements. These models have become known as MK-models. For the present discussion it is enough for the time being to simply know that these extensions exist; in any case, the main discussion can be focused on the three-dimensional case.

\subsection{First Critics}
\markboth{The Alleged Nullification of the Kochen-Specker Theorem}{First Critics}
The number of articles that criticize Meyer's paper is substantial. Apparently, new hope for hidden variables is not warmly welcomed. This came as a bit of a surprise to me, since many of these critics do agree that there are fundamental problems with quantum mechanics. It seems as though some people are afraid that some problems actually might have a possible solution. I will only discuss a selection of these articles, hoping not to leave out too many interesting comments.

The claim of Meyer that the Kochen-Specker Theorem has been nullified leads to the question what it is exactly that the Kochen-Specker Theorem states. A common notion is that the theorem states that (at any given time) not all observables can be assigned definite values that are independent of the measuring context. The necessary assumption to arrive at this conclusion is that observables obey natural functional relationships. The way this is proved, is by selecting a specific set of observables and showing that any assignment of values to those observables contradicts the functional relationships between those observables; see Section \ref{Kochen-Specker-sectie}.

\subsubsection{Non-Linearity of the MK-Models}

Cabello's first comment \cite{Cabello99} is quite straightforward, but seems to rely on a misinterpretation of the MK-models. In his view, each self-adjoint operator or each projection operator corresponds with a possible measurement. The Kochen-Specker Theorem then states that it is impossible ``to conceive a hidden variable model in which the outcomes of all measurements are pre-determined''.

From this point of view, Cabello is right. But clearly, the whole point of the finite precision discussion is that the 1-to-1 correspondence between observables and self-adjoint operators is not a necessary one. Cabello only focuses briefly on this viewpoint, but arrives at the conclusion that this cannot be what is meant in the MK-models since ``this loophole would have very weird consequences.'' For one, it would imply that the superposition principle is no longer valid. Indeed, in the three-dimensional case (in the Meyer model) the operators $\sigma^2_x$ and $\sigma^2_y$ correspond to observables. But the operator $\sigma^2_{x+y}$ doesn't, since the $x+y$-axis intersects $S^2$ in the points $\pm\frac{1}{\sqrt{2}}(1,1,0) \notin S^2\cap\mathbb{Q}^3$. However, the superposition principle\footnote{Usually the superposition principle is only postulated for states. However, each of these observables can be associated with a one-dimensional projection $\een-\sigma^2_r$. Since there is a one-to-one correspondence between (equivalence classes of pure) states and one-dimensional projections, one may therefore argue that the superposition principle should also hold for these operators.}, although a noticable aspect of quantum theory, cannot be a necessary criterion for a hidden variables theory (as is also noted in \cite{Barrett-Kent04}). One may even argue that it is a rather vague aspect of quantum mechanics altogether. 

\subsubsection{Contextuality of the MK-Models}\label{MerminNullify}

Also Mermin \cite{Mermin99} recognizes that a good understanding of the Kochen-Specker Theorem is necessary in order to get the discussion going. To relate this theorem to finite precision measurements, he restates it in the following way.\footnote{Here, somewhat more mathematical terms are used than in \cite{Mermin99} in order to obtain a better view.} 

Suppose $\Omega_{KS}=\{\A_1,\ldots,\A_n\}$ is a finite uncolorable set of observables (each corresponding to an operator with finite spectrum) and let $\mathcal{C}_{KS}$ be the set of all subsets of $\Omega_{KS}$ whose elements correspond to mutually commuting observables. That is, for each set $\{\A_{i_1},\ldots,\A_{i_k}\}\in\mathcal{C}_{KS}$ the observables $\A_{i_1},\ldots,\A_{i_k}$ can be measured simultaneously according to quantum theory. The set of all definite value assignments is again given by
\begin{equation}
	\Lambda=\{\lambda:\Omega_{KS}\to\mathbb{R}\:;\:\lambda(\A_i)\in\sigma(A_i),i=1,\ldots,n\}.
\end{equation}
Mermin then argues that the Kochen-Specker Theorem implies that for each probability measure $\pee$ on this space\footnote{For a $\sigma$-algebra one may take the power set.} there is a subset $\mathcal{C}_{KS}$ of mutually measurable observables and a value assignment $\lambda$ such that $\pee(\lambda)>0$ and such that $\lambda$ restricted to the set $\mathcal{C}_{KS}$ gives a valuation that is in conflict with the supposed functional relationships between these observables. In terms of the three-dimensional case: there is always a finite probability to measure the squared spin in three orthogonal directions and not find one of the outcomes (1,1,0), (1,0,1) or (0,1,1) (for a particular choice of the three directions). 

Certainly, this is a legitimate restatement of the Kochen-Specker Theorem, but already its opening assumption is susceptible to the criticism of Meyer and Kent, who would deny $\Omega_{KS}$ to be a set of observables. However, there are sets of observables that are arbitrarily close (in some sense) to the objects in the set $\Omega_{KS}$, and this is where Mermin seeks a loophole in the argument of Meyer and Kent. In his own words:
\begin{quote}
	``\ldots the KS theorem is not nullified by the finite precision of real experimental setups because of the fundamental physical requirement that probabilities of outcomes of real experiments vary only slightly under slight variations in the configuration of the experimental apparatus, and because the import of the theorem can be stated in terms of whether certain outcomes never occur, or occur a definite nonzero fraction of the time in a set of randomly chosen ideal experiments.'' \cite[p. 3]{Mermin99}
\end{quote}
Here Mermin has smuggled in a new assumption that was not part of the original Kochen-Specker Theorem, namely, that the probability of finding a certain measurement result depends continuously on the experimental setup. Thus, if anything, Mermin has only showed that \textit{given his continuity assumption} the Kochen en Specker theorem is not nullified by Meyer and Kent. In the case of the spin-1 particle, this condition would probably look something like this:
\begin{aanname}\label{Merminaanname}
Let $\sigma_r^2$ denote the squared spin along the $r$-axis, with $\|r\|=1$. Then Mermin's continuity assumption implies that for each state of the system (characterized by the probability measure $\pee$), and for each $r\in S^2$, for each $\epsilon>0$, there is a $\delta>0$ such that
\begin{equation}\label{mermincont}
	\left|\pee[\sigma_r^2=1]-\pee[\sigma_{r'}^2=1]\right|<\epsilon
\end{equation}
for all $r'\in S^2$ with $\|r'-r\|<\delta$.
\end{aanname}  
This assumption is indeed satisfied in quantum mechanics. It remains to be investigated if the Meyer and Kent models can satisfy this assumption and whether or not this is a `silly' assumption. 

Mermin states that without his continuity assumption, any physical theory capable of dealing with finite-precision measurements would be quite useless. It should be noted, however, that the assumption as stated in equation (\ref{mermincont}) is of a purely \textit{theoretical} nature, which is not susceptible to any experimental investigation. Indeed, when trying to determine whether the continuity criterion is satisfied, $r$ and $r'$ have to be chosen arbitrarily close to each other. A measurement of $\sigma_r^2$ then becomes indistinguishable from a measurement of $\sigma_{r'}^2$ and there is no way to compare the frequencies with which $[\sigma_r^2=1]$ or $[\sigma_{r'}^2=1]$ occur. Certainly, pragmatically speaking, any investigation of this condition leads to a verification of this condition and therefore one may consider it to be an a priori condition of scientific theories. But it is a metaphysical question if this experimental verification is a result of the condition actually being true (since the condition does not satisfy the falsification principle). 

It follows that outcomes of measurements of observables arbitrarily close to those in an uncolorable set must statistically resemble the outcomes predicted for the uncolorable observables predicted by quantum mechanics. But one does not need this assumption, since quantum mechanics already predicts certain probabilities for the colorable observables. The only assumption needed is that the hidden variables theory reproduces those statistics, in which case the continuity criterion will automatically follow for the observables in the MK-models. In this sense the Meyer-Kent models are incomplete, since no statistical behavior is specified in these models.\footnote{At this point, notice that the discussion is diverting from the possible existence of non-contextual hidden variables (which is claimed to be impossible by the Kochen-Specker Theorem) to the question if such a hidden variables theory can reproduce the statistics of quantum mechanics (about which the Kochen-Specker Theorem per se is silent).} However, incomplete as they are, Mermin claims that is in principle impossible for the Meyer-Kent models to reproduce the statistics of quantum mechanics without re-introducing contextuality. 

To state his argument for this claim, consider again an uncolorable set $\Omega_{KS}=\{\A_1,\ldots,\A_n\}$. For each observable $\A_i$ in this set, there is an observable $\A'_i$ in the colorable set that is empirically indistinguishable from $\A_i$. This gives a set $\Omega_{KS}'=\{\A_1',\ldots,\A_n'\}$. Mermin correctly notices that it is impossible to construct this set in such a way that for each set $C=\{\A_{i_1},\ldots,\A_{i_k}\}\in\mathcal{C}_{KS}$, the set $C'=\{\A'_{i_1},\ldots,\A'_{i_k}\}$ is again a set of observables corresponding to mutually \textit{commuting} operators. Indeed, if this were possible, any coloring of $C'$ would automatically yield a coloring of $C$, which is assumed to be impossible. His conclusion is the following:
\begin{quote}
	``[This] deficiency makes the MK set [$\Omega_{KS}'$] useless for specifying preassigned non-contextual values
agreeing with quantum mechanics for the outcomes of every one of the slightly imperfect experiments that corresponds to measuring a mutually commuting subset [some $C\in\mathcal{C}_{KS}$] of observables from the ideal KS uncolorable set [$\Omega_{KS}$].'' \cite[p. 2]{Mermin99}
\end{quote}

It seems a strange assumption that each time one intends to measure the observable $\A_i$, in fact the same observable $\A_i'$ is actually measured. Within the finite precision of measurement, however, there are countably many observables $\A_i'$ that may be measured if one intends to measure $\A_i$. The hidden assumption that Mermin makes is that the Meyer-Kent models have the property that whenever one attempts to measure $\A_i$, the observable actually measured is always the same $\A_i'$. But it is very likely that the measurement of $\A_i'$ that is actually performed will fluctuate in time within the boundaries of the finite precision with which the measurement is set up. However, against this line of reasoning Mermin brings forth the following objection:
\begin{quote}
	``If one tries to bridge this gap in the argument by associating more than a single nearby MK colorable observable
with some of the observables in the ideal uncolorable set, one sacrifices the non-contextuality of the value assignments.'' \cite[p. 2]{Mermin99}
\end{quote}
From a quantum-mechanical perspective, this is indeed true. If $\A$ appears in two measurement contexts $C_1$ and $C_2$, in the MK-model it may be associated with several different observables $\A_1'\in C_1'$ and $\A_2'\in C_2'$, depending on the context in which one wishes to measure $\A$. However, from the point of view of the MK-model, $\A_1'$ and $\A_2'$ are two distinct observables and there is no reason why they should be assigned the same value, or even values close to each other. Not even an appeal to Mermin's continuity assumption helps at this point, for the only requirement is that the distributions over the value assignments for these two observables resemble each other.


\markboth{The Alleged Nullification of the Kochen-Specker Theorem}{The Statistics of MKC-Models}
\subsection{The Statistics of MKC-Models}

As noted, the critique of Mermin focuses on the question whether or not the statistics of quantum mechanics can be reproduced by an MK-model. Clifton and Kent \cite{CliftonKent99} recognized this shortcoming and presented a modified non-contextual hidden-variable theory that is supposed to be able to reproduce the right statistical behavior. These modified models are known as MKC-models. 

The purpose of Clifton and Kent is to find a subset $\mathcal{P}_{CK}(\h)\subset\mathcal{P}(\h)$ that is colorable (i.e. there exists a valuation function) and dense in $\mathcal{P}(\h)$, where $\h$ is a finite-dimensional Hilbert space. That is, for each $P\in\mathcal{P}(\h)$ and for each $\epsilon>0$, there is a $P'\in\mathcal{P}_{CK}(\h)$ such that $\|P-P'\|<\epsilon$, where $\|.\|$ denotes the operator norm. Furthermore, it is required that the set of resolutions of the identity\footnote{A resolution of the identity is a sequence $(P_i)$ of pairwise orthogonal projection operators such that $\sum_iP_i=\een$.} generated by $\mathcal{P}_{CK}(\h)$ is dense in the set of all resolutions of the identity. That is, for each resolution $\{P_1,\ldots,P_n\}$ of the identity and each $\epsilon>0$, there is a resolution of the identity $\{P'_1,\ldots,P'_n\}\subset\mathcal{P}_{CK}(\h)$ such that $\|P_i-P_i'\|<\epsilon$ for all $i=1,\ldots,n$. 

This is sufficient to ensure that for any self-adjoint operator $A$, an observable in the MKC-model can be found that is empirically indistinguishable from $\A$. Indeed, for a finite-dimensional Hilbert space $\h$ of dimension $n$, each self-adjoint operator $A$ has a spectral decomposition of the form \begin{equation}
	A=\sum_{a\in\sigma(A)}aP_a,
\end{equation} 
where $\#\sigma(A)\leq n$ and $(P_a)_{a\in\sigma(A)}$ is a set of pairwise orthogonal projection operators that sum up to the unit operator $\een$. Then, for each $\epsilon>0$, there are pairwise orthogonal projection operators $P_a'\in\mathcal{P}_{CK}(\h)$, $a\in\sigma(A)$ such that $\|P_a-P_a'\|<\epsilon/n\sup\{|a|\:;\:a\in\sigma(A)\}$. The operator\footnote{If $\sup\{|a|\:;\:a\in\sigma(A)\}=0$, then $A=\nul$ and one may simply take $A'=\epsilon P'$ for any $P'\in\mathcal{P}_{CK}(\h)$.} $A':=\sum_{a\in\sigma(A)}aP_a'$ may then be seen as an observable\footnote{Throughout the remainder of this section, no distinction in notation between observables and their corresponding operators will be made.} in the MKC-model, and it satisfies $\|A'-A\|<\epsilon$.

In order to define statistical behavior, it is not enough that the set $\mathcal{P}_{CK}(\h)$ is colorable; in order to reproduce the predictions of quantum mechanics, it also needs to allow sufficiently many different colorings. To explain this, some notation will be introduced. 

For an orthonormal basis $\langle e_i\rangle_{i=1}^n$, let 
\begin{equation}
	\mathcal{P}_1\left(\langle e_i\rangle\right)=\{P_{e_1},\ldots,P_{e_n}\}
\end{equation} 
denote the set of one-dimensional projections on the basis vectors and let $\mathcal{P}\left(\langle e_i\rangle\right)$ be the set of all projections that project on subspaces of $\h$ spanned by some subset of $\{e_1,\ldots,e_n\}$ (i.e. $\mathcal{P}\left(\langle e_i\rangle\right)$ is the Boolean algebra generated by $\mathcal{P}_1\left(\langle e_i\rangle\right)$). Note that $\#\mathcal{P}_1\left(\langle e_i\rangle\right)=n$ and $\#\mathcal{P}\left(\langle e_i\rangle\right)=2^n$. A valuation function on $\mathcal{P}\left(\langle e_i\rangle\right)$ is completely determined by assigning the value 1 to precisely one of the $P_{e_i}$ (and hence, it always exists). 

\begin{definitie}\label{TotalIncompatibleDef}
Two (orthonormal) bases $\langle e_i\rangle_{i=1}^n$ and $\langle e'_i\rangle_{i=1}^n$ are called \underline{totally incompatible} if $[P,P']\neq0$ for all $P\in\mathcal{P}\left(\langle e_i\rangle\right)$ and $P'\in\mathcal{P}\left(\langle e'_i\rangle\right)$ with $P,P'\notin\{\nul,\een\}$. That is, for all $P\in\mathcal{P}\left(\langle e_i\rangle\right)$ and $P'\in\mathcal{P}\left(\langle e'_i\rangle\right)$, $P\h\subset P'\h$ implies $P=\nul$ or $P'=\een$.
\end{definitie}

The central theorem of Clifton and Kent \cite{CliftonKent99} can then be formulated as follows.
\begin{stelling}\label{CliftonKentTheorem}
For each finite-dimensional Hilbert space $\h$, there is a countable set of orthonormal bases $\{\langle e^{(1)}_i\rangle_{i=1}^n,\langle e^{(2)}_i\rangle_{i=1}^n,\langle e^{(3)}_i\rangle_{i=1}^n,\ldots\}$ that are pairwise totally incompatible, such that
\begin{equation}
	\mathcal{P}_{CK}(\h)=\bigcup_{m=1}^\infty \mathcal{P}\left(\langle e^{(m)}_i\rangle\right)
\end{equation}
is dense in $\mathcal{P}(\h)$ and such that the set of resolutions of the identity generated by $\mathcal{P}_{CK}(\h)$ is dense in the set of all resolutions of the identity.
\end{stelling}

The proof given by Clifton and Kent is not very illuminating and therefore I omit it.\footnote{Especially because they do not construct $\mathcal{P}_{CK}(\h)$ explicitely but rely on an existence proof.} The following lemma states that this set is in fact colorable.\footnote{The set of natural numbers $\mathbb{N}$ is here taken to exclude 0, i.e., $\mathbb{N}=\{1,2,3,\ldots\}$.} 
\begin{lemma}\label{CliftonKentColoring}
	For each $f:\mathbb{N}\to\{1,\ldots,n\}$, the function $\lambda_f:\mathcal{P}_{CK}(\h)\to\{0,1\}$ given by
	\begin{equation}\label{evalf} 
		\lambda_f(P)=\begin{cases}
							1, & \text{if }P_{e^{(m)}_{f(m)}}\leq P;\\
							0, & \text{otherwise}
						\end{cases}\quad\text{for }P\in\mathcal{P}\left(\langle e^{(m)}_i\rangle\right),\quad m\in\mathbb{N}
	\end{equation}
	is well-defined and defines a valuation function. Moreover, all valuation functions on $\mathcal{P}_{CK}(\h)$ are of this form and the correspondence is bijective.
\end{lemma}
\noindent
\textit{Proof:}\hspace*{\fill}\\
To see that the function is well-defined, note that for each $P\in\mathcal{P}_{CK}(\h)$ (unequal to $\nul$ and $\een$) there is exactly one $m$ such that $P\in\mathcal{P}\left(\langle e^{(m)}_i\rangle\right)$ because the bases are pairwise totally incompatible. 
\indent

Recall that a valuation function satisfies the finite sum rule\footnote{See Corollary \ref{eindigesomregel}.} for all projection operators in $\mathcal{P}_{CK}(\h)$. In fact, it is sufficient to only show the the finite sum rule is satisfied in order to show that $\lambda_f$ is a valuation function (see also \cite{Fine78}).
Now suppose $\{P_1,\ldots,P_k\}$ is a subset of $\mathcal{P}_{CK}(\h)\backslash\{\nul,\een\}$ such that $\sum_{i=1}^kP_i=\een$. Then all these projection operators commute. To see this, let $P_i$ and $P_j$ be any two projections in this set and let $P_h=\sum_{\substack{l=1\\l\neq i,j}}^kP_l$. Then
\begin{equation}
\begin{split}
	P_h+P_i+P_j
	&=\een=\een^2=(P_h+P_i+P_j)^2\\
	&=P_h^2+P_i^2+P_j^2+P_h(P_i+P_j)+(P_i+P_j)P_h+P_iP_j+P_jP_i\\
	&=P_h+P_i+P_j+P_h(P_i+P_j)+(P_i+P_j)P_h+P_iP_j+P_jP_i.
\end{split}
\end{equation}
Subtracting $P_h+P_i+P_j$ from both sides leads to
\begin{equation}
	P_h(P_i+P_j)+(P_i+P_j)P_h+P_iP_j+P_jP_i=\nul.
\end{equation}
Because $P_i+P_j=\een-P_h$, this results in $P_iP_j=-P_jP_i$. Finally, it follows that
\begin{equation}
	P_iP_j=P_iP_jP_j=-P_jP_iP_j=P_jP_jP_i=P_jP_i.
\end{equation}
Therefore, since all bases are totally incompatible, there is a unique $m\in\mathbb{N}$ such that $\{P_1,\ldots,P_k\}\subset\mathcal{P}\left(\langle e^{(m)}_i\rangle\right)$.
Then, since $\lambda_f$ restricted to each $\mathcal{P}\left(\langle e^{(m)}_i\rangle\right)$ is a valuation function, it follows that the finite sum rule holds. This shows that $\lambda_f$ is indeed a valuation function on $\mathcal{P}_{CK}(\h)$.

To show that all valuation functions are of this form and the correspondence is one-to-one, let $\lambda$ be an arbitrary valuation function. For each basis $\langle e_i^{(m)}\rangle_{i=1}^n$ it holds that $\sum_{i=1}^nP_{e_i^{(m)}}=\een$. Therefore, because of the finite sum rule, $f_\lambda(m)$ may be defined to be the \textit{unique} element $k$ of $\{1,\ldots,n\}$ with $\lambda\left(P_{e_k^{(m)}}\right)=1$. Because of the finite sum rule, one must have that if $P\leq P'$, then $\lambda(P)\leq \lambda(P')$. Therefore, (\ref{evalf}) is automatically satisfied and $\lambda=\lambda_{f_\lambda}$.
\hfill $\square$\\[0.5ex]

In the corresponding hidden-variable theory, the set of pure states $\Lambda$, is the set of all valuation functions on $\mathcal{P}_{CK}(\h)$. The above lemma establishes that $\Lambda\simeq \{1,\ldots,n\}^{\mathbb{N}}$. These pure states, the hidden variables, are suspected not to be empirically accessible. Instead, the system is described by selected probability distributions on the pure states. 
Each projection operator $P\in\mathcal{P}_{CK}(\h)$ reappears in the hidden-variable theory as a stochastic variable $\hat{P}:\Lambda\to\{0,1\}$ given by
\begin{equation}
	\hat{P}(\lambda):=\lambda(P).
\end{equation}
To show that the hidden-variable model can reproduce the statistical behavior of quantum theory, 
the following theorem must be proven.\footnote{The claim of this theorem is not stated very clearly in \cite{CliftonKent99} and an actual proof is lacking. Possibly it is one of the sources of confusion in the later criticisms on the MKC-models.}

\begin{stelling}\label{nc-statistics}
For each density operator $\rho$ on the Hilbert space $\h$, there is a probability measure $\pee_{\rho}$ on $\Lambda$ such that \begin{equation}\label{kansmaat}
	\pee_{\rho}[P=1]=\Trace(\rho P)
\end{equation}
for each $P\in\mathcal{P}_{CK}(\h)$, where $[P=1]=\{\lambda\in\Lambda\:;\:\lambda(P)=1\}=\hat{P}^{-1}(\{1\})$.
\end{stelling}

\noindent
\textit{Proof:}\hspace*{\fill}\\
In order to prove the existence of the probability measure, $\Lambda\simeq \{1,\ldots,n\}^{\mathbb{N}}$ has to be turned into a measurable space first. For a finite sequence $t_1,\ldots,t_k$ of natural numbers (not necessarily in increasing order) and a sequence $B_1,
\ldots,B_k$ of subsets of $\{1,\ldots,n\}$, define 
\begin{equation}
	S(t_1,\ldots,t_k;B_1,\ldots,B_k):=\{\lambda_f\in\Lambda\:;\:f(t_1)\in B_1,\ldots,f(t_k)\in B_k\}.
\end{equation}
For a fixed sequence $t_1,\ldots,t_k$, let $\Sigma(t_1,\ldots,t_k)$ be the $\sigma$-algebra generated by all these so-called cylinder sets.
Further, let $\Sigma$ be the smallest $\sigma$-algebra that contains all these $\sigma$-algebras, i.e.
\begin{equation}
	\Sigma=\sigma\left(\Sigma(t_1,\ldots,t_k)\:;\:\{t_1,\ldots,t_k\}\subset\mathbb{N}\right).
\end{equation}
Note that $\Sigma(t_1,\ldots,t_k)$ has in fact only finite many elements and that it is isomorphic (as a set) to the power set of $\{1,\ldots,n\}^{\{t_1,\ldots,t_k\}}$. Therefore, a probability measure on the space $(\Lambda,\Sigma(t_1,\ldots,t_k))$ is completely defined by its action on the sets that are equivalent to a singleton subset of $\{1,\ldots,n\}^{\{t_1,\ldots,t_k\}}$, i.e. the sets of the form
\begin{equation}
	 s(t_1,\ldots,t_k;j_1,\ldots,j_k):=\{\lambda_f\in\Lambda\:;\:f(t_i)=j_i\text{ for }i=1,\ldots,k\}.
\end{equation}
\indent

Now let a density operator $\rho$ be given. For each finite sequence $t_1,\ldots,t_k$ define a probability measure $\pee_{\rho,t_1,\ldots,t_k}$ on the space $(\Lambda,\Sigma(t_1,\ldots,t_k))$ by
\begin{equation}
\pee_{\rho,t_1,\ldots,t_k}[s(t_1,\ldots,t_k;j_1,\ldots,j_k)]
	:=
	\prod_{i=1}^k\Trace\left(\rho P^{(t_i)}_{e_{j_i}}\right),
\end{equation}
for each sequence $j_1,\ldots,j_k$ in $\{1,\ldots,n\}$.

It is easy to see that these probability measures satisfy the following consistency criteria:
\begin{enumerate}
\item For any finite sequence $t_1,\ldots,t_k$ and all permutations $(t_1',\ldots,t_k')=(t_{\pi(1)},\ldots,t_{\pi(k)})$ one has 
\begin{equation}
	\pee_{\rho,t'_1,\ldots,t'_k}[s(t'_1,\ldots,t'_k;j_{\pi(1)},\ldots,j_{\pi(k)})]
	=\pee_{\rho,t_1,\ldots,t_k}[s(t_1,\ldots,t_k;j_1,\ldots,j_k)].
\end{equation}
\item For each finite sequence $t_1,\ldots,t_k,t_{k+1}$, the measure $\pee_{\rho,t_1,\ldots,t_k,t_{k+1}}$ acts as $\pee_{\rho,t_1,\ldots,t_k}$ on every set in $\Sigma(t_1,\ldots,t_k)$, i.e. $\pee_{\rho,t_1,\ldots,t_k,t_{k+1}}(S)=\pee_{\rho,t_1,\ldots,t_k}(S)$ for all $S\in\Sigma(t_1,\ldots,t_k)$. 
\end{enumerate}
Then, according to the Kolmogorov's extension theorem (see for example Theorem 10.1 in \cite{Bhat07}), there exists a probability measure $\pee_{\rho}$ on the space $(\Lambda,\Sigma)$ such that $\pee_{\rho}$ acts as $\pee_{\rho,t_1,\ldots,t_k}$ on the sets in $\Sigma(t_1,\ldots,t_k)$, for each finite sequence $t_1,\ldots,t_k$.

The only thing left to show is that $\pee_{\rho}$ satisfies (\ref{kansmaat}). For one-dimensional projection this follows almost immediately:
\begin{equation}
	\pee_{\rho}[P_{e_l^{(m)}}=1]=\pee_{\rho}(\{\lambda_f\in\Lambda\:;\:f(l)=m\})
	=\pee_{\rho,l}(\{\lambda_f\in\Lambda\:;\:f(l)=m\})=\Trace(\rho P_{e_l^{(m)}}).
\end{equation}
And also for the zero and unit operator (\ref{kansmaat}) follows immediately since
\begin{equation}
	\pee_{\rho}[\nul=1]=\pee_{\rho}(\varnothing)=0\text{ and }\pee_{\rho}[\een=1]=\pee_{\rho}(\Lambda)=1.
\end{equation}
For all the other $P\in\mathcal{P}_{CK}(\h)$, there is exactly one $m\in\mathbb{N}$ with $P\in\mathcal{P}\left(\langle e^{(m)}_i\rangle\right)$. Therefore, 
\begin{equation}
\begin{split}
	[P=1]
	&=
	\{\lambda\in\Lambda\:;\:\lambda(P)=1\}
	=
	\bigcup_{i=1}^\infty\bigcup_{k=1}^n\left\{\lambda\in\Lambda\:;\:\lambda(P_{e_k^{(i)}})=1,P_{e_k^{(i)}}\leq P\right\}\\
	&=
	\bigcup_{k=1}^n\left\{\lambda\in\Lambda\:;\:\lambda(P_{e_k^{(m)}})=1,P_{e_k^{(m)}}\leq P\right\}
	=
	\bigcup_{\substack{k=1\\P_{e_k^{(m)}}\leq P}}^n[P_{e_k^{(m)}}=1],
\end{split}
\end{equation}
where in the third step I used the property that all the bases are totally incompatible. 
Since this is a union of disjoint sets, it follows that
\begin{equation}\label{KansEenProjectie}
\begin{split}
	\pee_{\rho}[P=1]
	&=
	\sum_{\substack{k=1\\P_{e_k^{(m)}}\leq P}}^n\pee_{\rho}[P_{e_k^{(m)}}=1]
	=
	\sum_{\substack{k=1\\P_{e_k^{(m)}}\leq P}}^n\Trace\left(\rho P_{e_k^{(m)}}\right)=
	\Trace\Bigl(\rho \sum_{\substack{k=1\\P_{e_k^{(m)}}\leq P}}^nP_{e_k^{(m)}}\Bigr)\\
	&=
	\Trace\left(\rho P\right).
\end{split}
\end{equation}
This completes the proof.
\hfill $\square$\\[0.5ex]

Putting the result of Theorem \ref{nc-statistics} back into the definition of the measure $\pee_\rho$, it follows that
\begin{equation}
	 \pee_\rho[P_{e_{f(i)}^{(i)}}=1\:\forall i=1,\ldots,\infty]=\pee_\rho(\{\lambda_f\})=\prod_{i=1}^{\infty}\pee_\rho[P_{e_{f(i)}^{(i)}}=1].
\end{equation}
Consequently, 
\begin{gevolg}\label{MKConafhankelijkheid}
	With repsect to any of the probability measures $\pee_{\rho}$, two projection operators $P_1,P_2\in\mathcal{P}_{CK}(\h)$ are independent as stochastic variables $\hat{P}_1,\hat{P}_2$ if and only if they do \emph{not} commute.
\end{gevolg}

It follows from Theorem \ref{nc-statistics} that the non-contextual hidden-variable theory defined by Lemma \ref{CliftonKentColoring} can indeed reproduce the statistical predictions of quantum theory with arbitrary precision. To see this, consider a quantum system in the state $\rho$ and let $A$ be a self-adjoint operator with spectral decomposition $A=\sum_{a\in\sigma(A)}aP_a$. The quantum-mechanical probability for finding the value $a$ is then given by $\Trace(\rho P_a)$. 

It may be that not all the $P_a$ lie in $\mathcal{P}_{CK}(\h)$. But for every $\epsilon>0$ there exists a (nonunique\footnote{It is of course not necessary to take $a\in\sigma(A)$. One may also take an other set $\sigma(A)'$ such that $a'$ is close to $a$ whenever $P_a'$ is close to $P_a$. However, this is not very convenient notation-wise and it is already sufficient to only consider taking $a\in\sigma(A)$.}) $A'=\sum_{a\in\sigma(A)}aP_a'$ such that $P_a'\in\mathcal{P}_{CK}(\h)$, $\sum_{a\in\sigma(A)}P'_a=\een$ and $\|A-A'\|<\epsilon$ (see the discussion just above Definition \ref{TotalIncompatibleDef}). Then, if one argues that a measurement of $A$ in fact comes down to the measurement of $A'$, one finds the value $a$ with a probability $\pee_{\rho}[P'_a=1]=\Trace(\rho P'_a)$ that is close to the probability predicted by quantum mechanics:
\begin{equation}\label{ContinueMaat}
	\left|\Trace(\rho P'_a)-\Trace(\rho P_a)\right|=
	\left|\Trace(\rho (P'_a-P_a))\right|\leq\|P'_a-P_a\|\Trace(\rho)<\epsilon.
\end{equation}
In fact, the more precise the measurements become (i.e., the smaller $\epsilon$ becomes), the more the probabilities obtained will resemble those predicted by quantum mechanics. This establishes, in particular, that the continuity criterion of Mermin (Section \ref{MerminNullify}) \textit{is} in fact satisfied by the MKC-models.


\markboth{The Alleged Nullification of the Kochen-Specker Theorem}{Further Criticism}
\subsection{Further Criticism}
\subsubsection{An Empirical Discrepancy with Quantum Mechanics (Part I)}
In \cite{Cabello02}, Cabello claims exactly the opposite of Clifton and Kent. Namely, he states that neither the model of Meyer, nor the models of Clifton and Kent can reproduce the statistics of quantum mechanics. It is not surprising that Meyer's model does not posses this property, for in the proof of Proposition \ref{MeyerLemma} only one possible coloring of the set $S^2\cap\mathbb{Q}^3$ was described. Without much use of imagination, this result can be extended to obtain three different colorings, but surely the richness of quantum stochastics cannot be reproduced by a probability space with only three elements. Also, it seems to me pointless to falsify a claim that was never made, certainly since the absence of a statistical part of the models of Meyer and Kent was one of the main motivations for Clifton and Kent to construct their models.

The proof Clifton and Kent gave to show that their models \textit{can} reproduce the statistics of quantum theory seems convincing, and it is a shame that Cabello did not point out what he thought was wrong with this proof. It seems that Cabello was probably puzzled when Clifton and Kent stated that 
\begin{quote}
``\ldots it should already be clear that the set of truth valuations [\ldots] will be sufficiently rich to recover the statistics of any quantum state by averaging over the values of the hidden variables that determine the various truth valuations.'' \cite[p. 6]{CliftonKent99}
\end{quote}
 This seemed indeed a puzzling statement, but I think it should have been clarified now by Theorem \ref{nc-statistics}. 

Instead, Cabello focuses on the statistical behavior of two specific observables. Consider again the sphere $S^2$ and let $\mathcal{D}$ be a dense subset for the MKC-model. That is, $\mathcal{D}$ is associated with the one-dimensional projections in $\mathcal{P}_{CK}(\mathbb{C}^3)$. Let $P_e$ denote the projection on the line spanned by some vector $e$ and consider $e_1=\tfrac{1}{\sqrt{3}}(1,1,1)$ and $e_2=\tfrac{1}{\sqrt{3}}(1,1,-1)$. Cabello now argues that for each probability measure $\pee$ on the set $\Lambda$ of pure states (i.e., colorings of the set $\mathcal{P}_{CK}(\mathbb{C}^3)$), the probability $\pee[P_{e'_1}=P_{e'_2}=1]$ can be made arbitrary small as long as $e'_1$ and $e'_2$ are taken close enough to $e_1$ and $e_2$. In more precise terms:

\begin{lemma}\label{Cabellolemma}
For each $\epsilon>0$, there exists a $\delta>0$, such that for all $P_{e'_1},P_{e'_2}\in\mathcal{P}_{CK}(\mathbb{R}^3)$ with $\|e_i-e'_i\|<\delta$ ($i=1,2$) one has
\begin{equation}
	\pee[P_{e'_1}=P_{e'_2}=1]<\epsilon
\end{equation}
for all probability measures $\pee$ on $\Lambda$.
\end{lemma}

Before turning to the proof, it is useful to ask why this property is supposed to be in conflict with quantum mechanics. In the usual interpretation, the observables associated with the operators $P_{e_1}$ and $P_{e_2}$ cannot be measured simultaneously (since the operators do not commute). The empirical result of finding the result 1 for both measurements observables must therefore be obtained by performing the experiments after each other, which calls for the use of the von Neumann postulate or some other postulate that describes the dynamics of measurements. To avoid this discussion, consider a system that is already described by the state $e_1$. If one then first measures $P_{e_1}$ and then $P_{e_2}$, the probability of finding for both the value 1 is given by the transition probability
\begin{equation}\label{Cabelloquantum}
	\pee_{QM}[P_{e_1}=P_{e_2}=1]=\langle e_1,P_{e_2}e_1\rangle=|\langle e_1,e_2\rangle|^2=\tfrac{1}{9}.
\end{equation}	
So if Lemma \ref{Cabellolemma} holds, this does seem a reasonable proof for the incapability of the MKC-models to reproduce quantum predictions. Hence a careful investigation of Cabello's proof is called for.

\noindent
\textit{Proof of Lemma \ref{Cabellolemma}:}\hspace*{\fill}\\
Consider the following two orthonormal bases of $\mathbb{C}^3$: 
\begin{equation}
\begin{aligned}
	f_1&=\frac{1}{2}\sqrt{2}(0,1,-1), & f_2&=\frac{1}{2}\sqrt{2}(0,1,1), & f_3&=(1,0,0); \\
	g_1&=\frac{1}{2}\sqrt{2}(1,0,-1), & g_2&=\frac{1}{2}\sqrt{2}(1,0,1), & g_3&=(0,1,0).
\end{aligned}
\end{equation}
These satisfy $f_1\perp e_1\perp g_1$, $f_2\perp e_2\perp g_2$ and $f_3\perp g_3$. Consequently, for every $\epsilon',\epsilon''>0$ two orthonormal bases $(f'_1,f'_2,f'_3), (g'_1,g'_2,g'_3)$ can be chosen that satisfy the following criteria:
\begin{itemize}
\item $P_{f'_i},P_{g'_i}\in\mathcal{P}_{CK}(\mathbb{R}^3)$ for $i=1,2,3$.
\item $\|P_{f'_1}P_{e_1}\|<\epsilon'$, $\|P_{g'_1}P_{e_1}\|<\epsilon'$, $\|P_{f'_2}P_{e_2}\|<\epsilon'$, $\|P_{g'_2}P_{e_2}\|<\epsilon'$.
\item $\|P_{f'_3}P_{g'_3}\|<\epsilon''$.
\end{itemize}
Indeed, this is done by choosing the bases close enough to the bases $(f_1,f_2,f_3)$ and $(g_1,g_2,g_3)$.
\indent

Now, in the MKC-model, an attempted measurement of $P_{e_1}$ leads to an actual measurement of $P_{e'_1}$ and similarly, an attempted measurement of $P_{e_2}$ leads to an actual measurement of $P_{e'_2}$ with $\|e_i-e'_i\|<\delta$, where $\delta>0$ denotes the precision of the measurement. Since $\delta$ is controlled by the experimenter, it can be taken arbitrarily small. 

Consider now the set of all $\lambda\in\Lambda$ that satisfy $\lambda(P_{e'_1})=\lambda(P_{e'_2})=1$. Cabello argues that $\epsilon'$ and $\delta$ can be taken small enough such that for any probability measure $\pee$ that is supposed to describe a quantum state, one has
\begin{equation}
\begin{gathered}
	\pee[P_{f'_1}=1|P_{e'_1}=P_{e'_2}=1]<\frac{1}{8}\epsilon;\quad
	\pee[P_{g'_1}=1|P_{e'_1}=P_{e'_2}=1]<\frac{1}{8}\epsilon;\\
	\pee[P_{f'_2}=1|P_{e'_1}=P_{e'_2}=1]<\frac{1}{8}\epsilon;\quad
	\pee[P_{g'_2}=1|P_{e'_1}=P_{e'_2}=1]<\frac{1}{8}\epsilon,
\end{gathered}
\end{equation}   
because each of the pairs $(f'_1,e'_1)$, $(g'_1,e'_1)$, $(f'_2,e'_2)$ and $(g'_2,e'_2)$ are arbitrarily close to being orthogonal (by choosing $\epsilon'$ and $\delta$ sufficiently small). Therefore, for `almost all'\footnote{This is to be considered with respect to $\pee$, but not in the usual measure-theoretic sense. Cabello's argument becomes a bit obscure at this point.} $\lambda\in\Lambda$ with $\lambda(P_{e'_1})=\lambda(P_{e'_2})=1$, one should have $\lambda(P_{f'_1})=\lambda(P_{f'_2})=\lambda(P_{g'_1})=\lambda(P_{g'_2})=0$. For these $\lambda$ one has $\lambda(P_{f'_3})=\lambda(P_{g'_3})=1$. But, if $\epsilon''$ is taken small enough, one should have $\pee[P_{f'_3}=P_{g'_3}=1]<\epsilon/2$, because $f'_3,g'_3$ are almost orthogonal.

To recapitulate, one has  
\begin{equation}
\begin{split}
	\pee[P_{e'_1}=P_{e'_2}=1]
	=&
	\pee[P_{e'_1}=P_{e'_2}=1,P_{f'_3}=P_{g'_3}=1]\\
	&+\pee[P_{e'_1}=P_{e'_2}=1,P_{f'_3}=0\text{ or }P_{g'_3}=0]\\
	\leq&
	\pee[P_{f'_3}=P_{g'_3}=1]
	+\pee[P_{f'_3}=0\text{ or }P_{g'_3}=0|P_{e'_1}=P_{e'_2}=1]\\
	\leq&
	\frac{1}{2}\epsilon+\pee[P_{f'_1}=1|P_{e'_1}=P_{e'_2}=1]+
	\pee[P_{g'_1}=1|P_{e'_1}=P_{e'_2}=1]\\&+
	\pee[P_{f'_2}=1|P_{e'_1}=P_{e'_2}=1]+
	\pee[P_{g'_2}=1|P_{e'_1}=P_{e'_2}=1]\\
	\leq& \frac{1}{2}\epsilon+\frac{1}{8}\epsilon+\frac{1}{8}\epsilon+\frac{1}{8}\epsilon+\frac{1}{8}\epsilon=\epsilon.
\end{split}
\end{equation}
\hfill $\square$\\[0.5ex]

This proof however, relies on an implicit assumption of continuity. Although the allowed probability distributions in the MKC-models (i.e. the ones of Theorem \ref{nc-statistics}) are continuous in a certain sense, the pure states are not. Indeed, in the previous section it was shown that for every $P_e\in\mathcal{P}_{CK}(\h)$, for every $\epsilon>0$ there is a $\delta>0$ such that
\begin{equation}\label{cabello1}
	\left|\pee[P_{e}=1]-\pee[P_{e'}=1]\right|<\epsilon
\end{equation}
for all $e'$ with $\|e-e'\|<\delta$ (this resembles the continuity criterion of Mermin). However, this does not imply that 
\begin{equation} 
	\left|\pee[P_e=P_{e'}=1]-\pee[P_e=1]\right| 
\end{equation}
becomes small as $e'$ approaches $e$. To be more explicit, in Cabello's proof it is assumed that if $(e_n)$ is a sequence that approaches a vector $e$ that is orthonormal to $f$, then
\begin{equation}\label{ContinuityCabello}
	\lim_{n\to\infty}\pee[P_{e_n}=P_f=1]=\pee[P_e=P_f=1]=0 
\end{equation}
for allowed probability distributions. This is simply not true. In fact, the discontinuity of the pure states is an essential part of the MKC-models (see below).

Cabello's argument (although implicitly) does point out a feature of the MKC-models that is not understood clearly yet, namely, the possible measurement of observables whose corresponding operators do not commute. It is no wonder that the models are silent about this topic, since it is also an unclear point in quantum theory itself. However, in the case considered in Lemma \ref{Cabellolemma}, it can easily be seen that the MKC-model \textit{does} reproduce equation (\ref{Cabelloquantum}) because of Theorem \ref{nc-statistics}. Indeed, in this case one has $\rho=P_{e_1}$ and using Corollary \ref{MKConafhankelijkheid} one finds 
\begin{equation}
\begin{split}
	\pee_{\rho}[P_{e'_1}=P_{e'_2}=1]
	&=
	\pee_{\rho}[P_{e'_1}=1]\pee_{\rho}[P_{e'_2}=1]=\Trace(P_{e_1}P_{e'_2})\Trace(P_{e_1}P_{e'_1})\\
	&\simeq 
	\Trace(P_{e_1}P_{e_2})\Trace(P_{e_1}P_{e_1})=\tfrac{1}{9}.
\end{split}
\end{equation}
So, not only does the proof of Cabello contain a flaw, the asserted lemma isn't true either. However, implicitly Cabello has shown that it is a necessary condition for the MKC-models that they violate (\ref{ContinuityCabello}).

\subsubsection{Non-Classicality}\label{ApplebyClassic}
The most extended criticism the MKC-models have attracted is due to Appleby \cite{Appleby00}, \cite{Appleby01}, \cite{Appleby02} and \cite{Appleby05}. From Appleby's investigation, three main objections against the MKC-models may be distinguished:
\begin{enumerate}
\item The MKC-models are not classical in some sense.
\item The MKC-models are contextual in some sense.
\item The MKC-models are not robust with regard to the finite precision of measurements.
\end{enumerate}
I chose to use the expression ``in some sense'', because in my opinion the validity of the first two claims is partly a matter of taste. That is, they depend on what one is supposed to expect from a hidden-variable theory. 
Appleby states that, although the MKC-models, strictly speaking, do nullify the Kochen-Specker Theorem\footnote{In the later two articles \cite{Appleby01} and \cite{Appleby05} (the first version of \cite{Appleby02} was written before \cite{Appleby01}) Appleby seems to doubt this earlier claim. I will return to this point in the next section.}, they do not nullify the essential point made by the Kochen-Specker Theorem, which, according to Appleby, is that
\begin{quote}
	``\ldots quantum mechanics (whether relativistic or not) is inconsistent with classical notions of physical reality.'' \cite[p. 1]{Appleby00}
\end{quote} 
This is, of course, nothing new. It was already clear in the founding days of quantum mechanics that this new theory was inconsistent with classical notions due to the wave-particle duality and the superposition principle.
So naturally, the question arises what is meant with these ``classical notions'' in this case. In \cite{Appleby00} three criteria are introduced:
\begin{itemize}
\item[\textbf{1)}] ``To each observable quantity characterizing a system there corresponds an objective physical quantity, which has a determinate value at every instant.''
\item[\textbf{2)}] ``An ideal, perfectly precise measurement gives, \textit{with certainty}, a value which \textit{exactly coincides} with the value which the quantity being measured objectively did possess, immediately before the measurement process was initiated.''
\item[\textbf{3)}] ``A non-ideal, approximate measurement gives, \textit{with high probability}, a value which is \textit{close} to the value which the quantity being measured objectively did possess, immediately before the measurement process was initiated.''
\end{itemize}
Criterion \textbf{1)} is the only one on which the original Kochen-Specker Theorem focuses. The fact that the MKC-models do satisfy it brings the alleged nullification. It may be good to dispose of a possible misconception as to why this criterion is satisfied. For each finite-dimensional Hilbert space, the MKC-models provide a function that assigns to each self-adjoint operator in a dense subset of all the self-adjoint operators a value in its spectrum in such a way that the FUNC-rule is satisfied. By the finite precision of measurement, it is then argued that every time one intends to measure an observable, actually an observable corresponding to a self-adjoint operator in the colorable set is being measured.
The other self-adjoint operators simply do not correspond to \textit{any observable} at all. I regard this as one of the few nice features of these models, since it implies that there are only (at most) countably many observables. 

It may also be noted that the MKC-models are in fact models about \textit{infinitely precise} measurements. Each of the observables in the theory can be measured with arbitrary precision and a measurement of an observable gives precisely the result it possesses. The only reason for introducing the notion of imprecise measurements is for arguing that the theory of quantum mechanics and the theory of MKC (both theories about precise measurements) are \textit{empirically} indistinguishable. Theoretically, however, they are fundamentally different (as they must be according to the Kochen-Specker Theorem). In fact, if one were able to perform infinitely precise measurements, one would be able to distinguish the MKC-models from quantum theory simply by trying to measure one of the observables that are supposed to exist according to quantum theory, but which is non-existent according to the MKC-models. It follows that the MKC-models do in fact satisfy the criteria 1) and 2) (as is also acknowledged by Appleby).

The rest of the article \cite{Appleby00} is devoted to proving that the MKC-models cannot satisfy criterion 3). However, a similar conclusion can already be drawn from the previous section. Indeed, if it is held that 3) should imply (\ref{ContinuityCabello}), it follows from Lemma \ref{Cabellolemma} that the MKC-models cannot satisfy 3) (this is also noted in \cite{Appleby01} and \cite{Appleby05}). 

In \cite{Appleby05} a more direct proof of the violation of 3) is given, motivated by looking at the coloring of Meyer (Proposition \ref{MeyerLemma}). The function $f$ introduced there is densely discontinuous, i.e., every non-empty open subset of $S^2\cap\mathbb{Q}^3$ contains both elements that are assigned the value 1 by $f$, as elements that are assigned the value 0. Appleby calls this feature ``patholologically'' discontinuous. It is taken to imply that any imprecise measurement of an observable $P_e$ (for some $e\in S^2\cap\mathbb{Q}^3$) might just as well yield the value $1-f(e)$ instead of the value $f(e)$ (since an imprecise measurement of $P_e$ will result in revealing the value of some $P_{e'}$ with $e'$ close to $e$). It is not a priori clear that the pure states in the MKC-models should also exhibit such densely discontinuous features. However, Appleby has proven that for any valuation function $f$ on any dense subset of $\mathcal{P}(\mathbb{C}^3)$, there is a non-empty open set on which $f$ is densely discontinuous. It therefore turns out that 3) is indeed necessarily violated by the MKC-models.

Is criterion 3) actually necessary for a hidden-variable theory? I don't see why this should be the case, and neither do Barrett and Kent \cite{Barrett-Kent04}.\footnote{It should be noted that \cite{Barrett-Kent04} was written after \cite{Appleby05}, despite its earlier publication in the same journal.} Appleby also seems to have some doubts about whether or not 3) should be a necessary condition, since a great deal of the papers \cite{Appleby01} and \cite{Appleby05} are devoted to arguing that the MKC-models in fact violate a criterion that is stronger than 3), namely, that measurements in the MKC-models do not furnish any information about the system at hand:
\begin{quote}
	``MKC focus on the point that, in their models, a measurement does always reveal the pre-existing value of \textit{something}. [\ldots] What they overlook is that, [\ldots] although the experimenter learns a value, s/he has no idea what it is a value \textit{of}. Consequently, the experimenter does not acquire any actual knowledge.'' \cite[p. 4]{Appleby05}
\end{quote}
This line of reasoning is of course true, but the conclusion drawn from it isn't. Although the experimenter doesn't know of which observable he/she has required the value, he/she does know \textit{approximately} of which observable it is the value. Thus an (imprecise) measurement doesn't provide an imprecise result of the observable one intended to measure but, instead, provides a precise result of an imprecise observable. In my opinion, this counts as actual knowledge. Moreover, performing the same measurement on an ensemble of systems, an experimenter does acquire knowledge about the macro state of the system. In that setting, it is no longer of importance what the actual observable being measured is (by the continuity of the probability distributions).  

Appleby goes on to argue that
\begin{quote}
	``What emerges from this is that PMKC\footnote{The P here stands for Pitowsky who was, in fact, the first to show that a densely defined valuation function on $S^2$ exists \cite{Pitowsky83}. 
	This model is non-constructive and relies on an unorthodox view on probability theory. 
	However, the main reason for omitting it here, is that I didn't find that it contributed much to the story.} have been asking the wrong question. 
	The important question is not: ``How much of $S^2$ [\ldots] can be coloured \textit{at all}?'' 
	But rather: ``How much of $S^2$ can be coloured in such a way that the colours are \textit{empirically knowable}?'' '' \cite[p. 4]{Appleby05}
\end{quote}
It seems to me that here it is Appleby who is asking the wrong question. By demanding that `the colors' should be empirically knowable, Appleby implicitly demands that the hidden variables $\lambda$ of the MKC-models should be empirically knowable. That is, the hidden variables are not allowed to be hidden. From a realist point of view this misses the point. The investigation of hidden-variable theories is based on the question whether or not quantum mechanics could be completed. That is, the question is if it is \textit{in principle} possible to assign definite values to all observables.\footnote{Where it remains a topic of discussion what should be considered to be the set of all observables.}  

In conclusion, I would say that it is a matter of taste whether one considers the MKC-models to be sufficiently classical or not and to what extent one requires a hidden-variable theory to be classical. The only hard fact is that the models violate Appleby's criterion 3). However, as it turns out, this is not the end of the story.

\subsubsection{An Empirical Discrepancy with Quantum Mechanics (Part II)}\label{ApplebyExperiment}

 In another attempt to prove that the MKC-models do not satisfy 3) (and that they are contextual in some sense), Appleby (in \cite{Appleby00}) proved a more disturbing feature of the MKC-models: if one takes into account the finite precision of measurement \textit{correctly}, the MKC-models empirically violate the predictions made by quantum mechanics. This is what is meant with the non-robustness of (iii).

Recall that (as argued above) neither the MKC-models, nor quantum mechanics define theories that incorporate the finite precision of measurements explicitly. So one has to keep in mind that what Appleby in fact shows is that a \textit{certain modification of the MKC-models} is incompatible with a \textit{certain modification of quantum theory}. Indeed, a great deal of the paper \cite{Appleby00} is devoted to presenting a modification of quantum theory that does allow one to compare measurement results of imprecisely measured (and therefore possibly incompatible) observables. Although of interesting nature, I will not discuss it here; the results of this modification that are necessary for this discussion will be presented along the way.

The argument takes place in the setting of the spin-1 particle. Consider, again, the measurement of the squared spin along some axis $r$. Although only the operators $\sigma^2_r\in\mathcal{P}_{CK}$ correspond to actual observables, in an actual experiment one may of course also intend to measure $\sigma^2_{r'}$ for some $\sigma^2_{r'}\notin \mathcal{P}_{CK}$ (especially since it may not be known to the experimentalist what the set $\mathcal{P}_{CK}$ exactly is).
So which values of $r'$ are acceptable? Appleby assumes that \textit{all} values of $r'$ are acceptable as long as they are ``finitely specifiable''. One may think of computable real numbers.\footnote{Roughly stated, this is the set of numbers that can be estimated up to arbitrary precision using an algorithm on a Turing machine. It is a countable set that not only incorporates all rationals, but also numbers like $\sqrt{2}$.} It doesn't actually matter much which set is taken, as long as three criteria are satisfied: 
\begin{itemize}
\item The set contains all the $r$ for which $\sigma^2_r\in\mathcal{P}_{CK}$ (let's denote this set $S_2^{CK}$). 
\item The set contains a subset $\{r_1,\ldots,r_n\}$ that is uncolorable. 
\item The set is countable.
\end{itemize}
Let $S_2^A$ be a set that satisfies the above three criteria. An object $\sigma_{r'}^2$ for $r'\in S_2^A\backslash S_2^{CK}$ may be called a \underline{pseudo-observable}.

Now, the intended measurement of a pseudo-observable $\sigma^2_{r'}$ necessarily results in the exposure of the value of $\sigma_r^2$ for some $r\in S_2^{CK}$. Appleby assumes that this direction $r$ is selected by some probability measure $\mu_{r'}$ on the set $S_2^{CK}$. For given $\lambda$, define
\begin{equation}
	p_\lambda(r'):=\mu_{r'}\{r\in S_2^{CK}\:;\:\lambda(\sigma_r^2)=1\},
\end{equation}
the probability of finding the value 1 if one intends to measure $\sigma_{r'}^2$, given that the particle is in the state $\lambda$. Of course, such a measure $\mu_{r'}$ may be assumed to exist for all $r'\in S_2^A$ (one may even take $\mu_{r'}$ to be the Dirac measure for the set $\{r'\}$ whenever $r'\in S_2^{CK}$). Then each state $\lambda$ induces a generalized state $\tilde{\lambda}:\{\sigma^2_{r'}\:;\:r'\in S_2^A\}\to\{0,1\}$ by
\begin{equation}
	\tilde{\lambda}(\sigma_{r'}^2):=
	\begin{cases}
		0,&\text{if }p_\lambda(r')<\frac{1}{2};\\
		1,&\text{if }p_\lambda(r')\geq\frac{1}{2}.
	\end{cases}
\end{equation}
It is not easy to give a clear interpretation of these generalized states. They don't contain any more information than that $\tilde{\lambda}(\sigma_{r'}^2)$ gives the most probable outcome of the measurement of $\sigma^2_{r'}$, given that the state is $\lambda$. Any extra information given by the measure $\mu_{r'}$ is thrown away. Also, note that in general for $r\in S_2^{CK}$ one does not have $\lambda(\sigma_r^2)=\tilde{\lambda}(\sigma_r^2)$.

Since $S_2^A$ is uncolorable, there exists a set of orthonormal directions $e_1,e_2,e_3\in S_2^A$ such that 
\begin{equation}\label{AppleSpecker}
 \tilde{\lambda}(\sigma_{e_1}^2)+\tilde{\lambda}(\sigma_{e_2}^2)+\tilde{\lambda}(\sigma_{e_3}^2)\neq2.
\end{equation}
Next, consider an experiment in which one actually intends to measure these pseudo-observables and let $\mathcal{M}_1,\mathcal{M}_2,\mathcal{M}_3$ denote the corresponding measurement results. In this setting, the following lemma can be proven.\footnote{In \cite{Appleby00} this result is only stated and a proof is omited.}

\begin{lemma}\label{impreciseApple}
	Consider a spin-1 particle in the state $\lambda$ and let $\sigma_{e_1}^2,\sigma_{e_2}^2,\sigma_{e_3}^2$ be a triple of pseudo-observables such that $e_1,e_2,e_3$ is an orthonormal basis and the relation (\ref{AppleSpecker}) holds. Assume that the measures $\mu_{e_1},\mu_{e_2},\mu_{e_3}$ are independent. Then the probability that $\mathcal{M}_1+\mathcal{M}_2+\mathcal{M}_3\neq2$ is greater than $\tfrac{1}{2}$.
\end{lemma}
\noindent
\textit{Proof:}\hspace*{\fill}\\
Consider the event $\mathcal{M}_1+\mathcal{M}_2+\mathcal{M}_3=2$. Since the measures $\mu_{e_1},\mu_{e_2},\mu_{e_3}$ are independent, the probability of this event can simply be calculated by using the product measure $\mu_{e_1}\times\mu_{e_2}\times\mu_{e_3}$ on the space $S_2^{CK}\times S_2^{CK}\times S_2^{CK}$. One then has
\begin{equation}
\begin{split}
	\pee\left[\mathcal{M}_1+\mathcal{M}_2+\mathcal{M}_3=2\right]
	=&
	\pee\left[\mathcal{M}_1=0,\mathcal{M}_2=\mathcal{M}_3=1\right]
	+\pee\left[\mathcal{M}_2=0,\mathcal{M}_1=\mathcal{M}_3=1\right]\\
	&+\pee\left[\mathcal{M}_3=0,\mathcal{M}_1=\mathcal{M}_2=1\right]\\
	=&
	(1-p_\lambda(e_1))p_\lambda(e_2)p_\lambda(e_3)
	+p_\lambda(e_1)(1-p_\lambda(e_2))p_\lambda(e_3)\\
	&+p_\lambda(e_1)p_\lambda(e_2)(1-p_\lambda(e_3)).
\end{split}
\end{equation}  
For the sake of notational convenience, set $p_i=p_\lambda(e_i)$ for $i=1,2,3$.
By symmetry, one may assume that $p_1\leq p_2\leq p_3$.
Using relation (\ref{AppleSpecker}), the following cases can be distinguished. Either 1) $ p_1\geq\tfrac{1}{2}$, or 2) $p_3<\tfrac{1}{2}$ or 3) $p_2<\tfrac{1}{2}$ and $p_3\geq \tfrac{1}{2}$. For the remainder of the proof, recall that for any $p\in[0,1]$, one has $p(1-p)\leq\tfrac{1}{4}$.
\indent
 
Case 2) is easy. One has
\begin{equation}
\begin{split}
	\pee\left[\mathcal{M}_1+\mathcal{M}_2+\mathcal{M}_3=2\right]
	=&
	(1-p_1)p_2p_3+p_1(1-p_2)p_3+p_1p_2(1-p_3)\\
	<&
	\frac{1}{2}p_2+p_1(1-p_3)p_3+p_1p_3(1-p_3)\\
	\leq&
	\frac{1}{2}p_2+p_1\frac{1}{4}+p_1\frac{1}{4}\leq p_2<\frac{1}{2}.
\end{split}
\end{equation}
For case 1), set $f(p_2,p_3)=(1-p_2)p_3+p_2(1-p_3)$. It follows that
\begin{multline}
	\pee\left[\mathcal{M}_1+\mathcal{M}_2+\mathcal{M}_3=2\right]
	=
	(1-p_1)p_2p_3+p_1(1-p_2)p_3+p_1p_2(1-p_3)\\
\begin{split}
	=&
	(1-p_1)\left(p_2p_3-f(p_2,p_3)\right)	+f(p_2,p_3)\\
	\leq&
	\sup\left\{\frac{1}{2}\left(p_2p_3-f(p_2,p_3)\right)
	+f(p_2,p_3)\:;\:p_2p_3-f(p_2,p_3)\geq0,p_2,p_3\in[\frac{1}{2},1]\right\}\\
	&\vee
	\sup\left\{f(p_2,p_3)\:;\:p_2p_3-f(p_2,p_3)<0,p_2,p_3\in[\frac{1}{2},1]\right\}\\
	\leq&
	\sup\left\{\frac{1}{2}\left(p_2p_3+f(p_2,p_3)\right)\:;\:p_2,p_3\in[\frac{1}{2},1]\right\}
	\vee
	\sup\left\{f(p_2,p_3)\:;\:p_2,p_3\in[\frac{1}{2},1]\right\}\\
	&=
	\frac{1}{2}\vee\frac{1}{2}=\frac{1}{2}.
\end{split}
\end{multline}
Finally, for case 3) one has
\begin{equation}
	\begin{split}
	\pee\left[\mathcal{M}_1+\mathcal{M}_2+\mathcal{M}_3=2\right]
	=&
	(1-p_1)p_2p_3+p_1(1-p_2)p_3+p_1p_2(1-p_3)\\
		=&
		p_2\left(p_1(1-p_3)+(1-2p_1)p_3\right)+p_1p_3.\\
	\end{split}
\end{equation}
Note that both $(1-p_3)\geq0$ as $(1-2p_1)\geq0$ so the factor for $p_2$ is always positive. Therefore
\begin{equation}
	\begin{split}
	\pee\left[\mathcal{M}_1+\mathcal{M}_2+\mathcal{M}_3=2\right]
	<&
		\frac{1}{2}\left(p_1(1-p_3)+(1-2p_1)p_3\right)+p_1p_3\\
	=&
		\frac{1}{2}\left(p_1+p_3-p_1p_3\right)\leq\frac{1}{2}.
	\end{split}
\end{equation}
 
So for each of the three possible cases one has
\begin{equation}
 	\pee\left[\mathcal{M}_1+\mathcal{M}_2+\mathcal{M}_3\neq2\right]=
 	1-\pee\left[\mathcal{M}_1+\mathcal{M}_2+\mathcal{M}_3=2\right]\geq\frac{1}{2}.
\end{equation}
\hfill $\square$\\[0.5ex]

Appleby then concludes that if the finite precision of measurement is taken into account correctly (namely by introducing the measures $\mu_{r}$), the MKC-models provide a non-negligible probability ($\geq\tfrac{1}{2}$) to find a measurement result that contradicts the FUNC-rule, no matter how precise the measurement becomes (i.e., no matter how much the measures $\mu_r$ approach the appropriate Dirac measures). On the other hand, Appleby's theory of imprecise measurements for quantum mechanics predicts that in that theory this probability goes to zero as measurements become more precise. He then reaches his conclusion:
\begin{quote}
	``This establishes that, if the alignment errors are random, and statistically independent, then the model must exhibit a form of contextuality: for it means that the probable outcome of an approximate measurement must, in general, be strongly dependent, not only on the observable which is being measured, but also on the particular way in which the measurement is carried out. It follows that, if the stated assumptions are true, the model fails to satisfy clause 3)\ldots'' \cite[p. 17]{Appleby00}
\end{quote}

Although Appleby focuses his objections on the statement that the MKC-models cannot satisfy clause 3), I think the direct consequence of Lemma \ref{impreciseApple}, stating that the MKC-models make predictions that are empirically different from the predictions made by quantum theory, is the strongest objection that can be made against these models. It is indeed a bit strange that Appleby doesn't come back to this point extensively in any of his later articles.  

The reply of Barrett and Kent to this objection isn't very extensive either.
They simply state that the assumption that imprecise measurements may lead to simultaneous measurements of observables that do not correspond to commuting observables is wrong. In their own words:
\begin{quote}
	``\ldots in a CK model for projective measurements, the projectors actually measured are always commuting (assuming that they are measured simultaneously)--this is one of the axioms of the theory that relate its mathematical structure to the world\ldots'' \cite[p. 163]{Barrett-Kent04}
\end{quote}
This defense works only as long as the pseudo-observables $\sigma_{e_1}^2,\sigma_{e_2}^2$ and $\sigma_{e_3}^2$ are measured simultaneously, for then the entire procedure can be regarded as the measurement of a single pseudo-observable (which then results in the measurement of a single real observable). In this case, the measures $\mu_{e_1},\mu_{e_2},\mu_{e_3}$ in Lemma \ref{impreciseApple} are not independent.

However, quantum mechanics predicts the same results irrespective of whether the pseudo-observables are measured simultaneously or sequentially. Appleby argues that in the case of sequential measurements, requiring that the measures $\mu_{e_1},\mu_{e_2},\mu_{e_3}$ are dependent leads to complications.  
He thinks of a measurement where one selects three directions $e_1,e_2,e_3$ that are very close to being orthogonal but with the property that one is still able to measure that they are not. That is, he tries to trick nature by acting as if the precision of measurement is not as good as it actually is. Then, since the observables that are actually measured must commute, either it must be physically impossible to set up this experiment, or, when the measurement is performed, there is some force that actually changes the alignment of the set up (in a way noticeable for the experimenter). Either way, the model provides a definite empirical prediction (which is distinguishable from predictions made by quantum mechanics). 

As they stand, the MKC-models do not provide an obvious way to escape these difficulties. It therefore seems that they are indeed not robust with regard to the finite precision of measurements, in a very drastic way. But, as noted earlier, the only thing proven is that a certain \textit{modification} of the MKC-models is in conflict with predictions made by quantum mechanics. The question is left open of whether every modification should posses this property. I will show in the next section that this is not the case, by constructing an explicit counterexample.


\subsection[A Modification of the MKC-Models for Imprecise Measurements]{A Modification of the MKC-Models for Imprecise\\ Measurements}\label{MyModification}
\markboth{The Alleged Nullification of the Kochen-Specker Theorem}{A Modification of the MKC-Models for Imprecise Measurements}

To examine the behavior of the MKC-models for sequential measurements, it is good to look at the way quantum mechanics deals with this. Considering the situation of Lemma \ref{impreciseApple}, after the first measurement of $\sigma_{e_1}^2$, the von Neumann postulate implies that the state of the system changes instantaneously in such a (discontinuous) way that the measurement results will satisfy $\mathcal{M}_1+\mathcal{M}_2+\mathcal{M}_3=2$ with high probability.\footnote{Other mechanisms that preserve the empirical relations like $\mathcal{M}_1+\mathcal{M}_2+\mathcal{M}_3=2$ have been suggested to prevent the use of the von Neumann postulate. However, that is not important for this discussion.} The new state also makes the system robust with regard to the finite precision of measurements. It may be possible to introduce a similar postulate in the MKC-models. Barrett and Kent state that this is indeed the case:
\begin{quote}
	``If the projectors are measured sequentially, then the rules of the model stipulate that the hidden state changes discontinuously after the measurement and Appleby's analysis no longer applies.'' \cite[p. 163]{Barrett-Kent04}
\end{quote}
This discontinuous change is supposed to take place in such a way
\begin{quote}
	``that the probability distribution of the post-measurement hidden variables correspond to that defined by the post-measurement quantum mechanical state vectors.'' \cite[p. 160]{Barrett-Kent04}
	\end{quote}
However, it is not clear how this discontinuous change should effect the pure state (of the hidden variable model) of a single system and if this can be done in such a way that Lemma \ref{impreciseApple} can no longer be applied. Indeed, the following comment by Appleby still seems to apply:
\begin{quote}
	``The models discussed by MKC are incomplete, since they do not include a specification of the dynamics. It is a highly non-trivial question, as to whether there exists a dynamics which, in every situation, gives rise to the probability distribution having the desired properties--not only in situations where the alignment errors arise ``naturally'', but also in situations where the errors are adjusted ``by hand'' (in the manner described in the last paragraph) [i.e., Appleby's procedure to trick nature]'' \cite[p. 17]{Appleby00}
\end{quote} 

The difficulty that arises when trying to consider a possible modification is that an immediate change of the pure state $\lambda$ is required to ensure robustness, whilst the pure states themselves have no direct relation with quantum mechanics. 
In line with Appleby, one may assume that for each (imprecise) measurement of the (pseudo-)observable $P$, there is a probability measure $\mu_P$ on $\mathcal{P}_{CK}(\h)$ that selects the actual observable of which the value is revealed upon measurement. 
To investigate what properties the immediate state change should obey, the following definition is introduced.

\begin{definitie}
For $\epsilon>0$, an imprecise measurement of the pseudo-observable $P$ is called \underline{$\epsilon$-precise} if 
\begin{equation}
	\mu_P\{P'\in\mathcal{P}_{CK}(\h)\:;\:\|P'-P\|>\epsilon\}<\epsilon.
\end{equation}
\end{definitie}

Now consider the sequential measurement of two pseudo-observables $P_1,P_2$ with $\|P_2P_1\|<\delta$ for some $\delta>0$. Suppose the first measurement of $P_1$ yielded the result 1 and let $\lambda'$ denote the state of the system after the measurement. If all measurements are $\epsilon$-precise, $\lambda'$ is required to satisfy the following relations:
\begin{equation}
\begin{gathered}
	\mu_{P_1}\{P'\in\mathcal{P}_{CK}(\h)\:;\:\lambda'(P')=1\}\to1,\quad\text{as }\epsilon,\delta\to0;\\
	\mu_{P_2}\{P'\in\mathcal{P}_{CK}(\h)\:;\:\lambda'(P')=1\}\to0,\quad\text{as }\epsilon,\delta\to0.
\end{gathered}
\end{equation}
Equivalently, if the measurement of $P_1$ yielded the result 0:
\begin{equation}
	\mu_{P_1}\{P'\in\mathcal{P}_{CK}(\h)\:;\:\lambda'(P')=1\}\to0,\quad\text{as }\epsilon,\delta\to0.\\
\end{equation}
The most natural way to establish this would be to require that after the measurement of $P_1$ yielded the result 1, the state changes to a state $\lambda'$ that assigns the value 1 to all $P\in P_{CK}(\h)$ in some neighborhood of $P_1$ and the value 0 to all $P\in P_{CK}(\h)$ that are in the neighborhood of some $P'$ perpendicular to $P_1$. Similarly if the measurement yielded the result 0.

This seems an impossible task, since it was shown by Appleby that \textit{every} $\lambda$ is densely discontinuous in a certain region $D_\lambda$. However, Appleby made no specific claim about the regions where $\lambda$ can be continuous. Indeed, Appleby did not (extensively) answer his own question about how much of $S^2$ can be colored in such a way that the colors are empirically knowable. It turns out that the MKC-models are sufficiently flexible to allow colorings that are empirically knowable.
	
\begin{propositie}\label{ContinuousMKC}
For every finite-dimensional Hilbert space $\h$, for every unit vector $e\in\h$ with $P_e\in\mathcal{P}_{CK}(\h)$, there are open subsets $U,U'$ of $\mathcal{P}(\h)$ (with respect to the norm on $B(\h)$) and a non-empty subset $\Lambda_e\subset\Lambda$ such that
\begin{enumerate}
\item $P_e\in U$ and $P_{e'}\in U'$ for all $e'\perp e$.
\item For all $\lambda\in\Lambda_e$ $\lambda(P)=1$ for all $P\in U\cap\mathcal{P}_{CK}(\h)$ and $\lambda(P)=0$ for all $P\in U'\cap\mathcal{P}_{CK}(\h)$.
\end{enumerate}
\end{propositie}
\noindent
\textit{Proof:}\hspace*{\fill}\\
Let $e$ be given and define $U$ and $U'$ as follows:
\begin{equation}
\begin{gathered}
	U:=\left\{P\in\mathcal{P}(\h)\:;\: \|PP_e\|>\frac{1}{2}\right\},\\
	U':=\left\{P\in\mathcal{P}(\h)\:;\: \|PP_e\|<\frac{1}{n}\right\},
\end{gathered}
\end{equation}
where $n$ is the dimension of $\h$. It is easy to see that these sets are open.\footnote{Indeed, if $P\in U$ choose $0<\epsilon<\|PP_e\|-\tfrac{1}{2}$. Then $P'\in U$ whenever $\|P'-P\|<\epsilon$ (for all $P'\in\mathcal{P}(\h)$):
\begin{equation*}
	\|PP_e\|\leq\|P-P'\|\|P_e\|+\|P'P_e\|<\epsilon+\|P'P_e\|,
\end{equation*}
thus $\|P'P_e\|>\|PP_e\|-\epsilon>\frac{1}{2}$. A similar argument shows that $U'$ is open.}
\indent

Recall that each $\lambda$ is completely defined by its action on the one-dimensional projections (by Lemma \ref{CliftonKentColoring}). Now, define 
\begin{equation}
	\Lambda_e:=\left\{\lambda_f\in\Lambda\:;\:\substack{f(m)=j\text{ whenever }P_{e_j^{(m)}}\in U,\\ f(m)\neq j\text{ whenever }P_{e_j^{(m)}}\in U'}\right\}.
\end{equation}
It follows directly from this definition that $\Lambda_e$ satisfies criterion (ii). To show that it is not empty, the following two statements have to be proven for every orthonormal basis $e_1,\ldots,e_n$ of $\h$:
\begin{itemize}
\item[1)] If $P_{e_1}\in U$, then $P_{e_j}\notin U$ for all $j>1$ (i.e., at most one of the $P_{e_i}$ lies in $U$).
\item[2)] If $P_{e_1},\ldots,P_{e_{n-1}}\in U'$, then $P_{e_n}\notin U'$.
\end{itemize}

To prove 1) and 2), it is useful to note that for every projection $P$ one has
\begin{equation}
\begin{split}
	\|PP_e\|
	&=
	\sup\{\|PP_e\psi\|\:;\:\psi\in\h,\:\|\psi\|=1\}
	=
	\sup_{\{\psi\:;\:\|\psi\|=1\}}\sqrt{\langle PP_e\psi,PP_e\psi\rangle}\\
	&=
	\sup_{\{\psi\:;\:\|\psi\|=1\}}\sqrt{\langle P_e\psi,PP_e\psi\rangle}
	=
	\sup_{\{\psi\:;\:\|\psi\|=1\}}\sqrt{\langle \psi,e\rangle\langle e,Pe\rangle\langle e,\psi\rangle}
	=
	\|Pe\|.
\end{split}
\end{equation}
Consequently, whenever $P$ and $P'$ are orthogonal projections, one has
\begin{equation}\label{LineaireNorm}
	\|(P+P')P_e\|=\|PP_e\|+\|P'P_e\|.
\end{equation}
Now let $e_1,\ldots,e_n$ be any orthonormal basis. Suppose $P_{e_1}\in U$. Then for all $j>1$
\begin{equation}
	1\geq \|(P_{e_1}+P_{e_j})P_e\|=\|P_{e_1}P_e\|+\|P_{e_j}P_e\|>\frac{1}{2}+\|P_{e_j}P_e\|,
\end{equation}
hence $\|P_{e_j}P_e\|<\tfrac{1}{2}$. Thus $P_{e_j}\notin U$.

To prove the second statement, suppose $P_{e_i}\in U'$ for all $i=1,\ldots,n-1$. Using (\ref{LineaireNorm}), it follows that
\begin{equation}
	\|P_{e_n}P_e\|
	=
	\|\een P_e\|-\|\sum_{i=1}^{n-1}P_{e_i}P_e\|
	=
	1-\sum_{i=1}^{n-1}\|P_{e_i}P_e\|
	>
	1-(n-1)\frac{1}{n}=\frac{1}{n}.
\end{equation}
This proves the proposition.
\hfill $\square$\\[0.5ex]

Using this proposition, it may be argued that after the measurement of any (pseudo-)observable $P$, the following procedure takes place: let $\lambda$ denote the state of the system before the measurement and let $P'$ be the observable whose value was actually displayed by the measurement. If $\lambda(P')=1$, the state changes to some $\lambda'\in\Lambda_e$ for some unit vector $e$ with $P_e\leq P'$. If $\lambda(P')=0$, the state changes to some $\lambda'\in\Lambda_e$ for some unit vector $e$ with $P_e\leq P'^\perp$.
This procedure ensures that the MKC-models are robust with regard to sequential imprecise measurements of almost commuting pseudo-observables. As a consequence, Appleby's Lemma \ref{impreciseApple} no longer applies to this modification.

The above-described procedure is a very drastic one to ensure robustness. For example, the sets $U$ and $U^\perp$ of Proposition \ref{ContinuousMKC} could have been taken a lot smaller. But there is also a more natural way to select a new $\lambda'$. 
Consider a system in the macro state $\pee_{\rho}$. If the measurement of the (pseudo-)observable $P$ on an individual system yielded the result 1, a new state $\lambda'$ may be selected at random in accordance with the measure $\pee_{P\rho P}$. Equivalently, if the result was 0, a new state may be selected in accordance with the measure $\pee_{P^\perp\rho P^\perp}$. 

This scheme enables one to define the entire dynamics for the models. Indeed, one may consider that the actual state of an individual system $\lambda(t)$ changes rapidly and stochastically in time, precisely in such a way that for every $t$ one has
\begin{equation}
	\pee[\lambda(t)\in\Delta]=\pee_{\rho(t)}[\Delta],\quad\forall\Delta\subset\Lambda,
\end{equation}
where $\rho(t)$ is given by the Schr\"odinger postulate (when no measurement is performed) and the von Neumann postulate (when a measurement is performed). More formally, the structure of the non-contextual hidden variable theory may be defined by the following postulates:
\begin{itemize}
\item[\textbf{1.}] \textbf{State Postulate:} For every physical system there is a set of pure states $\Lambda$, defined as the set of valuation functions on $\mathcal{P}_{CK}(\h)$ (which is well-defined by Theorem \ref{CliftonKentTheorem} and Lemma \ref{CliftonKentColoring}) , where $\h$ denotes the finite-dimensional Hilbert space associated with the system in quantum mechanics. The evolution of the state is given by a function $\omega\in\Omega=\Lambda^{\mathbb{R}}$. At each time $t$ the pure state $\omega(t)\in\Lambda$ gives a complete description of the system.
\item[\textbf{2.}] \textbf{Observable Postulate:} With each physical observable $\A$, there is associated a function $\hat{A}:\Lambda\to\mathbb{R}$ that assigns to each state the value possessed by the observable $\A$ in that state. A measurement of the observable $\A$ at the time $t$ yields the value $\hat{A}(\omega(t))$.
\item[\textbf{3.}] \textbf{Dynamics Postulate:} The evolution of the state of a single system is described by a stochastic process $(X_t)_{t\in\mathbb{R}}$ with filtration $(\mathcal{F}_t)_{t\in\mathbb{R}}$ on the space $(\Omega,\Sigma,\pee)$ where 
\begin{equation}
	X_t(\omega):=\omega(t),
\end{equation}
and $\mathcal{F}_t$ is the smallest $\sigma$-algebra generated by $(X_{t'})_{t'\leq t}$ and $\Sigma$ is the smallest $\sigma$-algebra containing $\cup_{t\in\mathbb{R}}\mathcal{F}_t$. At each time $t$, the probability that $\lambda\in\Delta$ for any subset $\Delta\subset\Lambda$ is given by $\pee[X_t^{-1}(\Delta)]$.
\end{itemize}
These three postulates indeed imply that the theory at hand is a stochastic non-contextual theory, which in principle admits a realist interpretation. The following postulates only serve to ensure empirical equivalence with quantum mechanics.
\begin{itemize}
\item[\textbf{4.}] \textbf{Extended Observable Postulate:} With each physical observable $\A$, there is also associated a self-adjoint operator $A$ for which the spectral decomposition $A=\sum_{a\in\sigma(A)}aP_a$ satisfies $P_a\in\mathcal{P}_{CK}(\h)$ $\forall a\in\sigma(A)$. The relation between $\hat{A}$ and $A$ is given by
\begin{equation}
	\hat{A}(\lambda)=\sum_{a\in\sigma(A)}a\lambda(P_a).
\end{equation}
\item[\textbf{5.}] \textbf{Extended Dynamics Postulate:} The probability measure $\pee$ on the space $(\Omega,\Sigma)$ satisfies
\begin{equation}
	\pee[X_t^{-1}(\Delta)]=\pee_{\rho(t)}[\Delta],\quad\forall\Delta\subset\Lambda,
\end{equation}
where the right-hand side is defined by Theorem \ref{nc-statistics} and $\rho(t)$ is the quantum-mechanical state of the system whose time-evolution is given by the Schr\"odinger and von Neumann postulates.
\end{itemize}
This last axiom ensures that the probability of finding a value in $\Delta$ upon measurement of the observable $\A$ at the time $t$ is given by
\begin{equation}
\begin{split}
	\pee\left[X_t^{-1}\left(\hat{A}^{-1}(\Delta)\right)\right]
	&=
	\pee\left[X_t^{-1}\left(\cup_{a\in\Delta}\{\lambda\:;\:\lambda(P_a)=1\}\right)\right]\\
	&=
	\pee_{\rho(t)}\left[\cup_{a\in\Delta}\{\lambda\:;\:\lambda(P_a)=1\}\right]\\
	&=
	\sum_{a\in\Delta}\pee_{\rho(t)}\left[\{\lambda\:;\:\lambda(P_a)=1\}\right]\\
	&=
	\sum_{a\in\Delta}\Trace(\rho(t)P_a)=\Trace(\rho(t)\mu_A(\Delta)),
\end{split}
\end{equation}	
where equation (\ref{KansEenProjectie}) is used to arrive at the last line. As noted earlier, the continuity property of the probability measures ensures that an imprecise measurement of an observable $\A$ yields, with approximately the same probability, approximately the same result (see (\ref{ContinueMaat})). This is the sincerest form of robustness one may expect for a \textit{stochastic} theory and in this sense, Appleby's objection (iii) (Section \ref{ApplebyClassic}) no longer applies. 

As a final remark it may be noted that the theory defined by the above five postulates is in fact fundamentally indeterministic. Indeed, the actual probability measure $\pee$ is only determined up to the time $t$ since $\rho(t)$ is not determined for all $t$. That is, unless one argues that it is also actually determined which experiments will be performed by experimenters, the von Neumann postulate implies that $\rho$ actually evolves indeterministically.


\subsection{Non-Locality of the MKC-Models}\label{NonLocalMKC}
\markboth{The Alleged Nullification of the Kochen-Specker Theorem}{Non-Locality of the MKC-Models}
An interesting light may be shed on the MKC-Models when studying them in the context of the EPRB-experiment (Example \ref{EPRB}) and the Bell inequality for this experiment (Section \ref{The Bell Inequality}). The original MKC-model is extremely non-local for this system: it turns out that almost every measurement on the first particle also interacts with the second particle and vice versa. This can be seen as follows.  The Hilbert space associated with the pair of spin-$\tfrac{1}{2}$ particles is $\mathbb{C}^4$. In quantum mechanics, an observable for the first particle corresponds with an operator of the form $A\otimes\een$, where $A$ is some self-adjoint operator acting on the space $\mathbb{C}^2$. But even if $A$ corresponds to an observable in the MKC-model, $A\otimes\een$ will most likely be a pseudo-observable. For suppose $A\otimes\een$ and $\een\otimes A'$ are two observables with $A=a_1P_1+a_2P_2$, $A'=a_1'P_1'+a_2'P_2'$. Then 
\begin{equation}
	P_1\otimes P_1',\quad P_1\otimes P_2',\quad P_2\otimes P_1',\quad P_2\otimes P_2'
\end{equation}
forms a resolution of the identity. Since each projection operator is only allowed to appear in one resolution of the identity, there can be no other observables of the form $B\otimes\een$ or $\een\otimes B'$ unless $[A,B]=0$ or $[A',B']=0$. Consequently, a measurement of a (pseudo-)observable $A\otimes\een$ will result in the measurement of some observable $(A\otimes\een)'$ close to $A\otimes\een$, which is most likely not of the form $A'\otimes\een$ (and probably not even of the form $A'\otimes\een'$).\footnote{The exact form of these observables depends, of course, on how one exactly constructs $\mathcal{P}_{CK}(\mathbb{C}^4)$. It is an open question if it can be constructed in such a way that it respects the structure of $\mathbb{C}^2\otimes\mathbb{C}^2$.} In general, it is not easy to find an interpretation of an observable that is not factorizable (like $(A\otimes\een)'$), but it is generally agreed that it is a non-local observable, that is, a measurement of the observable requires an interaction with both particles.\footnote{A similar sort of reasoning can be found in \cite{Appleby02} to argue for a certain form of contextuality. But even if the (in my opinion somewhat vague) argument presented there appeals to the reader, it may be disposed of by the modified MKC-models as I will indicate later.} 

Despite this extreme non-local property, it does not automatically follow that the MKC-model should actually violate the Bell inequality. This is because the EPRB-experiment is generally expected to involve two subsequent measurements (which procedure is not yet well-understood in the original MKC-models). In fact, if one would use an approach similar to the one used by Appleby in Section \ref{ApplebyExperiment}, it would follow that the MKC-model does satisfy the Bell inequality. That is, unless it is assumed that the spins of the two particles are measured simultaneously every time. In this case the Bell inequality is violated in a somewhat peculiar way. Each measurement reveals the value of some observable $(\sigma_{r_1}\otimes\sigma_{r_2})'$ in the neighborhood of $\sigma_{r_1}\otimes\sigma_{r_2}$. But whereas in quantum mechanics $\sigma_{r_1}\otimes\sigma_{r_2}$ can be viewed as the product of two observables $\sigma_{r_1}\otimes\een$ and $\een\otimes\sigma_{r_2}$, it is in general not the case that $(\sigma_{r_1}\otimes\sigma_{r_2})'$ can be seen as the product of two observables in the neighborhood of $\sigma_{r_1}\otimes\een$ and $\een\otimes\sigma_{r_2}$. This factorizability is indeed a necessary condition to derive a Bell inequality (see for example \cite{Bub97}). Consequently, the estimates made in (\ref{BellAfschatting1}) and (\ref{BellAfschatting2}) no longer hold. In other words, the measurement of $\sigma_{r_1}\otimes\sigma_{r_2}$  can no longer be viewed as measuring the two spins and then taking their product in the MKC-model.

The modified MKC-models introduced in the previous section can violate the Bell inequality by the discontinuous state change upon measurement. It is easy to see that this entitles a violation of OILOC (while CILOC is left intact). But although a new form of non-locality appears, one may argue that the extreme form of non-locality can be dropped in the modified models. 

The question is actually how a system consisting of several subsystems should be described in the context of the MKC-model. 
In a straightforward approach, for a system consisting of two subsystems, the set of states would be given by the set of all valuation functions on $\mathcal{P}_{CK}(\h_1\otimes\h_2)$, where $\h_1$ and $\h_2$ are the two Hilbert spaces associated with the subsystems in quantum mechanics. This approach leads to the extreme non-locality discussed above. However, instead of first trying to define the states, it is better to first look for a definition of appropriate observables. It seems reasonable to assume that any measurement on the composite system first requires an interaction with one subsystem and then with the other. For example, a measurement of $A\otimes B$ consists of a measurement of $A$ on system 1, a measurement of $B$ on system 2, and finally of taking the product of the results. Since quantum mechanics predicts that the result will be independent of the order in which $A$ and $B$ are measured, one can unambiguously speak of the measurement of $A\otimes B$. It therefore seems reasonable that each observable for the composite system should be a function of observables for the individual systems. The set of observables obtained in this way will in general no longer be colorable. Consider for example the following set of observables for the pair of two spin-$\tfrac{1}{2}$ particles, which is uncolorable\footnote{This example is taken from \cite{Mermin93}.}:
\begin{equation*}
\xymatrix{
		\sigma_x\otimes\een \ar@{-}[r]\ar@{-}[d] &
		\een\otimes\sigma_x \ar@{-}[r]\ar@{-}[d] &
		\sigma_x\otimes\sigma_x \ar@{-}[d] \\ 
		\een\otimes\sigma_y \ar@{-}[r]\ar@{-}[d] &
		\sigma_y\otimes\een \ar@{-}[r]\ar@{-}[d] &
		\sigma_y\otimes\sigma_y \ar@{-}[d] \\ 
		\sigma_x\otimes\sigma_y \ar@{-}[r] &
		\sigma_y\otimes\sigma_x \ar@{-}[r] &
		\sigma_z\otimes\sigma_z}
\end{equation*}
It is plausible that there exists an orthonormal basis $x,y,z$ such that all these operators would actually correspond to observables in this theory. However, in the scheme illustrated above the only way to measure $\sigma_y\otimes\sigma_y$ is by a measurement of $\sigma_y\otimes\een$ and a measurement of $\een\otimes\sigma_y$. A measurement of the pair $\sigma_y\otimes\sigma_y,\sigma_x\otimes\sigma_x$ would actually involve measurements of all four of the `local' observables that appear in the square. A disturbance due to the von Neumann postulate would therefore render this example meaningless. Indeed, for the pure states the functional relations (i.e., the FUNC rule) are only required to be satisfied locally, whereas the von Neumann postulate ensures that the relations are also satisfied for non-local measurements. 

The above discussion can be summarized as follows: for a composite system, consisting of two subsystems, the set of pure states is given by $\Lambda_1\times\Lambda_2$, where $\Lambda_i$ is the set of all valuation functions on $\mathcal{P}_{CK}(\h_i)$. The evolution of the state of the composite system ($\omega:\mathbb{R}\to\Lambda_1\times\Lambda_2$) may now be viewed as described by two coupled stochastic processes guided by the density operators $\rho_1$ and $\rho_2$ that are derived from the density operator $\rho$ on the space $\h_1\otimes\h_2$ and the use of the partial trace, i.e. $\rho_1=\Trace_2(\rho)$ where $\Trace_2:B(\h_1\otimes\h_2)\to B(\h_1)$ is the unique linear operator that satisfies
\begin{equation}
	\Trace_2(A_1\otimes A_2)=\Trace(A_2)A_1,\quad\forall A_1\in B(\h_1),A_2\in B(\h_2),
\end{equation}
and similarely $\rho_2=\Trace_1(\rho)$. The procedure to cope with these partial traces becomes clearer in an example.

Let $\A$ be an observable for the first subsystem and $\mathcal{B}$ an observable for the second system, and suppose one wants to measure their product $\A\mathcal{B}$. The self-adjoint operator associated with this observable is denoted by $A\otimes B$, and the corresponding stochastic variable is given by
\begin{equation}
	\widehat{AB}:\Lambda_1\times\Lambda_2\to\mathbb{R},\quad \widehat{AB}(\lambda_1,\lambda_2):=\hat{A}(\lambda_1)\hat{B}(\lambda_2).
\end{equation}
Suppose a measurement of $\A\mathcal{B}$ first entails an interaction with subsystem 2 and then with subsystem 1. Let $t$ be the time at which the measurement was initiated and let $t'$ be the time at which the interaction with subsystem 2 has taken place ($t$ and $t'$ may be assumed arbitrarily close to each other). The probability that the measurement yields the result $x$ may then be calculated as follows:
\begin{equation}
\begin{split}
	\pee[\widehat{AB}\in\{x\}]
	&=
	\sum_{b\in\sigma(B)}\pee_{\rho_1(t')}\left[\hat{A}\in\{x/b\}|\hat{B}\in\{b\}\right]\pee_{\rho_2(t)}\left[\hat{B}\in\{b\}\right]\\
	&=
	\sum_{b\in\sigma(B)}1_{\sigma(A)}(x/b)
	\frac{\Trace\left((\een\otimes P_b\rho(t)\een\otimes P_b)P_{x/b}\otimes\een\right)}{\Trace\left(\een\otimes P_b\rho(t)\een\otimes P_b\right)}
	\Trace(\rho(t) \een\otimes P_b)\\
	&=
	\sum_{b\in\sigma(B)}1_{\sigma(A)}(x/b)\Trace\left(\rho(t)P_{x/b}\otimes P_b\right)\\
	&=
	\sum_{\substack{(a,b)\in\sigma(A)\times\sigma(B)\\ ab=c}}\Trace\left(\rho(t)P_{a}\otimes P_b\right),
	\end{split}
\end{equation}
which is exactly the probability predicted by quantum mechanics. It follows that this probability is independent of whether the measurement apparatus first interacts with subsystem 1 or with subsystem 2. The requirement that this interaction between the two subsystems (i.e. the conditionalizing of the probabilities) must take place no matter how close $t'$ is to $t$, in fact leads to a violation of OILOC. This violation thus comes as a blessing for this theory since it enables one to avoid the extreme non-locality discussed at the beginning of this section.


\subsection{Conclusion}
\markboth{The Alleged Nullification of the Kochen-Specker Theorem}{Conclusion}
In my opinion, the modified version of the MKC-models is a non-contextual hidden-variable theory. If it is taken to be the sole statement of the Kochen-Specker Theorem that theories that assign definite values to all observables in a \textit{non-contextual} way (consistent with the algebraic relations holding in quantum mechanics) are impossible to construct, one might say that the theorem has indeed been `nullified'. However, the term `nullification' seems to imply that the entire theorem may be rendered useless. With this statement I disagree, and instead I tend to agree with Appleby that
\begin{quote}
	``The PMKC models do not nullify the Bell-KS theorem. Instead, they give us a deeper and more accurate insight into what the theorem is telling us.'' \cite[p. 22]{Appleby05}
\end{quote}
Appleby states that what the Kochen-Specker Theorem is telling us, is actually a deeper statement than the one that there are no non-contextual hidden-variable theories. He adopts Bell's view on the theorem that the main conclusion to be drawn is that
\begin{quote}
	``\ldots the result of the measurement does not actually tell us about some property previously possessed by the system\ldots'' \cite{Bell71}
\end{quote}   
But, as argued before, I think this statement is too strong and does not apply to the MKC-models. Instead, the main statement may be taken that point 3) of Section \ref{ApplebyClassic} must be violated by any hidden-variable theory, and it is a matter of taste if one is willing to allow this. It may also be emphasized that the full scope of the Kochen-Specker Theorem is restored as soon as there appears to be a method to perform infinitely precise measurements (if only for a finite uncolorable set of observables). Although unlikely and unimaginable, this may be a possibility.

But even if the Kochen-Specker Theorem happens to have been nullified, there are still theorems stating that every hidden-variable theory must be non-local. Indeed, the MKC-models turn out to be no exception (I will also come back to this point in the next Chapter). For me, this is enough to rule out the possibility of acceptable hidden-variable theories.

For a closing comment, I concur with the one made by Barrett and Kent:
\begin{quote}
	``We would like to emphasize that neither the preceding discussion nor earlier contributions to this debate [\ldots] are or were intended to cast doubt on the essential importance and interest of the Kochen-Specker Theorem. As we have stressed throughout, our interest in examining the logical possibility of non-contextual hidden variables simulating quantum mechanics is simply that it \textit{is} a logical--if scientifically highly implausible--possibility, which demonstrates interesting limitations on what we can rigorously infer about fundamental physics.'' \cite[p. 174]{Barrett-Kent04}
\end{quote}

\clearpage

\markboth{The Free Will Theorem Stripped Down}{Introduction}
\section{The Free Will Theorem Stripped Down}\label{FWThoofdstuk}

\begin{flushright}
\begin{minipage}[300pt]{0.6\linewidth}
\textit{There is no effective scientific test for free will. You can't run the universe again, with everything exactly as it was, and see if a different choice can be made the second time round.}
\end{minipage}
\end{flushright}
\begin{flushright}
-- T. Pratchett, I. Stewart \& J. Cohen
\end{flushright}

\subsection{Introduction}
In \cite{ConwayKochen06} and \cite{ConwayKochen09} Conway and Kochen present a very remarkable theorem. In their own words 
\begin{quote}
``[i]t asserts, roughly, that if indeed we humans have free will, then elementary particles already have their own small share of this valuable commodity.'' \cite{ConwayKochen09}
\end{quote}
This is a very strong and strange statement, and it seems unlikely that such a strong philosophical statement can be proven by means of mathematics alone (along with some physical axioms). While discussing the theorem, I'll try to lay bare where the philosophical assumptions turn into mathematical ones and where the mathematical conclusions are transformed back to philosophical conclusions. The stronger the claim of a theorem, the more important such transformations become. But of course, it is easy to criticize such transformations. The simplest argument will be that philosophical notions simply have a different nature from mathematical ones and therefore, any translation from the one into the other cannot be flawless (see also the discussion at the end of Section \ref{The Bell Inequality}). But I will also try to give some critique while accepting these transformations, weakening the so-called ``free will theorem'' even for the believers. 

The theorem that will be discussed here is in fact the ``The Strong Free Will Theorem'' \cite{ConwayKochen09}, which seems a bit more transparent than the original Free Will Theorem \cite{ConwayKochen06}. I will try to explain the theorem in such a way that it becomes clear that it doesn't so much depend on either quantum mechanics or relativity. At best, the theorem can be understood by anyone with a decent knowledge of mathematics and philosophy, with the hope that also people from these disciplines can enter the discussion. This should certainly be the aim of any theorem that is supposed to make a statement about free will.

\markboth{The Free Will Theorem Stripped Down}{The Axioms}
\subsection{The Axioms}
Three axioms are introduced named SPIN, TWIN and MIN. I will state them here in the form as they appear verbatim in \cite{ConwayKochen09}. 

\begin{quote}\textbf{SPIN:} Measurements of the squared (components of) spin of a spin-1 particle in three orthogonal directions always give the answers 1,0,1 in some order.
\end{quote}
This axiom is of course derived from quantum theory. But this theory is not necessarily needed to make it an acceptable axiom. It is in fact an experimentally testable prediction.
The term `squared spin of a spin-1 particle along some axis' may be directly associated with some associated procedure to measure this observable. Such a procedure (or an appropriate equivalence class thereof) may even be taken as the definition of the observable.\footnote{The thesis that observables only have meaning with reference to the way they are measured is common in Copenhagen interpretations of quantum mechanics. See for example \cite[Ch. III]{Heisenberg58}.} That is, once the axiom has been experimentally verified\footnote{This can of course only be done up to certain precision; I will come back to this point in Section \ref{RobustnessFWT}. It may also be noted that the known elementary spin-1 particles (the $W^\pm$ or $Z^0$ boson) aren't the easiest ones to manipulate for experimental testing. However, properties similar to the ones assumed in SPIN may be accomplished by working with coupled spin-$\tfrac{1}{2}$ particles. For example, something similar was done in \cite{Howell02}. That is, one doesn't necessarily need \textit{elementary} spin-1 particles to test SPIN.}, it becomes an axiom that must be derivable from any theory that describes the associated experiment.

From this point of view, the axiom can be explained in the following way: for each choice of three orthogonal directions $e_1,e_2,e_3$, one can construct a measuring context that always gives three numbers $x_1,x_2,x_3$ with the special property that two of these numbers are equal to 1, and the other is equal to 0. 
It is also possible to unambiguously associate each of the obtained numbers $x_i$ with exactly one of the chosen directions $e_i$. The exact procedure involved is irrelevant for the argument made by the Free Will Theorem. 

Explained in this way, the SPIN axiom is a very modest one. There are, of course, many other procedures to obtain three numbers $x_1,x_2,x_3$ with the special property that two of these numbers equal 1 and the other equals 0. One may, for example, put three balls in a bag, one of them blue and the other two red, then take out the balls one at the time and take a number for each ball; 0 if the ball is blue and 1 if it is red.\footnote{Of course, the statistical behavior of spin-1 particles is much richer, as predicted by quantum mechanics. But it is not demanded here that quantum mechanics is true.} The reason for taking spin-1 particles only becomes clear in the second axiom.

\begin{quote}\textbf{TWIN:} For twinned spin-1 particles, suppose experimenter A performs a triple experiment of measuring the squared spin component of particle $a$ in three orthogonal directions $x,y,z$, while experimenter B measures the twinned particle $b$ in one direction, $w$. Then if $w$ happens to be in the same direction as one of $x,y,z$, experimenter B's measurement will necessarily yield the same answer as the corresponding measurement by A.
\end{quote}

The above formulation of this axiom by Conway and Kochen is a bit unfortunate. In fact, it is not an axiom at all, but rather a definition of the term ``twinned spin-1 particles''. The axiom should then be taken to be the statement that twinned spin-1 particles exist. Or, more precisely, that it is possible to set up two systems, conveniently named particles $a$ and $b$, such that a measurement on system $b$ necessarily yields the same value as one of the obtained values from a measurement on system $a$ if certain criteria are met. My point becomes clearer in a story.

The TWIN axiom states that two experimenters A (Alice) and B (Bob) can come together to set up part of an experiment. 
Then they split the experiment into two and each of them takes one part to his/her own home laboratory. 
Once at home, each of the experimenters can choose to perform a measurement on their part of the experiment at any time they like. In advance, Alice and Bob have agreed on two sets of possible measurements from which they are allowed to pick one. The experiments Bob is allowed to perform are parameterized by a three-dimensional direction $w$. 
Suppose that he is allowed to choose one of the 33 directions from the Peres proof of the Kochen-Specker Theorem (see table \ref{Peresvectors}). Each of the possible experiments has 0 and 1 as the only possible outcomes i.e., with each of the 33 directions there is associated a physical quantity with possible values 0 and 1. The experiments can be thought of as the ones appearing in the SPIN axiom, with the extra procedure that two of the three results are thrown away.
Alice, on the other hand, is allowed to choose from 40 experiments, each of them associated with one of the triads that appear in the Peres proof (table \ref{Perestriads}) and corresponding to the experiment from the SPIN axiom. 

Thus far, this story is merely a consequence of the SPIN axiom. The new assumption is that when Alice and Bob are still together, they can arrange their experiment in such a way that if the direction Bob chooses at home coincides with one of the directions in the triad that Alice chooses, then their associated measurement results will also coincide \textit{no matter at what time either of the experimenters chooses to do their experiment}. In short, they can prepare a pair of twinned spin-1 particles.
Again, this is an experimentally testable axiom; the fact that quantum mechanics predicts this behavior is remarkable, though, of secondary importance. 

The third axiom actually incorporates two different axioms. One of these declares the free will of the experimenters and the other states a certain notion of locality. 

\begin{quote}
\textbf{MIN:} Assume that the experiments performed by A and B are space-like separated. Then experimenter B can freely choose any one of the 33 particular directions $w$, and $a$'s response is independent of this choice. Similarly and independently, A can freely choose any one of the 40 triples $x,y,z$, and $b$'s response is independent of that choice.
\end{quote}

The axiom firstly states that both Alice and Bob have a form a free will that allows them the ability of free choice. In particular, they both have the ability to choose whether or not they let their choice depend on the choice of the other. Their choices are free in the sense that their history (up to the point of the experiment) does not determine the settings.

Secondly, the notion of locality used in the axiom, is that the \textit{choice} of the experiment of any experimenter does not influence the \textit{outcome} of the experiment performed by the other. Since the experimenters themselves choose the time at which they perform their experiment, it may be arranged that neither does Alice's experiment lie in the causal future of Bob, nor does Bob's experiment lie in the causal future of Alice. This is possible under the assumption that instantaneous influences are prohibited and that there is no absolute frame of reference.

\markboth{The Free Will Theorem Stripped Down}{The Theorem}
\subsection{The Theorem}

Conway and Kochen stated their Strong Free Will Theorem as follows \cite{ConwayKochen09}:

\begin{stelling}\label{freewilltheorem}
The axioms SPIN, TWIN and MIN imply that the response of a spin 1 particle to a triple experiment is free -- that is to say, is not a function of properties of that part of the universe that is earlier than this response with respect to any given inertial frame.
\end{stelling}

Obviously, the second part of the statement makes the first part even more mystifying. Being mathematicians, Conway and Kochen focus their proof of the theorem primarily on mathematical considerations. As a result, the philosophical considerations have become obscured. In an attempt to give both sides of the story the attention they deserve, I'll first present the main mathematical reasoning in a lemma.

\begin{lemma}\label{freewilltheoremmath}
It is not possible to simultaneously assign definite values to the outcomes of all possible experiments on twinned spin-1 particles, without violating the SPIN or TWIN axiom.
\end{lemma}

\noindent
\textit{Proof:}\hspace*{\fill}\\
The assignment of `definite values to the outcomes of all possible experiments' means that with each experiment, one associates a unique number (or set of numbers) that denotes the outcome of the experiment. Let $E_B$ denote the set of the 33 possible experiments for Bob and let $E_A$ denote the set of the 40 possible experiments for Alice. A definite value assignment that meets the SPIN axiom then consists of two parts, namely a function $\theta_B:E_B\to\{0,1\}$ and a function $\theta_A:E_A\to\{(0,1,1),(1,0,1),(1,1,0)\}$. At this point there are still $2^{33}\cdot3^{40}\simeq10^{29}$ possible assignments.

For any triad $(x_1,x_2,x_3)\in E_A$, let $\theta_A^j(x_1,x_2,x_3)$ denote the $j$-th component of the triplet $\theta_A(x_1,x_2,x_3)$ for $j=1,2,3$. For any fixed $\theta_B$, the TWIN axiom implies that for every $x\in E_B$ the following condition should hold: if $x$ also appears as the $j$-th component in the triad $(x_1,x_2,x_3)\in E_A$, then
\begin{equation}
	\theta_B(x)=\theta_A^j(x_1,x_2,x_3).
\end{equation}
Note that $E_B$ and $E_A$ have been constructed in such a way that if $x$ appears both as the $j$-th component in $(x_1,x_2,x_3)\in E_A$ and as the $k$-th component in $(x'_1,x'_2,x'_3)\in E_A$, then $x\in E_B$. Consequently, if $x$ appears both as the $j$-th component in $(x_1,x_2,x_3)\in E_A$ and as the $k$-th component in $(x'_1,x'_2,x'_3)\in E_A$, then
\begin{equation}\label{non-context-free-will}
	\theta_A^j(x_1,x_2,x_3)=\theta_A^k(x'_1,x'_2,x'_3).
\end{equation}
This allows one to construct a function $\theta:E_B\to\{0,1\}$ in the following way:
\begin{equation}
	\theta(x):=\theta_A^j(x_1,x_2,x_3)\text{ if }x\text{ is the }j\text{-th component in the triad }(x_1,x_2,x_3).
\end{equation}
Since each element $x\in E_B$ appears in at least one triad in $E_A$, this function is actually defined for all $x\in E_B$, and because of (\ref{non-context-free-will}) it is also well-defined, i.e., the value assignment is independent of the choice of the triad in which $x$ appears.
Furthermore, the function $\theta$ satisfies the special property that for each orthogonal triple $x,y,z\in E_B$ one has
\begin{equation}
	\theta(x)+\theta(y)+\theta(z)=2.
\end{equation}
But according to Lemma \ref{KS3dim}, such a function does not exist. Since each choice of $\theta_B$ together with the TWIN axiom leads to the construction of an impossible function, the assignment of a definite value to the outcomes of all possible experiments is not possible without violating either the SPIN or the TWIN axiom (or both).
\hfill	$\square$\\[0.5ex]

\noindent
\textit{Proof of Theorem \ref{freewilltheorem}:}\hspace*{\fill}\\
The structure of the proof is as follows. First, it will be shown that the assumption of MIN, SPIN and TWIN together with the assumption that the responses of spin-1 particles to a triple experiment is not free implies the existence of an assignment of definite values to the outcomes of all possible experiments. It then follows from Lemma \ref{freewilltheoremmath} that the axioms SPIN and TWIN cannot both hold. Since all three the axioms SPIN, TWIN and MIN are assumed to hold, a contradiction is established, leading to the conclusion that the response of a spin-1 particle to a triple experiment cannot be free. It then follows that its response must indeed be free. 

From a logical point of view, Conway and Kochen avoid (possibly improper) use of the double negation elimination by giving free will a negative definition:
\begin{definitie}\label{particlefreewill}
	The response of the particle is said to be \underline{free} if it is \textit{not} determined by what has happened at earlier times (in any inertial frame).
\end{definitie}
This is an adaption of Conway and Kochen's defintion of the free \textit{choice} of the experimenter:
\begin{quote}
	``To say that A's choice of $x,y,z$ is free means more precisely that it is not determined by (i.e., it is not a function of) what has happened at earlier times (in any inertial frame).'' \cite{ConwayKochen09}
\end{quote}
There is no mentioning in their articles of how their definition exactly corresponds to the definition of the free \textit{response} of the particles, but I think Definition \ref{particlefreewill} is the most obvious one.

Now first consider particle $b$. Since the response of the particle is assumed to be deterministic, a definite value can be assigned to the outcome of the measurement that is actually being performed by Bob.\footnote{Here, one may raise the question whether the (mathematical) notion of value definiteness is a necessary condition for the (philosophical) concept of determinism to hold. I think it is.} Since Bob (by the MIN axiom) has a free choice which experiment he wishes to perform (i.e. the actual measurement he will perform is not determined), the particle must have a deterministic response for each possible choice of Bob. Thus at any time point $t_B$ (defined in the reference frame of Bob) there is a definite value assignment $\theta_B^{t_B}:E_b\to\{0,1\}$, where $\theta_B^{t_B}(x)$ denotes the pre-determined measurement result if Bob chooses to perform the measurement $x$ at time $t_B$ (here, the SPIN axiom has been used).
Using the same sort of reasoning, at any time point $t_A$ (defined in the reference frame of Alice) there is a definite value assignment $\theta_A^{t_A}:E_A\to\{(0,1,1),(1,0,1),(1,1,0)\}$, where $\theta_A^{t_A}(x_1,x_2,x_3)$ denotes the pre-determined measurement result if Alice chooses to perform the measurement $(x_1,x_2,x_3)$ at time $t_A$.

In principle, the TWIN axiom only has to be satisfied for the actual pair of experiments performed by Alice and Bob. If $t_B<t_A$ in all inertial frames\footnote{This is a fancy way of saying that events happening at the time $t_B$ are allowed to have a causal influence on the events happening at time $t_A$.}, $\theta_A^{t_A}$ can be arranged to satisfy the TWIN axiom for the specific outcome $\theta_B^{t_B}(x)$ for the specific choice of $x$ made by Bob. Vice versa, $\theta_B^{t_B}$ can be arranged to satisfy the TWIN axiom for the specific outcome $\theta_A^{t_A}(x_1,x_2,x_3)$ for the specific choice of $(x_1,x_2,x_3)$ made by Alice if $t_A<t_B$ in all inertial frames.

However, $t_A$ and $t_B$ may be chosen in such a way that neither $\theta_A^{t_A}$ is allowed to influence $\theta_B^{t_B}$, nor is $\theta_B^{t_B}$ allowed to influence $\theta_A^{t_A}$, that is, they are independent in the sense of the MIN axiom. Therefore, since all possible choices for experiments are allowed for both Alice and Bob, the TWIN axiom must actually be satisfied for all these possible choices. But this is impossible according to Lemma \ref{freewilltheoremmath}.

It then follows that if one maintains the axioms SPIN, TWIN and MIN, the responses of the particles $a$ and $b$ to a triple experiment are not both determined. By De Morgan's law, it then follows that at least one of the particles must have a `free will' in the sense of Definition \ref{particlefreewill}. 
\hfill	$\square$\\[0.5ex]


\markboth{The Free Will Theorem Stripped Down}{Discussion}
\subsection{Discussion}
\subsubsection{Free Will and Determinism}
Most readers will find the Free Will Theorem disturbing when they encounter it the first time. Indeed, most people have a certain notion of free will that is unlikely to be ascribable to particles. This notion will most likely also divert from the one given by Conway and Kochen (quoted just below Definition \ref{particlefreewill}), because having free will is usually seen as a \textit{positive} statement; it is a property someone may possess. Conway and Kochen however, define free will as a negative statement (and not even as a property): it is defined as the \textit{absence} of some property that someone's actions may have. The reason why they use this definition is rather strange. In their own words:
\begin{quote}
	``Our provocative ascription of free will to elementary particles is deliberate, since our freedom asserts that if experimenters have a certain freedom, then particles must have exactly the same kind of freedom. Indeed, it is natural to suppose that this latter freedom is the ultimate explanation of our own.'' \cite{ConwayKochen09}
\end{quote}  
I think it's questionable to state that this sort of freedom explains what free will is, because the assumption that they actually use in the proof of their theorem (under the guise of free will) is not that experimenters have free will, but that some of their choices are not pre-determined. Their definition also allows other forms of free will that are not explored in their paper. 

It seems natural that the freedom of choice possessed by the experimenter entails a certain form of indeterminism in the world, i.e. it is a common notion that some of our actual choices are not determined. However, the reasoning the other way around does seem to be a problem. Typically, our freedom of choice is accompanied by the feeling of being in control of the situation. Making a choice is an actual \textit{act}, whereas indeterminism can be entirely \textit{passive}. The notion of free will to which Conway and Kochen appeal in the MIN axiom is appealing, not because the choices made by Alice and Bob are \textit{not pre-determined}, as they claim, but, in my opinion, because they can actually \textit{make} these choices.\footnote{Indeed, if the choices made by the experimenters were only due to some stochastic process, there would be no motivation to assign definite values to the outcomes of \textit{all} possible experiments, since in that case one may argue that the stochastic process that selects the experiment also selects the outcome of the experiment (and thus doesn't have to select outcomes for all possible experiments).} It seems unreasonable to me to assume that the indeterministic behavior of the particles is similarly due to an active `choice' made by these particles.

But even within the context of the free will theorem, one may argue that the freedom of the particles cannot be ``exactly the same kind of freedom'' that experimenters have. It is clearly assumed by Conway and Kochen that Alice and Bob have an \textit{independent} form of free will. However, the particles in the story certainly do not. Once one particle has decided what value it wishes to bring about at the performance of a measurement, the other particle is immediately robbed of its free will. In fact, adopting the term `free will' to the behavior of the particles seems to introduce a new form of non-locality into the story, since the particles have to agree on what choice they make, using communicating skills that violate the law of causality. 
Secondly, it is plausible that the freedom to choose which experiment to perform is not the only freedom experimenters have. Besides being experimenters, most experimenters are also known to be normal human beings who sometimes choose \textit{not} to perform a measurement and rather have a cup of coffee.\footnote{Some may even consider throwing it across the room \cite{Conway06}.} This is also a freedom particles do not have. If the experimenter \textit{chooses} to perform a measurement, the particle is \textit{forced} to produce a measurement result. It seems that experimenters have an \textit{active} form of free will, whereas particles (if indeed they have some sort of free will) merely have a \textit{passive} form of free will.

Thus far I have only argued that the term `free will' is an unfortunate one to describe the peculiar behavior of the particles in the Free Will Theorem. It seems more natural to describe this behavior as indeterministic, which is usually defined as the denial of determinism. One often finds the following description of determinism by Laplace:
\begin{quote}
	``We ought then to regard the present state of the universe as the effect of its anterior state and as the cause of the one which is to follow. Given for one instant an intelligence which could comprehend all the forces by which nature is animated and the respective situation of the beings who compose it--an intelligence sufficiently vast to submit these data to analysis--it would embrace in the same formula the movements of the greatest bodies of the universe and those of the lightest atom; for it, nothing would be uncertain and the future, as the past, would be present to its eyes.'' \cite{Laplace02}
\end{quote}
More loosely put, given a complete account of the current situation of the entire universe, there is only one possible future.

But one does not need the Free Will Theorem to recognize that this form of determinism is excluded if experimenters have free will. One only needs to think of the following insipid example. If I hold a ball, I can choose to do with it whatever I want, making the path of the ball indeterministic. In more general terms, if there are indeterministic events present in a theory that have a causal influence on other events (which seems a necessary consequence of the form of free will of experimenters introduced by Conway and Kochen), these other events will necessarily also have indeterministic aspects. 

In Newtonian mechanics, the situation may be considered to be even worse. Here, one does not even have to assume the free will of experimenters to encounter a form of indeterminism.\footnote{The following example is taken from \cite{Landsman08}, which is in turn inspired by \cite{Earman07}.} 
Consider a particle with mass 1 in one dimension. Supposing it is subjected to the force $F(x,t)=6\mathrm{sgn}(x)|x|^{\frac{1}{3}}$ with initial position $x(0)=0$ and initial momentum $p(0)=0$, the equation of motion
\begin{equation}
	\frac{\dee^2 x}{\dee t^2}= 6\mathrm{sgn}(x)|x|^{\frac{1}{3}}
\end{equation}
does not have a unique solution, i.e. the motion of the particle is not determined.\footnote{Possible solutions are for example $x(t)=t^3$ or $x(t)=-t^3$. It should be noted that this indeterminacy can only occur when the force does not satisfy the Lipschitz condition. Sometimes this condition is added as an axiom for Newtonian mechanics to ensure that the equation of motion always has a unique solution.} 
The example is even robust against a more general definition of determinism that allows that an entire account of the present state alone may not be enough to determine the future:
\begin{quote}
	``Determinism is the thesis that the past and the laws of nature together determine, at every moment a unique future.'' \cite{Inwagen08}
\end{quote}
For suppose it is known that $x(t)=0$ and $p(t)=0$ for all $t\leq0$, then still $x(t)=t^3$, $x(t)=-t^3$ and $x(t)=0$ for $t>0$ are possible solutions.

The definition of determinism given by van Inwagen has my preference over the definition by Laplace. It has always puzzled me why determinism is often presented with the extra condition that the present renders the past as irrelevant for determining future events; which may perhaps conveniently be named the ``deterministic Markov property''.\footnote{The term may be even more appropriately related to this discussion than it seems at first sight. In \cite{Basharin04} it is argued that Markov came up with the idea of Markov chains partly to show that it is not true that independence is a necessary condition for the law of large numbers to hold. This erroneous assumption was used by Nekrasov to argue in favor of Christianity and the existence of free will, which was a source of annoyance for Markov (being an atheist).}

The above examples point out that value definiteness for all observables (a property of classical physics) does not necessarily imply determinism. Indeed, determinism does not only require that certain observables have a definite value, but also that the way these values change in time is uniquely given. If the Free Will Theorem is not to be rendered trivial, it must be shown that there is a crucial difference between indeterminism with value definiteness and indeterminism without value definiteness (and not only from a realist point of view). 

I don't believe it was the main intention of Conway and Kochen to show that determinism and free will are incompatible. Presumably, they did assume some form of compatibilism that is possible in classical physics, but is necessarily violated by quantum mechanics. Although I am by no means an expert on the philosophical problems involved\footnote{\cite{Inwagen08} provides a clear presentation of the problem with free will in philosophy. Roughly stated, the problem is that free will is incompatible both with determinism and indeterminism, but nonetheless seems to exist.}, I will try to give a definition of determinism that might appeal to Conway and Kochen:\footnote{I do not dare to claim that this definition makes free will compatible with determinism, but I do think that if compatibility is possible, this may be a possible step in the right direction. It is at least an attempt to alter our view on determinism instead of our view on free will, which seems the more customary approach to compatibilism.}
\begin{definitie}
	The response of a particle to an experiment is called \underline{locally deterministic} if it is completely determined by its past, \textit{given} the choice of the actual experiment performed, for \textit{all} possible choices of the experiment.
\end{definitie}
In a certain sense this is a milder form of determinism than the ones given earlier, since it only requires that the \textit{immediate} future is determined rather than the \textit{entire} future. This is what the term `locally' in the definition expresses; it refers to locality in time, rather than in space. Indeed, it is a form of determinism that is obeyed in the classical examples just given,\footnote{More generally, it is obeyed in any theory in which observables have definite values at all times that change continuously in time. In that case, the value assigned to some observable $\A$ at time $t$ is given by the limit $\lim_{t'\to t}\A(t')$ which is completely determined by what happened at earlier times.} but it also seems that it is exactly a violation of this form of determinism that appears in the proof of the Free Will Theorem. 
A possible objection against this definition of determinism is that it's also a more vague one, since it relies on the notion of `experiment'. However, the use of this notion in the definition seems necessary if Conway and Kochen's Definition \ref{particlefreewill} is to make any sense, since that in turn relies on the notion of `response'.
 At least, the definition is useful in the present discussion and it can be used to propose a new formulation of the Free Will Theorem that is sharper and emphasizes what is really proven:
\begin{stelling}\label{freewilltheorem2}
The axioms SPIN, TWIN and MIN imply that the response of a spin-1 particle to a triple experiment is not locally deterministic.
\end{stelling}
Or, even more precisely,\footnote{This last formulation avoids the use of the de Morgan law at the end of the proof of the Free Will Theorem, which is not generally accepted (e.g. in intuitionistic logic). Indeed, what is actually proven is that the assumption that the responses of \textit{both} particles are locally deterministic leads to a contradiction. So, for at least \textit{one} of the particles the assumption isn't true. However, the proof doesn't provide a way for us to decide for which particle this is the case.}
\begin{stelling}\label{freewilltheorem3}
The axioms SPIN, TWIN and MIN imply that the response of a spin-1 particle to a triple experiment cannot be locally deterministic for all particles.
\end{stelling}
Regardless if one accepts the radical definition of free will for particles introduced by Conway and Kochen, I think the new formulation gives a more precise formulation of what is really stated by the theorem (or rather, by its proof). 

Overall, I think it is a pity that Conway and Kochen presented their theorem in the way they did. If it had been formulated in the proposed way of Theorems \ref{freewilltheorem2} or \ref{freewilltheorem3}, probably more people would appreciate what is really accomplished by the theorem. Of course, the extension of free will to elementary particles is a witty notion, but I was surprised to see that even in \cite{ConwayKochen09-2} Conway states that the Free Will Theorem says that ``if we human beings do indeed have free will, then so do elementary particles have their own very small quantity of free will.'' Einstein's valuable lesson that ``a good joke should not be repeated too often'' seems to apply here.

\subsubsection{The Possibility of Absolute Determinism}
In my reformulation of the Free Will Theorem (Theorems \ref{freewilltheorem2} and \ref{freewilltheorem3}), the whole term `free will' has been eliminated. However, the term `Free Will Theorem' of course still applies, since it assumes the MIN axiom. For a true determinist (i.e. one who believes in a form of determinism \`a la van Inwagen or Laplace), the notion of free will as imposed by the MIN axiom is, obviously, appalling. Indeed, if one accepts this form of determinism, the choices made by Alice and Bob are of course also pre-determined, i.e., they don't have an actual choice. It is a shame that there do not seem to be a lot of philosophers who are determinists that mingle in the discussion. Fortunately, 't Hooft is both a respectable physicist and an outspoken determinist. In \cite{Hooft07} he argues in favor of determinism and against the notions of free will proposed in the articles on determinism and quantum mechanics. In fact, he easily disposes of the motivations for free will. About Tumulka's argument that
\begin{quote}
	``[w]e should require a physical theory to be non-conspirational, which means here that it can cope with arbitrary choices of the experimenters, as if they had free will [\ldots]. A theory seems unsatisfactory if somehow the initial
conditions of the universe are so contrived that EPR pairs [e.g. the correlated pairs of particles which are postulate by the TWIN axiom] always know in advance which magnetic fields the experimenters will choose.'' \cite{Tumulka06}, 
\end{quote}
't Hooft states that this form of conspirational aspects, like any conspirational aspects, are difficult to object to from a deterministic point of view. Indeed, the feeling of a conspiracy seems unavoidable if one accepts determinism, since in that case one may always answer the question why things happen in a certain way by saying that it was simply determined to be that way.\footnote{One may, for instance, think that it is conspirational that the sun comes up every day. From a determinist point of view, what seems conspirational or not is just a matter of opinion.}

Besides disposing of the arguments in favor of free will, 't Hooft also argues against the notion of free will ``meaning a modification of our actions without corresponding modifications of our past'' \cite{Hooft07}. From a deterministic point of view this is obvious, since every possible past implies a unique present, a modification of the present must imply a modification of the past. Remarkably, 't Hooft also states that this sort of free will is already prohibited in the structure of quantum mechanics:
\begin{quote}
	``Suppose we let an [annihilation] operator $a_i(\vec{x},t)$ act on a state, which means, more or less, that we remove a particle at the point $\vec{x}$, at time $t$. A different state is then obtained, \textit{in which both the future and the past development of operators look different from what they were in the old state}!'' \cite{Hooft07}
\end{quote}    
That is, any modification made in the present influences the past also in the structure of quantum mechanics. 
Or, put more strongly, according to 't Hooft quantum mechanics itself is in conflict with the MIN axiom. 
This is a dubious argument in my opinion. In the axioms of orthodox quantum mechanics, the role of the observer is described explicitly and the only alteration to the wave function that, according to these axioms, can be made by observers is its collapse. 
The act of annihilating particles is one that is not directly related to the axioms, and the question if the mathematical description of this act (about which 't Hooft speaks) has a direct relation to performing the act in real life (I do not know what this means exactly) is not easy to answer. 
But, to turn things around, if acts performed in the present by means of free will (e.g. turning on some magnetic field) \textit{do} retroactively influence the probabilities of finding certain measurement results in the past, it would be possible to perform measurements that conflict with these probabilities in the present, since the wrong wave function would be used to describe the experiment (because one does not know if it will be altered by some act performed in the future). I am not sure if I am meeting the objections made by 't Hooft, since his arguments are often clouded by the use of mathematical symbols and technical terminology. But hopefully, I have given some hint about why I think that quantum mechanics is not incompatible with free will \textit{in principle}, just like classical mechanics doesn't seem incompatible with free will \textit{in principle}.  

Despite 't Hooft's determinism, he doesn't entirely reject the idea of free will. Instead, he proposes a modified axiom of free will. To understand this modification, one first needs to understand what is expected from a deterministic theory. In his own words:
\begin{quote}
	``All we should demand is that the model in question obeys the most rigid requirements of internal logic. Our model should consist of a \textit{complete description} of its physical variables, the values they can take, and the laws they obey while evolving. The notion of time has to be introduced if only to distinguish \textit{cause} from \textit{effect}: cause must always precede effect. If we would not have such a notion of time, we would not know in which order the `laws of nature' that we might have postulated, should be applied.'' \cite{Hooft07}
\end{quote}

The notion of completeness used here is presumably the same as the one required by Einstein, Podolsky and Rosen. Being a deterministic model, it predicts what will happen in the future with certainty. However, exact calculations are assumed to be of such a complex nature that ``one will be forced to make crude approximations.'' On the other hand, to explain the present situation, ``numerous guesses concerning the past'' have to be made. This is where the free will of 't Hooft enters the discussion; that, although the model is about only one possible development of events, it should also describe other possible developments. In his own words, the modified axiom of free will states that
\begin{quote}
	``we must demand that our model gives credible scenarios for a universe \textit{for any choice of the initial conditions}!'' \cite{Hooft07}
\end{quote}
Thus, although one is not allowed to freely choose the experiment to perform, one is allowed to choose several initial conditions to check the theory. In particular, the theory should be \textit{defined} and make predictions for any such choice. Of course, this choice is not entirely free since an actual free performance of calculations using pen and paper will effect the entire past. In fact, one is only allowed to check the theory for the initial conditions one is determined to check by this same theory. In other words, there is no real notion of `any choice of the initial conditions' in the deterministic world. Another possible objection to this axiom is that it presumes the notion of `initial state of the universe', which, in general, cannot be given a sensible meaning but perhaps only in the supposed deterministic model.
Although I don't think that 't Hooft succeeds in showing that a sense of free will is still possible in a deterministic world, he does make an important point\footnote{Although it is by no means a new point.}: absolute determinism \textit{is} a logical possibility. 

But 't Hooft claims more than just the logical possibility of determinism; he also considers it possible that a deterministic theory of the universe might actually exist. I have reasonable doubts about this claim. 
Since such a theory would predict every action performed, a way to falsify it would be to find some of the experiments Alice and Bob are prohibited to perform and let them try to perform these experiments. This implies that the deterministic theory would actually make people aware of choices they are prohibited to make, which seems to me a very dubious situation (but indeed a logical possibility). A possible way out is that it is predetermined that such calculations (to determine prohibited experiments) will never be performed, but this seems unlikely because we've already come this far to discuss plans of performing these calculations (i.e., it will lead to a similarly dubious form of self-awareness).
The other option is that the calculations are simply too difficult to perform. In fact, the laws of nature must be of such a character that there doesn't exist any numerical method to calculate the directions of choice for Alice and Bob even within some reasonable precision. This appears to be the situation as 't Hooft envisages it. It therefore seems that 't Hooft has taken it to be his task to construct a theory that nobody understands and nobody can use. I'd like to note that this remark is not intended as a sneer but rather as a warning. I do believe that research on a possible deterministic theory is a noble cause, but I just don't see how it can succeed without ending in vague ambiguous technical terms. Hopefully, determinists will be aware of these lurking dangers. 

Finally, it should be noted that the Free Will Theorem in this case does establish that a potential determinist theory cannot possibly improve on the predictions of quantum mechanics for the experiments concerning spin-1 particles. The choice between free will and determinism therefore doesn't affect the way these experiments are described. But at least quantum theory takes the notion of experiment seriously, whereas from the deterministic point of view I can't help feeling a sense of defeat if possible experiments that would falsify the theory are prohibited even if these experiments can be performed in principle (i.e., they are not explicitly prohibited by the theory itself).


\subsubsection{Robustness}\label{RobustnessFWT}
The free will theorem may be seen as a modification of the Kochen-Specker Theorem. Both theorems use an argumentation of the form: 
\begin{quote}
`A certain type of theory should have certain properties. If these properties are satisfied, then one can construct a certain function on the unit sphere $S^2$ in $\mathbb{R}^3$. Since such a function does not exist, a theory of such a type cannot exist.'
\end{quote}

As seen in Chapter \ref{NullificatieStuk}, the Kochen-Specker Theorem is susceptible to loopholes if one takes the finite precision of measurement into account. Basically, the argument of Meyer \cite{Meyer99} was that for that `certain type of theory' the `certain function' doesn't need to be defined on the \textit{entire} set $S^2$. Since both the Kochen-Specker Theorem and the Free Will Theorem are based on the non-existence of the same function, it seems likely that Meyer's argument can also be used in this case. Although Conway and Kochen provide some discussion on the robustness\footnote{Robust under taking the finite precision of measurements into account.} of their theorem \cite{ConwayKochen06}, they do not explicitly go into the argument of Meyer (in fact, they don't even mention the MKC models). I will argue here that Meyer's argument in fact does not apply to the free will theorem. 

The original Kochen-Specker argument is set up so as to render realist theories that reproduce quantum-mechanical predictions impossible. From a realist point of view, each observable must be assigned a definite value such that one can unambiguously speak about actual properties possessed by a single system. The argument then runs that with each point on the unit sphere $S^2$ in $\mathbb{R}^3$ (more precisely, to each ray in $\mathbb{R}^3$) there must be an observable associated with this point. Given that the definite value assignment should satisfy certain rules to reproduce the quantum-mechanical predictions, one obtains a contradiction. In fact, it is sufficient to consider only a finite subset of points in $S^2$ to obtain a contradiction. The loophole in this argument is that if one takes into account the finite precision of measurement, one can no longer maintain the argument that indeed \textit{every} point in $S^2$ should coincide with an observable. In fact, it is enough to consider some dense countable subset.

The Free Will Theorem is different in spirit from the Kochen-Specker Theorem in that it does not start from the assumption of a realist approach toward nature. Instead, it focuses on the possibility of determinism and only assumes that measurements have outcomes. So no assignment of definite values to \textit{observables} is required, but instead an assignment of definite values to \textit{measurement results} is called for. Then, although not every point in the uncolorable subset $S'$ of $S^2$ may correspond to an actual observable, such points \textit{do} denote possible experiments that can be performed, and hence a definite value assignment to them is required. 

As seen, the MKC models respond to this situation in a peculiar way. Each time a measurement is performed, corresponding to a triad in the uncolorable subset, they claim that what in fact is measured are the values of some observables that together correspond to a triad close to the original triad one set out to measure. The indeterminacy of quantum mechanics is then retrieved by a probability distribution on the hidden states and hence on the values possessed by the actual observables. For an uncolorable subset $S'$, there are triads $x,y,z$ and $x,y',z'$ such that the observable actually measured if one wishes to measure $\sigma_x^2$ is necessarily a different one in each of the triads possessing a \textit{different} value. This is how the MKC-models satisfy the SPIN axiom. However, to satisfy the TWIN axiom, the models introduce a form of non-locality. Indeed, after Bob has performed his measurement, the state of the \textit{entire} system changes in such a way that Alice's measurement will automatically satisfy the TWIN and MIN axioms. At least, this is the case for the modified MKC-models proposed in Section \ref{MyModification}. In the original models, a discrepancy arises similar to the ones encountered in Section \ref{ApplebyExperiment} and Section \ref{NonLocalMKC}.

This form of non-locality cannot be sidestepped, since the TWIN axiom requires that the value obtained upon measurement of $\sigma_x^2$ is the same for both triads $x,y,z$ and $x,y',z'$. That is, in the view of the MKC-models, it is required that \textit{different} observables, which are close to each other, must be assigned the \textit{same} definite value. Since the definite value assignments are highly discontinuous at certain (non-negligible) regions (as they must be, as proven in \cite{Appleby05}), the MKC-models cannot satisfy this criterion.

This result is not surprising. Clifton and Kent \cite{CliftonKent99} already noticed that their models should be non-local. The ensuing action at a distance allows nature to select a triad that satisfies the quantum-mechanical predictions. It is, however, interesting to see that the argument of non-locality turns from the requirement that `each observable must be assigned a definite value independent of the measuring context' to the requirement that `different observables must be assigned the same definite value'. In a certain sense this also explains why the Free Will Theorem is robust. If locality is taken into account, the SPIN and TWIN axioms \textit{require} that most observables close to each other are assigned the \textit{same value}, which entails robustness.

A more extensive treatment of why the free will theorem is robust is also made in \cite{ConwayKochen06}. However, the argumentation used there appears to be a bit vague. The starting assumption is that it may of course be the case that the axioms SPIN and TWIN are in fact only approximately true and are more likely to be of the form: 
\begin{quote}\textbf{SPIN$_{FP}$:} Measurements of the squared (components of) spin of a spin-1 particle in three nominally orthogonal directions give the answers\footnote{One may assume that there are intervals $I_1$, $I_0$ containing 1 resp. 0 such that a measurement result in $I_1$ (resp. $I_0$) may be interpreted as the result $1$ (resp. $0$).} 1,0,1 in some order with a frequency of at least $1-\epsilon_S$.
\end{quote} 
\begin{quote}\textbf{TWIN$_{FP}$:} For twinned spin-1 particles, suppose experimenter A performs a triple experiment of measuring the squared spin component of particle $a$ in three nominally orthogonal directions $x,y,z$, while experimenter B measures the twinned particle $b$ in one direction, $w$. Then if $w$ happens to be nominally in the same direction as one of $x,y,z$, experimenter B's measurement will yield the same answer as the corresponding measurement by A with a frequency of at least $1-\epsilon_T$.
\end{quote}
Then, assuming an upper bound for the finite precision of measurement, an upper bound is derived for $3\epsilon_T+\epsilon_S$ using quantum theory. This confuses me, for why would one want to derive an upper bound for $3\epsilon_T+\epsilon_S$ using an assumption (quantum mechanics) that already states that $3\epsilon_T+\epsilon_S=0$? However, the upper bound derived by Conway and Kochen will still be useful for the following discussion.

Experimental investigations of the SPIN and TWIN axiom will provide upper bounds for the parameters $\epsilon_S$ and $\epsilon_T$. These upper bounds in turn provide an upper bound $f_{max}$ for the frequency with which the conjunction of SPIN and TWIN will be violated.  
On the other hand, for deterministic theories an estimate for the lower bound $f_{min}$ of the frequency with which experimental violations of SPIN and TWIN will appear can be made. Then, if $f_{max}<f_{min}$, the Free Will Theorem shows that a deterministic theory is not possible, even if the weakened versions SPIN$_{FP}$ and TWIN$_{FP}$ of SPIN and TWIN are assumed to be true.

As to the bound $f_{min}$, note that there are in fact $1320=\#(E_A\times E_B)$ possible experiments that Alice and Bob together can choose from. If each of these experiments is assigned a definite outcome, at least one of these assignments must conflict with either the SPIN or the TWIN axiom. Consequently,\footnote{Conway and Kochen derive here the frequency $1/40$. Somehow, they come to the conclusion that a violation of SPIN$\wedge$TWIN necessarily results in a violation of SPIN. I don't see why this should be true.}
\begin{equation}
	f_{min}\geq\frac{1}{1320}.
\end{equation}
The estimate for $f_{max}$ is a bit more tricky; here, an assumption on the frequency with which Alice and Bob choose their experiments is required. For the sake of simplicity, it is assumed that each of them independently chooses an experiment at random. Let $x,y,z$ denote the triad selected by Alice an $m$ the direction chosen by Bob. The corresponding measurement results will be denoted by $\mathcal{M}_x,\mathcal{M}_y,\mathcal{M}_z$ and $\mathcal{M}_m$. Whenever $m\in\{x,y,z\}$ (nominally),\footnote{It is reasonable to assume that Alice and Bob have agreed upon when this conclusion may be drawn and when not.} $\mathcal{M}_{m_A}$ denotes the measurement result obtained by Alice for the direction $m$. Furthermore, let $[101]$ denote the set $\{(0,1,1),(1,0,1),(1,1,0)\}$. Using this notation, one finds
\begin{equation}
\begin{split}
	f_{max}
	=&
	\pee[\neg\text{TWIN}\vee\neg\text{MIN}]\\
	=&
	\pee\left[\left((\mathcal{M}_x,\mathcal{M}_y,\mathcal{M}_z)\notin[101]\right)\vee
	\left(\mathcal{M}_m\neq\mathcal{M}_{m_A} \wedge m\in\{x,y,z\}\right)\right]\\
	\leq&
	\pee\left[(\mathcal{M}_x,\mathcal{M}_y,\mathcal{M}_z)\notin[101]\right]
	+
	\pee\left[\mathcal{M}_m\neq\mathcal{M}_{m_A}| m\in\{x,y,z\}\right]\pee\left[m\in\{x,y,z\}\right]\\
	\leq&
	\epsilon_S+\epsilon_T\pee\left[m\in\{x,y,z\}\right]\\
	=&
	\epsilon_S+\epsilon_T\left(\frac{3}{33}\frac{16}{40}+\frac{2}{33}\frac{24}{40}\right)=\epsilon_S+\frac{4}{55}\epsilon_T,
\end{split}
\end{equation}
where I used (implementing the randomness of the settings) that 16 of the 40 triads contain 3 directions that may be chosen by Bob, and the other 24 only contain 2 directions that may be chosen by Bob. It follows that, if it can be derived from experiment that $\epsilon_S$ and $\epsilon_T$ are actually so small that
\begin{equation}\label{FWTExpTest}
	\epsilon_S+\frac{4}{55}\epsilon_T<\frac{1}{1320},
\end{equation}
the Free Will Theorem holds. Here, the calculation of Conway and Kochen comes in. They derived (assuming the quantum mechanics is true) that it is reasonable to assume that actual experimental tests of SPIN and TWIN will yield
\begin{equation}
	3\epsilon_T+\epsilon_S\leq\frac{1}{2900000},
\end{equation}
which is clearly sufficiently small to entail (\ref{FWTExpTest}). But of course this is all just theoretical sophistry, which only becomes relevant when actual experimental tests of SPIN and TWIN will be performed.


\subsubsection{What does the Free Will Theorem add to the Story?}

\paragraph{The Bell Inequality}

Roughly stated, the claim made by the Free Will Theorem seems to be that determinism is a logical impossibility if one accepts certain forms of free will, locality, and some quantum-mechanical predictions. This claim strongly resembles the main conclusion of Bell's impossibility proof for contextual hidden variables. At first sight, the conclusions drawn from Bell's inequalities seem to be much stronger, since also certain types of \textit{indeterministic} hidden-variable theories appear to be excluded. However, the assumptions made to derive a Bell-type inequality turn out to be stronger too (or at least different). A comparison may be made by translating the Bell-type argument in the language of the Free Will Theorem. 

In the derivation of the Bell inequality for coupled spin-$\tfrac{1}{2}$ particles, certain analogues of the following axioms may be recognized:
\begin{quote}
	\textbf{SPIN':} Measurements of the spin of a spin-$\frac{1}{2}$ particle in any direction always gives one of the answers -1 or 1.
\end{quote}
\begin{quote}
	\textbf{TWIN':} For twinned spin-$\frac{1}{2}$ particles, suppose experimenter A performs an experiment of measuring the spin of particle $a$ in the direction $x_a$, while experimenter B measures the spin of the twinned particle $b$ in one direction, $x_b$. Then if $x_b$ happens to be the same direction as $x_a$, experimenter B's measurement will necessarily yield the opposite answer from the corresponding measurement by A.
\end{quote}
The MIN-axiom for this system would read as follows:
\begin{quote}
\textbf{MIN':} Assume that the experiments performed by A and B are space-like separated. Then experimenter B can freely choose any direction $x_b$, and $a$'s response is independent of this choice. Similarly and independently, A can freely choose any direction $x_a$, and $b$'s response is independent of that choice.
\end{quote}
However, it should be noted that this doesn't directly translate to the notions of locality CILOC and OILOC and the notion of free will assumed to derive the Bell inequality (see the discussion at the end of Section \ref{The Bell Inequality}). This is partly because MIN (and therefore MIN') uses very general terms like `independent' and `freely' whose meaning only becomes apparent by its use in the proof. On the other hand, the axioms CILOC and OILOC are of such a mathematical nature that it isn't quite apparent what they express \textit{exactly}. In fact, their meaning is only mathematically evident within the framework of Kolmogorov's probability theory. Indeed, it seems to be necessary for the derivation of the Bell inequality to assume that the measure-theoretic approach towards probability theory is suitable for describing actual probabilities of events in the real world. Among physicists this has become widely accepted, but many logicians would beg to differ.\footnote{\cite{Hajek01} gives a nice overview of logical probability. It also notes some of the axioms made by Kolmogorov that may be relaxed.} In fact, one may even consider the violation of Bell inequalities by quantum mechanics as an indication that the measure-theoretic approach is indeed flawed.

Overall, I think that the physical meaning of the various assumptions in the derivation of the Bell inequality is somewhat clouded by mathematical terms. For example, as seen in Section \ref{NonLocalMKC}, the notion of factorizability plays a crucial role, but it is not easily motivated without the use of the mathematical framework of quantum mechanics. However, it should be mentioned that the passage of time has given people the opportunity to find implicit assumptions in the derivation of the Bell inequality. It is likely that in the future similar hidden assumptions will be found for the Free Will Theorem.

\paragraph{Coupled Spin-1 Particles}

The main thrust of Lemma \ref{freewilltheoremmath} was known long before Conway and Kochen came with their theorem \cite{HeywoodRedhead83}, \cite{Stairs83}, \cite{BrownSvetlichny90}, and it has even been stated that the Free Will Theorem indeed expresses nothing that wasn't already known \cite{Goldstein09}. At first glance, this does indeed seem to be the case. However, a closer inspection shows that there are important differences between the Free Will Theorem and these earlier articles. 

To obtain a clear view of these differences, it is good to first consider the similarities. In all articles one may distinguish three assumptions AS1, AS2, AS3. The argument may then be arranged to have the following abstract form.
\begin{enumerate}
\item Because of AS1, all $40\times33=1320$ experiments must be assigned a definite outcome.
\item Because of AS2, each of the 1320 individual outcomes must be in accordance with quantum-mechanical predictions.
\item Because of AS3, the assignments must be non-contextual.
\item Because of the Kochen-Specker Theorem, this is not possible and therefore, the three assumptions AS1, AS2 and AS3 cannot all be true.
\end{enumerate}
In the Free Will Theorem, the assumptions are as follows:
\begin{itemize}
\item[AS1:] MIN and (local) determinism.
\item[AS2:] MIN, SPIN and TWIN.
\item[AS3:] MIN.
\end{itemize}
In the articles \cite{HeywoodRedhead83}, \cite{Stairs83} and \cite{BrownSvetlichny90} the assumptions are not formulated in this form, and it takes some more work to recover this line of reasoning from these articles. Also, none of the articles refer to any \textit{specific} set of experiments, which makes it harder to construct actual experiments to test the theorems. In \cite{HeywoodRedhead83} and \cite{Stairs83}, (i) is partly motivated by an appeal to realism:
\begin{quote}
	``We shall be concerned with the sort of realism which asserts at least that at all times and in all states every physical magnitude which pertains to the system has some value'' \cite[p. 482]{HeywoodRedhead83}
\end{quote}
And also in \cite{Stairs83} it is noted that value definiteness is motivated by `classical realism'.\footnote{Part of the investigation of Stairs concerns the question whether or nor value definiteness is a necessary condition for realism. A form of realism in which this condition may be relaxed is termed `quantum realism' by Stairs.} It is assumed that the correlation between possessed (existential) values and outcomes of experiments is given through the rule of Faithful Measurement:
\begin{quote}
	\textbf{FM:} ``Any measurement of a physical magnitude Q reveals the value which Q had immediately prior to the measurement'' \cite[p. 483]{HeywoodRedhead83}
\end{quote}
Without such a rule, a theorem can only exclude certain \textit{existential} behavior for a theory, which is of course of not much interest if it is not related to any \textit{empirical} behavior. Stairs is mainly concerned with existential behavior and therefore doesn't address the notion of experimental outcomes. Brown and Svetlichny have a more direct approach to (i), which is similar to the one followed by Conway and Kochen:
\begin{quote}
	``In a deterministic h.v. theory, once all the values of the relevant parameters are fixed at $t$, the predicted outcome of a measurement of any observable $A$--we shall denote it by $[A]^t$-- is uniquely determined.'' \cite[p. 1380]{BrownSvetlichny90}
\end{quote}
However, Brown and Svetlichny are not very clear about their motivation for stating that the outcome for \textit{any} observable should be determined. In fact, in a deterministic theory this is only required to hold for the observables that are determined to be measured. The only two reasons I can think of for requiring that the claim should also hold for other observables, is an appeal to either realism, or free will.\footnote{Sometimes, a more direct assumption is used called `counterfactual definiteness', which simply means that unperformed measurements also have (potential) outcomes. However, on its own, counterfactual definiteness doesn't seem a convincing assumption to me if it isn't motivated by either realism or free will.} Since Brown and Svetlichny state that 
\begin{quote}
	``it is not required that $[A]^t$, for every observable $A$, be interpreted to represent an objective element of reality [\ldots] associated with $A$, which the measurement process somehow faithfully reveals at $t$.'' \cite[p. 1384]{BrownSvetlichny90},
\end{quote}
I take it that they implicitly rely on a notion of free will. 

For step (ii), it is necessary to make assumptions on what constitutes (a part of) the set of observables if it is assumed that measurements reveal the values of observables. In all three articles \cite{HeywoodRedhead83, Stairs83, BrownSvetlichny90} it is assumed that observables can be associated with self-adjoint operators on a Hilbert space, at least, for the particular system of two spin-1 particles. Brown and Svetlichny actually (implicitly) require that all self-adjoint operators can be associated with an observable by relying on Gleason's lemma in their proof. The other proofs rely on the Kochen-Specker Theorem and therefore only require a finite subset of self-adjoint operators to correspond to observables. 

In \cite{HeywoodRedhead83}, (ii) is motivated by assuming functional relationships between the observables and the Value Rule:
\begin{quote}
	\textbf{VR:} If for a quantum-mechanical state $\psi$ it holds that 
	\begin{equation}
		\langle\psi,\mu_A(\{a\})\psi\rangle=0,
	\end{equation}
	then $\lambda_{\mathcal{C}}(\A)\neq a$ for all measuring contexts $\mathcal{C}$, for all $\lambda$ that are supposed to occur in the state determined by $\psi$.
\end{quote}
Together with FM this implies that individual outcomes of the 1320 experiments are in accordance with quantum-mechanical predictions. This seems trivial at first sight, but it should be noted that actually a modified version of FUNC is assumed to hold for the observables, which allows contextuality. I will formulate this here in terms that concur with the ones I used throughout this thesis (see Section \ref{The Bell Inequality}).
\begin{quote}
\textbf{FUNC*:} Let $\A_1,\A_2$ and $\A_3$ be three observables corresponding to operators $A_1,A_2$ and $A_3$ such that
\begin{equation}
	A_1=f_1(A_3),\quad A_2=f_2(A_3)\quad\text{and}\quad A_1=g(A_2)
\end{equation}
for certain functions $f_1,f_2$ and $g$. Assume that $A_3$ is a maximal operator (i.e., its spectral decomposition consists only of 1-dimensional projections), then 
\begin{equation}
	\lambda_{\{\A_1,\A_3\}}(\A_1)=g\left(\lambda_{\{\A_2,\A_3\}}(\A_2)\right)
\end{equation}
should hold for all $\lambda$.
\end{quote}

For a single spin-1 particle, the operator 
\begin{equation}\label{SpinHamilton}
	S=s_1\sigma_x^2+s_2\sigma_y^2+s_3\sigma_z^2
\end{equation}
is maximal for any triple of distinct real numbers $s_1,s_2,s_3$. Consequently, FUNC* implies that in a context in which $S$ can be measured, the observables $\sigma_x^2,\sigma_y^2,\sigma_z^2$ should be assigned values in accordance with the FUNC rule. However, for the system of two spin-1 particles, $S$ is no longer maximal, but the functional relationship can still be derived under the following extra assumption\footnote{The assumption stated here is not entirely the same as the one used by Heywood and Redhead, which is known as ontological locality. However, the differences are extremely minor and irrelevant for the present discussion.}:
\begin{quote}
	\textbf{Local Contextuality (LOCC):} If $\h$ and $\h'$ are the Hilbert spaces for two spatially separated systems and $\mathcal{S}$ is an observable associated with the operator $S\otimes\een$, where $S$ is a maximal operator on $\h$, then
	\begin{equation}
		\lambda_{\{\mathcal{S},\A_1\}}(\mathcal{S})=\lambda_{\{\mathcal{S},\A_2\}}(\mathcal{S})
	\end{equation}
	should hold for all hidden states $\lambda$ and for all observables $\A_1,\A_2$ corresponding to maximal operators $A_1,A_2$.
\end{quote}
This LOCC assumption\footnote{It is here stated as a notion of locality, but it may also be viewed as a notion of global non-contextuality i.e., it states that contextuality is only allowed locally.} can also be read in the MIN axiom of the Free Will Theorem. The FUNC* assumption, however, has no direct counterpart. In fact, one may argue that it is not well motivated at all in \cite{HeywoodRedhead83} or \cite{Stairs83}. Surely, if the FUNC relation is not satisfied for observables that can be measured simultaneously, the hidden-variable theory will lead to conflicting predictions whenever these observables are being measured. But this only motivates the idea that FUNC should hold for the observables \textit{that will actually be measured}. Extending this to \textit{all} observables may, again, only be motivated by appealing to the free choice of the experimenters. Indeed, realism can only motivate the requirement that the observables have an actual value, but cannot imply that these values should satisfy the FUNC rule (unless one adopts a many worlds interpretation).

In any case, Brown and Svetlichny replace the FUNC* assumption with the following one\footnote{A similar approach is followed by Stairs.}:
\begin{quote}
	\textbf{Excluded Joint Events (EJE):} If, for a system described by the quantum-mechanical state $\psi_t$, one has
	\begin{equation}
		\langle\psi_t,\mu_{A_1\otimes\een}(\{a_1\})\mu_{\een\otimes A_2}(\{a_2\})\psi_t\rangle=0,
	\end{equation}
	i.e., the probability of finding the value $a_1$ when measuring $A_1\otimes\een$ and the value $a_2$ when measuring $\een\otimes A_2$ is zero, where $A_1$ and $A_2$ are maximal operators on the individual Hilbert spaces, then either $[A_1\otimes\een]^t\neq a_1$ or $[\een\otimes A_2]^t\neq a_2$, or both.
\end{quote}
It is sufficient for them to only consider observables that are locally maximal, because they view the triple experiments of Alice as the measurement of a single observable of the form of (\ref{SpinHamilton}), which indeed is maximal. The measurements performed by Bob are not taken to be measurements of the squared spin but of the spin itself (i.e. $\sigma_w$ in stead of $\sigma_w^2$), which is also associated with a maximal operator.
Besides this assumption, they also rely on a notion of locality that is similar to the notion of LOCC. However, as with the FUNC* assumption, the EJE assumption for \textit{all} $A_1$ and $A_2$ may only be motivated with the aid of the assumption of free choice of the experimenters. Indeed, events that are assigned probability zero in quantum mechanics may be determined never to occur in a hidden-variable theory. But this doesn't imply that either $[A_1\otimes\een]^t\neq a_1$ or $[\een\otimes A_2]^t\neq a_2$, or both. It would be sufficient to require that $A_1\otimes\een$ and $\een\otimes A_2$ will not both be \textit{measured} whenever $[A_1\otimes\een]^t= a_1$ and $[\een\otimes A_2]^t= a_2$ (and this would indeed imply EJE if free will is assumed).

In \cite{BrownSvetlichny90}, the third step (iii) now follows almost immediately. Indeed, using EJE it follows that the values assigned to the measurement outcomes for Bob do not depend on the experiment performed by Alice. It also is derived from EJE that the value assignments for Bob should satisfy the FUNC rule (which is excluded by the Kochen-Specker Theorem and by Gleason's lemma). The argument by Heywood and Redhead is quite lengthy and relies on yet another assumption of locality, called Environmental Locality. It is not very interesting to discuss the details here.

From the above discussion, several conclusions may be drawn about the differences between the Free Will Theorem and the earlier articles mentioned above.
\begin{enumerate}
\item Heywood and Redhead and Stairs rely on a notion of realism to motivate value definiteness for outcomes of experiments, whereas Brown and Svetlichny don't motivate value definiteness that well at all. The Free Will Theorem relies on a strong notion of free will and the idea that measurements have outcomes to motivate value definiteness for all possible experiments.
\item All earlier articles rely on the theoretical structure of quantum mechanics. That is, the theorems are all formulated and proven in terms of the mathematics of Hilbert spaces, the tensor direct product of Hilbert spaces, local maximal self-adjoint operators, etc. By the use of this language it isn't very clear if the proven statements actually apply to \textit{all} possible hidden-variable theories.
\item The earlier articles rely on strong abstract assumptions like FUNC* and EJE, which actually imply the clearer and weaker assumptions SPIN and TWIN.
\item The abstract assumptions FUNC* and EJE are not well motivated and may perhaps only be well motivated by the same strong notion of free will assumed by Conway and Kochen.\footnote{Actually, an appeal to this form of free will may already be found in \cite{EPR} where Einstein, Podolsky and Rosen use it to motivate the incompleteness of quantum mechanics. However, there is no real consensus if this notion of free will is necessary to run their argument \cite{Fine09}.}
\end{enumerate}

Even if the above conclusions don't convince the reader that the Free Will Theorem actually states something new and adds something to the foundations of quantum mechanics, I think that the Free Will Theorem does present a clearer argument. In each of the articles \cite{HeywoodRedhead83, Stairs83, BrownSvetlichny90}, either the starting assumptions are not stated very clearly, or it is not very clear from the proof in which steps these assumptions are used or why the other steps are not based on extra implicit assumptions. It should be noted, however, that on this point a lot of work remains to be done for the Free Will Theorem as well, especially for the ways it was formulated and proven in \cite{ConwayKochen06} and \cite{ConwayKochen09} (as also noted throughout this chapter).

\clearpage

\markboth{The Strangeness and Logic of Quantum Mechanics}{The (In)completeness of Quantum Mechanics (Part II)}
\section{The Strangeness and Logic of Quantum Mechanics}\label{StrangenessSection}

\begin{flushright}
\begin{minipage}[300pt]{0.6\linewidth}
\textit{De kwantumtheorie leidt tot een logica waarin plaats is voor niet-weten.}
\end{minipage}
\end{flushright}
\begin{flushright}
-- G. Vertogen
\end{flushright}
\subsection{The (In)completeness of Quantum Mechanics (Part II)}
The discussion in the previous chapters was rather about hidden variables than about quantum mechanics itself. The main conclusion that can be drawn from this discussion is that it seems impossible to resolve the Einstein-Podolsky-Rosen paradox (of Example \ref{EPRB}) in the way envisaged by those same people without the use of dubious assumptions (i.e. either non-locality or absolute determinism).
The appropriate marginal note should always be that this conclusion is based on abstract mathematical considerations making use of numerous explicit and implicit assumptions. However, as it stands it seems to me that even after more than seventy years since \cite{EPR}, the standard realist approach has provided no satisfactory way for understanding quantum mechanics. To use a metaphor; it is raining in the realist quantum world. And as long as this is the case, one might as well go for a stroll in the Copenhagen quantum garden. Indeed, throughout the development of quantum mechanics Bohr maintained that the theory is in fact complete. Perhaps the conflict between Bohr and Einstein can be resolved by reaching an understanding of why Bohr came to this conclusion.

Bohr's first response to the Einstein-Podolsky-Rosen paradox quoted in Section \ref{The (In)completeness of Quantum Mechanics (Part I)} has puzzled many minds over the years. In terms of Example \ref{EPRB}, it seems that Bohr is saying that the phenomenon ``spin along the $z$-axis of the second particle'' before the measurement on the first particle is a phenomenon different from the one after that measurement, since the experimental context has changed. This seems somewhat reasonable, since the prediction of the value of the spin along the $z$-axis of the second particle \textit{requires} the measurement on the first particle. Note that this does not have to be seen as altering the observed \textit{system}. That is, one may still think that the second particle remains undisturbed under the measurement of the first particle. What merely changes is the measuring context. But then, if the system remains unchanged, why is it described as being in a different state (as a consequence of the von Neumann postulate)? What is the notion of a `state of a system' according to Bohr? Bohr answers these considerations in the following way:
\begin{quote}
	``In fact the paradox finds it complete solution within the frame of the quantum-mechanical formalism, according to which no well-defined use of the concept of ``state'' can be made as referring to the object separate from the body with which it has been in contact, until the external conditions involved in the definition of this concept are unambiguously fixed by a further suitable control of the auxiliary body.'' \cite[p. 21]{Bohr39}
\end{quote}
Hence it seems that according to Bohr the quantum-mechanical state of a system only has a meaning within a specified measuring context. In the quantum-mechanical formalism one should therefore not speak of \textit{the} state of a system, but instead, of the state $\psi$ relative to each possible measuring context. But isn't this bluntly accepting that the quantum-mechanical description \textit{is} incomplete? If the quantum-mechanical formalism \textit{does not} provide one with the state of a system, shouldn't one search for a theory that does? But according to Bohr, this would be asking the wrong question. To say that there would be such a thing as the state of the system would be to say that different measuring contexts could be compared with each other in an unambiguous way. The impossibility of doing so may be seen as the core lesson Bohr draws from quantum mechanics. In his own words:
\begin{quote} 
``[T]he impossibility of subdividing the individual quantum effects and of separating a behavior of the objects from their interaction with the measuring instruments serving to define the conditions under which the phenomena appear implies an ambiguity in assigning conventional attributes to atomic objects which calls for a reconsideration of our attitude toward the problem of physical explanation.'' \cite[p. 317]{Bohr48}
\end{quote}

Thus phenomena are only well-defined within a specified measuring context. 
Bohr considers this peculiarity of quantum mechanics to be fundamental, and this leads him to introduce a new philosophical concept: phenomena that \textit{require} different experimental setups may be defined as \textit{complementary}. Bohr states that information about the same object obtained by different experimental arrangements is complementary. It should be noted that Bohr places this philosophical concept prior to empirical experience. For Bohr, the fact that different measuring devices are required to measure different aspects of the same object is only considered empirical evidence for complementarity: 
\begin{quote}
``Such empirical evidence exhibits a novel type of relationship, which has no analogue in classical physics and which may conveniently be termed ``complementarity'' in order to stress that in the contrasting phenomena we have to do with equally essential aspects of all well-defined knowledge about the objects.'' \cite[p. 314]{Bohr48}
\end{quote}
But as a philosophical concept it stands on its own:
\begin{quote}	
	``[C]omplementarity presents itself as a rational generalization of the very ideal of causality.'' \cite[p. 317]{Bohr48}
\end{quote}

However, Bohr does not explain why the notion of complementarity is obvious and necessary. The vagueness of this notion also doesn't help much to see why quantum mechanics is complete and if so, in what sense. Especially the necessity of complementarity remains unclear, and at first sight it just seems yet another strange aspect of quantum mechanics that may be removed in a possible succeeding theory.   
In the end, it seems as if Bohr just tries to resolve the strangeness of quantum mechanics by ``conveniently terming it complementarity''. This unsatisfied feeling is not new, of course. In 1963 even Bohr's ally Rosenfeld had the following to say:
\begin{quote}
	``Complementarity is no system, no doctrine with ready-made precepts. There is no via regia to it; no formal definition of it can even be found in Bohr's writings, and this worries many people.'' \cite[p. 85]{WheelerZurek83}
\end{quote}

Personally, I think that it is only \textit{after} the search for a hidden-variable theory has failed that some more meaning can be found in the notion of complementarity, though not necessarily in favor of Bohr's interpretation.
 The principle of complementarity may only shed some light on the matter if one could find a philosophical argument for this principle, preferably based on considerations outside quantum mechanics. If this can be done, it will be most likely that a found notion of complementarity will differ from what Bohr had in mind. In the remainder of this Chapter, I will undertake a short investigation of the strangeness of quantum mechanics to finally come to a proposal for a philosophical foundation of complementarity.


\subsection{Quantum Mechanics as a Hidden-Variable Theory}\label{QMHVT}
\markboth{The Strangeness and Logic of Quantum Mechanics}{Quantum Mechanics as a Hidden-Variable Theory}

It remains fascinating that science works in such a way that a theory like quantum mechanics may emerge, differing fundamentally from \textit{a priori} notions of what a theory should be. The structure and requirements imposed on a hidden-variable theory that lead to impossibility proofs seem so reasonable that it is remarkable that they cannot be met. In the words of Bell \cite{Bell66}: ``That so much follows from such apparently innocent assumptions leads us to question their innocence.'' In fact, one may wonder if quantum mechanics avoids the objections raised against possible hidden-variable theories (and if so, how). Can quantum mechanics itself be interpreted as a hidden-variable theory?\footnote{The idea of this section was inspired by a small paragraph in \cite{SeevinckLN}, a more formal investigation of this question was carried out in \cite{Beltrametti95}. My discussion on contextuality is based on some remarks made in \cite{Mermin93}.}

In quantum mechanics the pure states are given by vectors $\psi$ in a Hilbert space $\h$. These states do not determine the value of all observables, but instead they assign a probability to possible measurement outcomes for each observable in such a way that with each observable one can associate a probability space in the following way:

\begin{lemma}\label{quantumprob}
For every observable $\A$ corresponding to the operator $A$ with spectrum $\sigma(A)$ and every state $\psi$, the triple $(\sigma(A),\Sigma_A,\pee_{\psi,A})$ is a probability space, where $\Sigma_A$ is the $\sigma$-algebra of all Borel subsets of $\sigma(A)$ and $\pee_{\psi,A}$ is defined by the Born postulate: 
\begin{equation}
	\pee_{\psi,A}(\Delta)=\frac{1}{\|\psi\|^2}\langle\psi,\mu_A(\Delta)\psi\rangle,\quad\forall\Delta\in\Sigma_A.
\end{equation}
\end{lemma} 

The proof is straightforward and well-known and therefore I omit it here.
Although this may be an unsurprising result, it is significant for the present discussion. Apparently, \textit{for each observable separately}, quantum mechanics behaves like a hidden variable-theory. That is, one may choose one observable (or, more generally, a set of observables whose corresponding operators mutually commute) and implement it as corresponding to an element of physical reality in the sense of Einstein-Podolsky-Rosen. One may then adopt the view that this specific observable does have a definite value at all times and adopt an ignorance interpretation towards the quantum-mechanical state of the system. This view plays an important role in so-called modal interpretations of quantum mechanics \cite{DicksonDieks09}. As far as I know, none of these interpretations solve all the problems encountered in the earlier Chapters in a satisfactory way. For example, (the modal interpretation of) Bohmian mechanics (in which position is the special selected observable) is still non-local.  

The Kochen-Specker Theorem shows that it is impossible to embed all the probability spaces in Lemma \ref{quantumprob} for all $A$ into one all-embracing classical probability space in a satisfactory way.\footnote{It was shown in Chapter \ref{NullificatieStuk} that this is possible if one restricts to a certain dense subset of all self-adjoint operators. However, the present discussion focusses on quantum mechanics itself and from this point of view there is no motivation to deny the reversal of the observable postulate and the Kochen-Specker Theorem applies.} At least, not without resorting to contextuality. Surprisingly enough, even though the Bohrian interpretation emphasizes the incompatibility of different measuring contexts, quantum mechanics itself is not contextual in a certain sense. 
As seen in Lemma \ref{quantumprob}, for a fixed observable the probabilities assigned to different events do \textit{not} depend on the measuring context. More specifically, if $\A_1,\A_2$ and $\A_3$ are three observables, corresponding to the operators $A_1,A_2,A_3$, such that $[A_1,A_3]=[A_2,A_3]=0$ but $[A_1,A_2]\neq0$, then (using the notation of Section \ref{The Bell Inequality}) one has
\begin{equation}
	\pee_{\{\A_1,\A_3\}}(\Delta)=\pee_{\{\A_2,\A_3\}}(\Delta),\quad\forall \Delta\in\Sigma_{A_3}.
\end{equation}
This is a consequence of the fact that probabilities assigned to events in quantum mechanics are in a certain sense blind to the actual observable considered: that is, for any pair of observables $\A_1,\A_2$ corresponding to the operators $A_1,A_2$ with $\Delta_1\in\Sigma_{A_1}$ and $\Delta_2\in\Sigma_{A_2}$ such that $\mu_{A_1}(\Delta_1)=\mu_{A_2}(\Delta_2)$, for any state $\psi$ one has 
\begin{equation}
	\pee_\psi[\A_1\in\Delta_1]=\pee_\psi[\A_2\in\Delta_2],
\end{equation} 
irrespective of whether or not $A_1$ and $A_2$ commute. Indeed, in quantum mechanics propositions of the form $\A\in\Delta$ fall into equivalence classes:
\begin{equation}\label{EquivPropQM}
	(\A_1\in\Delta_1)\sim(\A_2\in\Delta_2) \desda \mu_{A_1}(\Delta_1)=\mu_{A_2}(\Delta_2).
\end{equation}
Of course, this non-contextuality cannot be extended to the actual results of a measurements if $\A_1$ and $\A_2$ are incompatible. However, this cannot be considered a contextual aspect of the theory, but is rather a consequence of the indeterministic character of the theory. 

Thus far, quantum mechanics seems to behave pretty decently as a stochastic hidden-variable theory. But certainly quantum mechanics violates the Bell inequality and it must therefore be non-local. To investigate this, quantum theory has to be translated further into the language of stochastic hidden variables (Section \ref{The Bell Inequality}). Although usually the pure states are given by the one-dimensional projections, in this setting the pure states are given by the density operators. Indeed, to each observable $\A$ a density operator $\rho$ assigns a probability measure on the space $(\sigma(A),\Sigma_A)$ for
each measuring context $\mathcal{C}$ with $\A\in\mathcal{C}$ in the following way:
\begin{equation}
	\Sigma_A\ni\Delta\mapsto\pee_{\mathcal{C}}[\A\in\Delta|\rho]=\Trace(\rho\mu_A(\Delta)),
\end{equation}
which is indeed a probability measure (this follows from Lemma \ref{quantumprob}). The rule for conditionalizing is given by the von Neumann postulate. In this case, equation (\ref{conditionering}) becomes 
\begin{equation}
	\pee_{\mathcal{C}}[\A_2\in\Delta_2|\A_1\in\Delta_1,\rho]=
		\pee_{\mathcal{C}}\left[\A_2\in\Delta_2|\rho'\right],
\end{equation}
where\footnote{In case it worries the reader that $\Trace(\rho\mu_{A_1}(\Delta_1))$ may equal zero, note that in that case the probability of finding the result $\A_1\in\Delta_1$ given the state $\rho$ is zero, too. If one interprets events with probability zero as events that cannot occur, the situation where $\Trace(\rho\mu_{A_1}(\Delta_1))=0$ will never occur. In other cases, one may find it satisfactory to assign the value zero to $\pee_{\mathcal{C}}[\A_2\in\Delta_2|\A_1\in\Delta_1,\rho]$.}
\begin{equation}
	\rho'=\frac{1}{\Trace(\rho\mu_{A_1}(\Delta_1))}\sum_{a\in\Delta_1}\mu_{A_1}(\{a\})\rho\mu_{A_1}(\{a\}).
\end{equation}
The macro states for this hidden-variable theory may be considered all to be of the form $\mu=\delta_\rho$ in the sense of equation (\ref{SHVTDirac}). 

The notation introduced here allows one to check if quantum mechanics satisfies the locality conditions OILOC and CILOC. CILOC is clearly satisfied, since quantum mechanics is non-contextual, as seen above. Therefore, since quantum mechanics does violate the Bell inequality, it must violate OILOC. This is indeed the case. Consider again the state $\psi=\frac{1}{\sqrt{2}}(0,1,-1,0)$ in the Hilbert space $\mathbb{C}^2\otimes\mathbb{C}^2$ as in Example \ref{EPRB} and the measuring context $\mathcal{C}_{12}$ from Lemma \ref{SHVT-Bell}. One then has
\begin{equation}
	\pee_{\mathcal{C}_{12}}[\sigma_{r_1}=1|P_{\psi}]=\Trace(P_{\psi}(P_{r+}\otimes\een))=\frac{1}{2},
\end{equation}
but
\begin{equation}
\begin{split}
	\pee_{\mathcal{C}_{12}}[\sigma_{r_1}=1|\sigma_{r_2}
	&=1,P_{\psi}]=\Trace\left(
	\frac{(\een\otimes P_{r+})P_{\psi}(\een\otimes P_{r+})}{\Trace(P_{\psi}
	(\een\otimes P_{r+}))}(P_{r+}\otimes\een)\right)=0\\
	&=2\Trace(P_{\psi}(P_{r+}\otimes P_{r+}))=0.
\end{split}
\end{equation}
It would therefore seem that quantum mechanics is indeed non-local in this specific sense. But that conclusion seems be a bit rash. What does a violation of OILOC imply? 
There is in fact much discussion on the issue whether or not a violation of OILOC implies non-locality, which I will not discuss here.\footnote{\cite{Fine82}, \cite{Jarrett84}, \cite[Ch. 4]{Fraassen91}, \cite{Shimony93}, \cite[Ch. 4]{Maudlin94} is a (too) short list of relevant publications.} Instead, I will only make a few remarks.

From an information-theoretic point of view it would seem logical that new information alters the probabilities one should assign to events. However, according to the state postulate, the state $\psi$ should already contain \textit{all} the information necessary to describe the system, since it is supposed a \textit{complete} description of the system. It can only be maintained that both the states before the measurement and after the measurement provide a complete description of the system if the system at hand has been changed under the influence of the measurement.
Since the description of both subsystems has been altered after the measurement, both subsystems must have been influenced by the measurement, which implies an action at a distance. It is in this sense that a violation of OILOC implies non-locality.

This line of reasoning is appears to be completely satisfactory, but there are a few loopholes. In fact, the Copenhagen interpretation provides a way out. One may consider the wave function to be a complete description of the system and maintain that the second particle remains undisturbed under the influence of a measurement, \textit{provided one no longer considers the wave function as a property possessed by the system, but as a description of the system}. Then, after the first measurement, what changes is not the system but our knowledge about the system.\footnote{I will later return to the question of what is actually meant with `the knowledge about the system'.} Indeed, after the first measurement the observer has gained new information about the system, but this new information did not have any ontological value before the measurement. 

This may sound like a dispute of the reality of the system and an encouragement of the view that the measurement brings into being certain properties of the system. This is indeed a form of the Copenhagen interpretation often heard, but it is not a necessary conclusion. All that is maintained is that certain notions used to describe a system (i.e. the wave function, but also observables like spin, position and momentum) are not to be considered \textit{possessed} properties of the system but rather aspects of the way the system is \textit{described}. That is, in contrast with Einstein's view, these aspects are not supposed to correspond to elements of reality. In the words of Bohr:
\begin{quote}
 ``The entire formalism [quantum mechanics] is to be considered as a tool for deriving predictions, of definite or statistical character, as regards information obtainable under experimental conditions described in classical terms and specified by means of parameters entering into the algebraic or differential equations of which the matrices or the wave-functions, respectively, are solutions. These symbols themselves [\ldots] are not susceptible to pictorial interpretation; and even derived real functions like densities and currents are only to be regarded as expressing the probabilities for the occurrence of individual events observable under well-defined experimental conditions.'' \cite[p. 314]{Bohr48}
\end{quote}

There are some naturally appealing features to this viewpoint. What is being questioned here is not so much the \textit{existence} of reality but rather the direct correspondence between reality and the way we humans \textit{perceive} reality. This correspondence is assumed explicitly in \cite{EPR} by introducing a sufficient condition for the existence of an element of physical reality (EOPR). However, such an assumption is of a metaphysical nature and naturally a physical theory may only describe the way we humans perceive reality. 

A few remarks are in place. First, the viewpoint sketched here may be termed `the realist interpretation of Bohr' which is also promoted by Folse in \cite{Folse85}. However, there are also analyses of the philosophy of Bohr that portray Bohr as an anti-realist. In \cite{Landsman06} both these viewpoints are considered. Second, from the viewpoint sketched above, the notion of completeness in \cite{EPR} no longer makes sense. Instead, there may be other notions of completeness such that quantum mechanics may be considered complete. In fact, Bohr maintained throughout the development of quantum mechanics that it is a complete theory. However, there is no consensus about what he meant by that. At least, the impossibility proofs establish that it is a non-trivial matter to extend quantum mechanics to a theory that would be `more complete' in a satisfactory way. If one accepts the impossibility of extending quantum theory, then the theory may be considered complete. In \cite{ElbyBrownFoster93} the notion of completeness is discussed in more detail.

Finally, I'd like to make a remark that is perhaps somewhat controversial. In the experiment, after Alice has performed her experiment, her description of the entire system has been altered by the von Neumann postulate. However, Bob still uses the `old' description. It seems that it is natural to assume that Bob's description is `flawed'. However, this conclusion seems to rely on the assumption that Alice actually describes \textit{the} state of the system. There are (at least) two ways out of this dilemma. The first is to state that, although Alice's description is not \textit{the} state of the system, it is \textit{the} objective \textit{description of the system}. From this point of view, Bob's description is indeed flawed. The other way out is to assume that both Alice and Bob are right and there is no such thing as \textit{the} objective description of a system, but only subjective descriptions. Indeed, from Bob's perspective, his state gives a complete description of the system, in the sense that it contains all the information available to him.\footnote{These two viewpoints are not necessarily mutually exclusive. In \cite{Myrvold02} a modification of the quantum-mechanical state is introduced in such a way that its behavior is somewhat like that of the electromagnetic field in special relativity; the notion of the state is objective, but the way it should be interpreted depends on the reference frame of the observer (as does its time-evolution). Roughly, the collapse of the state propagates with the speed of light. So directly after the first measurement, both Alice's and Bob's description of the system are correct descriptions.}

\markboth{The Strangeness and Logic of Quantum Mechanics}{Quantum Logic and the Violation of the Bell Inequality}
\subsection{Quantum Logic and the Violation of the Bell Inequality}\label{QLogicSection}

In the previous section it was established that, if one views quantum mechanics as a stochastic hidden-variable theory, the violation of the Bell inequality by quantum mechanics may be attributed to a violation of OILOC. If OILOC is considered as a notion of locality, it is clear what a violation of it means and that it is disturbing. But, as I argued, this viewpoint is not necessary. On the other hand, it is not clear what OILOC means precisely if it is not interpreted as a notion of locality and it is not at all clear how one should understand a violation of OILOC in that case. This problem is not easily solved, and I will not make an attempt here. What I will do is establish another proof of the Bell inequality based on logical consideration instead of philosophical ones. From this proof I will explain how I think Bohr would consider it to be possible to violate the inequality, how orthodox quantum mechanics violates this inequality, and how I think the violation should be interpreted. 

Before I continue, there is an important remark to be made. The Bell inequality derived in the following lemma is one of probabilistic logic and not one of probability theory. The first is a branch of logic, whereas the second is a branch of mathematics. The mathematical tools used in probability theory are therefore not used and hence are not necessarily assumed to be true. For example, it is not assumed that all probability functions can be associated with finite measures on some measurable space. Instead, in probabilistic logic a probability function $\pee$ is a rule that to each proposition $\texttt{A}$ assigns a value $\pee(\texttt{A})\in[0,1]$ that denotes the probability that the proposition is true. There are many forms of probabilistic logic, each assuming different rules that should hold for the function $\pee$. I will assume the following rules:
\begin{equation}\label{LogischeEisen}
\begin{gathered}
	\text{1) If }\texttt{A}\to\texttt{B},\text{ then }\pee(\texttt{A})\leq\pee(\texttt{B})\\
	\text{2) If }\texttt{A}\text{ and }\texttt{B}\text{ are mutually exclusive, i.e., }\texttt{A}\wedge\texttt{B}\to\bot,\\
	\text{ then }\pee(\texttt{A}\vee\texttt{B})=\pee(\texttt{A})+\pee(\texttt{B})
\end{gathered}
\end{equation}
It is further assumed that the propositions obey all the relations of \textit{classical logic}.

\begin{lemma}\label{LogicBellLemma}
	For each probability function $\pee$ that satisfies (\ref{LogischeEisen}), and for all propositions $\texttt{A}_1$, $\texttt{B}_1$, $\texttt{A}_2$ and $\texttt{B}_2$ one has
	\begin{equation}\label{BellIneqLogic}
		\pee(\texttt{A}_1\wedge\texttt{B}_1)\leq
		\pee(\texttt{A}_1\wedge\texttt{B}_2)+\pee(\texttt{A}_2\wedge\texttt{B}_1)+\pee(\neg\texttt{A}_2\wedge\neg\texttt{B}_2)
	\end{equation}
\end{lemma}

\noindent
\textit{Proof:}\hspace*{\fill}\\
Using the rules assumed earlier, the inequality follows by simply expanding:
\begin{equation}
\begin{split}
	\pee(\texttt{A}_1\wedge\texttt{B}_1)
	&=
	\pee(\texttt{A}_1\wedge\texttt{B}_1\wedge(\texttt{B}_2\vee\neg\texttt{B}_2))\\
	&=
	\pee((\texttt{A}_1\wedge\texttt{B}_1\wedge\texttt{B}_2)\vee(\texttt{A}_1\wedge\texttt{B}_1\wedge\neg\texttt{B}_2))\\
	&=
	\pee(\texttt{A}_1\wedge\texttt{B}_1\wedge\texttt{B}_2)+\pee(\texttt{A}_1\wedge\texttt{B}_1\wedge\neg\texttt{B}_2)
	\leq
	\pee(\texttt{A}_1\wedge\texttt{B}_2)+\pee(\texttt{B}_1\wedge\neg\texttt{B}_2)\\
	&=
	\pee(\texttt{A}_1\wedge\texttt{B}_2)+\pee(\texttt{B}_1\wedge\neg\texttt{B}_2\wedge(\texttt{A}_2\vee\neg\texttt{A}_2))\\
	&=
	\pee(\texttt{A}_1\wedge\texttt{B}_2)+
	\pee((\texttt{B}_1\wedge\neg\texttt{B}_2\wedge\texttt{A}_2)\vee(\texttt{B}_1\wedge\neg\texttt{B}_2\wedge\neg\texttt{A}_2))\\
	&=
	\pee(\texttt{A}_1\wedge\texttt{B}_2)+
	\pee(\texttt{B}_1\wedge\neg\texttt{B}_2\wedge\texttt{A}_2)+\pee(\texttt{B}_1\wedge\neg\texttt{B}_2\wedge\neg\texttt{A}_2)\\
	&\leq
	\pee(\texttt{A}_1\wedge\texttt{B}_2)+\pee(\texttt{A}_2\wedge\texttt{B}_1)+\pee(\neg\texttt{A}_2\wedge\neg\texttt{B}_2)
\end{split}
\end{equation}
\hfill $\square$\\[0.5ex]
\indent

Typical propositions that play a role in physics are of the form $[\A\in\Delta]$, where $\A$ is an observable and $\Delta$ some subset of $\mathbb{R}$ (preferably Borel). Here, $[\A\in\Delta]$, may be understood as a shorthand notation of the statement: ``The answer to the question ``Does the value of $\A$ lie in $\Delta$?'' is yes'' (see also the discussion just after the proof of Lemma \ref{KS3dim} in Section \ref{Kochen-Specker-sectie}).  

By considering such propositions, it is not hard to show that quantum mechanics violates the inequality (\ref{BellIneqLogic}). It is again sufficient to look at the pair of spin-$\tfrac{1}{2}$ particles in the state $\psi=\tfrac{1}{2}\sqrt{2}(0,1,-1,0)$. Note that every spin operator along some axis $r$ in the $xy$-plane is of the form
\begin{equation}
	\sigma_r=\begin{pmatrix} 0 & \cos(\theta_r)-i\sin(\theta_r) \\  \cos(\theta_r)+i\sin(\theta_r) & 0 \end{pmatrix}.
\end{equation} 
Now let $\sigma_a^{(1)}$ be the spin along the axis $a$ for one particle and $\sigma_b^{(2)}$ the spin along the axis $b$ for the second particle. Using the notation of Example \ref{EPRB} one finds
\begin{equation}\label{ConjKans1}
\begin{split}
	\pee_{\psi}[\sigma_a^{(1)}\in\{1\}\wedge\sigma_b^{(2)}\in\{1\}]
	=&
	\langle\psi, P_{a+}\otimes P_{b+}\psi\rangle\\
	=&
	\frac{\langle e_1\otimes e_2-e_2\otimes e_1,(P_{a+}\otimes P_{b+})e_1\otimes e_2-e_2\otimes e_1\rangle}{2}\\
	=&
	\frac{
	{P_{a+}}_{11}{P_{b+}}_{22}-{P_{a+}}_{12}{P_{b+}}_{21}+{P_{a+}}_{22}{P_{b+}}_{11}
	-{P_{a+}}_{21}{P_{b+}}_{12}}{2}\\
	=&
	\frac{1}{8}\Bigl(1-(\cos(\theta_a)-i\sin(\theta_a))(\cos(\theta_b)+i\sin(\theta_b))\\
	&+1-(\cos(\theta_a)+i\sin(\theta_a))(\cos(\theta_b)-i\sin(\theta_b)\Bigr)\\
	=&
	\frac{1}{4}\left(1-\cos(\theta_a-\theta_b)\right).
\end{split}
\end{equation}
Similarely, one finds
\begin{equation}\label{ConjKans2}
	\pee_{\psi}[\sigma_a^{(1)}\in\{-1\}\wedge\sigma_b^{(2)}\in\{-1\}]=\frac{1}{4}\left(1-\cos(\theta_a-\theta_b)\right).
\end{equation}
Now take
\begin{equation}\label{ThetaKeuze}
	\theta_{a_1}=0,\quad\theta_{a_2}=\frac{2\pi}{3},\quad\theta_{b_1}=\pi,\quad\theta_{b_2}=\frac{\pi}{3},
\end{equation}
and set
\begin{equation}
	\texttt{A}_j=[\sigma_{a_j}^{(1)}\in\{1\}],\quad\neg\texttt{A}_j=[\sigma_{a_j}^{(1)}\in\{-1\}],\quad
	\texttt{B}_j=[\sigma_{b_j}^{(2)}\in\{1\}],\quad\neg\texttt{B}_j=[\sigma_{b_j}^{(2)}\in\{-1\}],
\end{equation}
for $j=1,2$.
With these identifications and the expressions (\ref{ConjKans1}) and (\ref{ConjKans2}), the inequality (\ref{BellIneqLogic}) would now read
\begin{equation}
	\frac{1}{2}=\pee(\texttt{A}_1\wedge\texttt{B}_1)\leq
		\pee(\texttt{A}_1\wedge\texttt{B}_2)+\pee(\texttt{A}_2\wedge\texttt{B}_1)+\pee(\neg\texttt{A}_2\wedge\neg\texttt{B}_2)
	=\frac{1}{8}+\frac{1}{8}+\frac{1}{8},
\end{equation}
which is of course false.
 
The derivation of the Bell inequality (\ref{BellIneqLogic}) is not based on philosophical concepts like realism, causality and locality, but on purely classically logical considerations. Of course, the same inequality can also be derived assuming the structure of a classical probability space and associating with each proposition some subset of this space. However, since such assumptions were not made, the usual objections made against the derivation of the Bell inequality, or explanations of its violation, can no longer be applied in a direct manner. The easiest objection against the logical derivation is that it implicitly assumes non-contextuality. Indeed, it is assumed that the proposition $\texttt{A}_1$ is the same in both the contexts $\{\sigma_{a_1}^{(1)},\sigma_{b_1}^{(2)}\}$ and $\{\sigma_{a_1}^{(1)},\sigma_{b_2}^{(2)}\}$, which is denied in viable hidden-variable theories. But, as argued in Section \ref{QMHVT}, quantum mechanics is in a certain sense a non-contextual theory, at least in such a way that the argument of contextuality cannot be applied without making additional assumptions. Bohr's answer to the paradox would probably be of a more philosophical nature. The derivation of the inequality relies on introducing propositions which involve speaking about both observables $\sigma_{a_1}^{(1)}$ and $\sigma_{a_2}^{(1)}$ or both $\sigma_{b_1}^{(2)}$ and $\sigma_{b_2}^{(2)}$. However, such observables are complementary. A proposition like $\texttt{A}_1\wedge\texttt{B}_1\wedge\texttt{B}_2$ relies on an abuse of language and is therefore ambiguous. This makes the whole derivation ambiguous and hence incorrect. However, although this Bohrian line of reasoning may sound appealing, it is not based on quantum mechanics as it is. There is no mentioning of the notion of complementarity in any of the axioms of this theory and also, there exists no derivation of this notion from the axioms. In fact, orthodox quantum mechanics can violate the inequality without resorting to any philosophical considerations. Each proposition in the derivation of (\ref{BellIneqLogic}) can be associated with a mathematical object in the theory in a consistent way. This association was first conceived in 1936 by Birkhoff and von Neumann \cite{BirkhoffNeumann36}. In this article it is shown that quantum mechanics can be viewed so as to obey a form of logic noticeably different from the logic that is customary held to be true (i.e. classical logic). This new form of logic is usually termed quantum logic. I will give a short derivation of this logic based on the one given in \cite{Isham95}.   

As noted earlier, typical propositions in physics are of the form $[\A\in\Delta]$. In quantum mechanics, each such proposition can be associated with a projection operator, namely, $\mu_A(\Delta)$. By (\ref{EquivPropQM}), propositions fall into equivalence classes and there is a 1-1 correspondence between the set of equivalence classes and the set of projection operators $\mathcal{P}(\h)$. 
Following Isham in \cite{Isham95}, the following meaning is given to the statement that $P$ is true, for some $P\in\mathcal{P}(\h)$:
\begin{definitie}
	For a quantum system in the state $\psi$ each proposition associated with the projection $P$ is true if $\langle\psi,P\psi\rangle=1$, i.e., a proposition $[\A\in\Delta]$ is true if the probability of finding a value in $\Delta$ upon measurement of $\A$ is one.
\end{definitie}
Naturally, with each proposition one can associate the set of all states for which this proposition is true. For a proposition associated with the projection $P$ this set is precisely given by $P\h$. Notice that $P\h$ is a closed linear subspace of $\h$ and that in fact all closed linear subspaces are of this form. This identification can be used to introduce the logical connectives. 
\begin{itemize}
\item Disjunction: The statement $[\A\in\Delta\text{ or } \A\in\Delta']$ is associated with the set of all states with the property that a measurement of $\A$ will yield an element of $\Delta$ or an element of $\Delta'$ with probability one. This corresponding set of states will then be the closure of the set $\Sp(\mu_A(\Delta)\h,\mu_A(\Delta')\h)$, where $\Sp$ denotes the linear span. This closed linear subspace is in fact equal to $\mu_A(\Delta\cup\Delta')\h$. Generalizing this result for different operators yields the identification
\begin{equation}
	[\A_1\in\Delta_1]\Qvee[\A_2\in\Delta_2]\:\hat{=}\:\overline{\Sp(\mu_{A_1}(\Delta_1)\h,\mu_{A_2}(\Delta_2)\h)},
\end{equation}
where the overlining denotes the closure of the set.
\item Conjunction: The statement $[\A\in\Delta\text{ and } \A\in\Delta']$ is associated with the set of all states with the property that a measurement of $\A$ will yield a result that is both an element of $\Delta$ and an element of $\Delta'$ with probability one. This corresponding set of states will then be $\mu_A(\Delta)\h\cap\mu_A(\Delta')\h$. This is again a closed linear subspace given by $\mu_A(\Delta\cap\Delta')\h$. Generalizing this result for different operators yields the identification
\begin{equation}
	[\A_1\in\Delta_1]\Qwedge[\A_2\in\Delta_2]\:\hat{=}\:\mu_{A_1}(\Delta_1)\h\cap\mu_{A_2}(\Delta_2)\h.
\end{equation}
This is the set of states with the property that a measurement of $\A_1$ will yield an element of $\Delta_1$ with probability 1 and a measurement of $\A_2$ will yield an element of $\Delta_2$ with probability 1.
\item Negation: The statement $[\A\notin\Delta_A]$ is associated with the set of all states with the property that a measurement of $\A$ will yield a result that is not an element of $\Delta_A$ with probability one. This corresponding set of states is given by $(\mu_A(\Delta)\h)^\bot$. This is again a closed linear subspace, given by $\mu_A(\Delta^c)\h$, where $\Delta^c$ denotes the complement in the set $\sigma(A)$. So
\begin{equation}
	\Qneg[\A\in\Delta]\:\hat{=}\:\mu_A(\Delta^c)\h.
\end{equation}
\end{itemize}
With these identifications, one sees that every proposition can be associated with a unique projection, irrespective of whether it concerns an elementary proposition\footnote{In logic, one may probably use the term `atomic formula'.} of the form $[\A\in\Delta]$, or a proposition that is formed using several elementary propositions together with the connectives. The propositional calculus then inherits the lattice structure of the set of projection operators. The partial ordering is given by
\begin{equation}	
	P_1\leq P_2\desda P_1\h\subset P_2\h.
\end{equation}
The top element $\top$ is given by the unit operator $\een$. This is taken to correspond with `absolute truth'. Indeed, the set $\h$ corresponds with the set of states for which all propositions are true. Similarly, the bottom element $\bot$ is given by the zero operator $\nul$. Within this structure, the following rules hold for all propositions $\texttt{A}_1,\texttt{A}_2$ and $\texttt{A}_3$:
\begin{equation}
\begin{gathered}
	\texttt{A}_1\vee(\texttt{A}_2\vee\texttt{A}_3)=(\texttt{A}_1\vee\texttt{A}_2)\vee\texttt{A}_3 \\
	\texttt{A}_1\vee\texttt{A}_2=\texttt{A}_2\vee\texttt{A}_1 \\
	\texttt{A}_1\vee(\texttt{A}_1\wedge\texttt{A}_2)=\texttt{A}_1 \\
	\texttt{A}_1\vee\neg\texttt{A}_1=\top
\end{gathered}\quad
\begin{gathered}
	\text{Associativity} \\
	\text{Commutativity} \\
	\text{Absorption} \\
	\text{Complements} 
\end{gathered}\quad
\begin{gathered}
	\texttt{A}_1\wedge(\texttt{A}_2\wedge\texttt{A}_3)=(\texttt{A}_1\wedge\texttt{A}_2)\wedge\texttt{A}_3 \\
	\texttt{A}_1\wedge\texttt{A}_2=\texttt{A}_2\wedge\texttt{A}_1 \\
	\texttt{A}_1\wedge(\texttt{A}_1\vee\texttt{A}_2)=\texttt{A}_1 \\
	\texttt{A}_1\wedge\neg\texttt{A}_1=\bot
\end{gathered}
\end{equation}
However, the relations 
\begin{equation}
	\texttt{A}_1\vee(\texttt{A}_2\wedge\texttt{A}_3)=(\texttt{A}_1\vee\texttt{A}_2)\wedge(\texttt{A}_1\vee\texttt{A}_3)
	\quad\text{Distributivity}\quad
	\texttt{A}_1\wedge(\texttt{A}_2\vee\texttt{A}_3)=(\texttt{A}_1\wedge\texttt{A}_2)\vee(\texttt{A}_1\wedge\texttt{A}_3)
\end{equation}
no longer hold in general. In particular, the second and the sixth step in the proof of Lemma \ref{LogicBellLemma} do not hold for the particular case of the two spin-$\tfrac{1}{2}$ particles considered.

There has been some discussion through the years about the question how significant the discovery of this so-called quantum logic is for the interpretation of quantum mechanics. Probably one of the most progressive views was advocated by Putnam in \cite{Putnam69}. In this article, it is argued that most of the problems concerning quantum mechanics would cease to exist if one adopts the viewpoint that quantum logic is in fact the only `true' logic. Classical logic then re-emerges as a certain limit similar to the way Euclidean geometry appears as a special case of non-Euclidean geometry. In particular, Putnam argued that one could easily adopt a (non-classical) realist point of view towards systems, observables and measurements. 

It seems to me that history wasn't very kind to this viewpoint. There were many more people willing to criticize this viewpoint than willing to endorse it.\footnote{This simply seems to be the faith of controversial ideas as also seen earlier in the `nullification' discussion in Chapter \ref{NullificatieStuk}. Consequently, it seems to me that such ideas have a tendency to be forgotten before they are well understood. Indeed, Maudlin's ``Tale of Qunatum Logic'' \cite{Maudlin05} has a beginning \textit{and} an end. It should be noted that one of the factors that may have played a role is that Putnam later distanced himself from his own ideas.} However, one of the most interesting articles due to Stairs \cite{Stairs83}, exposes some of the difficulties that in light of Putnam's thesis may be formulated as follows. Although quantum logic allows one to see why the Bell inequality (\ref{BellIneqLogic}) may be violated, it does not help much in understanding why the Bell inequality in Section \ref{The Bell Inequality} can be violated without engaging in the philosophical discussion involving realism and locality. But even so, what quantum logic does is replace the mystery of ``Why can the Bell inequality be violated?'' with the mystery of ``Why can the law of distributivity be violated?'' This last mystery is extensively discussed in \cite{Dummett76} and it seems that there is no solution at hand.

The quantum logical approach to the explanation of the violation of the Bell inequality (\ref{BellIneqLogic}) appears to be more precise than the Bohrian approach, since it exactly indicates which steps of the proof of Lemma \ref{LogicBellLemma} are flawed. But this approach also seems to immediately arrive at a philosophical dead end: the violation of distributivity. Indeed, from a philosophical point of view the Bohrian approach seems more appealing, even though it relies on the vague notion of complementarity. It does seem to be the case that the proof of Lemma \ref{LogicBellLemma} relies on ambiguous manipulations of the propositions. In fact, I believe that this discussion may provide us with a way to turn the situation around. Indeed, one should not search for an understanding of the notion of complementarity to explain why the proof of Lemma \ref{LogicBellLemma} is flawed, but instead, one should search for an understanding of the flaw of the proof to reach an explanation of the notion of complementarity. 

Approaches similar to this have been carried out several times, of course. For example, in \cite{Heelan70} an explanation of complementarity is explored from a quantum-logical point of view. However, this is of course again replacing a mystery by another mystery. In the same way, one also sometimes hears explanations based on non-commutativity, which plays a role in the mathematical structure of quantum mechanics. But this idea is of course circular. For Bohr, position and momentum are identified with non-commuting operators because the observables are of a complementary nature, not the other way around. In fact, one may argue that explanations based on mathematical structures are hardly explanations at all, ever. It also seems to me that the quantum-logical approach doesn't capture the ideas of Bohr at all. The first abuse of language in the proof of Lemma \ref{LogicBellLemma} already appears in the first step, where the incompatible statements about the observables $\sigma_{b_1}^{(2)}$ and $\sigma_{b_2}^{(2)}$ are introduced by appealing to the innocent looking law of excluded middle. However, this logical law has been accused earlier of being applied `recklessly' by Brouwer (the founder of Intuitionism), and it's worth an investigation to find out whether or not something similar is going on in quantum mechanics.       

\subsection{Intuitionism and Complementarity}\label{IntuiCompl}
\markboth{The Strangeness and Logic of Quantum Mechanics}{Intuitionism and Complementarity}

To see if intuitionism can play a possible role in the interpretation of quantum mechanics it is better to first have a small look at the philosophy behind intuitionism. Intuitionism is a philosophy of mathematics, which opposes the Platonic idea that mathematical objects exist independently of the mathematician. Consequently, for Brouwer, there is no independent mathematical truth; propositions only become true when one experiences its truth i.e., if one has found a proof for them. For Brouwer, this is also what a proof is; a construction that one rehearses in one's head. It seems he never saw any reason to formalize this notion much further, but fortunately, his student Heyting did, thereby introducing intuitionistic logic \cite{Heyting30}. Independently, a few years earlier Kolmogorov also had come up with a formalization of some of Brouwer's ideas \cite{Kolmogorov25}. These works have led to what is now known as the Brouwer-Heyting-Kolmogorov (BHK) interpretation of logical connectives. These are as follows:
\begin{itemize}
\item A proof of $\texttt{A}\wedge\texttt{B}$ consists of a proof of $\texttt{A}$ and a proof of $\texttt{B}$.
\item A proof of $\texttt{A}\vee\texttt{B}$ consists of a proof of $\texttt{A}$ or a proof of $\texttt{B}$ together with a rule that tells one of which statement one has the proof i.e., one can \textit{decide} whether $\texttt{A}$ is true or $\texttt{B}$ is true.
\item A proof of $\texttt{A}\to\texttt{B}$ consists of a rule that converts every proof of $\texttt{A}$ to a proof of $\texttt{B}$.
\item A proof of $\neg\texttt{A}$ consists of a rule that converts every proof of $\texttt{A}$ to a proof of $0=1$ i.e., a proof of $\neg\texttt{A}$ is a proof of $\texttt{A}\to\bot$.
\end{itemize}
In particular, a proof of $\texttt{A}\vee\neg\texttt{A}$ consists of showing that either $\texttt{A}$ is true, or showing that $\texttt{A}$ leads to a contradiction, and it is therefore not a triviality (as it is in the Platonic view). 

Brouwer himself was not that impressed by the introduction of intuitionistic logic. He viewed logic merely as the study of regularities that appear in the use of language \cite{Brouwer08}. But for him, performing mathematics is a language-free act. The only role language plays in mathematics is in communication, where one may only hope that expressing a proof in words leads to a similar construction in the mind of the person to whom you are explaining the proof. So for both Brouwer and Bohr, language may be seen as a necessary evil for communication. However, whereas for Bohr communication is seen as a necessary ingredient of science, for Brouwer communication wasn't that important.  

Of course, the BHK interpretation does not yet establish what an actual proof is,\footnote{What actually counts as a proof and what does not is one of the general questions in the philosophy of mathematics. A friendly book that looks at this question from a historic perspective is \cite{Krantz07}.} but only states how one should read the logical connectives. However, what counts as a proof is irrelevant to the present discussion. What \textit{is} interesting, is to see if these interpretations of connectives can be adopted to apply to physical statements, instead of mathematical ones. In particular, I'll first take a closer look at the inequality of the previous section.

The abuse of language involved in Lemma \ref{LogicBellLemma} actually already appears in the inequality (\ref{BellIneqLogic}) itself. The four statements $\texttt{A}_1\wedge\texttt{B}_1,\texttt{A}_1\wedge\texttt{B}_2,\texttt{A}_2\wedge\texttt{B}_1$ and $\neg\texttt{A}_2\wedge\neg\texttt{B}_2$ all refer to different experimental situations, in each of which only two observables are being measured. At least, this is the assumption one must make if one wishes to apply equations (\ref{ConjKans1}) and (\ref{ConjKans2}). However, in practice, only one of these experimental situations can be actual, the others are hypothetical. This means that inequality (\ref{BellIneqLogic}) expresses a counterfactual statement, which may indeed be considered an abuse of language.\footnote{This may be compared to the counterfactual reasoning that appears in Example \ref{EPRB}.} 

In the proof of Lemma \ref{LogicBellLemma}, counterfactual reasoning already enters in the first step. The reasoning is: 
\begin{quote}
	``If instead of measuring $\sigma_{a_1}^{(1)}$ and $\sigma_{b_1}^{(2)}$ one would measure $\sigma_{b_2}^{(2)}$, one would find either the result 1 or -1 i.e., one would be able to conclude $\texttt{B}_2$ or $\neg\texttt{B}_2$.'' 
\end{quote}
However, in the exact situation where one is measuring $\sigma_{a_1}^{(1)}$ and $\sigma_{b_1}^{(2)}$, one can neither draw the conclusion $\texttt{B}_2$, nor the conclusion $\neg\texttt{B}_2$. Consequently, one cannot draw the conclusion $\texttt{B}_2\vee\neg\texttt{B}_2$ either, that is (using the terminology of Brouwer), the statement $\texttt{B}_2\vee\neg\texttt{B}_2$ is \textit{reckless}, unless one actually measures $\sigma_{b_2}^{(2)}$. So, the physical `proof' of a statement may be regarded to be the actual appearance of a measurement result.

The obvious way out would seem to be to also measure $\sigma_{b_2}$. Then, if one also assumes that $\sigma_{a_2}$ will be measured, every step in the proof of Lemma \ref{LogicBellLemma} is intuitionisticaly justified. But surely the inequality would still be violated by quantum mechanics? However, it turns out that this is no longer the case, since in this experimental setup, where on one side first $\sigma_{a_1}^{(1)}$ is measured and then $\sigma_{a_2}^{(1)}$, and on the other side first $\sigma_{b_1}^{(2)}$ is measured and then $\sigma_{b_2}^{(2)}$, the equations (\ref{ConjKans1}) and (\ref{ConjKans2}) will no longer hold in general. Although in the special case of (\ref{ThetaKeuze}) one may still show that
\begin{equation}
	\pee(\texttt{A}_1\wedge\texttt{B}_2)=\pee(\texttt{A}_2\wedge\texttt{B}_1)=\frac{1}{8},
\end{equation}
one now finds that
\begin{multline}
	\pee(\neg\texttt{A}_2\wedge\neg\texttt{B}_2)\\
	\begin{split}
		=&
		\pee(\neg\texttt{A}_2\wedge\neg\texttt{B}_2\wedge\texttt{A}_1\wedge\texttt{B}_1)+
		\pee(\neg\texttt{A}_2\wedge\neg\texttt{B}_2\wedge\neg\texttt{A}_1\wedge\texttt{B}_1)\\
		&+
		\pee(\neg\texttt{A}_2\wedge\neg\texttt{B}_2\wedge\texttt{A}_1\wedge\neg\texttt{B}_1)+
		\pee(\neg\texttt{A}_2\wedge\neg\texttt{B}_2\wedge\neg\texttt{A}_1\wedge\neg\texttt{B}_1)\\
		=&
		\langle\psi,(P_{a_1+}P_{a_2-}P_{a_1+}\otimes P_{b_1+}P_{b_2-}P_{b_1+})\psi\rangle+
		\langle\psi,(P_{a_1-}P_{a_2-}P_{a_1-}\otimes P_{b_1+}P_{b_2-}P_{b_1+})\psi\rangle\\
		&+
		\langle\psi,(P_{a_1+}P_{a_2-}P_{a_1+}\otimes P_{b_1-}P_{b_2-}P_{b_1-})\psi\rangle+
		\langle\psi,(P_{a_1-}P_{a_2-}P_{a_1-}\otimes P_{b_1-}P_{b_2-}P_{b_1-})\psi\rangle
	\end{split}\\
	=\frac{9}{32}+0+0+\frac{1}{32}=\frac{5}{16}.
\end{multline}
So in this case, the inequality (\ref{BellIneqLogic}) reads $\tfrac{1}{2}\leq\tfrac{1}{8}+\tfrac{1}{8}+\tfrac{5}{16}$, which is actually true.

This example strengthens the believe that, in quantum mechanics, a statement of the form $\texttt{A}\vee\neg\texttt{A}$ may only be considered to be true if an actual experiment is performed that will decide whether $\texttt{A}$ or $\neg\texttt{A}$ is true i.e., if one can find a `physics proof' of $\texttt{A}\vee\neg\texttt{A}$. In this light, Peres' statement that ``unperformed experiments have no result''\footnote{This is the title of Peres' article \cite{Peres78} in which he advocates against the use of counterfactual reasoning. This opinion is also reflected in \cite{Peres84} and \cite{Peres02}.} may be explained by saying that any statement $\texttt{A}$ is neither true nor false unless one has a proof of one or the other i.e., unproven statements are not true (nor are they not true). Now, two statements $\texttt{A}$ and $\texttt{B}$ may be called \underline{complementary} if one cannot \textit{simultaneously} prove $\texttt{A}\vee\neg\texttt{A}$ and $\texttt{B}\vee\neg\texttt{B}$. That is, two statements are complementary if they are not simultaneously decidable, ever. 
 In quantum mechanics, this is for example the case if $\texttt{A}$ is a statement about some observable $\A$, and $\texttt{B}$ a statement of some observable $\mathcal{B}$ whose corresponding operators $A$ and $B$ do not commute. This seems a triviality, but notice that in this way complementarity is defined in logical terms. Its intuitive use in quantum mechanics is merely a \textit{consequence} of this definition. To make things clearer, let me give a more intuitive example.

Consider the double slit experiment. The analysis given by Putnam in \cite{Putnam69} provides a clear view from a logician's perspective. Consider a photon that went through the barrier and let $\texttt{R}$ be the statement ``the photon is detected within a certain region $R$ on the photographic plate''. Further, let $\texttt{S}_1$ be the statement ``the photon went through slit 1'' and $\texttt{S}_2$ the statement ``the photon went through slit 2''. Classically, one would expect that the probability of finding the photon at a certain region $R$ given that both slits are open is given by 
\begin{equation}\label{ClassicInterf}
	\pee(\texttt{R}|\texttt{S}_1\vee\texttt{S}_2)=\frac{1}{2}\pee(\texttt{R}|\texttt{S}_1)+\frac{1}{2}\pee(\texttt{R}|\texttt{S}_2).
\end{equation}
This expression is easily derived if one assumes that one may arrange that $\pee(\texttt{S}_1)=\pee(\texttt{S}_2)$ and that photons cannot go through both slits at the same time. Indeed, one then finds
\begin{equation}
\begin{split}
	\pee(\texttt{R}|\texttt{S}_1\vee\texttt{S}_2)
	=&
	\frac{\pee(\texttt{R}\wedge(\texttt{S}_1\vee\texttt{S}_2))}{\pee(\texttt{S}_1\vee\texttt{S}_2)}
	=
	\frac{\pee((\texttt{R}\wedge\texttt{S}_1)\vee(\texttt{R}\wedge\texttt{S}_2))}{\pee(\texttt{S}_1\vee\texttt{S}_2)}\\
	=&
	\frac{\pee(\texttt{R}\wedge\texttt{S}_1)}{\pee(\texttt{S}_1\vee\texttt{S}_2)}
	+
	\frac{\pee(\texttt{R}\wedge\texttt{S}_2)}{\pee(\texttt{S}_1\vee\texttt{S}_2)}
	=
	\frac{\pee(\texttt{R}\wedge\texttt{S}_1)}{2\pee(\texttt{S}_1)}
	+
	\frac{\pee(\texttt{R}\wedge\texttt{S}_2)}{2\pee(\texttt{S}_2)}\\
	=&
	\frac{1}{2}\pee(\texttt{R}|\texttt{S}_1)+\frac{1}{2}\pee(\texttt{R}|\texttt{S}_2).
\end{split}
\end{equation}
However, it is well known that with both slits open an interference pattern emerges on the photographic plate, which is not what is obtained if one takes the sum of the patterns that emerge with one slit closed. Putnam tries to adopt a realist point of view and therefore states that each photon that goes through the barrier, goes through exactly one slit. He blames the use of the law of distributivity in the second step, which is not generally true in quantum logic. 

Another often-heard conclusion is that the particular photon behaves as a wave instead of a particle in case both slits are open, and can therefore go through both slits at the same time. Both views are rather mystifying. From the intuitionistic point of view the derivation may in fact be considered to be correct. However, in this case the condition $\texttt{S}_1\vee\texttt{S}_2$ means that one can actually decide whether $\texttt{S}_1$ is the case, or $\texttt{S}_2$. Experiments have shown that in cases where one actually can make this decision, the interference pattern dissolves and equation (\ref{ClassicInterf}) holds. Indeed, Putnam makes the starting assumption that in cases where a photon actually goes through the barrier, the statement $\texttt{S}_1\vee\texttt{S}_2$ is always true. Feynman actually got it right from this point of view. About a similar experiment for electrons he concludes:
\begin{quote}
	``What we must say (to avoid making wrong predictions) is the following. If one looks at the holes or, more accurately, if one has a piece of apparatus which is capable of determining whether the electrons go through hole 1 or hole 2, then one \textit{can} say that it goes either through hole 1 or hole 2. \textit{But}, when one does \textit{not} try to tell which way the electron goes, when there is nothing in the experiment to 
disturb the electrons, then one may \textit{not} say that an electron goes either through hole 1 or hole 2. If one does say that, and starts to make any deductions from the statement, he will make errors in the analysis. This is the logical tightrope on which we must walk if we wish to describe nature successfully.'' \cite[p. 37-9]{Feynman63} 
\end{quote}
Indeed, in experiments in which an interference pattern is found, nothing can be said about the path of the photons, whereas in experiments where the path of the photons is detected, the interference pattern dissolves. These two possible views on the behavior of photons are thus complementary. In connection with the logical definition of complementarity, one can say that statements about the specific slit through which a photon passes and statements about the wavelength of the photons (which can be derived from studying the interference pattern) are complementary; they are not simultaneously decidable. 

It is likely that the given logical definition of complementarity doesn't coincide with Bohr's ideas about complementarity and I do not dare to claim that this definition serves as a magical key to a better understanding of the ideas of Bohr. However, the given definition does provide a way to take a new look at quantum mechanics based on considerations that do not depend on quantum mechanics. In this light, quantum mechanics presents itself as a theory in which complementary statements naturally and necessarily arise (e.g. statement about the position and statements about the momentum of a particle are complementary). But quantum mechanics is only a specific example in which complementarity arises and other examples outside physics are quite likely to be possible, as also envisaged by Bohr. The present notion of complementarity also shows what care should be taken in counterfactual reasoning. Indeed, although one can envisage that in a certain situation $\texttt{A}$ may be decidable and in another $\texttt{B}$ may be decidable, one should be careful to note that one cannot, in general, draw the conclusion that both are decidable at the same time. That is, there is no problem with counterfactual reasoning as such, but one should take care not to confuse counterfactual reasoning with factual reasoning.

\subsection{Towards Intuitionistic Quantum Logic}
\markboth{The Strangeness and Logic of Quantum Mechanics}{Towards Intuitionistic Quantum Logic}

In the previous sections some motivation was given to adopt an intuitionistic point of view towards quantum mechanics, and it was argued that this approach seems to be in line with parts of the Copenhagen interpretation. However, all these considerations only seem to float around quantum mechanics, but nowhere do they appear in the mathematical formulation of the theory. This is in strong contrast with, for example, the Bohmian interpretation of quantum mechanics or the MKC-models and the proposed axioms for these models discussed in Chapter \ref{NullificatieStuk}. In fact, as it stands the mathematical structure of quantum mechanics rather advocates the adoption of quantum logic than of intuitionistic logic. It is not at all clear that a consistent intuitionistic interpretation of quantum mechanics is possible. Unfortunately, I cannot present a solution to this problem, but I will discuss some possible ideas that may lead to possible consistent interpretations in the future. The reader more familiar with intuitionism may be warned. The reasonings I use sometimes borrow ideas from intuitionism, but not consistently so. The mathematics used is primarily classical, but the interpretation of it is not always. 

Formally, the task is to find a new correspondence between propositions and mathematical objects in the theory such that these mathematical objects together form the structure of a Heyting algebra, which is the formal algebraic structure of intuitionistic logic.\footnote{A Heyting algebra is a bounded lattice $L$ such that for all $l_1,l_2\in L$ there is an element denoted $l_1\to l_2$ that is the greatest element that satisfies $(l_1\to l_2)\wedge l_1\leq l_2$. Negation (often named the pseudo-complement in this context) may then be defined by $\neg l:=l\to\bot$. A Heyting algebra is a proper generalization of the notion of a Boolean algebra: every Boolean algebra is a Heyting algebra and a Heyting algebra is Boolean if and only if $\neg\neg l=l$ for all $l\in L$.}   
A natural approach is to take a closer look at quantum logic.
The first step taken by Birkhoff and von Neumann, where they associate propositions of the form $[\A\in\Delta]$ with closed linear subspaces (or the projections on these spaces), seems a very natural one to hang on to, so the focus is on the derivation of the logical connectives.
It seems plausible that their derivation contains some conceptual flaws, since a non-classical logic is derived using considerations based on classical logic. Indeed, the same approach when applied to classical physics leads to the construction of a classical propositional lattice (in the form of a Boolean algebra).\footnote{See for example \cite[Ch. 4]{Isham95}.} The derivation of quantum logic has of course been criticized several times over the years. An interesting contribution is due to Popper, who concludes:
\begin{quote}
	``It is of interest that the kind of change in classical logic which would fit what Birkhoff and von Neumann suggest [\ldots] would be the rejection of the law of excluded middle [\ldots], as proposed by Brouwer, but rejected by Birkhoff and von Neumann.'' \cite{Popper68}
\end{quote}
This conclusion is based on the following consideration. It is not hard to find two statements $\texttt{A}$ and $\texttt{B}$ (both of the form $[\A\in\Delta]$) such that in certain states one has
\begin{equation}\label{PopperExample}
	\pee(\texttt{B}\wedge(\texttt{A}\vee\neg\texttt{A}))=\pee(\texttt{B})>0
	=\pee(\texttt{B}\wedge\texttt{A})+\pee(\texttt{B}\wedge\neg\texttt{A})=\pee((\texttt{B}\wedge\texttt{A})\vee(\texttt{B}\wedge\neg\texttt{A})).
\end{equation}
The right-hand side of this inequality may be interpreted as stating that $\texttt{B}$ is incompatible both with $\texttt{A}$ and $\neg\texttt{A}$. On the other hand, $\pee(\texttt{B})>0$ can be interpreted to imply that $\texttt{B}$ is not an absurdity and therefore is a third possibility (besides the possibilities $\texttt{A}$ and $\neg\texttt{A}$). 

From this point of view, it does indeed seem strange that $\texttt{A}\vee\neg\texttt{A}$ is associated with triviality. From an intuitionistic point of view, the peculiarity doesn't seem to arise from the definition of negation. This deserves some explanation. In the BHK interpretation each mathematical proposition is associated with a project, namely, that of finding a proof. In particular, Kolmogorov associated propositions with problems of which it is the task of the mathematician to solve them (\cite{Kolmogorov32} see also \cite{Coquand07}). Indeed, the negation of a proposition is not just simply the absence of a proof, but actually requires a proof on its own. The notion of proof, however, is quite an ambiguous one in physics. Indeed, in mathematics one often considers `truth' to mean `provability', but in physics the best one can do is falsify a proposition. That is, for any acceptable physical proposition one must be able to construct an experiment that may have as an outcome that the proposition is not true. The statement $[\A\in\Delta]$ is falsified if a measurement of $\A$ yields a result outside $\Delta$ and similarly, $\neg[\A\in\Delta]$ can be falsified if it is associated with the statement $[\A\in\Delta^c]$ (note that this wouldn't be the case if $\neg[\A\in\Delta]$ were associated with the set $\h\backslash\mu_A(\Delta)\h$).     
 
On the other hand, in quantum logic the disjunction of two propositions seems to include much more than the truth of either one the individual propositions. Indeed, consider the situation of equation (\ref{PopperExample}) and consider a state for which $\texttt{B}$ is true. Quantum logic states that for this state the proposition $\texttt{A}\vee\neg\texttt{A}$ is true. However, one cannot decide which of the two ($\texttt{A}$ and $\neg\texttt{A}$) is true and even worse: assuming either option leads to a contradiction. Thus it seems more natural to associate the set $\mu_A(\Delta)\h\cup\mu_A(\Delta^c)\h$ with the statement $[\A\in\Delta]\vee\neg[\A\in\Delta]$. This leads to an extension of the set of propositions that not only includes closed linear subspaces of the Hilbert space, but at least also incorporates finite unions of closed linear subspaces:
\begin{equation}
	L_1:=\left\{\bigcup_{j=1}^nK_j\:;\:n\in\mathbb{N},K_j\subset\h\text{ is a closed linear subspace }\forall j\right\}.
\end{equation}
This set forms a partially ordered set under inclusion i.e.,
\begin{equation}
	\bigcup_{j=1}^nK_j\leq\bigcup_{j=1}^mK'_j\:\desda\:\bigcup_{j=1}^nK_j\subset\bigcup_{j=1}^mK'_j.
\end{equation}
The top element $\top$ is given by $\h$ and the bottom element $\bot$ is the empty set $\varnothing$, taking an empty inclusion. The bottom element may be identified with the zero-dimensional subspace $\{0\}$, since the zero vector is not a state. This leads to a small adjustment of the set $L_1$ by introducing the equivalence relation
\begin{equation}
	l_1\sim l_2\:\desda\: l_1\backslash l_2 \cup l_2\backslash l_1 \subset \{0\},\quad l_1,l_2\in L_1,
\end{equation}
and then taking $L_1/\sim$. I will treat $L_1$ as if it is in fact $L_1/\sim$. A disjunction and a conjunction can be defined as follows:
\begin{equation}
\begin{gathered}
	\bigcup_{j=1}^nK_j\vee\bigcup_{k=1}^mK'_j=\bigcup_{j=1}^n\bigcup_{k=1}^m(K_j\cup K'_k)\in L_1,\\
	\bigcup_{j=1}^nK_j\wedge\bigcup_{k=1}^mK'_j=\bigcup_{j=1}^n\bigcup_{k=1}^m(K_j\cap K'_k)\in L_1.
\end{gathered}
\end{equation}
The definition of the disjunction is a straight forward generalization of the ideas that led to the definition of $L_1$. The definition of the conjunction is inspired by its use in quantum logic. It is easy to see that these definitions in fact coincide with the join and the meet that turn $L_1$ into a lattice.

In $L_1$, the definition of negation generalizes to
\begin{equation}
	\neg \bigcup_{j=1}^nK_j:=\left\{\psi\in\h\:;\:\langle\psi,\phi\rangle=0,\:\forall\phi\in\bigcup_{j=1}^nK_j\right\},
\end{equation}
which is a closed linear subspace and thus again an element of $L_1$.
This negation has some interesting properties. For example, for any element $l\in L_1$, its double negation $\neg\neg l$ is the smallest closed linear subspace that contains $l$. Consequently, the closed linear subspaces are precisely the regular elements of $L_1$ i.e., the elements for which the relation $\neg\neg l=l$ holds. The negation also behaves non-classically, since one has
\begin{equation}
	l\vee\neg l\neq\top\text{ and }\neg l\vee\neg\neg l\neq\top,\quad\forall l\in L_1\backslash\{\top,\bot\}.
\end{equation}
However, this negation is not a pseudo-complement (which is required for $L_1$ to be a Heyting algebra), since although one does have $l\wedge\neg l=\bot$ for all $l\in L_1$, $\neg l$ is not the greatest element that satisfies this property. In fact, there is no greatest element in $L_1$ that satisfies this criterion for any $l\in L_1\backslash\{\top,\bot\}$, and a pseudo-complement can therefore not be defined at all. To see this, consider the two-dimensional Hilbert space with orthonormal basis $e_1,e_2$. Then any element $l\in L_1$ that satisfies $P_{e_1}\h\nleq l$ satisfies $P_{e_1}\h\wedge l=\bot$, but the supremum over all these elements does not exist (unless one also allows arbitrary unions in $L_1$, in which case the pseudo-complement simply becomes the set-theoretic complement). From the impossibility to define a pseudo-complement it does not only follow that $L_1$ cannot be given the structure of a Heyting algebra, but also that it cannot conform to a broader class of algebras (such as the pseudo-complemented lattices). 

In conclusion, one cannot embed the quantum-logical structure in a Heyting algebra simply by redefining conjunction. But redefining negation is also not an option. Indeed, it is easy to show that one cannot define an implication on the lattice of projections with disjunction and conjunction defined as in Section \ref{QLogicSection}. To see this, consider again the two-dimensional Hilbert space and let $f:=\tfrac{1}{2}\sqrt{2}(e_1+e_2)$. The proposition $P_f\to P_{e_1}$ should be the supremum of all elements that satisfy $P\wedge P_f\leq P_{e_1}$. Thus $P_f\to P_{e_1}$ should be the supremum over all $P$ that satisfy $P_{e_1}\nleq P$, which does not exist. These considerations exclude every acceptable Heyting algebra in terms of subsets of the Hilbert space.\footnote{Remaining options are to find yet an other definition of the join, or to find an other definition of the partial order. Both options are not likely to result in any structure that also allows an interpretation.}

A way to break out of the structure of subsets of the Hilbert space may be found when looking once more at the particular proposition $[\A\in\Delta]\vee\neg[\A\in\Delta]$, where $\neg[\A\in\Delta]$ is interpreted as $[\A\in\Delta^c]$. According to quantum mechanics, a measurement of $\A$ will with certainty yield a result in the set $\Delta\cup\Delta^c$. Does this also mean that one can say that the result lies either in $\Delta$ or in $\Delta^c$? From an intuitionistic point of view the answer is not clear, since for an intuitionist the equality $\Delta\cup\Delta^c=\mathbb{R}$ does not hold in general. There may be numbers $x\in\mathbb{R}$ for which one cannot make this decision. However, in practice one may imagine that the set $\Delta$ is well-behaved and the set of all numbers $x\in\mathbb{R}$ for which one cannot decide $x\in\Delta$ or $x\in\Delta^c$ is most likely to be assigned probability zero in any quantum-mechanical state. What I want to get at, is that in the event of an actual measurement of $\A$, it is natural to assume that $[\A\in\Delta]\vee\neg[\A\in\Delta]$ is true. However, in the case of a measurement of an other observable, the proposition $[\A\in\Delta]\vee\neg[\A\in\Delta]$ is not likely to make any sense. What may be considered true and what may not, may depend on the measurement context. The information about measuring contexts is thrown away when propositions are identified with closed linear subspaces, and this may be seen as an explanation of why it is not possible to unambiguously associate propositions with subsets of the Hilbert space from an intuitionistic point of view. 

It was shown in \cite{CHLS09} that if one adopts a richer structure that also accounts for measuring contexts, it is possible to construct a Heyting algebra to obtain an intuitionistic quantum logic for systems associated with a finite-dimensional Hilbert space. Their results are obtained by looking at quantum mechanics from a topos-theoretic point of view. This may be seen as a difficult way of taking the easy way out. The easy part is that Heyting algebras arise naturally in topos theory. The difficult part is that the approach is very abstract from a mathematical point of view. It is not easy to find physical interpretations for the results obtained in this way, which is also my main concern with the Heyting algebra obtained in \cite{CHLS09}. To discuss these concerns, I'll first introduce the Heyting algebra under consideration.

Each finite-dimensional Hilbert space over $\mathbb{C}$ is isomorphic with $\mathbb{C}^n$. The set of operators then corresponds with the set of all $n\times n$ matrices $\mathfrak{A}=M_n(\mathbb{C})$. Measuring contexts are associated with Abelian unital sub-C*-algebras of $\mathfrak{A}$. The motivation for this is based on the FUNC' rule of Section \ref{Kochen-Specker-sectie}. For any observable $\A$ and any Borel function $f$, one can introduce the observable $f(\A)$ by applying $f$ to the result of a measurement of $\A$. So in any measurement context in which one can measure $\A$, one can also measure all observables of the form $f(\A)$. Then, if $\A$ is associated with the operator $A$, one may associate $f(\A)$ with $f(A)$. So the measurement context in which one can measure $\A$ may be associated with the set $\{f(A)\:;\:f\text{ Borel}\}$. These are precisely all the self-adjoint operators in the smallest unital sub-C*-algebra that contains $A$. This algebra is automatically Abelian. Furthermore, all Abelian unital sub-C*-algebras of $\mathfrak{A}$ are of this form. The set of all these sub-algebras will be denoted by $C(\mathfrak{A})$. 

As argued earlier, it is a problem in quantum logic that propositions are specified without referring to a measurement context. For example, the statement $[\A\in\Delta]\vee\neg[\A\in\Delta]$ may be considered to be true in any context in which one can measure $\A$ and not true in other contexts. The question is, of course, how these notions should be linked together. One suggestion is to associate propositions with functions 
\begin{equation}
	S:C(\mathfrak{A})\to\mathcal{P}(\mathfrak{A}),
\end{equation}
where $\mathcal{P}(\mathfrak{A})$ is the set of all projection operators in $\mathfrak{A}$. The idea is that for each measuring context a function $S$ specifies what is actually being measured in that context. In \cite{CHLS09}, the following restrictions are derived for these functions:
\begin{equation}\label{CHLSrooster}
	L_2:=\{S:C(\mathfrak{A})\to\mathcal{P}(\mathfrak{A})\:;\:S(\mathcal{C})\in\mathcal{P}(\mathcal{C}), 
	S(\mathcal{C})\leq S(\mathcal{D})\text{ if }\mathcal{C}\subset\mathcal{D},\:\forall\mathcal{C},\mathcal{D}\in C(\mathfrak{A})\},
\end{equation}     
where the partial order refers to the one from quantum logic. This set becomes a bounded lattice under the definition
\begin{subequations}
\begin{equation}\label{orderCHLS}
	S_1\leq S_2\quad\desda\quad S_1(\mathcal{C})\leq S_2(\mathcal{C}),\quad\forall\mathcal{C}\in C(\mathfrak{A}),
\end{equation}
such that its top and bottom element are given by
\begin{equation}\label{boundsCHLS}
	\top(\mathcal{C})=\een,\quad\bot(\mathcal{C})=\nul,\quad\forall\mathcal{C}\in C(\mathfrak{A}),
\end{equation}
and the join and meet are given by
\begin{equation}\label{veeCHLS}
	(S_1\vee S_2)(\mathcal{C})=S_1(\mathcal{C})\vee S_2(\mathcal{C}),\quad\forall\mathcal{C}\in C(\mathfrak{A});
\end{equation}
\begin{equation}\label{wedgeCHLS}
	(S_1\wedge S_2)(\mathcal{C})=S_1(\mathcal{C})\wedge S_2(\mathcal{C}),\quad\forall\mathcal{C}\in C(\mathfrak{A}).
\end{equation}
Note that the lattice operations on the right-hand side are those from quantum logic.
As it turns out, with the definition of implication given by
\begin{equation}
	(S_1\to S_2)(\mathcal{C}):=\bigvee\{P\in\mathcal{P}(\mathcal{C})\:;\:P\leq S_1(\mathcal{D})^\perp\vee S_2(\mathcal{D}),\quad\forall\mathcal{D}\supset\mathcal{C}\},
\end{equation}
\end{subequations}
the lattice becomes a Heyting algebra, which in general is non-Boolean (i.e., if one defines $\neg S:=S\to\bot$, there are elements $S$ for which $S\vee\neg S\neq\top$). From a mathematical point of view this is enough to speak of an intuitionistic logic of $M_n(\mathbb{C})$. However, without an interpretation this logic is meaningless from a physical point of view. In \cite{CHLS09}, any indication for a possible interpretation is missing. The question therefore becomes: ``What does a proposition $S$ actually state?'' And: ``What is the significance of the restrictions in (\ref{CHLSrooster})?''
The idea behind the first restriction may be that statements in a certain measuring context should make sense in that specific measuring context. The idea behind the second condition may be that in a larger measuring context more may be true. But these notions are, of course, a bit vague and so it may be better to investigate them in an explicit example.

Consider the system of a single spin-$\tfrac{1}{2}$ particle with associated Hilbert space $\mathbb{C}^2$. More specifically, consider a measurement of the spin along the $z$-axis. The associated operator and corresponding minimal measuring context are given by
\begin{equation}
	\sigma_z=\begin{pmatrix} 1 & 0 \\ 0 & -1\end{pmatrix},\quad
	\mathcal{C}_z=\left\{ \begin{pmatrix} \lambda_1 & 0 \\ 0 & \lambda_2\end{pmatrix}\:;\:\lambda_1,\lambda_2\in\mathbb{C}\right\}.
\end{equation}
The algebra $\mathcal{C}_z$ is the only measuring context in which $\sigma_z$ can be measured and the only measuring context properly contained in this one is $\mathbb{C}\een$. Now consider the proposition $[P_z+=1]$, which is generally read as ``the spin along the $z$-axis is up'' and which in quantum logic is associated with the projection operator $P_z+$. It is a fundamental proposition, since it may be the outcome of a measurement; any physical theory should at least incorporate such propositions. So it becomes the task to associate an element $S_{[P_z+=1]}$ of $L$ with this proposition. For its value in $\mathcal{C}_z$ two possibilities come to mind:
\begin{equation}
  S_{[P_z+=1]}(\mathcal{C}_z)=P_z+\quad\text{or}\quad S_{[P_z+=1]}(\mathcal{C}_z)=\een.
\end{equation}
The second option immediately leads to problems because it would imply $S_{[P_z-=1]}(\mathcal{C}_z)=\een$, which consequently results in $S_{[P_z+=1]}=S_{[P_z-=1]}$. The first option implies $S_{[P_z+=1]}(\mathbb{C}\een)=\nul$ because $S_{[P_z+=1]}$ must be an element of $L$. Thus the proposition $S_{[P_z+=1]}$ seems to state that in the measurement context in which one can only measure trivialities ($\mathbb{C}\een$), one finds a contradiction (which is the standard interpretation of the zero operator). This seems very puzzling to me to say the least, and I think the problem arises from the restriction $S(\mathcal{C})\leq S(\mathcal{D})$ whenever $\mathcal{C}\subset\mathcal{D}$.

Consider the following interpretation of a function $S:C(\mathfrak{A})\to\mathcal{P}(\mathfrak{A})$: given the proposition $S$, then $S(\mathcal{C})$ gives all the available information that is relevant for predictions concerning measurements in the measuring context $\mathcal{C}$. Now consider the given information $[P_z+=1]$. An associated proposition $S_{[P_z+=1]}$  should state that in the measuring context $\mathcal{C}_z$ one will only find results compatible with $[P_z+=1]$ i.e., $S_{[P_z+=1]}(\mathcal{C}_z)$ should be equal to $P_z+$. However, in the measuring context $\mathbb{C}\een$, the information $[P_z+=1]$ is completely useless; it is as good as no information at all. Without information one can only hang on to trivialities, which may best be associated with the projection $\een$. Opposite to trivialities, a contradiction may be interpreted as an excess of information, associated with the projection $\nul$. Indeed, the most natural contradiction arises from $\texttt{A}\wedge\neg\texttt{A}$ for any proposition $\texttt{A}$. 
Note that this line of reasoning may be applied to any measuring context that doesn't contain $\sigma_z$. Therefore, it seems natural to define
\begin{equation}
	S_{[P_z+=1]}(\mathcal{C}):=\begin{cases} P_z+,& P_z+\in\mathcal{C};\\ \een,& P_z+\notin\mathcal{C},\end{cases}\quad\forall\mathcal{C}\in C(\mathfrak{A}),
\end{equation}
which is no longer an element of $L_2$, since it violates the second condition.

From this point of view, it seems natural to introduce the set
\begin{equation}
	L_3:=\{S:C(\mathfrak{A})\to\mathcal{P}(\mathfrak{A})\:;\:S(\mathcal{C})\in\mathcal{P}(\mathcal{C}), 
	S(\mathcal{C})\leq S(\mathcal{D})\text{ if }\mathcal{C}\supset\mathcal{D},\:\forall\mathcal{C},\mathcal{D}\in C(\mathfrak{A})\}.
\end{equation}   
The new restriction, i.e. $S(\mathcal{C})\leq S(\mathcal{D})$ whenever $\mathcal{D}\subset\mathcal{C}$, may be motivated by the interpretation that the available information $S$ can be further specified in a broader measuring context.
The quantum propositional lattice $\mathcal{P}(\h)$ may be embedded into this set by taking
\begin{equation}
	S_P(\mathcal{C}):=\begin{cases} P,& P\in\mathcal{C};\\ \een,& P\notin\mathcal{C},\end{cases}\quad\forall\mathcal{C}\in C(\mathfrak{A}),\forall P\in\mathcal{P}(\h).
\end{equation}

This embedding may also be used to motivate the operations (\ref{orderCHLS}), (\ref{boundsCHLS}), (\ref{veeCHLS}) and (\ref{wedgeCHLS}) to turn $L_3$ into a lattice.\footnote{It is easy to see that $L_3$ with these operations is in fact a bounded lattice, and I will omit the proof here.} Information $S_1$ may be considered to be more precise then $S_2$ ($S_1\leq S_2$) if and only if the information is more precise in every measuring context ($S_1(\mathcal{C})\leq S_2(\mathcal{C})$ $\forall\mathcal{C}$). The top element corresponds with no information at all ($\top=S_{\een}$) and the bottom element corresponds with contradictory information ($\bot=S_{\nul}$). Now consider the information $S_{P_1}$ or $S_{P_2}$. In a measurement context containing both $P_1$ and $P_2$, one can then conclude that $P_1\vee P_2$ in this context. However, in any other context, one can draw no conclusion at all, since one doesn't know which information $P_1$ or $P_2$ is true.\footnote{This is a peculiarity even in the BHK interpretation. To prove $\texttt{A}\vee\texttt{B}$ one must prove at least one of the statements. Thus having obtained a proof of $\texttt{A}\vee\texttt{B}$, one has obtained a proof of $\texttt{A}$ or a proof of $\texttt{B}$. However, once the conclusion is converted from either $\texttt{A}$ is true to $\texttt{A}\vee\texttt{B}$ is true, or from $\texttt{B}$ is true to $\texttt{A}\vee\texttt{B}$ is true, one cannot recover which of the two is true from the proposition $\texttt{A}\vee\texttt{B}$. This is only recovered by looking at the proof of $\texttt{A}\vee\texttt{B}$, in which case one would rather have one of the propositions $\texttt{A}$, $\texttt{B}$ or $\texttt{A}\wedge\texttt{B}$. The same reasoning may be applied when thinking of $S_{P_1}\vee S_{p_2}$.} One then finds $(S_{P_1}\vee S_{P_2})(\mathcal{C})=S_{P_1}(\mathcal{C})\vee S_{P_2}(\mathcal{C})$ for all $\mathcal{C}$. This may be compared to the proposition $S_{P_1\vee P_2}$, which contains more information from this perspective, since $S_{P_1\vee P_2}\leq S_{P_1}\vee S_{P_2}$. Finally, consider the information $S_{P_1}$ and $S_{P_2}$. In a measurement context containing both $P_1$ and $P_2$, both pieces of information can be applied to obtain $P_1\wedge P_2$. In a measurement context containing only one of the $P_1,P_2$, only one part of information is applicable and one may only obtain the relevant $P_1$ or $P_2$. One therefore finds $(S_{P_1}\wedge S_{P_2})(\mathcal{C})=S_{P_1}(\mathcal{C})\wedge S_{P_2}(\mathcal{C})$ for all $\mathcal{C}$.

The lattice $L_3$ is turned into a Heyting algebra by taking
\begin{equation}\label{ImplicationNew}
	(S_1\to S_2)(\mathcal{C}):=\bigwedge\{S_1(\mathcal{D})^\perp\vee S_2(\mathcal{D})\:;\:\mathcal{D}\subset\mathcal{C}\}.
\end{equation}
Indeed, for each $\mathcal{C}$, the maximal element of $\mathcal{P}(\mathcal{C})$ that can be assigned to $S_1\to S_2$ such that $(S_1\to S_2)(\mathcal{C})\wedge S_1(\mathcal{C})\leq S_2(\mathcal{C})$ is $S_1(\mathcal{C})^\perp\vee S_2(\mathcal{C})$. However, since $S_1\to S_2$ also has to satisfy $(S_1\to S_2)(\mathcal{C})\leq (S_1\to S_2)(\mathcal{D})$ whenever $\mathcal{D}\subset\mathcal{C}$, the maximal element that can be assigned to $(S_1\to S_2)(\mathcal{C})$ such that $S_1\to S_2\in L_3$ is precisely the one given in (\ref{ImplicationNew}). This also shows that $L_3$ is indeed a Heyting algebra. The negation is then defined in the usual way by $\neg S:= S\to\bot$. This negation has some interesting features. Suppose $S(\mathbb{C}\een)=\een$. Then, by definition, 
\begin{equation}
	\neg S(\mathbb{C}\een)= (S\to\bot)(\mathbb{C}\een)= S(\mathbb{C}\een)^\bot\vee \bot(\mathbb{C}\een)=\nul.
\end{equation}
This implies that $\neg S=\bot$. On the other hand, if $S(\mathbb{C}\een)=\nul$, then $S=\bot$ and one has
\begin{equation}
\begin{split}
	\neg S(\mathbb{C}\een)
	=& 
	(S\to\bot)(\mathbb{C}\een)= \bigwedge\{S(\mathcal{D})^\perp\vee \bot(\mathcal{D})\:;\:\mathcal{D}\subset\mathcal{C}\}\\
	=& 
	\bigwedge\{\nul^\perp\vee \nul\:;\:\mathcal{D}\subset\mathcal{C}\}=\een.
\end{split}
\end{equation}
In conclusion
\begin{equation}
	\neg S=
	\begin{cases}
		\top, & S=\bot;\\ \bot, & S\neq\bot,
	\end{cases}
	\quad\forall S\in L_3.
\end{equation}
So the only regular elements of $L_3$ are $\bot$ and $\top$, which makes $L_3$ extremely non-Boolean. 

It seems to me that the Heyting algebra $L_3$ comes closer to describing an intuitionistic logic for $M_n(\mathbb{C})$ then $L_2$. But there is still a lot of work to be done if intuitionistic logic is to play an explicit role in the interpretation of quantum mechanics and it is not clear if $L_3$ is a step in the right direction. A significant feature of quantum mechanics is that propositions about future measurements can be assigned a probability of turning out to be true in the case of actual measurements. It is not clear what role these probabilities should play in the intuitionistic approach. Moreover, there is no generally accepted theory of intuitionistic probability, though there have been some interesting suggestions (see for example \cite{Fraassen81}, \cite{MorganLeblanc83}, \cite{MorganLeblanc83-2} and \cite{Weatherson03}). It is an open question if any of these suggestions can be applied to $L_3$ in a way that is consistent with the interpretation of the propositions.

A question that may play an important role is: ``What is the meaning of these probabilities in the Copenhagen interpretation?'' In the realist approach probabilities arise from lack of information about the actual state of the system, but from the Copenhagen point of view, the state of the system may be considered to be nothing but these probabilities. The information-theoretic approach I have used to discuss some problems in this thesis seems to advocate a Bayesian approach, in which the probability of an event expresses the degree of faith someone (usually called an agent in this context) has in that the event will actually occur based on the information available to that person. 

Investigations in a Bayesian approach to quantum mechanics are being carried out (\cite{Fuchs09} is one of the latest works and also refers to a lot of earlier work). One of the obstacles is to find a philosophical motivation for the Born axiom or rather an answer to the question: ``Why are the only admissible probabilities those that obey the structure postulated by the Born axiom?'' This is indeed a natural question. In orthodox quantum mechanics the probabilities are simply derived from the objective quantum state. But, as argued at the end of Section \ref{QMHVT}, the objective state does not make much sense from an information theoretic point of view. A difficulty that arises when trying to find an answer to the above question is that the Born axiom allows one to have connections between probabilities that apply to counterfactual situations, whereas the Copenhagen interpretation advocates the idea that one should not compare counterfactual situations. It seems to me quite possible that these difficulties may have more natural solutions if one adopts an intuitionistic point of view, since the intuitionistic point of view provides a way of dealing with counterfactual propositions, as shown in Section \ref{IntuiCompl}. 

The final problem that I want to discuss, which arises if one wants to adopt an intuitionist point of view on quantum mechanics, is the following. The theory of Hilbert spaces, being the mathematical foundation of quantum mechanics, is based on classical mathematics and relies on theorems that may not be true from an intuitionistic point of view. Attempts have been made to find a constructive foundation for quantum mechanics in \cite{Bridges99}, \cite{RichmanBridges99} and \cite{BridgesSvozil00} but it has never come so far as to find a constructive approach in which the axioms of quantum mechanics can be reformulated (at least, not as far as I know). However, one may also raise the question to what extent the Hilbert space formalism is truly necessary for quantum mechanics. Approaches have become more algebraic over the years and it may be possible to obtain a formulation in which it is easier to obtain an intuitionistic mathematical approach. But the problem may also find an other resolution. One may have an intuitionistic point of view towards physics while maintaining a classical point of view towards mathematics. However, it is a deep philosophical question if this view can be self-consistent.

\markboth{The Strangeness and Logic of Quantum Mechanics}{Epilogue}
\subsection{Epilogue}
The ideas expressed in Chapter \ref{StrangenessSection} may seem premature and perhaps even a bit farfetched. Although I do believe that intuitionistic ideas may play a possible role in understanding quantum mechanics, I should stress that I don't think that their role will ever be strictly necessary; this depends on what one expects from a physical theory. 

In \cite{Bell90}, Bell expressed his dislike for the standard interpretation of quantum mechanics (which is roughly quantum mechanics as presented by the six postulates in Section \ref{PostulatenZelf}). For example, he states that
\begin{quote}
	``the idea that quantum mechanics [\ldots] is exclusively about the results of experiments would remain disappointing.'' \cite[p. 34]{Bell90}
\end{quote}
But, like many physicists, Bell not only opposes the instrumentalist view on science, but he also advocates that a physical theory should actually be about Nature itself rather than about merely our observations of Nature. In other words, Bell argues that a satisfactory realist interpretation should be a necessary criterion for any physical theory.

The impossibility proofs discussed in this thesis are taken to imply some unsatisfactory features every realist interpretation of quantum mechanics necessarily possesses. However, none of these proofs seem loophole-free to me and I don't think they can be. The reason is quite simple: one never knows for sure if one has considered all possible loopholes. The discovery of the `finite precision'-loophole in the Kochen-Specker Theorem, which leads to the MKC-models, is a striking example, in my opinion.

No matter what exactly is proven by any of the impossibility proofs, it has been established that a realist interpretation is not possible without adding, removing or modifying any of the postulates. Indeed, both the von Neumann proof and the Kochen-Specker Theorem imply that any realist interpretation will encounter difficulties incorporating the notion of non-commutativity of observables,\footnote{It is usually understood to imply the notion of contextuality but as discussed in this thesis, this conclusion is not a necessary one. In the MKC-models, non-commutativity leads to statistical independence (see Corollary \ref{MKConafhankelijkheid}). Either way, it seems unlikely that the notion of non-commutativity will vanish completely in a realist interpretation.} whereas the Bell-type arguments and the Free Will Theorem show that any realist interpretation will meet difficulties incorporating the notion of entanglement. 

It is a peculiar aspect of quantum mechanics that the choice between realism and instrumentalism seems to strongly influence the way the theory should be formulated. But it may be even more peculiar thatmost of the realist interpretations known to date (the best-known probably being Bohmian mechanics, the GRW theory and Everett's many worlds interpretation) describe a significantly different reality. This may cast even more doubt on the realist view than the introduction of non-locality.
The most noticeabe advantage some of the present realist interpretations of quantum mechanics have is that they are easy to understand; they state what is actually being measured when a measurement is performed, and sometimes even describe what constitutes a measurement. In other words, they only seem compelling due to the vagueness and difficulty of the standard interpretation.

But when comprehension rather than truth becomes the decisive criterion for physical theories, there is no reason to adopt a realist view \textit{per se}. Comprehension may also be found using other approaches, and Bohr's views seem a good starting point to me. 
The Copenhagen interpretation is a departure from the materialist paradise upheld by classical physics.\footnote{This departure somewhat resembles the departure from Cantor's paradise that took place in mathematics from around 1900. Hilbert would have nothing of it: ``Aus dem Paradies, das Cantor uns geschaffen, soll uns niemand vertreiben k\"onnen.'' \cite[p. 170]{Hilbert26}. And many mathematicians still agree with this view to some extent.} The emphasis is on the subject (the experimenter) and how the subject interacts with the object (through experiments) without making any statements about the object by itself. In a certain sense, it is an anti-Copernican revolution; the Copenhagen interpretation puts the observer back into the center of the universe. 

It should be noted that the Bohrian approach is not really an instrumentalist approach. A true instrumentalist would be content with just the mathematical formulation of a theory along with some rules how to apply it. In fact, I think Bohr would agree with Bell that there's more to physics than just measurement results. It is the aim to acquire some understanding of Nature. The difference is that realists start by making assumptions about Nature itself whereas Bohr starts from the point of view of the observer, accepting the (likely) possibility that in this approach one may never come to a complete understanding of Nature itself.

This is roughly the point of view I had when I first came up with the idea of a possible role for intuitionistic logic in quantum mechanics. It was only later that I found some resemblances with the ideas expressed by Bohr.\footnote{Although Bohr stressed the necessity of the use of classical logic in physics: ``all well-defined experimental data [\ldots] must be expressed in ordinary language making use of common logic.'' \cite[p. 317]{Bohr48}.}
In a realist interpretation the law of excluded middle seems necessary; every statement about something that \textit{is}, is either true or false. But from the point of view of the observer this is not obvious. For example, a statement like `the particle went either through slit one or slit two' has no significant meaning if one does not perform an experiment that decides through which slit the particle went. 

Intuitionistic logic also seems the natural logic that is used when deriving theories from experimental data, since such theories naturally do not talk about unperformed measurements. From this point of view it may not be all that surprising that a theory (i.e. quantum mechanics) could have emerged at the beginning of the twentieth century that seemed to defy the notions of logic itself, since the methods used to construct the theory use a logic that differs from the one demanded for the theory itself.

\clearpage

\addcontentsline{toc}{section}{References}
\markboth{}{}
\bibliographystyle{rhalphanum}
\bibliography{referenties}

\end{document}